\documentclass[prx, 10pt, twocolumn, times, floatfix, superscriptaddress, notitlepage, footinbib]{revtex4-1}

\usepackage{times, mathrsfs, amsmath, amsfonts, graphics, graphicx, color, bbm, amssymb, epsdice}
\usepackage[english]{babel}
\usepackage[margin=.75in]{geometry}
\usepackage{subfigure}
\usepackage{verbatim}
\usepackage{makecell}
\usepackage{booktabs}
\usepackage{xfrac,relsize,float}
\usepackage{placeins}
\usepackage[math]{cellspace}
\newcommand{\ftnt}[1]{{\textsuperscript{\scriptsize\footnote{#1}}}}

\usepackage{tikz}
\newcommand{\Cross}{\mathbin{\tikz [x=1.4ex,y=1.4ex,line width=.2ex] \draw (0,0) -- (1,1) (0,1) -- (1,0);}}%

\usepackage[unicode=true, breaklinks=false, pdfborder={0 0 1}, backref=false, colorlinks=true, linkcolor=blue, citecolor=blue]{hyperref}
\newcommand\vcdice[1]{\vcenter{\hbox{\epsdice{#1}}}}
\newcommand\vcdiceb[1]{\vcenter{\hbox{\epsdice[black]{#1}}}}

\newcommand{\trps}{\text{T}}

\newcommand{\Rtype}{\texttt{R}}
\newcommand{\Ftype}{\texttt{F}}

\def\bra#1{\mathinner{\langle{#1}|}}
\def\ket#1{\mathinner{|{#1}\rangle}}
\def\braket#1{\mathinner{\langle{#1}\rangle}}

\def\bbra#1{\mathinner{\langle\!\langle{#1}|}}
\def\kket#1{\mathinner{|{#1}\rangle\!\rangle}}
\def\BraVert{\egroup\,\mid\,\bgroup}

\def\ketbra#1#2{|#1\rangle \langle#2|}
\def\kketbbra#1#2{|#1\rangle\!\rangle \langle\!\langle#2|}
\def\bbrakket#1#2{\langle\!\langle#1|#2\rangle\!\rangle}

\newcommand{\stxt}{S}
\newcommand{\stxtp}{S'}

\newcommand{\etxt}{E}
\newcommand{\setxt}{SE}

\newcommand{\tr}{{\rm tr}}
\newcommand{\tre}{{\rm tr}_\etxt}

\newcommand{\Acal}{\mathcal{A}}
\newcommand{\Bcal}{\mathcal{B}}
\newcommand{\Ccal}{\mathcal{C}}
\newcommand{\Dcal}{\mathcal{D}}
\newcommand{\Ecal}{\mathcal{E}}
\newcommand{\Fcal}{\mathcal{F}}
\newcommand{\Hcal}{\mathcal{H}}
\newcommand{\Ical}{\mathcal{I}}
\newcommand{\Jcal}{\mathcal{J}}

\newcommand{\Rcal}{\mathcal{R}}
\newcommand{\Scal}{\mathcal{S}}
\newcommand{\Tcal}{\mathcal{T}}
\newcommand{\Ucal}{\mathcal{U}}
\newcommand{\Vcal}{\mathcal{V}}
\newcommand{\Wcal}{\mathcal{W}}

\newcommand{\Pprob}{\mathbb{P}}

\def\Ucalt#1{\Ucal_{#1}}
\newcommand{\Asf}{\mathsf{A}}
\newcommand{\Bsf}{\mathsf{B}}
\newcommand{\Tset}{\mathbf{T}}
\newcommand{\Tsetk}{\mathbf{T}_k}
\newcommand{\Tsetl}{\mathbf{T}_\ell}
\newcommand{\Jset}{\mathbf{J}}
\newcommand{\Jsetk}{\mathbf{J}_{\Tset_k}}
\newcommand{\Jsetl}{\mathbf{J}_{\Tset_\ell}}
\newcommand{\xset}{\mathbf{x}}
\newcommand{\xsetk}{\mathbf{x}_{\Tset_k}}
\newcommand{\Aset}{\mathbf{A}}
\newcommand{\Asetk}{\mathbf{A}_{\xsetk}}

\newcommand{\Bsetk}{\mathbf{B}_{\xsetk}}

\newcommand{\Asetc}{\mathscr{A}}
\newcommand{\Asetck}{\mathscr{A}_{\xsetk}}

\newcommand{\Upsk}{\Upsilon_{\Tsetk}}
\newcommand{\Tcalk}{\Tcal_{\Tsetk}}
\newcommand{\Tcall}{\Tcal_{\Tsetl}}

\newcommand{\inp}{\texttt{i}}
\newcommand{\out}{\texttt{o}}
\newcommand{\ident}{\openone}
\newcommand*{\mc}{\mathcal}
\newcommand*{\markov}{{\scriptscriptstyle{(\mathrm{M})}}}
\newcommand{\sbt}{\,\begin{picture}(-1,1)(-1,-3)\circle*{2.5}\end{picture}\ }

\usepackage{fdsymbol}
\newcommand*{\f}{\frac}

\usepackage{ntheorem, theorem}

\newtheorem*{definition*}{Definition}
\newtheorem*{proposition*}{Proposition}
\newtheorem*{lemma*}{Lemma}
\newtheorem*{algorithm*}{Algorithm}
\newtheorem*{fact*}{Fact}
\newtheorem*{theorem*}{Theorem}
\newtheorem*{corollary*}{Corollary}
\newtheorem*{conjecture*}{Conjecture}
\newtheorem*{postulate*}{Postulate}
\newtheorem*{axiom*}{Axiom}
\newtheorem*{remark*}{Remark}
\newtheorem*{example*}{Example}
\newtheorem*{question*}{Question}

\definecolor{Blue}{rgb}{0,0,1}
\definecolor{Red}{rgb}{1,0,0}
\definecolor{Green}{rgb}{0,1,0}
\definecolor{darkgreen}{rgb}{0,.7,0}
\definecolor{Purp}{rgb}{.2,0,.2}
\definecolor{white}{rgb}{1,1,1}

\begin{document}
\title{Quantum stochastic processes and quantum non-Markovian phenomena}
\author{Simon Milz}
\email{simon.milz@oeaw.ac.at} 
\affiliation{Institute for Quantum Optics and Quantum Information, Austrian Academy of Sciences, Boltzmanngasse 3, 1090 Vienna, Austria}

\author{Kavan Modi}
\email{kavan.modi@monash.edu}
\affiliation{School of Physics and Astronomy, Monash University, Clayton, Victoria 3800, Australia}

\date{\today}

\begin{abstract}
The field of classical stochastic processes forms a major branch of mathematics. They are, of course, also very well studied in biology, chemistry, ecology, geology, finance, physics, and many more fields of natural and social sciences. When it comes to quantum stochastic processes, however, the topic is plagued with pathological issues that have led to fierce debates amongst researchers. Recent developments have begun to untangle these issues and paved the way for generalizing the theory of classical stochastic processes to the quantum domain without ambiguities. This tutorial details the structure of quantum stochastic processes, in terms of the modern language of quantum combs, and is aimed at students in quantum physics and quantum information theory. We begin with the basics of classical stochastic processes and generalize the same ideas to the quantum domain. Along the way, we discuss the subtle structure of quantum physics that has led to troubles in forming an overarching theory for quantum stochastic processes. We close the tutorial by laying out many exciting problems that lie ahead in this branch of science.
\end{abstract}

\maketitle
\tableofcontents

\section{Introduction}
\label{sec::Intro}

Many systems of interest, in both natural and social sciences, are not isolated from their environment. However, the environment itself is often far too large and far too complex to model efficiently and thus must be treated statistically. This is the core philosophy of open systems; it is a way to render the description of systems immersed in complex environments manageable, even though the respective environments are inaccessible and their full description out of reach. Quantum systems are no exception to this philosophy. If anything, they are more prone to be affected by their complex environments, be they stray electromagnetic fields, impurities, or a many-body system. It is for this reason that the study of quantum stochastic processes goes back a full century. The field of classical stochastic processes is a bit older, however, not by much. Still, there are stark contrasts in the development of these two fields; while the latter rests on solid mathematical and conceptual grounds, the quantum branch is fraught with mathematical and foundational difficulties.

The 1960s and 1970s saw great advancements in laser technology, which enabled isolating and manipulating single quantum systems. However, of course, this did not mean that unwanted environmental degrees of freedom were eliminated, highlighting the need for a better and formal understanding of quantum stochastic processes. It is in this era great theoretical advancements were made to this field. Still going half a century into the future from these early developments, there is yet another quantum revolution on the horizon; the one aimed at processing quantum information. While quantum engineering was advancing, many of the early results in the field of quantum stochastic processes regained importance and new problems have arisen requiring a fresh look at how we characterize and model open quantum systems.

Central among these problems is the need to understand the nature of memory that quantum environments carry. At its core, memory is nothing more than information about the past of the system we aim to model and understand. However, the presence of this seemingly harmless feature leads to highly complex dynamics for the system that require different tools for their description from the ones used in the absence of memory. This is of particular importance for engineering fault-tolerant quantum devices which are by design complex and the impact of memory effects will rise with increased miniaturization and read-out frequencies. Consequently, here, one aims to characterize the underlying processes with the hope to mitigate or outmaneuver complex noise and making the operation of engineered devices robust to external noise. On the other hand, there are natural systems that are immersed in complex environments that have functional or fundamental importance in, e.g., biological systems. These systems too undergo open quantum processes with memory as they interact with their complex environments. Here, in order to exploit them for technological development or to understand the underlying physics, one aims to better understand the mechanisms that are at the heart of complex quantum processes observed in nature.

For the reasons stated above, over the years many books have been dedicated to this field of research, e.g.~\cite{alicki_semi_1987, Gardiner, BreuerPetruccione, accardi_quantum_2002, Wiseman}. In addition, the progress, both in experimental and theoretical physics has been fast leading to many review papers focusing on different facets of open quantum systems~\cite{qprob2006, Rivas2014, RevModPhys.88.021002, deVega2017, Li2018, eplrev2, eplrev1} and the complex multilayered structure of memory effects in quantum processes~\cite{Li2018}. This tutorial adds to this growing literature and has its own distinct focus. Namely, we aim to answer two questions: how can we overcome the conceptual problems encountered in the description of quantum stochastic processes, and how can we comprehensively characterize multi-time correlations and memory effects in the quantum regime when the system of interest is immersed in a complex environment. 

A key aim of this tutorial is to render the connection between quantum and classical stochastic processes transparent. That is, while there is a well-established formal theory of classical stochastic processes, does the same hold true for open quantum processes? And if so, how are the two theories connected? Thus we begin with a pedagogical treatment of classical stochastic process centered around several examples in Sec.~\ref{sec:clexample}. Next, in Sec.~\ref{sec::ClFormal} we formalize the elements of the classical theory, as well as present several facets of the theory that are important in practice. In Sec.~\ref{sec::EarlyProg} we discuss the early results on the quantum side that are well-known. Here, we also focus on the fundamental problems in generalizing the theory of quantum stochastic processes such that it is on equal footing as its classical counterpart. Sec.~\ref{sec:QStochProc} begins with identifying the features of quantum theory that impose a fundamentally different structure for quantum stochastic processes than that encountered in the description of classical processes. We then go on to detail the framework that allows one to generalize the classical theory of stochastic processes to the quantum domain. Finally, in Sec.~\ref{sec::PropQuant} we present various features of quantum stochastic processes, like, e.g., the distinction between Markovian and non-Markovian processes. Throughout the whole manuscript, we give examples that build intuition for how one ought to address multi-time correlations in an open quantum system. We close with several applications. 

Naturally, we cannot possibly hope to do the vast field of open quantum system dynamics full justice here. The theory of classical stochastic processes is incredibly large, and its quantum counterpart is at least as large and complex. Here, we focus on several aspects of the field and introduce them rather by concrete example than aiming for absolute rigor. It goes without saying that there are countless facets of the field that will remain unexplored, and of what is known and well-established, we only scratch the surface in our presentation in this tutorial. We do, however, endeavor to present the intuition at the core of this vast field. While we aim to provide as many references as possible for further reading, we do so without a claim to comprehensiveness, and much of the results that have been found in the field will be left unsaid, and far too much will not even be addressed.

\section{\texorpdfstring{Classical Stochastic Processes \\ Some Examples}{}}
\label{sec:clexample}

A typical textbook on stochastic processes would begin with a formal mathematical treatment by introducing the triplet $(\Omega, \mathcal{S}, \omega)$ of a sample space, a $\sigma$-algebra, and a probability measure. Here, we are not going to proceed in this formal way. Instead, we will begin with intuitive features of classical stochastic processes and then motivate the formal mathematical language retrospectively. We will then introduce and justify the axioms underpinning the theory of stochastic processes and present several key results in the theory of classical stochastic processes in the next section. The principal reason for introducing the details of the classical theory is that, later in the tutorial, we will see that many of these key results cannot be imported straightforwardly into the theory of quantum stochastic processes. We will then pivot to provide resolutions of how to generalize the features and key ingredients of classical stochastic processes to the quantum realm.

\subsection{Statistical state}
\label{sec::StatState}

Intuitively, a stochastic process consists of sequences of measurement outcomes, and a rule that allocates probabilities to each of these possible sequences. Let us start with a motivating example of a simple process -- that of tossing a die -- to clarify these concepts. After a single toss, a die will roll to yield one of the following outcomes
\begin{gather}\label{eq:twowdice}
\Rtype_1 = \left\{\vcdice{1}, \vcdice{2}, \vcdice{3}, \vcdice{4}, \vcdice{5}, \vcdice{6}\right\}.
\end{gather}
Here, $\Rtype$ (for roll of the die) is called the event space capturing all possible outcomes. If we toss the die twice in a row then the event space is
\begin{gather}
\Rtype_2 = \left\{\vcdice{1} \vcdice{1}, \vcdice{1} \vcdice{2}, \dots, \vcdice{6} \vcdice{5}, \vcdice{6} \vcdice{6} \right\}.
\end{gather}
While this looks the same as a single toss of two dice
\begin{gather}\label{eq:twodice}
\Rtype_2 = \left\{\vcdice{1} \vcdiceb{1}, \vcdice{1} \vcdiceb{2}, \dots, \vcdice{6} \vcdiceb{5} , \vcdice{6} \vcdiceb{6} \right\},
\end{gather}
the two experiments -- tossing two dice in parallel, and tossing a single die twice in a row -- can, depending on how the die is tossed, indeed be different. However, in both cases the event spaces are the same and grow exponentially with the number of tosses. For example, for three tosses the event space $\Rtype_3$ has $6^3$ entries. 

While the event spaces for different experiments can coincide, the probabilities for the occurrence of different events generally differ. Any possible event $\mathbf{r}_{\textbf{K}} \in \Rtype_{\textbf{K}}$ has a probability 
\begin{gather}
\Pprob(\mathbf{R}_{\textbf{K}} = \mathbf{r}_{\textbf{K}}),
\end{gather}
where the boldface subscript $\textbf{K}$ denotes the number of times or the number of dice that are tossed in general, and $\mathbf{R}_{\textbf{K}}$ is the random variable corresponding to $\textbf{K}$ tosses. Throughout, we will denote the random variable at toss $k$ by $R_k$, and the specific outcome by $r_k$ and we will use boldface notation for sequences. Importantly, two experiments with the same potential outcomes and the same corresponding probabilities cannot be statistically distinguished. For example, tossing two dice in parallel, and hard tossing (see below) of one die twice in a row yield the same probabilities and could not be distinguished, even though the underlying mechanisms are different. Consequently, we call the allocation of probabilities to possible events the \textit{statistical state} of the die, as it contains all inferable information about the experiment at hand. In anticipation of our later treatment of quantum stochastic processes, we emphasize that this definition of state chimes well with the definition of quantum states, which, too, contain all statistical information that is inferable from a quantum system. 
Importantly, the respective probabilities not only depend on how the die is made, i.e., its bias, but also on how it is tossed. Since we are interested in the stochastic process and, as such, sequential measurements in time, we will focus on the latter aspects below.

\begin{figure}[t]
\centering
\includegraphics[width=0.95\linewidth]{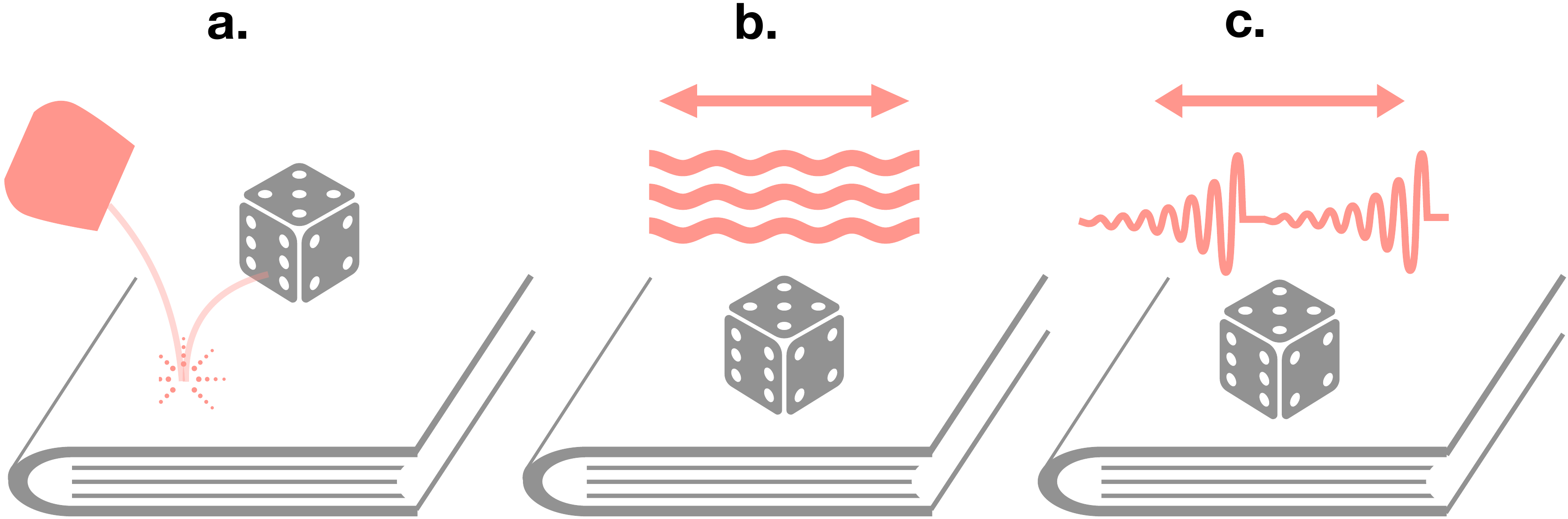}
\caption{\textbf{Classical die processes.} Panel (a) denotes fair toss; panel (b) denotes perturbed toss; and panel (c) the perturbation strength depends on the history.}
\label{fig::cldie}
\end{figure}

\subsection{Memoryless process}
\label{sec::Memless}
Let us now, to see how the probabilities $\Pprob_{\textbf{K}}$ emerge, look at a concrete `experiment', the case where the die is tossed hard. For a single toss of a fair die, we expect the outcomes to be equally distributed as
\begin{gather}
\label{eq:FairDie}
\Pprob(R_{1}= \vcdice{1}) = \ldots = \Pprob(R_{1} = \vcdice{6}) = 1/6. 
\end{gather}
Now, imagine this fair die is tossed `hard' successively. By hard, we mean that it is shaken in between tosses -- in contrast to merely being perturbed (see below). Then, importantly, the probability of future events does not depend on the past events; observing, say, $\vcdice{4}$, at some toss, has no bearing on the probabilities of later tosses. In other words, a hard toss of a fair die is a fully random process that has \textit{no} memory of the past. Consequently, this successive tossing of a single die at $k$ times is not statistically distinguishable from the tossing of $k$ unbiased dice in general. 

The memorylessness of the process is not affected if a biased die is tossed, e.g., a die with distribution
\begin{gather}
\begin{split}
&\Pprob(R = \left\{\vcdice{1}, \vcdice{2}, \vcdice{3}, \vcdice{4}, \vcdice{5}\right\}) = \tfrac{4}{25} \\ 
&\mbox{and} \qquad \Pprob(R = \vcdice{6}) = \tfrac{1}{5}.
\end{split}
\end{gather}
Here, while the bias of the die influences the respective probabilities, the dependence of these probabilities on prior outcomes solely stems from the way the die is tossed. Alternatively, suppose, we toss two identical dice with event space given in Eq.~\eqref{eq:twodice}. Now, if we consider the aggregate outcomes (sum of the outcomes of the two dice) $\{2,3,\dots,12\}$, they do not occur with uniform probability. Nevertheless, the process itself remains random as the future outcomes do not depend on the past outcomes. Processes without any dependence on past outcomes are often referred to as \textit{Markov order $0$} processes. We now slightly alter the tossing of a die to encounter processes with higher Markov order.

\subsection{Markov process}
\label{sec::MarkProc}
To introduce a dependence on prior outcomes, let us now ease the tossing and imagine placing the die on a book and then gently shaking the book horizontally for three seconds, see the depiction in Figure~\ref{fig::cldie}(b). We refer to this process as \textit{perturbed die}. The term `perturbed' here highlights that the toss is only a small perturbation on the current configuration. In this process, the probability to tip to any one side is $q$, rolling to the opposite side is highly unlikely\ftnt{The sum of the opposite ends of a die always equals 7.} (with probability $s$), while it is highly likely (with probability $p$) that the die stays on the same side. Concretely, suppose we start the die with $\vcdice{2}$, then the probability for the outcomes of the next roll will be
\begin{gather}\label{eq:conddist}
\Pprob(R_{k} | R_{k-1} = \vcdice{2}) = \left[q \ p \ q \ q \ s \ q \right]^\trps, 
\end{gather}
where $\trps$ denotes transpose, i.e., the probability distribution is a column vector. The perturbative nature of the toss means that $p>q \gg s$ and normalization gives us $p+4q+s=1$. Above, $R_{k}$ and $R_{k-1}$ are the random variables describing the die at the $k$-th and $(k-1)$-th toss, respectively. The conditional probabilities in Eq.~\eqref{eq:conddist} denote the probability for the outcomes $\{\vcdice{1}, \vcdice{2}, \vcdice{3}, \vcdice{4}, \vcdice{5}, \vcdice{6}\}$ at the $k$-th toss, given that the $(k-1)$-th toss yielded $\vcdice{2}$. For example, for the die to yield outcome $R_k = \vcdice{4}$ (i.e., to roll on its side) at the $k$-th toss, given that it yielded $r_{k-1} = \vcdice{2}$ in the previous toss, is $\Pprob(R_k = \vcdice{4} | R_{k-1} = \vcdice{2}) = q$.

A word of caution is needed. In the literature, conditional probabilities often carry an additional subscript to denote how many previous outcomes the probability of the current outcome is conditioned on. For example, $\Pprob_{1|k}$ would denote the probability of one (the current) outcome conditioned on the $k$ previous outcomes, while $\Pprob_k$ would represent a joint probability of $k$ outcomes. Here, in slight abuse of notation, we use the same symbol for conditional probabilities, as we used for one-time probabilities, e.g., in Eq.~\eqref{eq:FairDie}, and we omit additional subscripts. However, since the number of arguments always clarifies what type of probability is referred to, there is no risk of confusion, and we will maintain this naming convention also for the case of conditional probabilities that depend on multiple past outcomes.

In this example, even though the die may be unbiased, the toss itself is not and the distribution for the future outcomes of the die depends on its current configuration. As such, the process remembers the current state. However, for the probabilities at the $k$th toss, it is only the outcome at the $(k-1)$th toss that is of relevance, but none of the earlier ones. In other words, only the current configuration matters for future statistics, but the earlier history does \textit{not} matter. Such processes are referred to as \textit{Markov processes}, or, as they `remember' only the most current outcome, processes of Markov order $1$. Importantly, as soon as any kind of memory effects are present, the successive tossing of a die can be distinguished from the independent, parallel tossing of several identical dice, as in this latter case, the statistics of the $k$th die cannot depend on the $(k-1)$th die (or any other die). 

Again, we emphasize that this process will remain Markovian even if the die is replaced by two dice or by a biased die. Similarly, the above considerations would not change if the perturbation depended on the number of the toss $k$, i.e., if the parameters of Eq.~\eqref{eq:conddist} were functions $q(k),p(k),s(k)$. We will now discuss the case where this assumption is not satisfied, i.e., where the perturbation at the $k$-th toss can depend on past outcomes, and memory over longer periods of time starts to play a non-negligible role.

\subsection{Non-Markovian processes}\label{sec:clnm}

Let us now modify the process in the last example a bit by changing the perturbation intensity as we go. Above, we considered the process where the die was placed on a book, and the book was shaken for three seconds. Suppose that after the first shake the die rolls on its side, say $\vcdice{2} \mapsto \vcdice{3}$. The process is such that, after the number of pips changes, the next perturbation has unit intensity. If this intensity is low enough then we are likely to see $\vcdice{3} \mapsto \vcdice{3}$, and if that happens -- i.e., the number of pips is unchanged -- then the intensity is doubled the next shake; and we keep doubling the intensity until either die rolls to a new value or the intensity reaches the value of eight units (four times), which we assume to be equal to shaking the die so strongly that its initial value does not influence future outcomes. After this, the shaking intensity is reset to the unit level. We have depicted this process in Figure~\ref{fig::cldie}(c).

In this example, to predict the future probabilities we not only need to know the current number of pips the die shows, but also its past values. That is, the probability of observing an event, say $\vcdice{6}$, after observing two consecutive $\vcdice{3}$ outcomes is different than if one had previously observed $\vcdice{3}$ and $\vcdice{2}$, i.e.,
\begin{gather}
\Pprob(\vcdice{6} \ | \ \vcdice{3}, \vcdice{3}) \ne \Pprob(\vcdice{6} \ | \ \vcdice{2}, \vcdice{3}). 
\end{gather}

The necessity for remembering the past beyond the most recent outcomes makes this process \textit{non-Markovian}. On the other hand, here, we only have to remember the past four outcomes of the die due to the resetting protocol of the perturbation strength. Concretely, the future probabilities are independent of the past beyond four steps. For example, we have
\begin{gather}
\Pprob(\vcdice{6} \ | \ \vcdice{3}, \vcdice{3}, \vcdice{3}, \vcdice{3}, \vcdice{3} ) = 
\Pprob(\vcdice{6} \ | \ \vcdice{3}, \vcdice{3}, \vcdice{3}, \vcdice{3} , \vcdice{2}). 
\end{gather}
To be more precise, predicting the next outcome with correct probabilities requires knowing the die's configuration for the past four steps. That is, the future distribution is fully determined by conditional probabilities
\begin{gather}\label{eq:markovorder}
 \Pprob(R_k | R_{k-1},\ldots,R_{0}) = \Pprob(R_k | R_{k-1},\ldots,R_{k-4}),
\end{gather}
where we only need to know a part (here, the last four outcomes) of the history. 

As mentioned, the size of the memory is often referred to as the \textit{Markov order} or \textit{memory length} of the process. A fully random process -- like the hard tossing of a die -- has a Markov order 0, and a Markov process has an order of 1. A non-Markovian process has an order of 2 or larger. This, in turn, implies that the study of non-Markovian processes contains Markovian processes as well as fully random processes as special cases. Indeed, most processes in nature will carry memory, and Markovian processes are the -- well studied -- exception rather than the norm~\cite{vanKampen1998}. 

In general, the complexity of a non-Markovian process is higher than that of the Markov process in the last subsection; this is because there is more to remember. Put less prosaically, the process has to keep a ledger of the past outcomes to carry out the correct type of perturbation at each point. And, in general, the size of this ledger, or the complexity, grows exponentially with the Markov order $m$: for a process with $d$ different outcomes at each time (6 for a die), it is given by $d^m$. However, sometimes it is possible to compress the memory. For instance, in the above example, we only need to know the current configuration and the number of time steps it has remained unchanged; thus the size of the memory is linear in the Markov order for this example. Moreover, looking at histories larger than the Markov order will not reveal anything new and thus does not add to the complexity of the process.

\subsection{Stochastic matrix}
\label{sec::StochMat}

\begin{figure}
 \centering
 \includegraphics[width=0.95\linewidth] {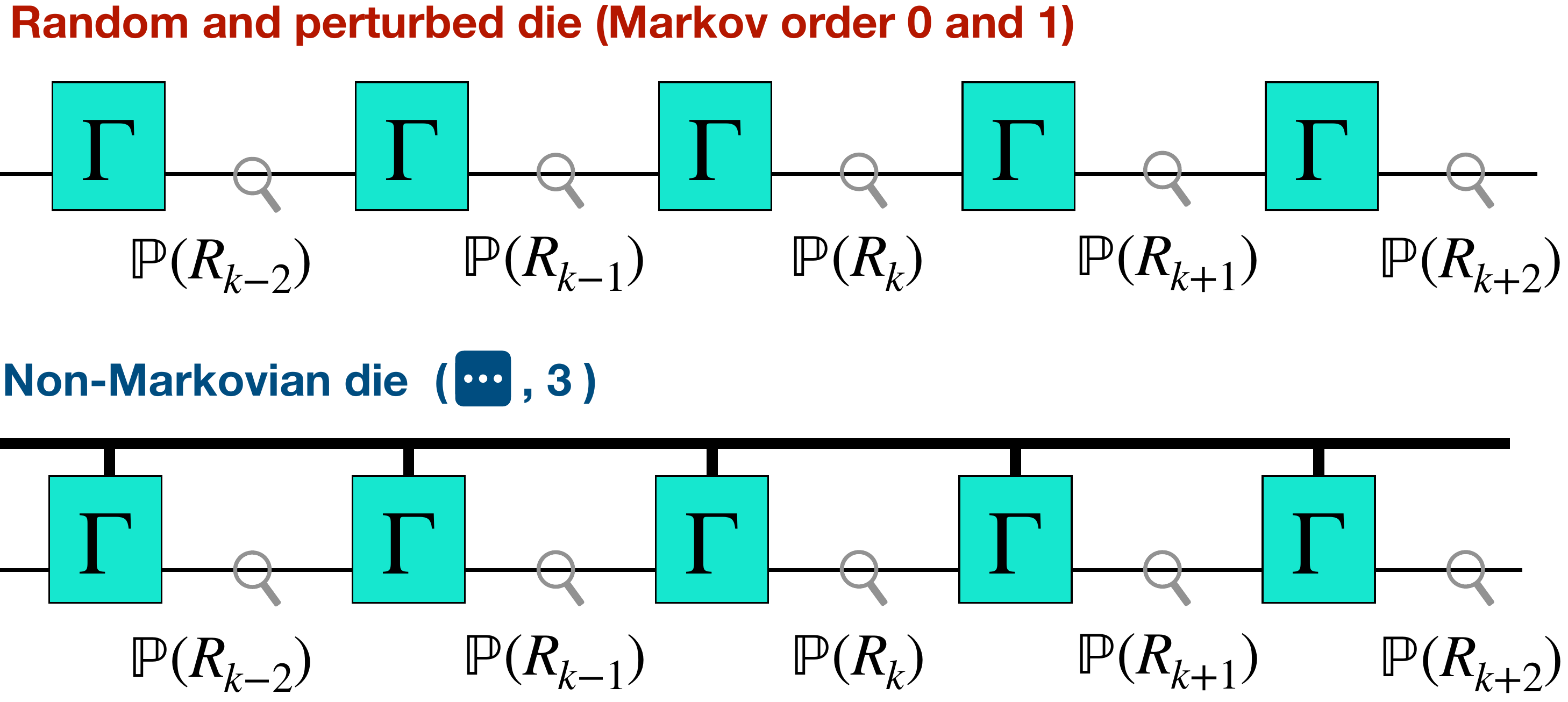}
 \caption{\textbf{Random, Markovian, and non-Markovian processes.} Top panel shows the circuits for random and Markovian die. In these cases, there are no extra lines of communication between the tosses (represented by boxes). Only the system carries the information forward for a Markov process. The bottom panel shows the non-Markovian die. Here, information is sent between tosses (represented by boxes) in addition to what the system carries, which is the memory of the past states of the system (die). This memory is then denoted by the thick line. The memory has to carry the information about the state of the die in the past four tosses to determine the intensity of the next perturbation.}\label{fig::markovdie}
\end{figure}

Having discussed stochastic processes and memory at a general level, it is now time to look in more detail at the mathematical machinery used to describe them. A convenient way to model stochastic processes is the stochastic matrix, which transforms the current state of the system into the future state. It also lends itself to a clear graphical depiction of the process in terms of a circuit, see, e.g., Figure~\ref{fig::markovdie} for circuits corresponding to the three examples above. In what follows, we will write down the stochastic matrices corresponding to the three processes above. The future states can then be computed by following the circuit and performing appropriate matrix multiplication.

\subsubsection{Transforming the statistical state}
\label{sec::TransStatState}

Before describing the process, let us write down the state of the system at time $k-1$. At any given time, the die has a probability of be in one of six states, not necessarily uniformly distributed. We can think of this distribution as the statistical state of the system:
\begin{gather}\label{eq:statstatedie}
 \Pprob(R_{k-1}) = \left[\Pprob(\vcdice{1}) \ \Pprob(\vcdice{2}) \ \Pprob(\vcdice{3}) \ \Pprob(\vcdice{4}) \ \Pprob(\vcdice{3}) \ \Pprob(\vcdice{6}) \right]^\trps.
\end{gather}
Here again, $\trps$ denotes transposition, i.e., the statistical state is a column vector. 

Suppose the die in the $(k-1)$-th toss rolls to $r_{k-1}$. Along with this, if we knew the conditional (or transition) probabilities $\Pprob(r_k|r_{k-1})$, the probability to find the die to rolls to $r_k$ in the $k$-th toss can be straight forwardly computed via 
\begin{gather}\label{eq:oneprop}
\Pprob(r_k) = \sum_{r_{k-1}} \Pprob(r_k|r_{k-1}) \Pprob(r_{k-1}).
\end{gather}
This can be phrased more succinctly as
\begin{gather}\label{eq:stochmat}
 \Pprob(R_k) = \Gamma_{(k:k-1)} \ \Pprob(R_{k-1}),
\end{gather}
where stochastic matrix $\Gamma_{(k:k-1)}$ is the mechanism by which the statistical state changes in time from time step $k-1$ to $k$. For brevity we will generally omit the subscript on $\Gamma$ (the time at which it acts will be clear from the respective arguments it acts on) unless it is required for clarity. The elements of the stochastic matrix are called transition probabilities as they indicate how two events at $k$ and $k-1$ are correlated.

Before examining the explicit stochastic matrices for the above examples of processes, let us first discuss their general properties. First, all entries of $\Gamma$ are positive, as they correspond to transition probabilities. Second, to ensure that the l.h.s of Eq.~\eqref{eq:stochmat} is a probability distribution, the columns of the stochastic matrix sum to one, which is a direct consequence of the identity $\sum_{r_k} \Pprob(r_k|r_{k-1}) = 1$ which holds for all $r_{k-1}$. On the other hand, the rows of $\Gamma$ do not have to add to unity, as generally we have $\sum_{r_{k-1}} \Pprob(r_k|r_{k-1}) \neq 1$ (this is also clear in Eq.~\eqref{eq:randomprocessdie} for a biased die below). In the case where the rows actually add to 1, the matrix is called bistochastic, and it has some nice properties and applications~\cite{marshall_inequalities_2011}, which we will not cover in detail in this tutorial; for example, any bistochastic matrix can be represented as a convex combination of permutation matrices, a fact known as Birkhoff's theorem.

\subsubsection{Random process}
\label{sec::RandProc}

Now, making the concept of stochastic matrices more concrete, we begin by constructing the stochastic matrix for the fully random process of the tossing of a die without memory. In this case, it does not matter what the current state of the die is, and the future state will be the one given in Eq.~\eqref{eq:statstatedie}. This is achieved by the following matrix
\begin{gather}\label{eq:randomprocessdie}
 \Gamma^{(0)} = \left(\begin{matrix}
 \Pprob(\vcdice{1}) \! & \! \Pprob(\vcdice{1}) \! & \! \Pprob(\vcdice{1}) \! & \! \Pprob(\vcdice{1}) \! & \! \Pprob(\vcdice{1}) \! & \! \Pprob(\vcdice{1}) \\
 \Pprob(\vcdice{2}) \! & \! \Pprob(\vcdice{2}) \! & \! \Pprob(\vcdice{2}) \! & \! \Pprob(\vcdice{2}) \! & \! \Pprob(\vcdice{2}) \! & \! \Pprob(\vcdice{2}) \\
 \Pprob(\vcdice{3}) \! & \! \Pprob(\vcdice{3}) \! & \! \Pprob(\vcdice{3}) \! & \! \Pprob(\vcdice{3}) \! & \! \Pprob(\vcdice{3}) \! & \! \Pprob(\vcdice{3}) \\
 \Pprob(\vcdice{4}) \! & \! \Pprob(\vcdice{4}) \! & \! \Pprob(\vcdice{4}) \! & \! \Pprob(\vcdice{4}) \! & \! \Pprob(\vcdice{4}) \! & \! \Pprob(\vcdice{4}) \\
 \Pprob(\vcdice{5}) \! & \! \Pprob(\vcdice{5}) \! & \! \Pprob(\vcdice{5}) \! & \! \Pprob(\vcdice{5}) \! & \! \Pprob(\vcdice{5}) \! & \! \Pprob(\vcdice{5}) \\
 \Pprob(\vcdice{6}) \! & \! \Pprob(\vcdice{6}) \! & \! \Pprob(\vcdice{6}) \! & \! \Pprob(\vcdice{6}) \! & \! \Pprob(\vcdice{6}) \! & \! \Pprob(\vcdice{6}) \\
 \end{matrix}\right).
\end{gather}
As stated above, a fully random process has Markov order of $0$, which we denote by the extra superscript $(0)$. Additionally, all the columns of the above $\Gamma^{(0)}$ add up to one, independent of whether or not the die is biased, while in general, i.e., when the die is biased, the rows do not add up to unity. 

It is easy to check that the above stochastic matrix indeed leads to the correct transitions; suppose the current state of the die is $\vcdice{6}$, i.e., $\Pprob(R_{k-1}) = \left[0 \ 0 \ 0 \ 0 \ 0 \ 1 \right]^\trps$. The statistical state after the roll will be the one given in Eq.~\eqref{eq:statstatedie}, i.e., 
\begin{gather}
\begin{split}
 \Pprob(R_{k}) &= \Gamma^{(0)}\ \Pprob(R_{k-1}) \\
 &= \left[\Pprob(\vcdice{1}) \ \Pprob(\vcdice{2}) \ \Pprob(\vcdice{3}) \ \Pprob(\vcdice{4}) \ \Pprob(\vcdice{3}) \ \Pprob(\vcdice{6}) \right]^\trps.
\end{split}
\end{gather}
Evidently, this process does not care about the current state -- the `new' probabilities at the $k$-th toss do not depend on the previous ones -- but it merely independently samples from the underlying distribution corresponding to the bias of the coin. As already mentioned, we could readily incorporate a temporal change of said bias, by making it dependent on the number of tosses. However, as long as this dependence is only on the number of tosses, and not on the previous outcomes, we would still consider this process memoryless (strictly speaking, the die along with a clock represents a memoryless process). To avoid unnecessary notational cluttering, we will always assume that the bias and/or the transition probabilities are independent of the absolute toss number but may depend on previous outcomes, as shown below.

For an unbiased die the above stochastic matrix will be simply
\begin{gather}\label{eq:fairrandomprocessdie}
 \Gamma^{(0)} = \tfrac{1}{6} \left(\begin{matrix}
 1 & 1 & 1 & 1 & 1 & 1 \\
 1 & 1 & 1 & 1 & 1 & 1 \\
 1 & 1 & 1 & 1 & 1 & 1 \\
 1 & 1 & 1 & 1 & 1 & 1 \\
 1 & 1 & 1 & 1 & 1 & 1 \\
 1 & 1 & 1 & 1 & 1 & 1 \\
 \end{matrix}\right),
\end{gather}
which is not only a stochastic, but a bistochastic map. Again, it is easy to check that the output is the uniform distribution 
\begin{gather}
\Gamma^{(0)} \ \Pprob(R_{k-1}) = \Pprob(R_{k}) = \tfrac{1}{6} \left[ 1 \ 1 \ 1 \ 1 \ 1 \ 1 \right]^\trps, 
\end{gather}
for \textit{any} $\Pprob(R_{k-1})$.

\subsubsection{Markov process}
\label{sec::MarkProcEx}

Let us now move to the perturbed die process, which we argued is a Markovian process. In this case the stochastic matrix has the form
\begin{gather}
 \Gamma^{(1)} = \left(\begin{matrix}
 \Pprob(\vcdice{1}|\vcdice{1}) & \Pprob(\vcdice{1}|\vcdice{2}) & \cdots & \Pprob(\vcdice{1}|\vcdice{6}) \\
 \Pprob(\vcdice{2}|\vcdice{1}) & \Pprob(\vcdice{2}|\vcdice{2}) & \cdots & \Pprob(\vcdice{2}|\vcdice{6}) \\
 \vdots & \vdots & \ddots & \vdots \\
 \Pprob(\vcdice{6}|\vcdice{1}) & \Pprob(\vcdice{6}|\vcdice{2}) & \cdots & \Pprob(\vcdice{6}|\vcdice{6}) \\
 \end{matrix}\right),
\end{gather}
where, again, we have used the superscript ${(1)}$ to signify that the underlying process is of Markov order $1$.

The hallmark of this matrix is that it gives us different future probabilities, depending on the current configuration; the probability $\Pprob(\vcdice{1}|\vcdice{6})$ to find the die showing $\vcdice{1}$ at the $k$-th toss, given that it showed $\vcdice{6}$ at the $k-1$-th toss generally differs from the probability $\Pprob(\vcdice{1}|\vcdice{2})$ to show $\vcdice{1}$ given that it previously showed $\vcdice{2}$. In contrast, for the fully random process above, both of these transition probabilities would be given by $\Pprob(\vcdice{1})$. 

Concretely, for the perturbed die process given in Eq.~\eqref{eq:conddist}, the stochastic matrix will have the form
\begin{gather}
 \Gamma^{(1)} = \left(\begin{matrix}
 p & q & q & q & q & s \\
 q & p & q & q & s & q \\
 q & q & p & s & q & q \\
 q & q & s & p & q & q \\
 q & s & q & q & p & q \\
 s & q & q & q & q & p \\
 \end{matrix}\right).
\end{gather}
Again, here the conditions $p > q \gg s$ and $p+4q+s=1$ are assumed, and we have $\Pprob(\vcdice{1}|\vcdice{6}) = s \neq q = \Pprob(\vcdice{1}|\vcdice{2})$. Again, it is easy to see, that the normalization of the conditional probabilities implies that the columns of the stochastic matrix add to one. Additionally, here, the rows of $\Gamma^{(1)}$ add up to one, too, making it a bistochastic matrix. 

For a Markov process, the state $\Pprob(R_k)$ is related to an earlier state $\Pprob(R_j)$, with $j<k$, by repeated applications of the stochastic matrix
\begin{gather}
 \Pprob(R_k) = \Gamma^{(1)}_{(k:k-1)} \cdots \ \Gamma^{(1)}_{(j+2:j+1)} \Gamma^{(1)}_{(j+1:j)} \Pprob(R_j).
\end{gather}
Alternatively, we may describe the process from $j$ to $k$ with the stochastic matrix
\begin{gather}
 \Gamma^{(1)}_{(k:j)} := \Gamma^{(1)}_{(k:k-1)} \cdots \ \Gamma^{(1)}_{(j+2:j+1)} \Gamma^{(1)}_{(j+1:j)}.
\end{gather}
This is clearly desirable as the above stochastic matrix is simply obtained by matrix multiplications, which is easy to do on a computer. Another way to compute the probability for two sequential events, say $r_k$ given we saw event $r_j$ at respective times, is by employing Eq.~\eqref{eq:oneprop}:
\begin{gather}\label{eq:chapman}
 \Pprob(r_k|r_j) = 
 \!\!\!\!\!\!\sum_{\{r_m\}_{m>j}^{k-1}}
 \prod_{i = j}^{k-1}
 \Pprob(r_{i+1}|r_{i}) \Pprob(r_j).
\end{gather}
This is known as the Chapman-Kolmogorov equation. Here, we have summed over all trajectories between event $r_j$ and event $r_k$, i.e., all possible sequences that begin with outcome $r_j$ at $t_j$ and end with outcome $r_k$ at $t_k$.

\subsubsection{Non-Markovian process}
\label{sec::NonMarkProc}

\begin{figure}[t]
\centering
\includegraphics[width=0.95\linewidth] {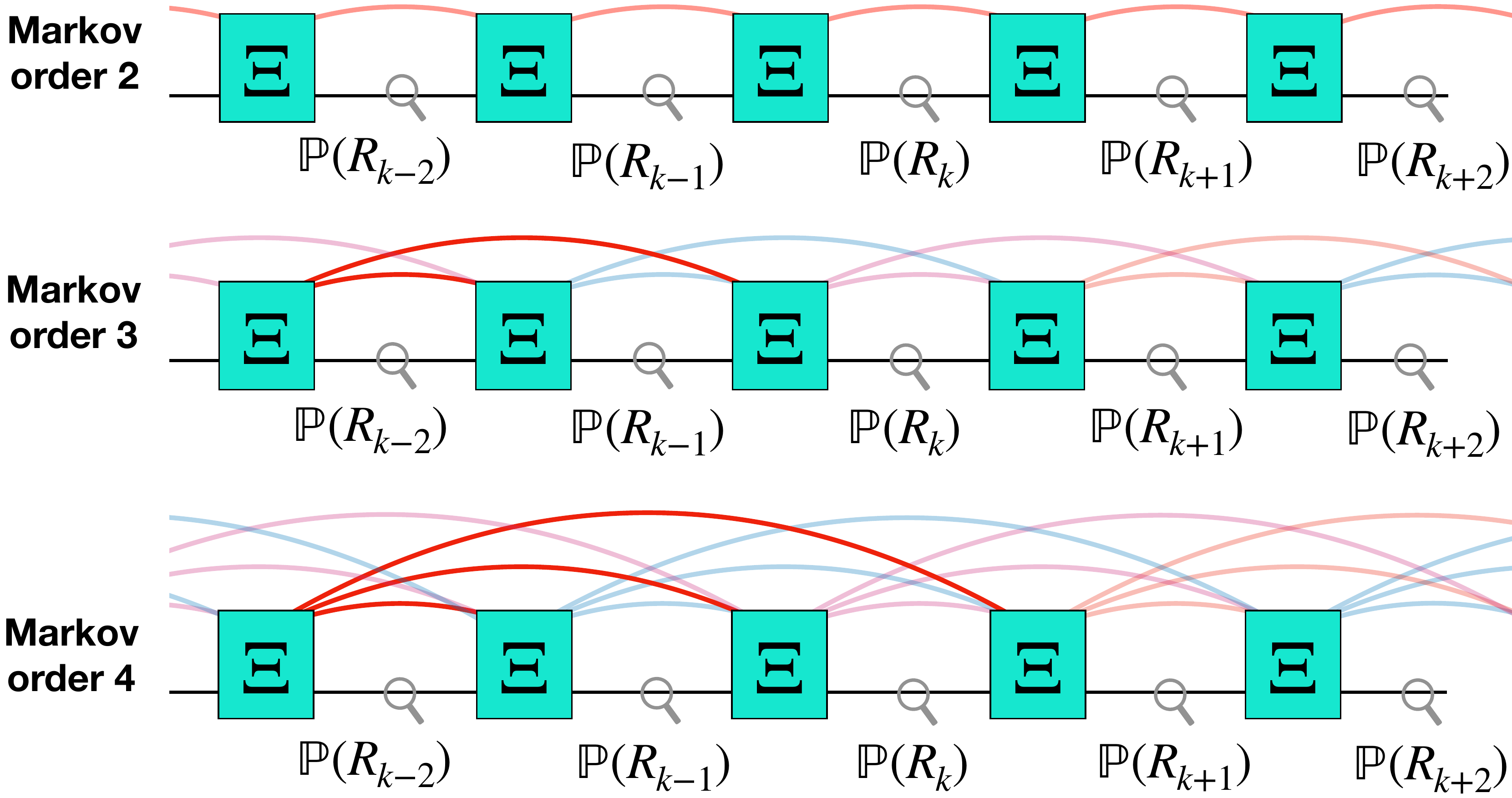}
\caption{\textbf{Memory in non-Markovian processes.} For processes with memory, besides the state of the system at a time/toss $k$, we need additional information -- depicted by the additional memory lines -- about the past to correctly predict future statistics. If only the probability of the next outcome is of interest, then a map $\Gamma^{(m)}$ of the form of Eq.~\eqref{eq:ProcTensClass} is sufficient, if all future probabilities are to be computed via the concatenation of a single map, then $\Xi$, given in Eq.~\eqref{eq:ximap}, is required. Together, the system and memory undergo Markovian dynamics.}
\label{fig::clnm}
\end{figure}

Above, we have required that the stochastic matrix $\Gamma$ maps the statistical state $\Pprob(R_{j})$ at a single time to another single-time statistical state $\Pprob(R_k)$. This was the correct way of computing future statistics, as they only depended on the current state of the system, but not on any additional memory. Now, turning our attention to non-Markovian dynamics, we will expand our view to consider processes that map multi-time statistical states, e.g., $\Pprob(R_{j-1}, R_{j-2},\dots,R_{j-m})$, to either a single-time state, e.g. $\Pprob(R_{k})$, or a multi-time state, e.g., $\Pprob (R_{k-1}, R_{k-2},\dots,R_{k-m})$, depending on what we aim to describe. This can be done in several ways, either by considering collections of stochastic map, or a single stochastic map that acts on a larger space. We briefly discuss both of these options.

First, let us consider the stochastic matrix for the non-Markovian process described in Sec.~\ref{sec:clnm}, where the perturbation intensity depended on the sequence of previously observed number of pips the die showed. As mentioned before, for this example we need to know the current state \textit{and} the number of times it has not changed -- which we will denote by $\mu$ -- to correctly predict future statistics. As the perturbation strength is reset after the die has shown the same number of pips three consecutive times, we have $\mu \in [0,1,2,3]$. For each $\mu$, we can then write the stochastic matrix as
\begin{gather}
\label{eq:NM_stochMat}
 \Gamma^{(\mu)} = \left(\begin{matrix}
 \Pprob^\mu(\vcdice{1}|\vcdice{1}) \!& \Pprob^\mu(\vcdice{1}|\vcdice{2}) \!& \cdots \!& \Pprob^\mu(\vcdice{1}|\vcdice{6}) \\
 \Pprob^\mu(\vcdice{2}|\vcdice{1}) \!& \Pprob^\mu(\vcdice{2}|\vcdice{2}) \!& \cdots \!& \Pprob^\mu(\vcdice{2}|\vcdice{6}) \\
 \vdots & \vdots & \ddots & \vdots \\
 \Pprob^\mu(\vcdice{6}|\vcdice{1}) \!& \Pprob^\mu(\vcdice{6}|\vcdice{2}) \!& \cdots \!& \Pprob^\mu(\vcdice{6}|\vcdice{6}) \\
 \end{matrix}\right),
\end{gather}
where the superscript on the transition probabilities and the stochastic matrices denotes that they depend on the number of times the outcome has not changed. For $\mu=3$, the perturbation strength is such that the process becomes the random process given in Eq.~\eqref{eq:randomprocessdie}, and $\mu=4$ is the same as $\mu=0$. Evidently, Eq.~\eqref{eq:NM_stochMat} defines four distinct stochastic matrices, one for each $\mu$ that leads to distinct future statistics. For any given $\mu$, $\Gamma^{(\mu)}$ allows us to correctly predict the probability of the next toss of the die.

It is always possible to write down a family of stochastic matrices for \textit{any} non-Markovian process. Given the current state and history, we make use of the appropriate stochastic matrix to get the correct future state of the system. In general, for Markov order $m$, there are at most $d^m$ distinct histories, i.e., $\mu \in \{0,\dots, d^{m-1}-1\}$; each such history (prior to the current outcome) then requires a distinct stochastic matrix to correctly predict future probabilities. This exponentially growing storage requirement of distinct pasts highlights the complexity of a non-Markovian process.

On the other hand, such a collection of stochastic matrices for a process of Markov order $m$ could equivalently be combined into one $d\times d^m$ matrix of the form 
\begin{gather}
\label{eq:ProcTensClass}
 \Gamma^{(m)} \!=\! \left(\begin{matrix}
 \Pprob(\vcdice{1}|\vcdice{1} \cdots \vcdice{1}) \!& \Pprob(\vcdice{1}|\vcdice{1} \cdots \vcdice{2}) \!& \cdots \!& \Pprob(\vcdice{1}|\vcdice{6}\cdots \vcdice{6}) \\
 \Pprob(\vcdice{2}|\vcdice{1} \cdots \vcdice{1}) \!& \Pprob(\vcdice{2}|\vcdice{1} \cdots \vcdice{2}) \!& \cdots \!& \Pprob(\vcdice{2}|\vcdice{6} \cdots \vcdice{6}) \\
 \vdots & \vdots & \ddots & \vdots \\
 \Pprob(\vcdice{6}|\vcdice{1} \cdots \vcdice{1}) \!& \Pprob(\vcdice{6}|\vcdice{1} \cdots \vcdice{2}) \!& \cdots \!& \Pprob(\vcdice{6}|\vcdice{6}\cdots \vcdice{6}) \\
 \end{matrix}\right),
\end{gather}
that acts on $d^m$-dimensional probability vectors 
\begin{gather}
\begin{split}
\Pprob(\mathbf{R}_{\mathbf{K}}) \!=\! \left[ 
\Pprob(\vcdice{1}\cdots \vcdice{2}) \cdots \Pprob(\vcdice{6} \cdots \vcdice{5}) \ \Pprob(\vcdice{6} \cdots \vcdice{6}) \right]^\trps,
\end{split}
\end{gather}
to yield the correct future statistics, i.e., $\Pprob(R_{k}) = \Gamma^{(m)} \Pprob(\mathbf{R}_{\mathbf{K}})$. 
Here, $\mathbf{K}$ denotes the last $m$ tosses and
thus by $\mathbf{R}_{\mathbf{K}}$ we denote the random variable corresponding to sequences of the last $m$ outcomes starting at the $(k-1)$-th toss. As before, $\Gamma^{(m)}$ is a stochastic matrix, as all of its entries are positive, and its columns sum to one. However, in contrast to the Markovian and the fully random case, it ceases to be a square matrix. We thus have to widen our understanding of a statistical `state' from probability vectors of outcomes at \textit{one} time/toss, to probability vectors of outcomes at \textit{sequences} of times/tosses. In quantum mechanics, this shift of perspective allows one to resolve many of the apparent paradoxes that appear to plague the description of quantum stochastic processes. In the following section, we will see a concrete example of this way of describing non-Markovian processes.

We have graphically depicted non-Markovian processes, with Markov orders 2, 3, and 4, in Figure~\ref{fig::clnm}. Here, the lines above the boxes denote the memory that is passed to the future and required to correctly predict future statistics. Each box simply has to pass the information about the current state -- which generally is a multi-time object -- to future boxes, which, again, can make use of this information. Considering Figure~\ref{fig::clnm}, we can already see that the description of stochastic processes with the memory provided above is somewhat incomplete. While $\Gamma^{(m)}$ allows us to compute the probabilities of the next outcome, given the last $m$ outcomes, it only yields a one-time state, not an $m$-time state. While this is sufficient if we are only interested in the statistics of the next outcome, it is not enough to compute statistics further in the future. Concretely, we cannot let $\Gamma^{(m)}$ act successively to obtain \textit{all} future statistics. Expressed more graphically, a map that allows us to fully compute statistics for a process of Markov order $m$ needs $m$ input and $m$ output lines (see Figure~\ref{fig::clnm}). Naturally, such a map, which we will denote as $\Xi$ can always be constructed from $\Gamma^{(m)}$, as we discuss in more detail in the next section. Importantly, its action looks just like a square stochastic matrix:
\begin{gather}\label{eq:ximap}
\Xi^{(1)} \ \Pprob(R_{k-1}, \dots, R_{k-m}) \!=\!
\Pprob(R_{k}, \dots, R_{k-m+1}),
\end{gather}
which allows us to simply compute statistics via the concatenation of $\Xi$ just like in the Markovian case and hence the superscript 1. In other words, we can think of any non-Markovian process as a Markovian process on a larger system, as depicted in the bottom panel. Graphically, this can already easily be seen in Figure~\ref{fig::clnm}, where the system of interest (the die) plus the required memory lines form a Markovian process. 

Returning to our discussion of the complexity of non-Markovian processes, usually, not all distinct pasts -- even within the Markov order -- lead to distinct futures, and memory can be compressed. This effect can already be seen for the perturbed coin above, where, instead of $6^3 = 216$ stochastic matrices, we can compute the correct future by means of merely $4$ stochastic matrices. We will not discuss the issue of memory compression in this tutorial, but details can be found in the vast literature on the so-called $\varepsilon$-machines, see, for example, Ref.~\cite{crutchfield_inferring_1989,shalizi_computational_2001,crutchfield_between_2012}. Finally, we emphasize that, while here we have been focusing on the underlying mechanisms through which the respective probabilities emerge, a stochastic process is also fully described once all joint probabilities for events are known. For example, considering a three-fold toss of a die, once the probabilities $\Pprob(R_2, R_1, R_0)$ are known, all probabilities for smaller sequences of tosses (say, for example, $\Pprob(R_2, R_0)$) as well as all conditional probabilities for those three tosses can be computed. Knowing the full joint distribution is thus equivalent to knowing the underlying mechanism.

\subsection{Hidden Markov model}
\label{sec::HidMark}

\begin{figure}[t]
\centering
\includegraphics[width=0.95\linewidth]{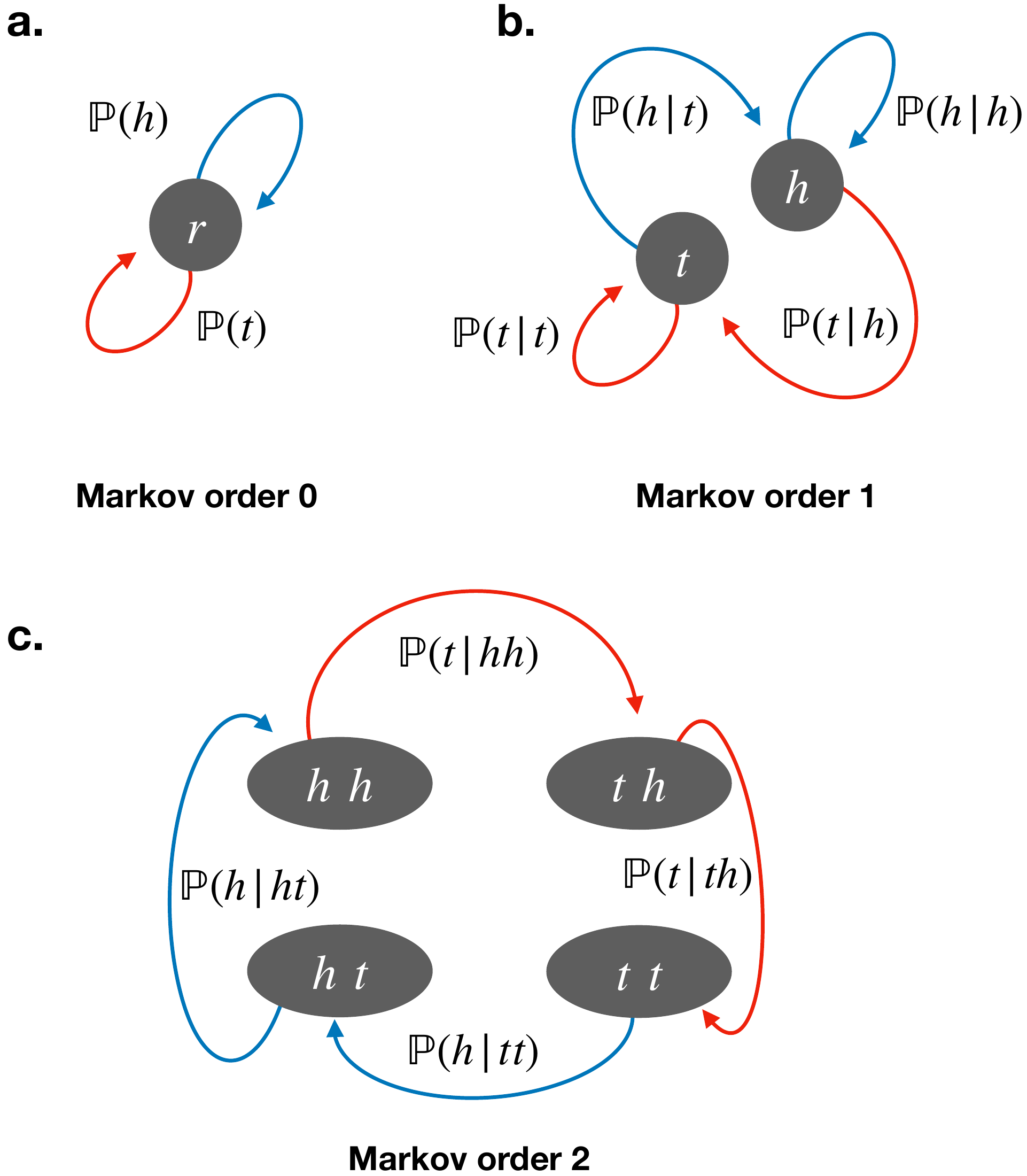}
\caption{\textbf{Markov chains.} Given a time series of coin flips we can deduce any of the above hidden Markov models. At memory length 2 we have a deterministic process and therefore longer memory will not yield any more information. In other words, the Markov order of the process in panel (c) is 2.}\label{fig::mchain}
\end{figure}

An important concept in many disciplines and one that is crucial to be able to deduce probabilities from sequences of measurement outcomes is that of stationarity. For stochastic processes, stationarity means time translation symmetry. That is, it does not matter when we flip a coin, we only need to consider its states up to the Markov order. This is useful because often we are interested in characterizing a process whose inner workings are hidden from us. In such a case, we can try to infer the inner working by noting the statistics of the state of the system in a \textit{time series}. For example, given a long sequence of coin-flip outcomes $\Ftype_1 = \{h,t\}$, we can determine the statistics for seeing `heads' $h$ and `tails' $t$, or any other sequence, say $hhttht$. Of course, this requires that the total data size is much larger than any sequence whose probability we wish to estimate. From this data we construct \textit{hidden Markov model} for the system that will reproduce the statistics up any desired Markov order~\cite{crutchfield_inferring_1989,shalizi_computational_2001, crutchfield_between_2012}.

An illuminating graphical representation for a stationary stochastic process is the so-called \textit{Markov chain}, which is associated with the stochastic matrix. For simplicity of the diagram let us consider a process dichotomic outcomes; e.g., a coin flip with the random variables $F_k$ (for flips). Again, this can be a fully random process, a Markov process, or a non-Markovian process, depending on how the coin is flipped. Suppose now that the coin is flipped a million times in succession and we are given the sequence of results. For simplicity, we assume stationarity, i.e., probabilities and conditional probabilities do not depend on the cardinal number of the coin toss. Under this assumption, from the observed results, we can compute how frequently one sees $h$ or $t$, which is quantified by $\Pprob(F_k)$. We might compute how often $h$ flips to $t$ or remain $h$ and so on; this is quantified by $\Pprob(F_k|F_{k-1})$ or the joint distributions $\Pprob(F_k|F_{k-1})$ for all $k$. Analogously, we may also compute the probability of seeing longer sequences, like $hhh$, $hht$, etc. With all of this, we can obtain conditional probabilities of the form $\Pprob(F_k|F_{k-1},F_{k-2})$ and $\Pprob(F_k|F_{k-1},F_{k-2},F_{k-3})$. Let us assume that both of these conditional probabilities coincide, which leads us to conclude that the Markov order of the process is 2 (technically, we should check that $\Pprob(F_k|F_{k-1},F_{k-2}) = \Pprob(F_k|F_{k-1}, \dots, F_{k-n})$ for all $n\ge2$, but as it is unlikely that \textit{only} longer memory exists, we consider this test for Markov order $2$ sufficient.).

In this case, following the ideas laid out below Eq.~\eqref{eq:ProcTensClass}, the probabilities of future outcomes can be described by a single stochastic matrix of the form 
\begin{gather}
 \Gamma^{(2)} \!=\! \left(\begin{matrix}
 \Pprob(h|hh) & 
 \Pprob(h|ht) &
 \Pprob(h|th) &
 \Pprob(h|tt)\\
 \Pprob(t|hh) & 
 \Pprob(t|ht) &
 \Pprob(t|th) & \Pprob(t|tt)\\
 \end{matrix}\right).
\end{gather}
This map will act on a statistical state that has the form 
\begin{gather}
\Pprob(F_{k},F_{k-1}) = \left[ \begin{matrix}\Pprob(hh) \! & \Pprob(ht) \! & \Pprob(th) \! & \Pprob(tt)\end{matrix}\right]^\trps.
\end{gather}
The action of the stochastic matrix on the statistical state gives us the probability for next flip.
\begin{gather}
\label{eq:MarkovEmbed}
\begin{split}
\Pprob(F_{k+1}) &= \Gamma^{(2)} \Pprob(F_{k},F_{k-1}) \\
&= \left[\begin{matrix}
\sum_{xy \in \{ht\}}\Pprob(h|xy) \Pprob(xy)\\
\sum_{xy \in \{ht\}}\Pprob(t|xy) \Pprob(xy)\\
\end{matrix}\right].
\end{split}
\end{gather}

Combining the probabilities for two successive outcomes into a single probability vector thus allows us to compute the probabilities for the next outcome in a Markovian fashion, i.e., by applying a single stochastic matrix to said probability vector. However, as already alluded to above, there is a slight mismatch in Eq.~\eqref{eq:MarkovEmbed}; while the random variables we look at on the r.h.s. are sequences of two successive outcomes, the random variable on the l.h.s. is a single outcome at the $k+1$-th toss. To obtain a fully Markovian model, one would rather desire a stochastic matrix that provides the transition probabilities from one sequence of two outcomes to another, i.e., a stochastic matrix $\Xi$ that yields
\begin{gather}
\label{eqn::HiddMarkovSand}
 \Pprob(F_{k+1},F_{k}) = \Xi^{(1)} \Pprob(\widetilde{F}_{k},F_{k-1}),
\end{gather}
where, for better bookkeeping, we formally distinguish between the random variables on the LHS and the RHS. Additionally, we give $\Xi$ an extra superscript to underline that it describes a process of Markov order one. To do so, it has to act on a larger space of random variables, namely, the combined previous two outcomes. 

Now, in our case, it is easy to see that the action of $\Xi^{(1)}$ can be simply computed from $\Gamma^{(2)}$ as
\begin{gather}
\begin{split}
 \Pprob(F_{k+1},F_{k}) &= \Xi^{(1)} \Pprob(\widetilde{F}_{k},F_{k-1}) \\
 &= \delta_{F_k\widetilde{F}_k} \Gamma^{(2)} \Pprob(\widetilde{F}_{k},F_{k-1}),
\end{split}
\end{gather}
where $\delta$ is the Kronecker function. This, in turn, implies that $\Xi^{(1)}$ and $\Gamma^{(2)}$ contain the same information, and the distinction between them is more of formal then of fundamental nature. Importantly though, $\Xi^{(1)}$ can be applied in succession, e.g., we have $\Pprob(F_{k+n},F_{k+n-1}) = (\Xi^{(1)})^n \Pprob(\widetilde{F}_{k},F_{k-1})$, while the same is not possible for $\Gamma^{(2)}$ due to the mismatch of input and output spaces.

Eq.~\eqref{eqn::HiddMarkovSand} then describes a Markovian model for the random variable $\textbf{F}_2$, which takes values $\{hh, ht, th, tt\}$. As knowledge of all relevant, i.e., within the Markov order, transition probabilities allows the computation of all joint probabilities, such an embedding into a higher dimensional Markovian process via a redefinition of the considered random variables is always possible. The corresponding Markovian model is often called \textit{hidden Markov model}. As a brief aside, we note that the amount of memory that needs to be considered in an experiment depends both on the intrinsic Markov order of the process at hand, as well as the amount of information an experimenter can or wants to store. If, for example, one is only interested in correctly recreating transition probabilities $\Pprob(R_k|R_{k-1})$ between adjacent times, but not necessarily higher-order transition probabilities, like, e.g., $\Pprob(R_k|R_{k-1}, R_{k-2})$, then a Markovian model without any memory is fully sufficient (but will not properly reproduce higher-order transition probabilities). 

Returning to our process, we have depicted the corresponding Markov chains for each case in Figure~\ref{fig::mchain}. For a fully random process, the Markov chain only has one state; after each flip, the state returns to itself, and the future probabilities do not change based on the past. For a Markov process, the chain has two states, and four transitions are possible. Finally, the non-Markovian process is chosen to be deterministic: $hh$ always goes to $th$, and so on. Note that, here, as mentioned above, if we only care about transition probabilities $\Pprob(F_k|F_{k-1})$, i.e., we only consider the last outcome and not the last two outcomes (i.e., we identify $hh$ and $ht$, and $th$ and $tt$), then the process of the panel (c) in Figure~\ref{fig::mchain} reduces to the simpler one in panel (b), but the information is lost.

All of the panels of Figure~\ref{fig::mchain} describe Markovian processes, however for different random variables. This is a general feature: any non-Markovian process can be represented by a \textit{hidden Markov model} or a Markov chain by properly combining the past into a `large enough' random variable~\cite{vanKampen1998} (for example, the random variable with values $\{hh, th, ht, tt\}$ in panel (c)). This intuition will come in handy when we move to the case of quantum stochastic processes. But first, we need to formalize the theory of classical stochastic process and show where lie the pitfalls when generalizing this theory to the quantum domains.

\subsection{(Some) mathematical rigor}
\label{sec::rig1}
As mentioned, in our presentation of stochastic processes, we rather opt for intuitive examples than full mathematical rigor. However, laying the fundamental concepts of probability theory in detail provides a more comprehensive picture of stochastic processes, and renders the generalizations needed to treat quantum processes mathematically straightforward.

The basic ingredient for the discussion of stochastic processes is the triplet $(\Omega, \mathcal{S}, \omega)$ of a sample space $\Omega$, a $\sigma$-algebra $\mathcal{S}$ and a probability measure $\omega$. Intuitively, $\Omega$ is the set of all events that can occur in a given experiment (for example, $\Omega$ could represent the myriad of microstates a die can assume or the possible numbers of pips it can show), $\mathcal{S}$ corresponds to all the outcomes that can be resolved by the measurement device (for the case of the die, $\Scal$ could, for example, correspond to the number of pips the die can show, or to the less fine-grained information `odd' or `even') and $\omega$ allocates a probability to each of these observable outcomes. 

More rigorously, we have the following definition~\cite{tao_introduction_2011}:
\begin{definition*}[$\sigma$-algebra]
Let $\Omega$ be a set. A $\sigma$-algebra on $\Omega$ is a collection $\Scal$ of subsets of $\Omega$, such that 
\begin{itemize}
 \item $\Omega \in \Scal$ and $\varnothing \in \Scal$.
 \item If $s \in \Scal$, then $\Omega\setminus s \in \Scal$.
 \item $\Scal$ is closed under (countable) unions and intersections, \textit{i.e.}, if $s_1, s_2, \dots \in \Scal$, then $\bigcup\limits_{j= 1}^\infty s_j \in \Scal$ and $\bigcap \limits_{j= 1}^\infty s_j \in \Scal$. 
\end{itemize}
\end{definition*}
For example, if the sample space is given by $\Omega = \{\vcdice{1},\dots,\vcdice{6}\}$ and we only resolve whether the outcome of the toss of a die is odd or even, the corresponding $\sigma$-algebra is given by $\{\{\vcdice{1},\vcdice{3},\vcdice{5}\}, \{\vcdice{2},\vcdice{4},\vcdice{6}\}, \varnothing, \Omega\}$, while in the case where we resolve the individual numbers of pips, $\Scal$ is simply the power set of $\Omega$. 

A pair $(\Omega,\Scal)$ is called a \textit{measurable space}, as now, we can introduce a probability measure for observable outcomes in a well-defined way: 

\begin{definition*}[Probability measure]
\label{def::ProbMeas}
 Let $(\Omega,\Scal)$ be a measurable space. A probability measure $\omega: \Scal \rightarrow \mathbbm{R}$ is a real-valued function that satisfies
 \begin{itemize}
 \item $\omega(\Omega) = 1$.
 \item $\omega(s) \geq 0$ for all $s \in \Scal$
 \item $\omega$ is additive for (countable) unions of disjoint events, \textit{i.e.}, $\omega\left(\bigcup_{j=1}^\infty s_j\right) = \sum_j^\infty \omega(s_j)$ for $s_j \in \Scal$ and $s_j \cap s_{j^{\, \prime}} = \varnothing$ when $j \neq j^{\, \prime}$. 
 \end{itemize}
\end{definition*}
The corresponding triplet $(\Omega, \Scal, \omega)$ is then called a \textit{probability space}~\cite{tao_introduction_2011}. As the name suggests, $\omega$ maps each event $s_j$ to its corresponding probability, and, using the convention of the previous sections, we could have denoted it by $\Pprob$, and will do so in what follows. Evidently, in our previous discussions, we already made use of sample spaces, $\sigma$-algebras and probability measures, without caring too much about their mathematical underpinnings. 

The mathematical machinery of probability spaces provides a versatile framework for the description of stochastic processes, both on finitely and infinitely many times (see Sec.~\ref{sec::RigII} for an extension of the above concepts to the multi-time case).

So far, we have talked about processes that are discrete both in time and space. It does not make much sense to talk about the state of a die when it is in mid-air; nor does it make sense to attribute a state of $4.4$ to a die. On the other hand, of course, there are processes that are both continuous in time and space. A classic example is \textit{Brownian motion}~\cite{lemons_introduction_2002}, which requires that time be treated continuously. If not, the results lead to pathological situations where the kinetic energy of the Brownian particle blows up. Moreover, in such instances, the event space is the position of the Brownian particle and can take uncountably many different real values. Nevertheless, the central object in the theory of stochastic processes does not change; it remains the joint probability distribution for all events, which in the case of infinitely many times is a probability distribution on a rather complicated, and not easy to handle $\sigma$-algebra. Below, we will discuss how due to a fundamental result by Kolmogorov it is sufficient to deal with finite distributions instead of distributions on $\sigma$-algebras on infinite Cartesian products of sample spaces. Finally, this machinery straightforwardly generalizes to positive operator valued measures (POVMs) as well as instruments, fundamental ingredients for the discussion of quantum stochastic processes.

\section{\texorpdfstring{Classical Stochastic Processes \\ Formal approach}{}}
\label{sec::ClFormal}

Up to this point, both in the examples we provided, as well as the more rigorous formulation, we have somewhat left open what exactly we mean by a stochastic process, and what quantity encapsulates it. We will do so now, and provide a fundamental theorem for the theory of stochastic processes, the Kolmogorov extension theorem (KET), which allows one to properly define stochastic processes on infinitely many times, based on finite time information. 

\subsection{What then is a stochastic process?}
\label{sec::What}

Intuitively, a stochastic process on a set of times $\Tset_k:=\{t_0, t_1, \dots, t_k\}$ with $t_i \leq t_j$ for $i\leq j$ is the joint probability distribution over observable events. Namely, the central quantity that captures everything that can be learned about an underlying process is 
\begin{gather}
 \Pprob_{\Tset_{k+1}}:= \Pprob(R_k, \ t_k; R_{k-1}, \ t_{k-1}; \ \dots; R_0, \ t_{0}), 
\end{gather}
corresponding to all joint probabilities 
\begin{gather}
 \{\Pprob(R_k = r_k, R_{k-1}=r_{k-1}, \ \dots, R_0=r_0,)\}_{r_k,\dots,r_0}
\end{gather}
to observe all possible realizations $R_k=r_k$ at time $t_k$, $R_{k-1}=r_{k-1}$ at time $t_{k-1}$ and so on. Evidently, the time label -- which we omit above and for most of this tutorial -- could also correspond to a label of the number of tosses, etc. We also adopt the compact notation of $\Pprob_{\Tset_{k+1}}$, as defined above, to denote a probability distribution on a set of $k+1$ times.

More concretely, suppose the process we have in mind is tossing a die five times in a row. This stochastic process is fully characterized by the probability of observing all possible sequence of events
\begin{gather}
 \begin{split}
&\{\Pprob (\vcdice{1}, \vcdice{1}, \vcdice{1}, \vcdice{1}, \vcdice{1}), \ \dots, \ \Pprob (\vcdice{1}, \vcdice{1}, \vcdice{1}, \vcdice{1}, \vcdice{6}), \dots\\
&\qquad \qquad \vdots \qquad\quad \, \qquad \qquad \qquad \ \ \vdots\\
&\phantom{\{} \Pprob (\vcdice{6}, \vcdice{6}, \vcdice{6}, \vcdice{6}, \vcdice{1}), \ \dots, \
\Pprob (\vcdice{6}, \vcdice{6}, \vcdice{6}, \vcdice{6}, \vcdice{6})\},
 \end{split}
\end{gather}
where, as before, we omit the respective time/tossing number labels. 

From the joint distribution for five tosses, one can obtain \textit{any} desired marginal distributions for fewer tosses, e.g. $\Pprob(R_3)$; or \textit{any} conditional distributions (for five tosses), such as, for example, the conditional probability $\Pprob(R_2=\vcdice{1}| R_1=\vcdice{6},R_0=\vcdice{6})$, to obtain outcome $\vcdice{1}$ at the third toss, having observed two $\vcdice{6}$ in a row previously; the conditional distributions in turn allows computing the stochastic matrices, which in turn allow casting processes as a Markov chain. Having the total distribution is enough to determine whether a process is fully random, Markovian, or non-Markovian. This statement, however, is contingent on the respective set of times. Naturally, without any further assumptions of memory length and/or stationarity, knowing the joint probabilities of outcomes -- and thus everything that can be learned -- on a set of times $\Tset_k$ does not provide knowledge about the corresponding process on a different set of times $\Tset_{k'}$. Consequently, we identify a stochastic process with the joint probabilities it displays with respect to a fixed set of times.

While joint probabilities contain all inferable information about a stochastic process, working with them is not always desirable because their number of entries grows exponentially. Nevertheless, they are the central quantity in the theory of classical stochastic processes. Our first aim when extending the notion of stochastic processes to the quantum domain will thus be to construct the analogy to joint distribution for time-ordered events. Doing so has been troubling for the same foundational reasons that make quantum mechanics so interesting. Most notably, quantum processes, in general, do not straightforwardly allow for a Kolmogorov extension theorem, which we discuss below. However, upon closer inspection, such obstacles can be overcome by properly generalizing the concept of joint probabilities to the quantum domain. Before doing so, we will first return to our more rigorous mathematical treatment and define stochastic processes in terms of probability spaces.

\begin{figure}
 \centering
 \includegraphics[width=0.95\linewidth] {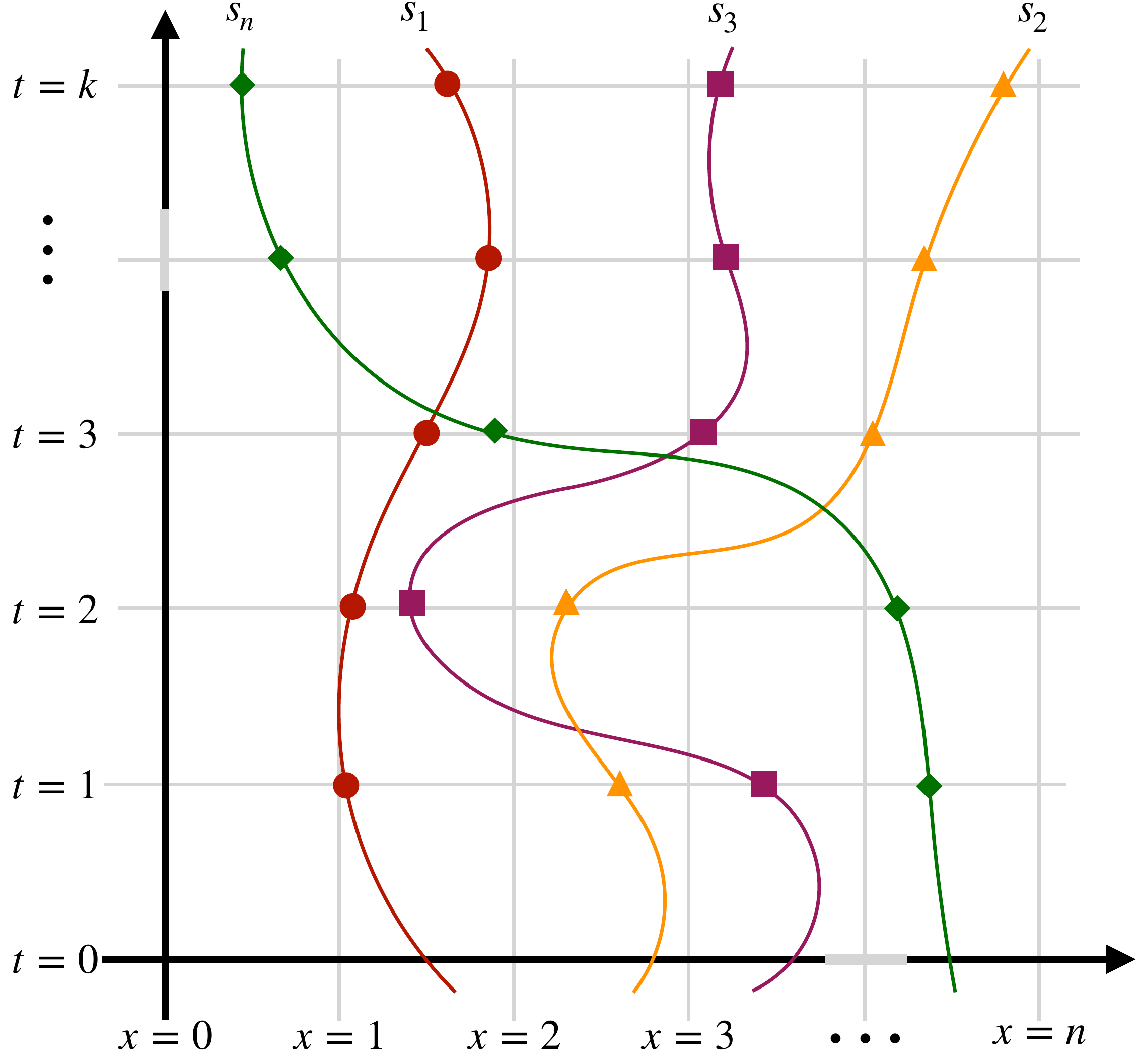}
 \caption{\textbf{Continuous process.} A stochastic process is a joint probability distribution over all times. From a physical perspective, we can think of it as the probability of observing a trajectory $s_k$. This is highly desirable when talking about the motion of a Brownian particle. However, this interpretation requires some caution as there are cases where trajectories may not be smooth or even continuous.}\label{fig:traj}
\end{figure}

\subsection{Kolmogorov extension theorem}
\label{sec::KET}

While, for the example of the tossing of a die, a description of the process at hand in terms of joint probabilities on finitely or countably many times/tosses is satisfactory, this is not always the case. For example, even though it can in practice only be probed at finitely many points in time, when considering Brownian motion, one implicitly posits the existence of an `underlying' stochastic process, from which the observed joint probabilities stem. Intuitively, for the case of Brownian motion, this underlying process should be fully described by a probability distribution that ascribes a probability to all possible trajectories the particle can take. Connecting the operationally well-defined finite joint probabilities a physicist can observe and/or model with the concept of an underlying process is the aim of the Kolmogorov extension theorem \textbf{(KET)}.

Besides not being experimentally accessible, working with probability distributions on infinitely many times has the additional drawback that the respective mathematical objects are rather cumbersome to use, and would make the modeling of stochastic processes a fairly tedious business. Luckily, the KET allows one to deduce the existence of an underlying process on infinitely many times, from properties of only finite objects. With this, modeling a proper stochastic process on infinitely many times amounts to constructing finite time joint probabilities that `fit together' properly.

To see what we mean by this last statement let $\Pprob_{\Tset_\ell}$ be the joint distribution obtained for an experiment for some fixed $\ell$ times. For now, we will stick with the case of Brownian motion, and $\Pprob_{\Tset_\ell}$ could correspond to the probability to find a particle at positions $x_0, \dots, x_{\ell-1}$ when measuring it at times $\Tset_\ell = \{t_0,\dots, t_{\ell-1}\}$. As mentioned before, $\Pprob_{\Tset_\ell}$ contains all statistical information for fewer times as marginals, i.e., for any subset $\Tset_{k} \subseteq \Tset_{\ell}$ we have
\begin{gather}
 \Pprob_{\Tset_{k}} = \sum_{\Tset_\ell \setminus \Tset_{k}} \Pprob_{\Tset_{\ell}} =: \Pprob_{\Tset_\ell}^{|\Tset_k},
\end{gather}
where we denote the sum over the times in the complement of the intersection of $\Tset_{k}$ and $\Tset_{\ell}$ by $\Tset_\ell \setminus \Tset_{k}$ and use an additional superscript to signify that the respective joint probability distribution is restricted to a subset of times via marginalization. For simplicity of notation, here and in what follows, we always denote the marginalization by a summation, even though, in the uncountably infinite case, it would correspond to an integration. 

For classical stochastic processes, all probabilities on a set of time can be obtained from those on a superset of times by marginalization. We will call this consistency condition between joint probability distributions of a process on different sets of times \textit{Kolmogorov consistency conditions}. Naturally, consistency conditions hold in particular if the finite joint probability distributions stem from an underlying process on infinitely many times $\Tset \supseteq \Tset_\ell \supseteq \Tset_k $, where we leave the nature of the corresponding probability distribution $\Pprob_\Tset$ somewhat vague for now (see Sec.~\ref{sec::RigII} for a more thorough definition).

Importantly, the KET shows, that satisfaction of the consistency condition on all finite sets $\Tset_k \subseteq \Tset_\ell \subseteq \Tset$ is already sufficient to guarantee the existence of an underlying process on $\Tset$. Specifically, the Kolmogorov extension theorem~\cite{kolmogorov_foundations_1956, feller_introduction_1968, tao_introduction_2011, BreuerPetruccione} defines the minimal properties finite probability distributions have to satisfy in order for an underlying process to exist:

\vspace{5pt}
\textbf{Theorem. (KET)} \textit{Let $\Tset$ be a set of times. For each finite $\Tset_k \subseteq \Tset$, let $\Pprob_{\Tset_k}$ be a (sufficiently regular) $k$-step joint probability distribution. There exists an underlying stochastic process $\Pprob_T$ that satisfies $\Pprob_{T_k} = \Pprob_{\Tset}^{|\Tset_k}$ for all finite $\Tset_k\subseteq \Tset$ iff $\Pprob_{\Tset_k} = \Pprob_{\Tset_\ell}^{|\Tset_k}$ for all $\Tset_k\subseteq \Tset_\ell \subseteq \Tset$.}
\vspace{5pt}

Put more intuitively, the KET shows that for a given family of finite joint probability distributions that satisfy consistency conditions,\ftnt{Besides consistency, the individual probability distributions also have to be inner regular. We will not concern ourselves with this technicality, see Ref.~\cite{tao_introduction_2011} for more details.} the existence of an underlying process, that contains all of the finite ones as marginals, is ensured. Importantly, this underlying process does not need to be known explicitly in order to properly model a stochastic process. 

We emphasize that, in the (physically relevant) case where $\Tset$ is an infinite set, the probability distribution $\Pprob_\Tset$ is generally not experimentally accessible. For example, in the case of Brownian motion, the set $\Tset$ could contain all times in the interval $[0,t]$ and each realization would represent a possible continuous trajectory of a particle over this time interval, see Figure~\ref{fig:traj}. While we assume the existence of these underlying trajectories (and hence the existence of $\Pprob_\Tset$) in experiments concerning Brownian motion, we often only access their finite time manifestations, \textit{i.e.}, $\Pprob_{\Tset_k}$ for some $\Tset_k$. The KET thus bridges the gap between the finite experimental reality and the underlying infinite stochastic process, in turn defining in terms of accessible quantities what one means by a stochastic process on infinitely many times. For this reason, many books on stochastic processes begin with the statement of KET.

In addition, the KET also enables the modeling of stochastic processes: Any mechanism that leads to finite joint probability distributions that satisfy a consistency condition is ensured to have an underlying process. For example, the proof of the existence of Brownian motion relies on the KET as a fundamental ingredient~\cite{wiener_differential_1966, Levy_1940, ciesielski_lectures_1966, bhattacharya_basic_2017}. 

Loosely speaking, the KET holds for classical stochastic processes, because there is no difference between `doing nothing' and conducting a measurement but `not looking at the outcomes' (\textit{i.e.}, summing over the outcomes at a time); otherwise, as we shall see in the discussion of quantum stochastic processes, Kolmogorov consistency conditions are not satisfied. Put differently, the validity of the KET is based on the fundamental assumption that the interrogation of a system does not, on average, influence its state. This assumption generally fails to hold in quantum mechanics, which makes the definition of quantum stochastic processes somewhat more involved, and their structure much richer than that of their classical counterparts. 

\begin{figure}
 \centering
 \includegraphics[width=0.95\linewidth]{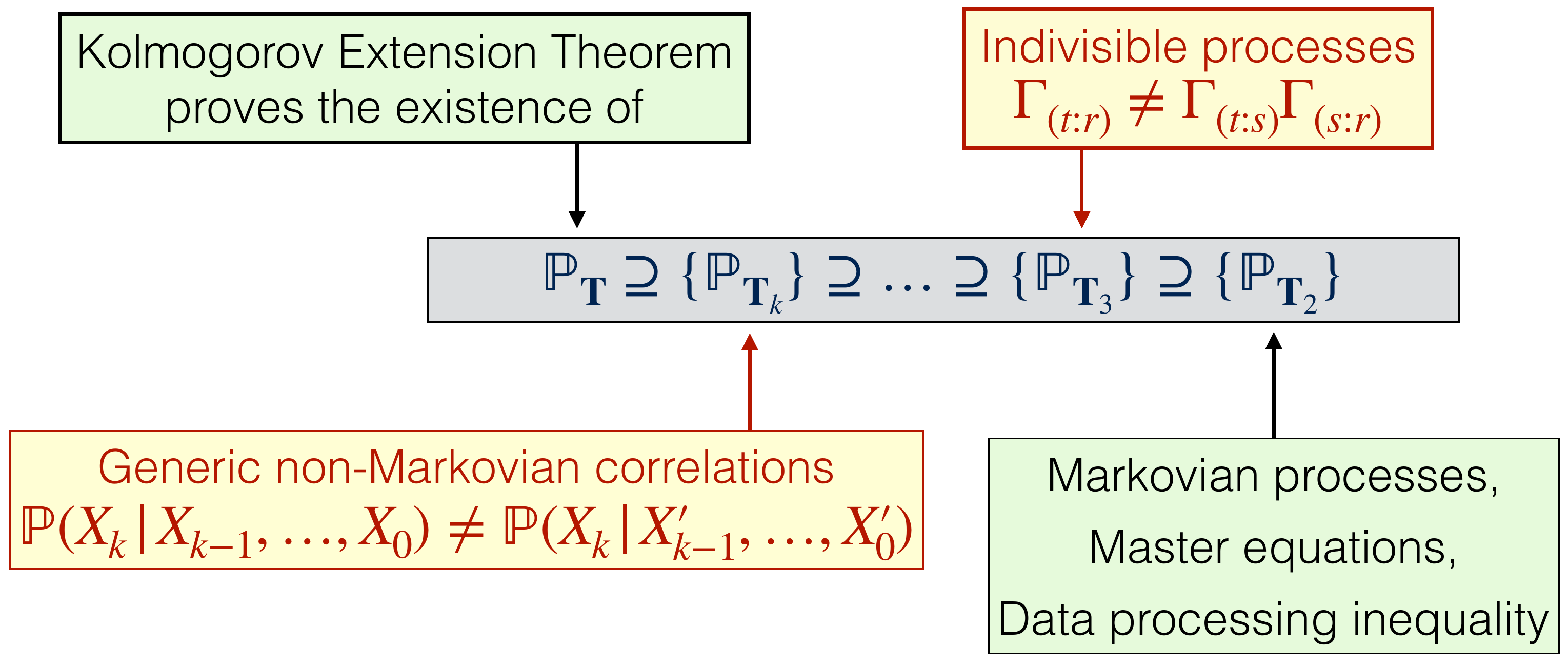}
 \caption{\textbf{Hierarchy of multi-time processes}. A stochastic process is the joint probability distribution over all times. Of course, in practice one looks only at finite time statistics. However, the set of all $k$-time probability distributions $\{\Pprob_{\Tset_k}\}$ contain, as marginals, all $j$-time probability distributions $\{\Pprob_{\Tset_j}\}$ for $j<k$. Moreover, the set of two and three time distributions play a significant roles in the theory of stochastic processes.}
 \label{fig:hierarchy}
\end{figure}

\subsection{Practical features of stochastic processes}
\label{sec::PracFeat}

Now that we have a formal definition of a stochastic process let us ask what it is useful for. It is worth saying that working with a probability distribution of a large number of random variables is not desirable as the complexity grows exponentially. However, for a given problem, what we care about is the structure of the stochastic process and what we may anticipate when we sample from this distribution. We depict the hierarchy of stochastic processes in Figure~\ref{fig:hierarchy}, and in this section focus on the short end of the hierarchy, i.e., Markovian processes or non-Markovian processes with low Markov order.

Naturally, the examples in Sec.~\ref{sec:clexample} and the formal theory in the last subsection only begin to scratch at the massive literature on stochastic processes. We, of course, cannot cover all facets of this field here. However, in practice, there are a few important topics that must be mentioned. Below we will discuss several common tools that one encounters in the field of stochastic processes. Here, we do so rather to provide a quick overview than a thorough introduction to the field. First among the tools used in the field are master equations, which are employed ubiquitously in the sciences, finance, and beyond. Next, we will briefly cover methods to differentiate between Markovian and non-Markovian processes, as well as quantify the memory using tools of information theory. While many of these examples only deal with two-time correlations, we do emphasize that there are problems that naturally require multi-time correlations.

\subsubsection{Master equations}
\label{sec::MastEq}

A master equation is a differential equation that relates the rate of change in probabilities with the current and the past states of the system. Put simply, they are equations of motion for stochastic processes and thus provide the underlying mechanism by which the transition probabilities we discussed in the previous section come about. There are of course many famous master equations in physics: Pauli, Fokker-Plank, Langevin, to name a few on the classical side. We will not delve into the details of this very rich topic here, and once again just begin to scratch the surface. We refer the reader to other texts for more in-depth coverage of master equations~\cite{StochProc, hanggi_stochastic_1982, lemons_introduction_2002}. 

It will suffice for our purpose that a master equation, in general, has the following form\ftnt{Some authors will not call this a master equation due to its temporal non-locality~\cite{StochProc}.}
\begin{gather}\label{eq:master}
 \frac{d}{d t} \Pprob(X_t) = \int_{s}^t \mathcal{G}(t,\tau)\Pprob(X_\tau) \ d\tau,
\end{gather}
where $\mathcal{G}(t,\tau)$ is a matrix operator. The time derivative of the state at $t$ depends on the previous states up to a time $s$, which is the memory length. If the memory length is infinite, then $s \to -\infty$. As mentioned before, such a master equation allows one, in principle, to compute the change of probabilities, given some information about the past of the system.

Since the master equation expresses the probabilities continuously in time it may be then tempting to think that a master equation is equivalent to a stochastic process as defined above by means of the KET. However, this is not the case because a master equation needs at most joint probabilities of two times or lower. Namely, the set of joint probability distributions, 
\begin{gather}\label{eq:triset}
\{\Pprob_{\Tset_2} \} := \{\Pprob(X_b,X_a) \}_{b > a} \quad \forall \ b>a>0
\end{gather}
is sufficient to derive Eq.~\eqref{eq:master}. The LHS can be computed by setting $b=t$ and $a=t-dt$. While the RHS can be expressed as a linear combination of product of stochastic matrices $\Gamma_{c:b} \Gamma_{b:a}$, with $c=t$, $b=\tau \ge s$, and $a=r<\tau$. In fact, the RHS is concerned with functions such as $\Gamma_{c:a} - \Gamma_{c:b} \Gamma_{b:a}$, which measure the temporal correlations between $a$ and $c$, given an observation at $b$. In any case, these stochastic matrices only depend on joint distributions of two times, as seen in Eqs.~\eqref{eq:oneprop} and \eqref{eq:stochmat}, and are not concerned with multi-time statistics. Thus, the family of distributions in Eq.~\eqref{eq:triset} suffices for the RHS. Formally, showing that the RHS can be expressed as a product of two stochastic matrices can be done by means of the Laplace transform~\cite{bassano1,vacchini} or the ansatz known as the \textit{transfer tensor}~\cite{CerrilloCao2014, Rosenbach2016, Pollock-quantum}. These technical details aside, master equations play an important practical role for the description of scenarios, where only two-time probabilities and/or the change of single time probabilities are required. By construction, they do not, however, allow for the computation of multi-time joint probabilities. In turn, this implies that they do not provide a full description of stochastic processes in the sense of the KET. Nonetheless, they constitute an important tool for the description of aspects of stochastic processes.

\subsubsection{Divisible processes}
\label{sec:cl-div}
To shed more light on the concept of master equations, let us consider a special case (which we will also encounter in the quantum setting). Specifically, let us consider a family of stochastic matrices that satisfy
\begin{gather}\label{eq:divisible}
 \Gamma_{(t:r)} = \Gamma_{(t:s)} \Gamma_{(s:r)} \quad \forall \ t>s>r.
\end{gather}
Processes described by such a family are called \textit{divisible}. Once the functional dependence of $\Gamma_{(t:r)}$ on $t$ and $r$ is known, one can build up the set of distributions contained in Eq.~\eqref{eq:triset}. It is easy to see that the family of stochastic matrices in Eq.~\eqref{eq:divisible} is a superset of Markovian processes. That is, any Markov process will satisfy the above equation. However, there are non-Markovian processes that also satisfy the divisibility property~\footnote{Due to this inequivalence of divisibility and Markovianity, the maps $\Gamma_{t:s}$ in Eq.~\eqref{eq:divisible} cannot always be considered as matrices containing conditional probabilities $\Pprob(R_t|R_s)$ -- as these conditional probabilities might depend on prior measurement outcomes -- but rather as mapping from a probability distribution at time $s$ to a probability distribution at time $t$~\cite{hanggi_time_1977, hanggi_stochastic_1982}. This breakdown of interpretation also occurs in quantum mechanics~\cite{Milz2019}. In the Markovian case, $\Gamma_{t:s}$ indeed contains conditional probabilities}. Nevertheless, checking for divisibility is often far simpler than checking for the satisfaction of the Markov conditions since the latter requires the collection of multi-time statistics, while the former can be decided based on two-time statistics only. Moreover, as we will see shortly, the divisibility of the process implies several highly desirable properties for the process.

A nice property of divisible processes is the corresponding master equation. Applying Eq.~\eqref{eq:divisible} to the LHS of Eq.~\eqref{eq:master} we get
\begin{gather}\label{eq:markov-master}
 \frac{\Pprob(X_t) - \Pprob(X_{t-dt})}{d t} = \frac{\Gamma_{(t:t-dt)}-\openone}{dt} \Pprob(X_{t-dt})
\end{gather}
where $\openone$ is the identity matrix. Taking the limit $dt \to 0$ yields the generator $\mathbf{G}_t := \lim_{dt \to 0} [\Gamma_{(t:t-dt)}-\openone]/dt$. This is a time-local master equation in the sense that the derivative of $\Pprob$ -- in contrast to the more general case of Eq.~\eqref{eq:master} -- only depends on the current time $t$, but not on previous times. In turn, the generator is related to the stochastic matrix as $\Gamma_{(t:t-dt)} = \exp(\mathbf{G}_t dt)$, which is obtained by integration. When the process is stationary, i.e., symmetric under time-translation, both $\Gamma$ and $\mathbf{G}$ will be time independent.

\textbf{A divisible Markovian process.} To make the above more concrete, let us consider a two level system that undergoes the following infinitesimal process
\begin{gather}\label{eq:cldivproc}
 \Gamma_{(t:t-dt)}= (1-\gamma dt) \left(\begin{matrix} 1 & 0 \\ 0 & 1 \end{matrix}\right) + \gamma dt
 \left(\begin{matrix} g_0 & g_0 \\ g_1 & g_1
 \end{matrix}\right).
\end{gather}
The first part of the process is just the identity process, and the second part is a random process. However, together they form a Markov process. Using Eq.~\eqref{eq:markov-master} we can derive the generator for the master equation. This process is very similar to the perturbed die in the last section, with the difference that here, we consider a process that is continuous in time; it takes any state $\Pprob(X_{t-dt})$ at $t-dt$ to 
\begin{gather}
\Pprob(X_{t}) = (1-\gamma dt) \Pprob(X_{t-dt}) + \gamma dt \ \mathbb{G},
\end{gather}
where $\mathbb{G} = [g_0 \ g_1]^\trps$. After some time $\tau = n dt$, i.e., after $n$ applications of the stochastic matrix, we have 
\begin{gather}
\Pprob(X_{\tau}) = (1-\gamma dt)^n \Pprob(X_{t}) + \gamma n dt \ \mathbb{G}.
\end{gather}
That is, the process relaxes any state of the system to the fixed $\mathbb{G}$ exponentially fast with a rate $\gamma$. Many processes, such as thermalization, have such a form. In fact, one often associates Markov processes with exponential decay. However, as already mentioned above, such an identification is not exact, since there are non-Markovian processes that satisfy a divisible master equation as we shall see now by means of two explicit examples (we will encounter an explicit example of this phenomenon in the quantum case in Sec.~\ref{sec::ExDivMark}.

\begin{figure}
 \centering
 \includegraphics[width=0.95\linewidth]{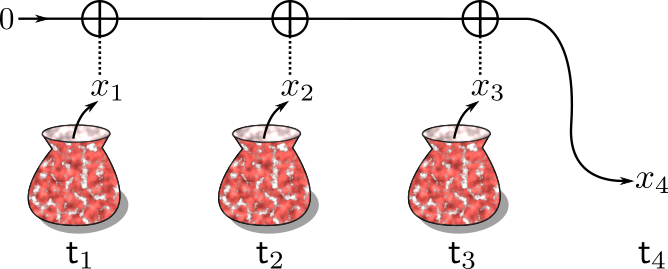}
 \caption{\textbf{Stroboscopic divisible non-Markovian process.} At each time $t_j$, each of the possible outcomes $0$ and $1$ occurs with probability $1/2$ (for example, they could be drawn from urns with uniform distributions). At the final time $t_4$, the observed outcome is equal to the sum (modulo 2) of the previous three outcomes. While the stochastic map between any two points in time is completely random -- and thus the process is divisible -- the overall joint probability distribution shows multi-time memory effects (as laid out in the text).}
 \label{fig::Four_Time_Class}
\end{figure}

\textbf{A stroboscopic divisible non-Markovian process.} As mentioned, divisibility and Markovianity do not coincide. To see this, we provide the following example which comes from Ref.~\cite{capela_monogamy_2020} and provides a stroboscopic -- in the sense that we only consider it at fixed points in time -- non-Markovian process that is divisible. Let us consider a single bit process with $x_j=0,1$ with probability $1/2$ for $j=1,2,3$. That is, the process yields random bits in the first three times. 
At the next time, we let $x_4=x_1+x_2+x_3$, where the addition is modulo 2 (see Figure~\ref{fig::Four_Time_Class}). It is easy to see that the stochastic matrix between \textit{any} two times will correspond to a random process, making the process divisible. However, $\Pprob(X_4,X_3,X_2,X_1)$ is not uniform; when $x_1 + x_2 + x_3 = x_4$ the probability will be $\tfrac{1}{8}$ and $0$ otherwise. Consequently, there are genuine four-time correlations, but there are not two or three time correlations.

\textbf{A process with long memory.}
Let us now consider a process where the probability of observing an outcome $x_t$ is correlated with what was observed some time ago $x_{t-s}$ with some probability
\begin{gather}
 \Pprob(X_t = x_t |X_{t-s} = x_{t-s} ) = p \ \delta_{x_t,x_{t-s}} + \frac{1-p}{d}
\end{gather}
Here $d$ is the size of the system. This process only has two-time correlations, but the process is non-Markovian as the memory is a long-range one. A master equation, of the type of Eq.~\eqref{eq:master}, for this process, can be derived by differentiating. For sake of brevity, we forego this exercise.

As mentioned, master equations are a ubiquitously used tool for the description of stochastic processes, both in the classical as well as the quantum (see below) domain. They allow one to model the evolution of the one-time probability distribution $\Pprob(X_t)$. However, they are not well-suited for the description of multi-time joint probabilities. This will be particularly true for the quantum case, where intermediate measurements -- required to collect multi-time statistics -- unavoidable influence the state of the system. For many real world applications though, knowledge about $\Pprob(X_t)$ is sufficient, making master equations an indispensable tool for the modeling of stochastic processes. On the other hand, in order to analyze memory length and strength in detail, one must -- particularly in the quantum case -- go beyond the description of stochastic processes in terms of master equations (see Sec.~\ref{sec::Multi-time}). This widening of the horizon beyond master equations then also enables one to carry over the intuition developed for stochastic processes in the classical case to the quantum realm, as well as a rigorous definition of quantum stochastic processes in the spirit of the KET.

At this stage, it is worth pointing out why Markov processes are of interest in many cases, and how they fit into the picture. Suppose we are following the trajectory of a particle at position $x$ at time $t$, which then moves to $x'$ at $t'$. If the difference in time is arbitrarily small, say $\delta t$, then for a physical process, $x'$ cannot be too different from $x$ due to continuity. Thus, it is natural to write down a master equation to describe such a process. Since the future state will always depend on the current position, the process will be at least Markovian. Still, the process may have higher-order correlations, but they are often neglected for simplicity. Importantly, if the process is indeed memoryless, then master equations actually allow for the computation of \textit{all} joint probability distributions and provide a complete picture of the process at hand. Due to their practical importance, we now provide some tools that are frequently used when dealing with memoryless processes, and to gauge deviation from Markovian statistics in an operationally accessible way. As before, this short overview is by no means intended to be comprehensive but merely aimed at providing a quick glimpse of possible ways to quantify memory.

\subsubsection{Data processing inequality}
\label{sec::DataProc}
Somewhat abstractly, a stochastic process can be understood as a state being processed in time. Memory, then, means that some information about the past of the state of the system at hand is stored and used at a later time to influence the future statistics of the system. Unsurprisingly, the mathematical means we use to make this intuition manifest and quantify the presence of memory are borrowed from information theory. Here, we introduce them, starting from the special case of divisible processes.

One of the most useful properties of Markov (and, more generally, divisible) processes is the satisfaction of the data processing inequality (\textbf{DPI}). Suppose we are able to prepare the system in two possible initial states $\Pprob(X_0)$ and $\mathbb{R}(X_0)$, and then subject each to a process $\Gamma_{(t:0)}$ to yield $\Pprob(X_t) =  \Gamma_{(t:0)} \Pprob(X_0)$ and $\mathbb{R}(X_t) = \Gamma_{t:0}\mathbbm{R}(X_0)$, respectively. The intuition behind DPIs is that the process has no mechanism to increase the distinguishability between two initial states unless it has some additional information.

For instance, a natural measure for distinguishing probability distribution is the so called the \textit{Kolmogorov distance} or the \textit{trace distance}
\begin{gather}
 \|\Pprob(X)- \mathbb{R}(X)\|_{1} := \frac{1}{2}\sum_{x} 
 | \Pprob(x)- \mathbb{R}(x)|.
\end{gather}
When two states are fully distinguishable, the trace distance will be 1, which is the maximal value it can assume. On the other hand, if the two distributions are the same then the trace distance will be 0. The DPI guarantees that the distance between distributions is non-increasing under the action of stochastic maps, i.e.,
\begin{gather}
 \|\Pprob(X_0)- \mathbb{R}(X_0)\|_{1} \ge \|\Pprob(X_t)- \mathbb{R}(X_t)\|_{1}
\end{gather}
for all times $t>0$ and equality (for all pairs of initial distributions) holds if and only if the process is reversible. Additionally, for Markov processes the DPI will hold for all times $t\ge s$, i.e.
\begin{gather}
\|\Pprob(X_s)- \mathbb{R}(X_s)\|_{1} \ge \|\Pprob(X_t)- \mathbb{R}(X_t)\|_{1}.
\end{gather}
Conversely, if the distinguishability between two distributions increases at any point of their evolution, then the underlying dynamics cannot be Markovian and stochastic maps $\Gamma_{t:s}$ between two points in time do not provide a full picture of the process at hand.

There are many metrics and pseudo-metrics that satisfy DPI, but not all. For instance, the Euclidean norm, $\|\Pprob(X)\|_2 := \sqrt{\sum_x \Pprob(x)^2}$, does not satisfy the DPI. As an example consider a two-bit process with initial states $\Pprob(X_0) \ \Pprob_u$ and $\mathbb{R}(X_0)\ \Pprob_u$, where the second bit's state is the uniform distribution $\Pprob_u$. If the process simply discards the second bit, then the final Euclidean distance is simply $\Pprob(X_0) - \mathbb{R}(X_0)$. However, the initial Euclidean distance is exactly $\tfrac{1}{2} (\Pprob(X_0) - \mathbb{R}(X_0))$. Thus the class of functions that are contractive under the action of a stochastic matrix are typically good candidates to formulate DPIs.

The DPI plays an important role in information theory because it holds for two important metrics, the mutual information and the Kullback-Leibler divergence (also known as relative entropy). For a random variable, the Shannon entropy is defined as 
\begin{gather}\label{eq:shannon}
 H(X):= -\sum_x \Pprob(x) \log[\Pprob(x)].
\end{gather}
The mutual information between two random variables $X$ and $Y$, that posses a joint distribution $\Pprob(X,Y)$, is defined as
\begin{gather}\label{eq:mutual}
 H(X:Y):= H(X)+H(Y)-H(XY).
\end{gather}
Here, $H(X)$ is computed from the marginal distribution $\Pprob(X) = \sum_y \Pprob(X,Y=y)$; and $H(Y)$ is computed from the marginal distribution $\Pprob(Y) = \sum_x \Pprob(X=x,Y)$. The corresponding DPI then is 
\begin{gather}\label{eq:mutualDPI}
 H(X_0:Y_0) \ge H(X_t:Y_t),
\end{gather}
under the action of a stochastic matrix. For Markov processes we have a stronger inequality
\begin{gather}\label{eq:MarkovmutualDPI}
 H(X_s:Y_s) \ge H(X_t:Y_t),
\end{gather}
for all times $t\ge s$.

The relative entropy between two distributions $\Pprob(X)$ and $\Pprob'(X)$ is defined as
\begin{gather}\label{eq:RE}
 H[\Pprob(X)\| \mathbb{R}(X)] := -\sum_x \Pprob(x) \log\left[\frac{\mathbb{R}(x)}{\Pprob(x)}\right].
\end{gather}
Note that this is not a quantity that is symmetric in its arguments. The relative entropy is endowed with an operational meaning as the probability of confusion~\cite{vedral2002role}; that is, if one is promised $\mathbb{R}(X)$ but given $\Pprob(X)$ instead, then after $n$ samples the confusion probability is quantitatively given by
$\Pprob(X)$ for $\mathbb{R}(X)$ 
\begin{gather}
\text{Pr}_\text{conf} = \exp(-n H[\Pprob(X)\| \mathbb{R}(X)]).
\end{gather}
We will see later that a similar expression can be employed in the quantification of memory effects in quantum stochastic processes (see Sec.~\ref{sec::RelEnt}). The corresponding DPI here has the form
\begin{gather}\label{eq:REdpi}
 H[\Pprob(X_0)\| \mathbb{R}(X_0)] \ge H[\Pprob(X_t)\| \mathbb{R}(X_t)],
\end{gather}
under a stochastic transformation. For Markov processes, we get the stronger version
\begin{gather}\label{eq:MarkovREdpi}
 H[\Pprob(X_s)\| \mathbb{R}(X_s)] \ge H[\Pprob(X_t)\| \mathbb{R}(X_t)],
\end{gather}
that holds for all $t\ge s$.

The behavior of relative entropy and the related pseudo-metric in quantum and classical dynamics is an ongoing research effort~\cite{PhysRevE.99.012120, PhysRevE.88.012112, rivas_strong_2020}. The meaning of all of these DPIs for Markov processes is that the system is progressively loosing information as time marches forward. This clearly, has implication on our understanding of the second law of thermodynamics and the arrow of time. There are still other inequalities that are being discovered, e.g. see Ref.~\cite{capela_monogamy_2020} for the so-called monogamy inequality. For detailed coverage of DPI see~\cite{yeung1, yeung2}. Moreover, recently, researchers have employed the so-called \textit{entropy cone}~\cite{Janzing, cone} to infer causality in processes, which is closely related to many of our interests in this tutorial. However, for brevity, we do not go into these details here. Here, we merely aimed to emphasize that metrics that satisfy DPI can be used as a herald for non-Markovian behaviour based on two-time distributions only.

\subsubsection{Conditional mutual information}
\label{sec::CondMut}

Naturally, we can go further in the investigation of the connection of memory and correlation measures from information theory. While Markov processes, i.e., processes with finite Markov order $1$, satisfy the DPI, a general process with finite Markov order (introduced in Sec.~\ref{sec:clnm}) has vanishing \emph{conditional mutual information} (\textbf{CMI}), mirroring the fact that such a process is conditionally independent of past outcomes that lie further back than a certain memory length (Markov order) of $\ell$. 

For ease of notation, we will group the times $\{t_k,\dots, t_0\}$ on which the process at hand is defined into three segments: the history $H = \{ t_{1}, \hdots, t_{k-\ell-1} \}$, the memory $M = \{t_{k-\ell}, \hdots, t_{k-1} \}$ and the future $F = \{ t_k , \hdots, t_n\}$. With this, the CMI of a joint probability distribution on past, memory and future is defined as
\begin{gather}
\label{eqn::CMI}
H(F\!:\!H|M)
\! = \! H(F|M) \!+ \! H( H|M) \!- \! H(F,H|M),
\end{gather}
where the conditional entropy is given by
\begin{gather}\label{eq:condent}
H(X|Y)=H(XY)-H(Y). 
\end{gather}
This latter quantity is the entropy of the conditional distribution $\Pprob(X|Y)$ and has a clear interpretation in information theory as the number of bits $X$ must send to $Y$ so the latter party can reconstruct the full distribution. 

Consequently, $H(F:H|M)$ is a measure of the correlations that persist between $F$ and $H$, once the outcomes on $M$ are known. Intuitively then, for a process of Markov order $\ell$, $H(F\!:\!H|M)$ should vanish as soon as $M$ contains more than $\ell$ times. This can be shown by direct insertion. Recall that by means of (the general form of) Eq.~\eqref{eq:markovorder}, we can write $\Pprob(F|M,H) = \Pprob(F|M)$ for a process of Markov order $\ell\leq |M|$, implying
\begin{gather}\label{eq:cmarkovcondindep}
 \Pprob(F,H|M) = \Pprob(F|M) \Pprob(H|M).
\end{gather}
This means that $H(F,H|M) = H(F|M) + H( H|M)$ and, consequently, the CMI in Eq.~\eqref{eqn::CMI} vanishes. Importantly, the CMI only vanishes for processes with finite Markov order (and $|M|\geq \ell$), but not in general. If the CMI vanishes, then the future is decoupled from the entire history given knowledge of the memory. Vanishing CMI can thus be used as an alternative, equivalent definition of Markov order. 

Following this interpretation, the Markov order then encodes the complexity of the process at hand, as it is directly related to the number of past outcomes that need to be remembered to correctly predict future statistics; if there are $d$ different possible outcomes at each time, then no more than $d^\ell$ different sequences need to be remembered. While, in principle, $\ell$ may be large for many processes, they can often be approximated by processes with short Markov order. This is, in fact, the assumption that is made when real-life processes are modeled by means of Markovian Master equations. 

Additionally, complementing the conditional independence between history and future, processes with vanishing CMI admit a so-called `recovery map' $\mathcal{R}_{M \to FM}$ that allows one to deduce $\Pprob(F, M, H)$ from $\Pprob(M, H)$ by means of a map that only acts on $M$ (but \textit{not} on $H$). Indeed, we have 
\begin{gather}
\label{eqn::RecoveryMap}
\begin{split}
\Pprob(F, M, H) &= \Pprob(F|M)\Pprob(M, H) \\
&=: \mathcal{R}_{M \to FM}[\Pprob(M, H)],
\end{split}
\end{gather}
where we have added additional subscripts to clarify what variables the respective joint probability distributions act on. In spirit, the recovery map is analogous to the map $\Xi^{(1)}$ we discussed in Sec.~\ref{sec::HidMark} in the context of hidden Markov models, with the important difference that, here, the input and output spaces of $\mathcal{R}_{M \to FM}$ differ.

While seemingly trivial, the above equation states that the future statistics of a process with Markov order $\ell$ can be recovered by only looking at the memory block. Whenever the memory block one looks at is shorter than the Markov order, any recovery map only approximately yields the correct future statistics. Importantly, though, the approximation error is bounded by the CMI between $F$ and $H$~\cite{Fawzi2015, Sutter2016}, providing an operational interpretation of the CMI, as well as quantifiable reasoning for the memory truncation of non-Markovian processes. 

While the treatment of concepts used to detect and quantify memory we provide here is necessarily cursory, there are two simple overall points that will carry over to the quantum case. On the one hand, in a process without memory, the distinguishability between distributions cannot increase, a fact mirrored by the satisfaction of the DPI. Put more intuitively, in a process without memory, information is leaked into the environment but never the other way round, leading to a `wash-out' of distributions and a decrease in their distunguishability. On the other hand, memory is generally a question of conditional independence between outcomes in the future and the past. One way to make this concept manifest is by means of the CMI.

As we will see in Sec.~\ref{sec::QMarkOrd}, many of these properties will also apply in some form to quantum processes of finite Markov order, with the caveat that the question of memory length possesses a much more layered answer in the quantum case than it does in the classical one.

\subsection{(Some more) mathematical rigor}
\label{sec::RigII}

In this section we discussed master equation as a means to model aspects of stochastic processes in a fashion that is continuous in time. This point of view is somewhat at odds with the discrete examples and definitions we discussed in the previous sections. As promised above, we shall now define what we mean by a stochastic process in more rigorous terms, and thus give a concrete meaning to the probability distribution $\Pprob_{\Tset}$ when $|{\Tset}|$ is infinite. 

Before advancing, a brief remark is necessary to avoid potential confusion. In the literature, stochastic processes are generally defined in terms of random variables~\cite{feller_introduction_1968, BreuerPetruccione}, and above, we have already phrased some of our examples in terms of them. However, both in the previous examples, as well as those that follow, explicit reference to random variables is not a necessity, and all of the results we present can be phrased in terms of joint probabilities alone. Thus, foregoing the need for a rigorous introduction of random variables and trajectories thereof, we shall phrase our formal definition of stochastic processes in terms of probability distributions only. For all intents and purposes, though, there is no difference between our approach and the one generally found in the literature.

To obtain a definition of stochastic processes on infinite sets of times, we will define stochastic processes -- first for finitely many times, then for infinitely many -- in terms of probability spaces, which we introduced in Sec.~\ref{sec::rig1}. This can be done by merely extending their definition to sequences of measurement outcomes at (finitely many) multiple times, like, for example, the sequential tossing of a die (with or without memory) we discussed above.
\begin{definition*}[Classical stochastic process]
\label{def::ClassProc}
A stochastic process on times $\alpha\in\Tset_k$ with $|\Tset_k|=k<\infty$ is a triplet $(\Omega_{\Tset_k},\Scal_{\Tset_k}, \Pprob_{\Tset_k})$ of a sample space
 \begin{gather}
 \Omega_{\Tset_k} = \Cross_{\alpha \in \Tset_k} \Omega_\alpha,
 \end{gather}
a $\sigma$-algebra $\Scal_{\Tset_k}$ on $\Omega_{\Tset_k}$, and a probability measure $\Pprob_{\Tset_k}$ on $\Scal_{\Tset_k}$ with $\Pprob_{\Tset_k}(\Omega_{\Tset_k}) = 1$.
\end{definition*}
The symbol $\times$ denotes the Cartesian product for sets. Naturally, as already mentioned, the set $\Tset_k$ the stochastic process is defined on does not have to contain times, but could, as in the case of the die tossing, contain general labels of the observed outcomes. Each $\Omega_\alpha$ corresponds to a sample space at $t_\alpha$, and the probability measure $\Pprob_{\Tset_k}: \Scal_{\Tset_k} \rightarrow [0,1]$ maps any sequence of outcomes at times $\{t_\alpha\}_{\alpha\in \Tset_k}$ to its corresponding probability of being measured. A priori, this definition of stochastic processes is not concerned with the respective mechanism that leads to the probability measure $\Pprob_{\Tset_k}$; above, we have already seen several examples of how it emerges from the stochastic matrices we considered. However, as mentioned, once the full statistics $\Pprob_{\Tset_k}$ are known, all relevant stochastic matrices can be computed. Put differently, once $\Pprob_{\Tset_k}$ is known, there is no more information that can be learned about a classical process on $\Tset_k$.

We now formally define a stochastic process on sets of times $\Tset$, where $|\Tset|$ can be infinite. Using the mathematical machinery we introduced, this is surprisingly simple:
\begin{definition*}
\label{def::ClassProc1}
A stochastic process on times $\alpha \in\Tset$ is a triplet $(\Pprob_{\Tset},\Omega_{\Tset},\mathcal{S}_{\Tset})$ of a sample space
 \begin{gather}
 \Omega_{\Tset} = \Cross_{\alpha \in \Tset} \Omega_\alpha,
 \end{gather}
a $\sigma$-algebra $\mathcal{S}_{\Tset}$ on $\Omega_{\Tset}$, and a probability measure $\Pprob_{\Tset}$ on $\mathcal{S}_{\Tset}$ with $\Pprob_{\Tset}(\Omega_{\Tset}) = 1$.
\end{definition*}

While almost identical to the analogous definition for finitely many times, conceptually, there is a crucial difference between the two. Notably, $\Pprob_{\Tset}$ is not an experimentally reconstructable quantity unless $|\Tset|$ is finite. Additionally, here, we simply posit the $\sigma$-algebra $\mathcal{S}_{\Tset}$. However, generally, the explicit construction of this $\sigma$-algebra from scratch is not straightforward, and starting the description of a given stochastic process on times $\Tset$ from the construction of $\mathcal{S}_{\Tset}$ is a daunting task, which is why, for example, the modeling of Brownian motion processes does not follow this route. Nonetheless, we often implicitly assume the existence of an `underlying' process, given by $(\Pprob_{\Tset},\Omega_{\Tset},\mathcal{S}_{\Tset})$ when discussing, for example, Brownian motion on \textit{finite} sets of times. Connecting finite joint probability distributions to the concept of an underlying process is the main achievement of the Kolmogorov extension theorem, as we will lay out in detail below.

\section{Early Progress on Quantum Stochastic Processes}
\label{sec::EarlyProg}

Our goal in the present section, as well as the next section, will be to follow the narrative presented in the last two chapters to obtain a consistent description of \textit{quantum} stochastic processes. However, the subtle structure of quantum mechanics will generate technical and foundational problems that will challenge our attempts to generalize the theory of classical stochastic processes to the quantum domain. Nevertheless, it is instructive to understand the kernel of these problems before we present the natural generalization in the next section. Thus we begin with the elements of quantum stochastic processes that are widely accepted. It should be noted that we assume a certain level of mastery of quantum mechanics from the reader. Namely, statistical quantum states, generalized quantum measurements, composite systems, and unitary dynamics. We refer the readers unfamiliar with these standard elements of quantum theory to textbooks on quantum information theory, e.g.~\cite{NielsenBook, bengtsson_geometry_2007, wilde2013quantum}. However, for completeness, we briefly introduce some of these elements in this section.

The intersection of quantum mechanics and stochastic processes dates back to the inception of quantum theory. After all, a quantum measurement itself is a stochastic process. However, the term \textit{quantum stochastic process} means a lot more than that a quantum measurement has to be interpreted probabilistically. Perhaps, the von Neumann equation (also due to Landau) is the first instance where elements of the stochastic process come together with those of quantum mechanics. Here, the evolution of a (mixed) quantum state is written as a master equation, though this equation is fully deterministic. Nevertheless, a few years after the von Neumann equation, genuine phenomenological master equations appeared to explain atomic relaxations and particle decays~\cite{Pauli}. Later, further developments were made as necessitated, e.g., Jaynes introduced what is now known as a random unitary channel~\cite{jaynes1957}. 

Serious and formal studies of quantum stochastic processes began in the late 1950s and early 1960s. Two early discoveries were the exact non-Markovian master equation due to Nakajima and Zwanzig~\cite{Nakajima1958, Zwanzig1960} as well as the phenomenological study of the maser and laser~\cite{lamb_theory_1964, weidlich_coherence-properties_1965, weidlich_quantummechanical_1967I, weidlich_quantummechanical_1967II}. It took another decade for the derivation of the general form of Markovian master equations~\cite{sudarshangorini, lindblad75}. In the early 1960s, Sudarshan et al.~\cite{SudarshanMatthewsRau61, jordan_dynamical_1961} generalized the notion of the stochastic matrix to the quantum domain, which was again discovered in the early 1970s by Kraus~\cite{kraus_general_1971}. 

Here, in a sense, we follow the historic route by not directly fully generalizing classical stochastic processes to the quantum domain, but rather doing it piecewise, with an emphasis on the problems encountered along the way. We begin by introducing the basic elements of quantum theory and move to quantum stochastic matrices (also called quantum channels, quantum maps, dynamical maps), and discuss their properties and representations. This then lays the groundwork for a consistent description of quantum stochastic processes that allows one to incorporate genuine multi-time probabilities.

\subsection{Quantum statistical state}
\label{subsec::Qstatstate}
As with the classical case we begin with defining the notion of quantum statistical state. A (pure) quantum state $\ket{\psi}$ is a ray in a $d$-dimensional Hilbert space $\Hcal_{\stxt}$ (where we employ the subscript $\stxt$ for \textit{system}). Just like in the classical case, $d$ corresponds to the number of perfectly distinguishable outcomes. Any such pure state can be written in terms of a basis:
\begin{gather}\label{eq:purestate}
 \ket{\psi} = \sum_{s=1}^d c_s \ket{s},
\end{gather}
where $\{\ket{s}\}$ is an orthonormal basis, $c_s$ are complex number, and we assume $d<\infty$ throughout this article. Thus the quantum state is a complex vector, which is required to satisfy the property $\braket{\psi|\psi}=1$, implying $\sum_{s} |c_s|^2 = 1$. It may be tempting to think of $\ket{\psi}$ as the quantum generalization of the classical statistical state $\Pprob$. However, as mentioned, a state that is represented in the above form is pure, i.e., there is no uncertainty about what state the system is in. To account for potential ignorance, one introduces density matrices, which are better suited to fill the role of quantum statistical states.

Density matrices are written in the form
\begin{gather}\label{eq:densityconvex}
 \rho = \sum_{j=1}^n p_j \ket{\psi_j}\!\bra{\psi_j},
\end{gather}
which can be interpreted as an ensemble of pure quantum states $\{\ket{\psi_j}\}_{j=1}^n $ that are prepared with probabilities $p_j$ such that $\sum_{j=1}^n p_j=1$. Such a decomposition is also called a convex mixture. Naturally, pure states are special cases of density matrices, where $p_j=1$ for some $j$. In other words, density matrices represent our ignorance about which element of the ensemble or the exact pure quantum state we possess. It is important though, to add a qualifier to this statement: seemingly, Eq.~\eqref{eq:densityconvex} provides the `rule', by which the statistical quantum state at hand was prepared. However, this decomposition in terms of the pure state is neither unique nor do the states $\{\ket{\psi_j}\}$ that appear in it have to be orthogonal. For any non-pure density matrix, there are infinitely many ways of decomposing it as a convex mixture of pure states~\cite{schrodinger_probability_1936,gisin_quantum_1984, hughston_complete_1993}. This is in stark contrast to the classical case, where any probability vector can be \textit{uniquely} decomposed as a convex mixture of perfectly distinguishable `pure' states, i.e., events that happen with unit probability. 

For a $d$-dimensional system, the density matrix is a $d\times d$ square matrix (i.e., an element of the space $\Bcal(\Hcal)$ of bounded operators on the Hilbert space $\Hcal$)
\begin{gather}
 \rho = \sum_{r,s=1}^d \rho_{rs} \ket{r}\!\bra{s} \quad \mbox{and} \quad \rho \in \Bcal(\Hcal).
\end{gather}
Due to physical considerations, like the necessity for probabilities to be real, positive, and normalised, the density matrix must be
\begin{itemize}
 \item Hermitian $\rho_{rs} = \rho_{sr}^*$,
 \item positive semidefinite $\braket{x|\rho|x} \geq 0$ for all $\ket{x}$, and
 \item unit-trace $\sum_{r} \rho_{rr} = 1$.
\end{itemize}

Throughout, we will denote semidefiniteness by $\rho \geq 0$. As noted above, the density matrix is really the generalization of the classical probability distribution $\Pprob$. In fact, a density matrix that is diagonal in the computational basis is just a classical probability distribution. Conversely, the off-diagonal elements of a density matrix, known as coherences, make quantum mechanics non-commutative and are responsible for the interference effects that give quantum mechanics its wave-like nature. However, it is important to realize that a density matrix is like the single time probability distribution $\Pprob$, in the sense that it provides the probabilities for any conceivable measurement outcome at a single given time to occur. It will turn out that the key to the theory of quantum stochastic process lies in clearly defining a multi-time density matrix.

There are many interesting properties and distinct origins for the density matrix. While we have simply heuristically introduced it as an object that accounts for the uncertainty about what current state the system at hand is in, there are more rigorous ways to motivate it. One way to do so is, e.g., Gleason's theorem, which is grounded in the language of measure theory~\cite{gleason_measures_1975,busch_quantum_2003}, and, basically, derives density matrices as the most general statistical object that provides `answers' to all questions an experimenter can ask. 

Concerning its properties, a density matrix is pure if and only if $\rho^2 = \rho$. Any (non-pure) mixed quantum state $\rho_\stxt$, of the system $\stxt$, can be thought of as the marginal $\rho_\stxt = \tr_{S'}[\ketbra{\psi}_{\stxt\stxtp}{\psi}]$ of a bipartite pure quantum state $\ket{\psi}_{\stxt\stxtp}$, which must be entangled. This fact is known as quantum purification and it is an exceedingly important property that we will discuss in Sec.~\ref{sec::PurDil}. Of course, the same state $\rho_\stxt$ can also be thought as a proper mixture of an ensemble of quantum states on the space $\stxt$ alone. However, quantum mechanics does not differentiate between proper mixtures and improper mixtures, i.e., mixedness due to entanglement (see~\cite{mixtures} for a discussion of these different concepts of mixtures). As mentioned, mixtures are non-unique. The same holds true for purifications; for a given density matrix $\rho_\stxt$, there are infinitely many pure states that have it as a marginal.

Finally, let us say a few words about the mathematical structure of density matrices. Density matrices are elements of the vector space of $d\times d$ Hermitian matrices, which is $d^2$-dimensional. Consequently, akin to the decomposition of pure state in Eq.~\eqref{eq:purestate} in terms of an orthonormal basis, a density matrix can also be cast in terms of a fixed set of $d^2$ orthonormal basis operators:
\begin{gather}\label{eq:paulibasis}
 \rho = \sum_{k=1}^{d^2} r_k \hat\sigma_k,
\end{gather}
where we can choose different sets $\{\hat{\sigma}_k\}$ of basis matrices.\ftnt{We denote operator (and later superoperator) basis elements with hats.} They can, for example, be Hermitian observables (e.g., Pauli matrices plus the identity matrix), in which case $\{r_k\}$ are real numbers. Also, $\{\hat\sigma_k\}$ can be non-Hermitian elementary matrices, in which case $\{r_k\}$ are complex numbers; In both cases, we may have the matrix orthonormality condition $\tr[\hat\sigma_j \hat\sigma_k^\dag] = N \delta_{jk}$, with $N$ being a normalization constant. However, in neither case, the matrices $\{\hat\sigma_k\}$ correspond to physical states, as there is no set of $d^2$ orthogonal $d\times d$ quantum states, since a $d$-dimensional system can only have $d$ perfectly distinguishable states.

\begin{figure}
 \centering
 \includegraphics[width=0.5\linewidth]{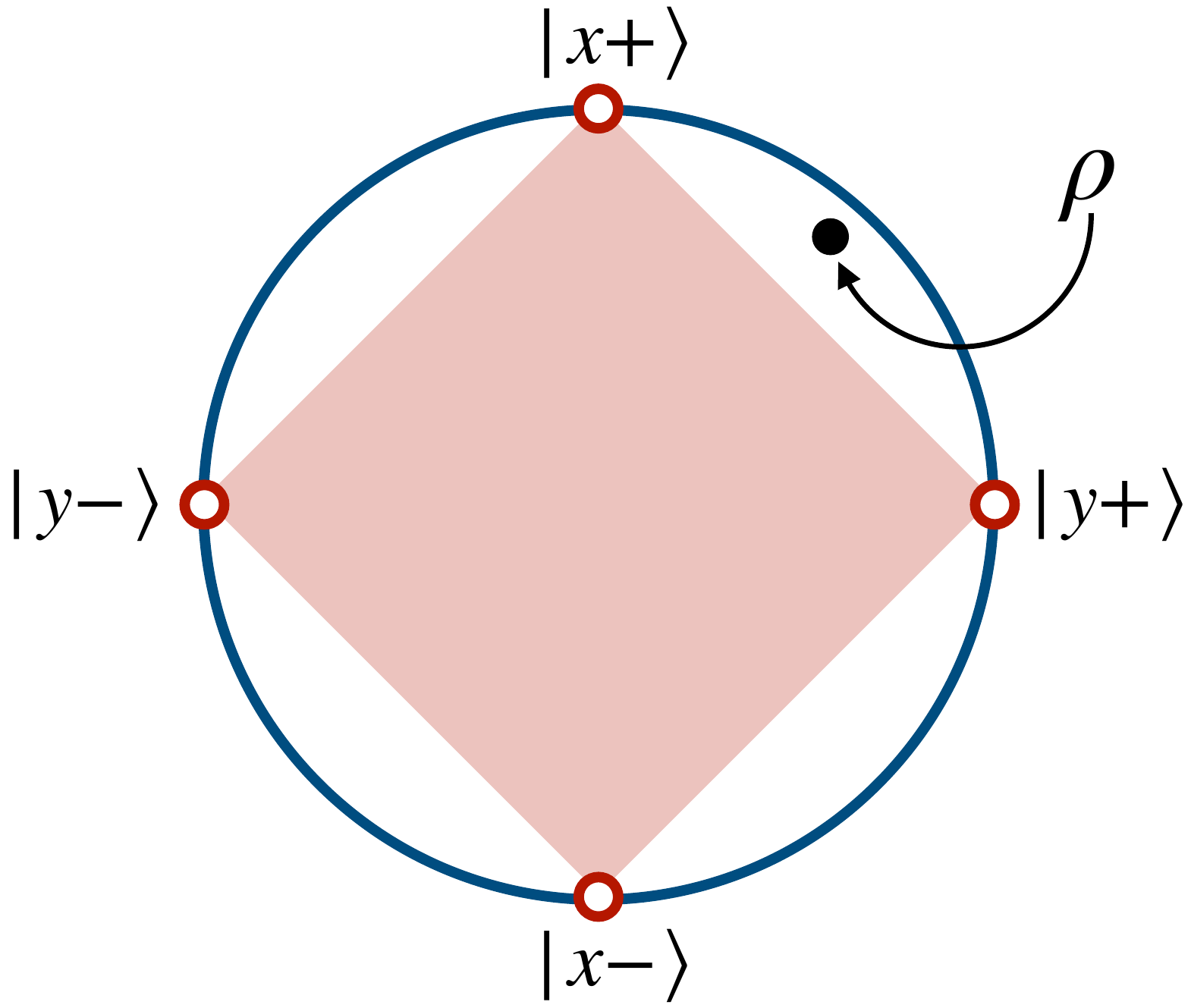}
 \caption{\textbf{Non-convex decomposition}. All states in the $x-y$ plane of the Bloch sphere, including the pure states, can be described by the basis state $\hat\varrho_1$, $\hat\varrho_2$, and $\hat\varrho_4$ in Eq.~\eqref{eq:statebasis}. However, only the states in shaded region will be convex mixtures of these basis states. Of course, no pure state can be expressed as a convex mixture.}
 \label{fig:convex}
\end{figure}

We can, however, drop the demand for orthonormality, and write \textit{any} density matrix as a linear sum of a fixed set of $d^2$ linearly independent density matrices $\{\hat{\varrho}_k\}$
\begin{gather}\label{eq:densitybasis}
 \rho = \sum_k q_k \hat{\varrho}_k.
\end{gather}
Here, $\{q_k\}$ will be real but generally not positive, see Figure~\ref{fig:convex}. This appears to be in contrast to Eq.~\eqref{eq:densityconvex}, where the density matrix is written as a convex mixture of physical states. The reason for this distinction is that in the last equation we have \textit{fixed} the basis operators $\{\hat{\varrho}_k\}$, which span the whole space of Hermitian matrices, and demand that any quantum state can be written as a linear combination of them, while in Eq.~\eqref{eq:densityconvex} the states $\{\ket{\psi_j}\}$ can be any quantum state, i.e., they would have to vary to represent different density matrices as convex mixtures. Understanding these distinctions will be crucial in order to grasp the pitfalls that lie before us as well as to overcome them.

\subsubsection{Decomposing quantum states}
\label{sec:statedecomp}
Let us illustrate the concept of quantum states with a concrete example for $d=2$, i.e., the qubit case. A generic state of one qubit can, for example, be written as
\begin{gather}
 \alpha = \frac{1}{2}\left(\hat\sigma_0 + a_1 \hat\sigma_1 + a_2 \hat\sigma_2 + a_3 \hat\sigma_3 \right)
\end{gather}
in terms of Pauli operators $\{\hat \sigma_1, \hat\sigma_2, \hat \sigma_3\}$ and the identity matrix $\hat \sigma_0$. The set $\{\hat \sigma_0, \hat \sigma_1, \hat \sigma_2, \hat \sigma_3\}$ forms an orthogonal basis of the space of $2\times 2$ Hermitian matrices, which implies $a_j \in \mathbbm{R}$, while positivity of $\alpha$ enforces $\sum a_j^2 \leq 1$. We can write the same state in terms of elementary matrices 
\begin{gather}
\alpha = \sum_{ij} e_{ij} \hat{\varepsilon}_{ij} \quad
\mbox{where} \quad
\hat{\varepsilon}_{ij} = \ketbra{i}{j},
\end{gather}
with complex coefficients $\{e_{00}, e_{01},
e_{10}, e_{11}\}$ being
\begin{gather}
\left\{\frac{1+a_3}{2}, \frac{a_1-ia_2}{2}, \frac{a_1+ia_2}{2}, \frac{1-a_3}{2} \right\}.
\end{gather}
The elementary matrices are non-Hermitian but orthonormal, i.e., $\tr[\hat{\varepsilon}_{ij} \hat{\varepsilon}_{kl}^\dag] = \delta_{ik}\delta_{jl}$.

These are, of course, two standard ways to represent a qubit state in terms of well-known orthonormal bases. On the other hand, we can expand the same state in terms of the following basis states
\begin{gather}\label{eq:statebasis}
\begin{split}
& \hat\varrho_1 = \ket{+x}\!\bra{+x}, \
 \hat\varrho_2 = \ket{+y}\!\bra{+y},\\
& \hat\varrho_3 = \ket{+z}\!\bra{+z}, \
 \hat\varrho_4 = \ket{-x}\!\bra{-x},
\end{split}
\end{gather}
where $\ket{\pm x}$, $\ket{\pm y}$, and $\ket{\pm z}$ are the eigenvectors of $\hat\sigma_1$, $\hat\sigma_2$, and $\hat\sigma_3$, respectively. With this, for any Hermitian matrix $\alpha$ we have $\alpha = \sum_k q_k \hat\varrho_k$. It is easy to see that the density matrices $\hat \varrho_k$ are Hermitian and linearly independent, but \textit{not} orthonormal. The real coefficients $\{q_1,q_2,q_3,q_4\}$ are obtained by following the inner product
\begin{gather}
\label{eqn::DualCoeff}
 q_k = \tr(\alpha \hat{D}_k^\dag) = \sum_j q_j \tr(\hat\varrho_j \hat{D}_k^\dag)
\end{gather}
where the set $\{\hat{D}_k\}$ is dual to the set of matrices in Eq.~\eqref{eq:statebasis} satisfying the condition $\tr(\hat\varrho_i \hat{D}_j^\dag) = \delta_{ij}$. We will see below that such dual matrices are a helpful tool for the experimental reconstruction of density matrices. See the Appendix in Refs.~\cite{modi_positivity_2012, arXiv:1708.00769} for a method for constructing the dual basis.

For example, for the set of density matrices in Eq.~\eqref{eq:statebasis} the dual set is
\begin{gather}\label{eq:dualbasis}
\begin{split}
& \hat{D}_1 = \frac{\hat\sigma_0 + \hat\sigma_1 -\hat\sigma_2 - \hat\sigma_3}{2}, \ \ \ 
 \hat{D}_2 = \hat\sigma_2,\\
& \hat{D}_3 = \hat\sigma_3, \ \ \ 
 \hat{D}_4 = \frac{\hat\sigma_0 - \hat\sigma_1 -\hat\sigma_2 - \hat\sigma_3}{2},
\end{split}
\end{gather}
Note that, even though the states $\hat \varrho_k$ are positive, this dual set $\{\hat D_k\}$ does not consist of positive matrices (all the duals of a set of Hermitian matrices are Hermitian, though~\cite{arXiv:1708.00769}). Nonetheless, it gives us the coefficient in Eq.~\eqref{eqn::DualCoeff} as
\begin{gather}
\left\{ \frac{1+a_1-a_2-a_3}{2}, a_2, a_3,
\frac{1-a_1-a_2-a_3}{2} \right\}.
\end{gather}

Interestingly, the dual set of a basis itself also forms a linear basis, and we can write any state $\alpha$ as
\begin{gather}\label{eq:statetom}
 \alpha = \sum_k p_k \hat{D}_k^\dag,
\end{gather}
where $p_k = \tr(\alpha\hat \varrho_k)$. Note that, if all basis elements $\hat \varrho_k$ are positive semidefinite, then $p_k\geq 0$, and we have $\sum_k p_k = 1$ if $\sum_k \hat \varrho_k = \ident$. This decomposition in particular lends itself nicely to experimental reconstruction of the state $\alpha$. Specifically, given many copies of $\alpha$, the value $\tr(\alpha\hat \varrho_k)$ is obtained by projecting $\alpha$ along directions $x,y,z$, i.e., measuring the observables $\hat \sigma_1, \hat \sigma_2,$ and $\hat \sigma_3$. The inner product $\tr(\alpha\hat \varrho_k)$ is then nothing more than a projective measurement along direction $k$ and $p_k$ is the probability of observing the respective outcomes. Importantly, as the duals $\{\hat D_k\}$ can be computed from the basis $\{\hat \varrho_k\}$, these probabilities then allow us to estimate the state via Eq.~\eqref{eq:statetom}. 

Intuitively, this procedure is not unlike the way in which one determines the classical state of a system; for example, in order to determine the bias of a coin, one flips it many times and records the respective outcome probabilities for heads and tails. The crucial difference is quantum mechanics is that one must measure in different directions to fully construct the state of interest. Algebraically, this fact is reflected by the dimension of the space of $d\times d$ Hermitian matrices, which is $d^2$ dimensional, thus, in order to fully determine a density matrix, one needs to know its overlap with $d^2$ linearly independent Hermitian matrices. If, however, one knows in which basis the state one aims to represent is diagonal -- as is the case in classical physics -- then the overlap with the $d$ projectors that make up its eigenbasis is sufficient.

The procedure to estimate a quantum state by measuring it is called \textit{quantum state tomography}~\cite{JModOpt.44.2455, PhysRevLett.113.190404, kalev}. There are many sophisticated methods to this nowadays, which we will only briefly touch on in this tutorial.

\subsubsection{Measuring quantum states: POVMs and dual sets}
\label{subsec::POVM}

As we have seen, a quantum state can be recronstructed experimentally, by measuring enough observables (above, the observables $\hat\sigma_1, \hat\sigma_2,$ and $\hat\sigma_3$ were used), and collecting the corresponding outcome probabilities. Performing pure projective measurement is not the only way in quantum mechanics to gather information about a state. More generally, a measurement is described by a positive operator valued measure (POVM), a collection $\Jcal=\{E_k\}_{k=1}^n$ of positive operators (here, matrices), that add up to the identity, i.e., $\sum_k E_k = \ident$ (we will comment on the physical realizability of POVMs below; for the moment, they can just be thought of as a natural generalization of projective measurements). Each $E_k$ corresponds to a possible measurement outcome, and the probability to observe said outcome is given by the Born rule: 
\begin{gather}
 \label{eqn::Born}
 p_k = \tr(\rho E_k).
\end{gather}
Projective measurements are then a special case of POVMs, where $E_k = \ketbra{k}{k}$ and $\{\ket{k}\}$ are the eigenstates of the measured observable. For example, when measuring the observable $\hat\sigma_3$, the corresponding POVM is given by $\Jcal=\{\ketbra{+z}{+z}, \ketbra{-z}{-z}\}$, and the respective probabilities are computed via $p_\pm = \braket{\pm z|\rho| \pm z} = \tr(\rho\ketbra{\pm z}{\pm z})$. 

A less trivial example on a qubit is the symmetric informationally complete (SIC) POVM~\cite{JModOpt.44.2455} $\Jcal = \{\hat{E}_k = \tfrac{1}{2} \ketbra{\phi_k}{\phi_k}\}_{k=1}^{4}$, where 
\begin{gather}
\begin{split}
 &\ket{\phi_1} = \ket{0}, \\
 &\ket{\phi_k} = \sqrt{\tfrac{1}{3}}\ket{0} + \sqrt{\tfrac{2}{3}}e^{i\tfrac{2(k-2)\pi}{3}}\ket{1} \ \text{for} \ k=2,3,4.
\end{split}
\end{gather}
While still pure (up to normalization), these POVM elements are not orthogonal. However, as they are linearly independent, they span the $d^2=4$-dimensional space of Hermitian qubit matrices, and every density matrix is fully characterized once the probabilities $p_k=\tr(\rho \hat{E}_k)$ are known. As this holds true in any dimension for POVMs consisting of $d^2$ linearly independent elements, such POVMs are called `informationally complete' (\textbf{IC}).\ftnt{It is easy to construct informationally complete POVMs, adding symmetric part is hard. For our purposes IC will be sufficient.} Importantly, using the ideas outlined above, an informationally complete POVM allows one to fully reconstruct density matrices. 

In short, to do so, one measures the system with an IC-POVM, whose operators $\{\hat{E}_k\}$ linearly span the matrix space of the system at hand. The POVM yields probabilities $\{p_k\}$, and the measurement operators $\{\hat{E}_k\}$ have a dual set $\{\hat{\Delta}_k\}$. The density matrix is then of the form (see also Eq.~\eqref{eq:statetom})
\begin{gather}
\label{eqn::stateDefPOVM}
 \rho = \sum_kp_k \hat{\Delta}_k^\dagger,
\end{gather}
which can be seen by direct insertion; the above state yields the correct probability with respect to each of the POVM elements $\hat{E}_k$. Concretely, we have 
\begin{gather}
    \tr(\rho \hat{E}_k) = \sum_\ell p_\ell \tr(\hat \Delta_\ell \hat{E}_k) = p_k\, ,
\end{gather}
where we have used $\tr(\hat \Delta_\ell \hat{E}_k) = \delta_{\ell k}$. As the POVM is informationally complete, this implies the state defined in Eq.~\eqref{eqn::stateDefPOVM} yields the correct probabilities with respect to every POVM. 

It remains to comment on the existence of IC-POVMs, and the physical realizability of POVMs in general, which, at first sight, appear to be a mere mathematical construction. Concerning the former, it is easy to see that there always exists a basis of $d^2\times d^2$ Hermitian matrices that only consists of positive elements. Choosing such a set $\{F_k\}_{k=1}^{d^2}$ of positive elements, one can set $F:= \sum_{k=1}^{d^2} F_k$. By construction, $F$ is positive semi-definite. Without much loss of generality, let us assume that $F$ is invertible (if it is not, then we could work with the pseudo-inverse in what follows). Then,
\begin{gather}
    \Jcal = \{E_k = F^{\scriptscriptstyle{-1/2}} F_k F^{\scriptscriptstyle{-1/2}}\}_{k=1}^{d^2}
\end{gather} constitutes a set of positive matrices that add up to $\ident$. To see that the matrices $\{E_k\}_{k=1}^{d^2}$ are linearly independent, let us assume the opposite and that, for example, $E_1$ can be written in terms of the remaining $E_k$, i.e., $E_1 = \sum_{k=2}^{d^2} a_k E_k$. Multiplying this expression from the left and the right by $F^{\scriptscriptstyle{1/2}}$ then yields $F_1 = \sum_{k=2}^{d^2} a_k F_k$, which contradicts the original assumption that the matrices $\{F_k\}_{k=1}^{d^2}$ are linearly independent. Consequently, the set $\{E_k\}_{k=1}^{d^2}$ is an IC POVM. More pragmatically, one could make one's life easier and sample $d^2-1$ positive matrices $\{E_k\}_{k=1}^{d^2-1}$ according to one's measure of choice. In general, these sampled matrices are linearly independent. Then, one chosses an $\alpha > 0$ such that $E_{d^{2}} := \ident - \sum_k\{E_k\}_{k=1}^{d^2-1} \geq 0$. With this, the set $\{E_{k}\}_{k=1}^{d^2}$ is a POVM by construction, and in general also informationally complete.

With respect to the latter, i.e., the physical realizability of POVMs, due to Neumark's theorem~\cite{neumark, peresQT, paulsen_completely_2003}, any POVM can be realized as a pure projective measurement in a higher-dimensional space, thus putting them on the same foundational footing as `normal' measurements in quantum mechanics. Without going any deeper into the theory of POVMs, let us emphasize the take-home message of the above sections: quantum states can be experimentally reconstructed in a very similar way as classical states, by simply collecting sufficient statistics; however, the number of necessary measurements is larger and their structure is richer. While this latter point seems innocuous, it actually lies at the heart of the problems one encounters when generalizing classical stochastic processes to the quantum realm, like the break-down of the KET; if all necessary measurements could be chosen to be diagonal in the same basis, then there would be no fundamental difference between classical and quantum processes.

\subsection{Quantum stochastic matrix}
\label{subsec::Qstochmat}

Our overarching aim is to generalize the notion of stochastic processes to quantum theory. Here, after having discussed quantum states and their experimental reconstruction in the previous section, we generalize the notion of classical stochastic matrices. In the classical case a stochastic matrix, in Eq.~\eqref{eq:oneprop}, is a mapping of a statistical state
from time $t_j$ to time $t_k$, i.e., $\Gamma_{(k:j)}: \Pprob(X_j) \mapsto \Pprob(X_k)$. As such, in clear analogy, we are looking for a mapping of the form $\Ecal_{(k:j)}: \rho(t_j) \mapsto \rho(t_k)$. While there are different representations of $\Ecal_{(k:j)}$ (see, for example, Ref.~\cite{arXiv:1708.00769}), we start with the one that most closely resembles the classical case, where a probability vector gets mapped to another probability vector by means of a matrix $\Gamma_{(k:j)}$. We have already argued that the density matrix is the quantum generalization of the classical probability distribution. Then, consider the following transformation that turns a density matrix into a vector: 
\begin{gather}\label{eq:vectorisation}
 \rho = \sum_{rs} \rho_{rs} \ket{r}\!\bra{s} \longleftrightarrow
 \kket{\rho}
 := \sum_{rs} \rho_{rs} \kket{rs},
\end{gather}
where we use the $\kket{\sbt \,}$ notation to emphasize that the vector originally stems from a matrix. This procedure is often called \emph{vectorization} of matrices, for details see Refs.~\cite{dariano_orthogonality_2000, havel03, bengtsson_geometry_2007, gilchrist_vectorization_2009}. 

Next, in clear analogy to Eq.~\eqref{eq:randomprocessdie}, we can define a matrix $\breve\Ecal$ that maps a density matrix $\rho$ (say, at time $t_j$) to another density matrix $\rho'$ (say, at time $t_k$); we have added the symbol $\breve{}$ to distinguish the map $\Ecal$ from its matrix representation $\breve \Ecal$. Using the above notation, this matrix can be expressed as
\begin{gather}
\label{eqn::OuterMatrix}
\breve\Ecal := \sum_{r's',rs} \breve\Ecal_{r's',rs}
\kket{r's'}\!\bbra{rs}
\end{gather}
and the action of $\Ecal$ can be written as
\begin{gather}
\label{eqn::ActionA}
\kket{\Ecal[\rho]} = \breve\Ecal \kket{\rho} = \sum_{r's'rs} \breve\Ecal_{r's',rs}\rho_{rs} \kket{r's'} = \kket{\rho'}
\end{gather}
Here, $\breve\Ecal$ is simply a matrix representing the map $\Ecal: \Bcal (\Hcal^{\inp}) \rightarrow \Bcal (\Hcal^{\out})$,\ftnt{One should think of the map as an abstract object, and should be distinguished from its representation, hence we have $\Ecal$ versus $\breve{\Ecal}$.} very much like the stochastic matrix, that maps the initial state to final state. For better book-keeping, we explicitly distinguish between the input (\inp) Hilbert space  $\Hcal^{\inp}$ and output (\out) Hilbert space $\Hcal^{\out}$ and denote the space of matrices on said spaces by $\Bcal(\Hcal^\texttt{x})$. While for the remainder of this tutorial, the dimensions of these two spaces generally agree, in general, the two are allowed to differ, and even in the case where they do not, it proves advantageous to keep track of the different spaces.

It was with the above intuition Sudarshan \textit{et al.} called $\breve\Ecal$ the quantum stochastic matrix~\cite{SudarshanMatthewsRau61}. In today's literature, it is often referred to as a quantum channel, quantum dynamical map, etc. Along with many names, it also has many representations. We will not go much into these details here (see Ref~\cite{arXiv:1708.00769} for further information). We will, however, briefly discuss some of its important properties. Note, that we stick to discrete level systems and do not touch the topics of Gaussian quantum information~\cite{arXiv:quant-ph/0505151, RevModPhys.84.621}.

\textbf{Amplitude damping channel.} Before that, let us  quickly provide an explicit examples of a quantum stochastic matrix. Consider a relaxation process that takes any input quantum state to the ground state. Such as process is, for example, described by the so-called amplitude damping channel
\begin{gather}
\label{eq:ADchannel}
 \breve\Ecal^{\text{AD}}_{(t:0)} = \begin{pmatrix} 
 1 & 0 & 0 & 1-p(t)\\
 0 & \sqrt{p(t)} & 0 & 0\\
 0 & 0 & \sqrt{p(t)} & 0\\
 0 & 0 & 0 & p(t)\\
 \end{pmatrix}.
\end{gather}
This matrix acts on a vectorized density matrix of a qubit, i.e., $\kket{\rho(0)} = [\rho_{00}, \rho_{01}, \rho_{10}, \rho_{11}]^\trps$ to yield $\kket{\rho(t)} = [\rho_{00}+(1-p(t))\rho_{11}, \sqrt{p(t)}\rho_{01}, \sqrt{p(t)}\rho_{10}, p(t)\rho_{11}]^\trps$. When $p(t) = \exp\{-\gamma t\}$, we get relaxation exponentially fast in time, and for $t\rightarrow \infty$, any input state will be mapped to $[1, 0, 0, 0]^\trps$. This example is very close in spirit of the classical example in Eq.~\eqref{eq:cldivproc}.

Here, already, it is easy to see that the matrices $\breve\Ecal$, unlike their classical counterparts, do not possess nice properties, like Hermiticity, or stochasticity (note that, for example, neither the rows not the columns of $\breve\Ecal^{\text{AD}}_{(t:0)}$ sum to unity). However, these shortcomings can be remedied in different representations of $\breve\Ecal$. Also note that, here, we actually have a family of quantum stochastic matrices  parameterized by time (for each time $t$ we have in general a different map). When we speak of a family of maps we will label them with subscript $(t:0)$. However, often the stochastic matrix only represents a mapping from the initial time to a final time. In such cases, we will omit the subscript and refer to the initial and final states as $\rho$ and $\rho'$, respectively.

\subsubsection{Linearity and tomography}
\label{sec::LinTom}
Having formally introduced quantum maps, the generalization of stochastic matrices in the classical case, it is now time to discuss the properties they should display. We begin with one of the most important features of quantum dynamics. The quantum stochastic map, like its classical counterpart, is a linear map:
\begin{gather}
 \Ecal [\alpha A + \beta B] = 
 \alpha \Ecal[A] + \beta \Ecal[B].
\end{gather}
This is straightforwardly clear for the specific case of the quantum stochastic matrix $\breve\Ecal$ because the vectorization of a density matrix itself is also a linear map, i.e., $\kket{A+B}=\kket{A} + \kket{B}$. Once this is done, the rest is just matrix transformations, which are linear.

The importance of linearity cannot be overstated; we will exploit this property over and over and, in particular, the linearity of quantum dynamics plays a crucial role in defining an unambiguous set of Markov conditions in quantum mechanics and, as we have seen, it is the fundamental ingredient in the experimental reconstruction of quantum objects. Due to linearity, a quantum channel is fully defined once its action on a set of linearly independent states is known. From a practical point of view, this is important for experimentally characterizing quantum dynamics by means of a procedure known as \textit{quantum process tomography}~\cite{JModOpt.44.2455, poyatos} (see, \textit{e.g.}, Refs.~\cite{ringbauer_quantum_2017, wood_thesis} for a more in-depth discussion). 

To see how this works out in practice, let us prepare a set of linearly independent input states, say $\{\hat{\varrho}_k\}$, and determine their corresponding outputs $\{\Ecal[\hat{\varrho}_k]\}$ by means of quantum state tomography (which we discussed above). The corresponding input-output relation then fully determines the action of the stochastic map on any density matrix
\begin{gather}
 \Ecal [\rho] = \sum_k q_k \Ecal[\hat{\varrho}_k],
\end{gather}
where we have used Eq.~\eqref{eq:densitybasis}, i.e, $\rho = \sum_k q_k \hat \varrho_k$. The above equation highlights that, once the output states for a basis of input states are known, the action of the entire map is determined.

\begin{figure}
 \centering
 \includegraphics[width=0.98\linewidth]{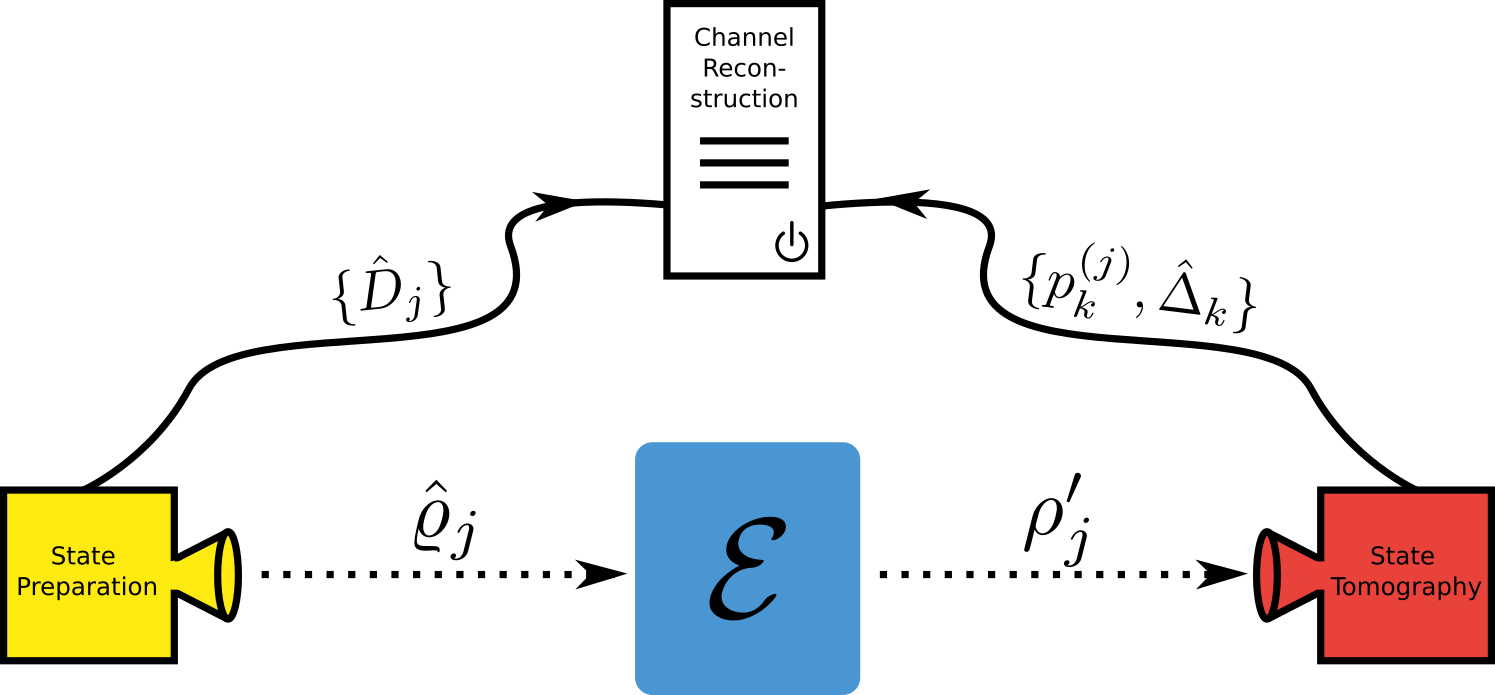}
 \caption{\textbf{Quantum process tomography.} Any quantum channel $\Ecal$ can be reconstructed by preparing a basis of input states and measuring the corresponding output states states $\rho_j' = \Ecal[\hat \varrho_j]$ with an informationally complete POVM. The corresponding duals $\{\hat D_j\}$, $\{\hat \Delta_k\}$ and the outcome probabilities then allow for the reconstruction of $\Ecal$ according to Eq.~\eqref{eqn::TomoEq}.} 
 \label{fig::ChannelTomo}
\end{figure}

Using ideas akin to the aforementioned tomography of quantum states we can also directly use linearity and dual sets to reconstruct the matrix $\breve\Ecal$. Above, we saw that a quantum state is fully determined, once the probabilities for an informationally complete POVM are known. In the same vein, a quantum map is fully determined, once the output states for a basis $\{\hat\varrho_j\}_{j=1}^{d^2}$ of input states are known. Concretely, setting $\rho_j' = \Ecal[\hat\varrho_j]$, and denoting the dual set of $\{\hat\varrho_j\}_{j=1}^{d^2}$ by $\{\hat D_k\}_{k=1}^{d^2}$, we have
\begin{gather}
\label{eqn::TomoVec}
 \breve\Ecal = \sum_{j=1}^{d^2} \kketbbra{\rho_j'}{\hat D_j}.
\end{gather}
Indeed, it is easy to see that $\bbrakket{A}{B} = \tr(A^\dagger B)$, implying that, with the above definition, we have $\breve\Ecal\kket{\hat \varrho_j} = \kket{\rho'_j}$ for all basis elements. Due to linearity, this implies that $\breve\Ecal$ yields the correct output state for \textit{any} input state. Measuring the output states for a basis of input states is thus sufficient to perform \textit{process tomography}. Specifically, if the output states $\rho_j'$ are measured by means of an informationally complete POVM $\{E_k\}$ with corresponding dual set $\{\Delta_k\}_{k=1}^{d^2}$, then $\rho_j' = \sum_k p_k^{(j)} \hat \Delta_k^\dagger$, with $p_k^{(j)} = \tr(\hat \varrho_j E_k)$ and Eq.~\eqref{eqn::TomoVec} reads 
\begin{gather}
\label{eqn::TomoEq}
 \breve\Ecal = \sum_{j,k=1}^{d^2} p_k^{(j)}\kketbbra{\hat \Delta_k}{\hat D_j}.
\end{gather}
As the experimenter controls the states they prepare at each run (and, as such, the dual $\hat D_j$ corresponding to each run), as well as the POVM they use, determining the probabilities $p_j^{(j)}$ thus enables the reconstruction of $\breve \Ecal$ (see Figure~\ref{fig::ChannelTomo} for a graphical representation). While we have not discussed the reconstruction of classical stochastic maps in detail, it is clear that it works in exactly the same vein. The above arguments only hinge on linearity as their main ingredient, implying that, analogous to the quantum case, a classical stochastic matrix $\Gamma_{(t:0)}$ is determined once the resulting output distributions $\Gamma_{(t:0)}[\hat \Pprob_j]$ for a basis of input distributions $\hat \Pprob_j$ are known.

\subsubsection{Complete positivity and trace preservation}
\label{sec::CPTP}
While linearity is a crucial property of quantum channels, it is naturally not the only pertinent one. A classical stochastic matrix maps probability vectors to probability vectors. As such, it is \textit{positive}, in the sense that it maps any vector with positive semi-definite entries to another positive semi-definite vector. In the same vein, quantum channels need to be positive, as they have to map all density matrices to proper density matrices, i.e., positive semi-definite matrices to positive semi-definite matrices. 

One crucial difference between classical stochastic maps and their quantum generalization is the requirement of \textit{complete} positivity. A positive stochastic matrix is guaranteed to map probabilities into probabilities even if it acts non-trivially only on a subpart (implying that only some but not all degrees of freedom undergo the stochastic process at hand), i.e., 
\begin{gather}
 \Gamma_A \Pprob_A = \mathbb{R}_A \geq 0 \Leftrightarrow (\Gamma_A \otimes \openone_B) \Pprob_{AB} = \mathbb{R}_{AB} \geq 0,
\end{gather}
for all $\Pprob_A$ and $\Pprob_{AB}$ where $A$ and $B$ are two different spaces and $\openone_B$ is the identity process on $B$. Here, $\Pprob \geq 0$ means that all the entries of the vector are positive semi-definite, and we have given all objects additional subscripts to denote the spaces they act/live on.

The same is not true in quantum mechanics. Namely, there are maps that take all density matrices to density matrices on a subspace, but their action on a larger space fails to map density matrices to density matrices, i.e.,
\begin{gather}
 \Ecal_A [\rho_A] \ge 0 \qquad \mbox{but} \qquad \Ecal_A \otimes \mathcal{I}_B [\rho_{AB}] \ngeq 0,
\end{gather}
where $\mathcal{I}_B$ is the identity map on the system $B$, i.e., $\mathcal{I}_B[\rho_B] = \rho_B$ for all $\rho_B$, and $\rho \geq 0$ means that all eigenvalues of $\rho$ are positive semidefinite. These maps are called positive maps, and they play an important role in the theory of entanglement witnesses~\cite{guhne_entanglement_2009, friis_entanglement_2019}. The most prominent of a positive, but not completely positive map is the transposition map $\rho \rightarrow \rho^\trps$. It is easy to show that positivity breaks only when the map $\Ecal$ acts on an entangled bipartite state (which is why positivity and complete positivity coincide in the classical case). Of course, giving up positivity of probability is not physical, and as such a positive map that is not also positive when acting on a part of a state is not physical.

One thus demands that physical maps must take \textit{all} density matrices to density matrices, even when only acting non-trivially on a part of them, i.e., 
\begin{gather}
 \Ecal_A \otimes \mathcal{I}_B [\rho_{AB}] \geq 0 \qquad \forall \ \rho_{AB} \geq 0.
\end{gather}
Maps, for which this is true for the arbitrary size of the system $B$ are called completely positive (\textbf{CP}) maps~\cite{alicki_semi_1987} and are the only maps we will consider throughout this tutorial (for a discussion of non-completely positive maps and their potential physical relevance -- or lack thereof -- see, for example, Refs.~\cite{pechukas, Alicki95,pechukas2,jordan:052110, StelmachovicBuzek01,arXiv:1708.00769}).

In addition to preserving positivity, i.e., preserving the positivity of probabilities, quantum maps also must preserve the trace of the state $\rho$, which amounts to preserving the normalization of probabilities. This is the natural generalization of the requirement on stochastic matrices that their columns sum up to $1$. Consequently, for a quantum channel, we demand that it satisfies
\begin{gather}
 \tr(\Ecal[\rho]) = \tr(\rho) \qquad \forall \rho.
\end{gather}
If a map $\Ecal_A$ is trace-preserving, then so is $\Ecal_A \otimes \mathcal{I}_B$. We will refer to completely positive maps that are also trace-preserving as \textbf{CPTP} maps, or quantum channels. Importantly, while the physicality of non-completely positive maps is questionable, we will frequently encounter maps that are CP, but only trace non-increasing instead of trace-preserving. Such maps are the natural generalizations of POVM elements and will play a crucial role when modeling quantum stochastic processes. Together, linearity, complete positivity, and trace preservation fully determine the set of admissible quantum channels.

\subsubsection{Representations}
\label{subsec::Representation}

In the classical case, stochastic matrices map vectors to vectors, and are thus naturally expressed in terms of matrices. In contrast, here, quantum channels map density matrices to density matrices, raising the question of how to represent their action for concrete calculations. We will not discuss the details of different representations here in much depth and only provide a rather basic introduction; we refer the reader to Ref.~\cite{zyczkowski_duality_2004, bengtsson_geometry_2007, arXiv:1708.00769} for further reading. 

Above, we have already seen the matrix representation $\breve\Ecal$ of the quantum stochastic map in Eq.~\eqref{eqn::OuterMatrix}. This representation is rather useful for numerical purposes, as it allows one to write the action of the map $\Ecal$ as a matrix multiplication. However, it is not straightforward to see how complete positivity and trace preservation enter into the properties of $\breve\Ecal$. When we add these two properties to the mix, there are two other important and useful representations that prove more insightful. First is the so-called Kraus representation of completely positive maps: 
\begin{gather}
\label{eqn::Kraus}
 \Ecal[\rho]
 = \sum_{j} K_j \, \rho \, K^\dag_j , 
\end{gather}
where the linear operators $\{K_j\}$ are called Kraus operators~\cite{kraus_general_1971,kraus_states_1983} (though, this form was first discovered in~\cite{SudarshanMatthewsRau61}). For the case of input and output spaces of the same size, Kraus operators are simply $d\times d$ matrices, hence are just operators on the Hilbert space. For this reason $\Ecal$ is often called a superoperator, i.e, an operator on operators. Denoting -- as above -- the space of matrices on a Hilbert space $\Hcal$ by $\Bcal(\Hcal)$, it can be shown that a map $\Ecal:\Bcal(\Hcal^\inp) \rightarrow \Bcal(\Hcal^\out)$ is CP iff it can be written in the Kraus form~\eqref{eqn::Kraus} for some $d\times d$ matrices $\{K_j\}$. For the `if' part, note that for any $\rho_{AB} \geq 0$, we have 
\begin{gather}
    \sum_{j} (K_j \otimes \ident_B) \rho_{AB} (K_j^\dagger \otimes \ident_B) = \sum_{j} M_j M_j^\dagger,
\end{gather} 
where $M_j := (K_j \otimes \ident_B) \sqrt{\rho_{AB}}$. Since any matrix that can be written as $M_jM_j^\dagger$ is positive, we see that a map $\Ecal$ that allows for a Kraus form maps positive matrices of any dimension onto positive matrices, making $\Ecal$ completely positive. The `only if' part is discussed below after introducing a second important representation of quantum channels.

Finally, the CP map $\Ecal$ is trace-preserving, iff the Kraus operators satisfy $\sum_j K_{j}^\dagger K_j = \ident$, which can be seen directly from $\tr(\sum_{j}K_j \rho K_j^\dagger) = \tr(\sum_{j}K_j^\dagger K_j \rho)$.

\begin{figure}[t]
 \centering
 \includegraphics[width=0.7\linewidth]{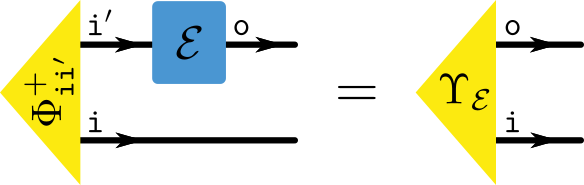}
 \caption{\textbf{Choi-Jamio{\l}kowski isomorphism.} A map $\Ecal:\Bcal(\Hcal^\inp) \rightarrow \Bcal(\Hcal^\out)$ can be mapped to a matrix $\Upsilon_\Ecal \in \Bcal(\Hcal^\out \otimes \Hcal^\inp)$ by letting it act on one half on a maximally entangled state. Note that, for ease of notation, here, we let $\Ecal$ act on $\Bcal(\Hcal_{\inp'})$, such that $\Upsilon_\Ecal\in \Bcal(\Hcal^\out \otimes \Hcal^\inp)$. As $\Hcal^{\inp'} \cong \Hcal^{\inp}$, this is merely a relabeling and not of conceptual relevance.}
 \label{fig::Choi}
\end{figure}

\textbf{Depolarizing channel.} Let us consider a concrete example of the Kraus representation. A common quantum map that one encounters in this representation is the depolarizing channel on qubits:
\begin{gather}
\label{eqn::DP}
 \Ecal^{\text{DP}}[\rho] = \sum_{j=0}^{3} p_j \, \sigma_j \, \rho \, \sigma_j \quad \mbox{with} \quad p_j \ge 0, \ \sum_j p_j = 1,
\end{gather}
where $\{\sigma_j\}$ are the Pauli operators. This map is an example of a random unitary channel~\cite{wolfunital}, i.e., it is a probabilistic mixture of unitary maps. When the $p_j$ are uniform the image of this map is the maximally mixed state for all input states. It is straightforward to see that the above map is indeed CPTP, as it can be written in terms of the Kraus operators $\{K_j=\sqrt{p_j}\sigma_j\}$, and we have $\sum_j{K_j^\dagger K_j} = \sum_jp_j \sigma_j \sigma_j = \ident$.

Another important representation that nicely incorporates the properties of complete positivity and trace preservation is that in terms of so-called Choi matrices. For this representation of $\Ecal$, consider its action on one part of an (unnormalized) maximally entangled state $\ket{\Phi^+} := \sum_{k=1}^{d^\inp} \ket{kk}$:
\begin{gather}\label{eqn::Choi}
\begin{split}
\Upsilon_{\Ecal} :=& \Ecal \otimes \Ical \left[\ketbra{\Phi^+}{\Phi^+}\right] \\=& \sum_{k,l=1}^{d^\inp} \Ecal\left[\ketbra{k}{l}\right]\otimes \ketbra{k}{l},
\end{split}
\end{gather}
where $\{{\ket{k}}\}$ is an orthonormal basis of $\Hcal^{\inp}$. See Figure~\ref{fig::Choi} for a graphical depiction. The resultant matrix $\Upsilon_{\Ecal} \in \Bcal(\Hcal^\out \otimes \Hcal^\inp)$ is isomorphic to the quantum map $\Ecal$. This can easily be seen by noting that in the last equation $\Ecal$ is acting on a complete linear basis of matrices, i.e., the elementary matrices $\{\hat\varepsilon_{kl}:=\ket{k}\!\bra{l}\}$. Consequently, $\Upsilon_\Ecal$ contains all information about the action of $\Ecal$. In  principle, instead of $\Phi^+$, any bipartite vector with full Schmidt rank could be used for this isomorphism~\cite{dariano_imprinting_2003}, but the definition we use here is the one encountered most frequently in the literature. In the form of~\eqref{eqn::Choi} it is known as the \emph{Choi-Jamio{\l}kowski isomorphism} (\textbf{CJI})~\cite{pillis_linear_1967,jamiolkowski_linear_1972, choi75}. It allows one to map linear maps, $\Ecal: \Bcal (\Hcal^{\inp}) \rightarrow \Bcal(\Hcal^{\out})$ to matrices $\Upsilon_{\Ecal} \in \Bcal(\Hcal^{\out})\otimes \Bcal(\Hcal^{\inp})$.

Usually, $\Upsilon_{\Ecal}$ is called the \emph{Choi matrix} or \emph{Choi state} of the map $\Ecal$. We will mostly refer to it by the latter. Given $\Upsilon_{\Ecal}$, the action of $\Ecal$ can be written as 
\begin{gather}
\label{eqn::ChoiAction}
\Ecal[\rho] = \tr_\inp\left[\left(\openone^\out \otimes \rho^\trps\right) \Upsilon_\Ecal \right],
\end{gather}
where $\tr_\inp$ is the trace over the input space $\Hcal^\inp$ and $\ident^\out$ denotes the identity matrix on $\Hcal^\out$. The validity of \eqref{eqn::ChoiAction} can be seen by direct insertion of~\eqref{eqn::Choi}:
\begin{gather}
\begin{split}
    \tr_\inp\left[\left(\openone^\out \!\otimes\! \rho^\trps\right) \Upsilon_\Ecal \right] 
    =& \sum_{k,\ell = 1}^{d^\inp} \braket{\ell|\rho^\trps|k} \Ecal\left[\ketbra{k}{l}\right] \\
    =&\sum_{k,\ell=1}^{d^\inp} \Ecal\left[\braket{k|\rho|\ell} \ketbra{k}{\ell}\right] \\
    =& \Ecal[\rho],
\end{split}
\end{gather}
where we have used the linearity of $\Ecal$. A related decomposition of the Choi state is
\begin{gather}\label{eq:tomrepmap}
\Upsilon_\Ecal = \sum_k \rho'_j \otimes \hat{D}^*_j, 
\end{gather}
where $\{\rho_j' = \Ecal[\hat\rho_j]\}$ are the output states corresponding to a basis of input states. The above equation is simply a reshuffling of Eq.~\eqref{eqn::TomoVec} from vectors to matrices. As was the case for Eq.~\eqref{eqn::TomoVec}, its validity can be checked by realizing that the Choi state $\Upsilon_\Ecal$ yields the correct output state for a full basis of input states, which can be seen by direct insertion of Eq.~\eqref{eq:tomrepmap} into~\eqref{eqn::ChoiAction}.

For quantum maps, $\Upsilon_\Ecal$ has particularly nice properties. Complete positivity of $\Ecal$ is equivalent to $\Upsilon_\Ecal \geq 0$, and it is straightforward to deduce from Eq.~\eqref{eqn::ChoiAction} that $\Ecal$ is trace-preserving iff $\tr_{\out}(\Upsilon_\Ecal) = \ident^\inp$ (see below for a quick justification of these statements). These properties are much more transparent, and easier to work with than, for example, the properties that make $\breve\Ecal$ a matrix corresponding to a CPTP map. Additionally, Eq.~\eqref{eqn::ChoiAction} allows one to directly relate the representation of $\Ecal$ in terms of Kraus operators to the Choi state $\Upsilon_\Ecal$, and, in particular, the minimal number of required Kraus operators to  the rank of $\Upsilon_\Ecal$. Specifically, in terms of its eigenbasis, $\Upsilon_\Ecal$ can be written as $\Upsilon_\Ecal = \sum_{\alpha=j}^r \lambda_j \ketbra{\Phi_j}{\Phi_j}$, where $r = \mathrm{rank}(\Upsilon_\Ecal)$ and $\lambda_j \geq 0$. Inserting this into Eq.~\eqref{eqn::ChoiAction}, we obtain
\begin{align}
\notag    \Ecal[\rho] \!&=\! \sum_{j=1}^r \! \left(\! \sqrt{\lambda_j} \! \sum_{\alpha = 1}^{d^\inp} \! \braket{\alpha|\Phi_j}\!\bra{\alpha} \! \right) \rho \left(\! \sqrt{\lambda_j} \! \sum_{\beta=1}^{d^\inp} \! \ket{\beta}\braket{\Phi_j|\beta}\! \right)\\
\label{eqn::ChoitoKraus}
    &=: \sum_{j=1}^r K_j \rho K_j^\dagger,
\end{align}
where $\{\ket{\alpha/\beta}\}$ is a basis of $\Hcal^\inp$. The above equation provides a Kraus representation of $\Ecal$ with the minimal number of Kraus operators (for more details on this connection between Choi matrices and Kraus operators, see, for example, Ref.~\cite{verstraete_quantum_2002}). Eq.~\eqref{eqn::ChoitoKraus} directly allows us to provide the missing properties of the Kraus and Choi representations that we alluded to above. Firstly, if $\Ecal$ is CP, then naturally, $\Upsilon_\Ecal$ is positive, which can be seen from its definition in Eq.~\eqref{eqn::Choi}. On the other hand, if $\Upsilon_\Ecal$ is positive, then Eq.~\eqref{eqn::ChoitoKraus} tells us that it leads to a Kraus form, implying that the corresponding map $\Ecal$ is completely positive. In addition, this means that any completely positive map admits a Kraus decomposition, a claim we made several paragraphs above. Finally, from Eq.~\eqref{eqn::ChoiAction} we see directly that $\tr_\out(\Upsilon_\Ecal) = \ident^\inp$ holds for all for trace preserving maps $\Ecal$.
Naturally, \textit{all} representations of quantum maps can be transformed into one another; details how this is done can be found, for example, in Refs.~\cite{wood_tensor_2011, arXiv:1708.00769}; however, for our purposes, it will prove very advantageous to predominantly use Choi states.

\textbf{Depolarizing Channel.} For concreteness, let us consider the above case of the depolarizing channel $\Ecal^{DP}$ and provide its Choi state. Inserting Eq.~\eqref{eqn::DP} into Eq.~\eqref{eqn::Choi}, we obtain
\begin{gather}
 \Upsilon_{\Ecal}^{DP} = \begin{pmatrix} 
 p_0 + p_3 & 0 & 0 & p_0-p_3\\
 0 & p_1+p_2 & p_1-p_2 & 0\\
 0 & p_1-p_2 & p_1+p_2 & 0\\
 p_0-p_3 & 0 & 0 & p_0 + p_3\\
 \end{pmatrix}. 
\end{gather}
The resulting matrix $\Upsilon_{\Ecal}^{DP}$ is positive semidefinite (with corresponding eigenvalues $\{2p_0,2p_1,2p_2,2p_3\}$), satisfies $\tr_\out\Upsilon_{\Ecal}^{DP} = \ident^\inp$, and $\tr \Upsilon_{\Ecal}^{DP} = 2 = d^\inp$.

Besides its appealing mathematical properties, the CJI is also of experimental importance. Given that a (normalized) maximally entangled state can be created in practice, the CJI enables another way of reconstructing a representation of the map $\Ecal$; letting it act on one half of a maximally entangled state and reconstructing the resulting state via state tomography directly yields $\Upsilon_\Ecal$. While this so-called \emph{ancilla-assisted process tomography}~\cite{PhysRevLett.90.193601, PhysRevLett.86.4195} requires the same number of measurements as the input-output procedure, it can be -- depending on the experimental situation -- easier to implement in the laboratory.

Additionally, since they simply correspond to quantum states with an additional trace condition, Choi states straightforwardly allow for the analysis of correlations in time -- in the same way as quantum states do in space. Consequently, below, when analyzing pertinent properties of quantum stochastic processes, like, for example, quantum Markov order, many of the results will be most easily be phrased in terms of Choi states and we will make ample use of them.

\begin{figure}[t]
\centering
\includegraphics[width=0.9\linewidth]{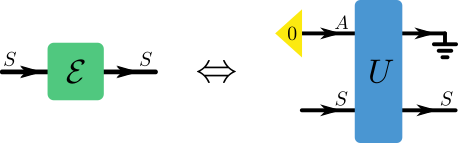}
\caption{
\textbf{Stinespring Dilation.} Any CPTP map on the system $S$ can be represented in terms of a unitary on a larger space $SA$ and final discarding of the additional degrees of freedom. The corresponding unitary can be computed by means of Eq.~\eqref{eqn::Unitary_Stine}. Here, for simplicity, we drop the explicit distinction between input and output spaces we normally employ.}
\label{fig::InputOutput}
\end{figure}

\subsubsection{Purification and Dilation}
\label{sec::PurDil}
At this point, after having pinned down the properties of statistical objects in quantum mechanics, it is worth taking a short detour and comment on the origin of stochasticity in the quantum case, and how it differs from the classical one. Importantly, in quantum mechanics any mixed state $\rho_{\stxt}$ can be thought of as the marginal of a pure state $\ket{\Psi}_{\stxt\stxtp}$ in a higher dimensional space that. I.e., for any $\rho_{\stxt}$, there exists a pure state $\ket{\Psi}_{\stxt\stxtp}$ such that $\tr_{\stxtp} (\ket{\Psi}_{\stxt\stxtp} \! \bra{\Psi}) = \rho_\stxt$. The state $\ket{\Psi}_{\stxt\stxtp}$ is then called a \textit{purification} of $\rho_\stxt$. The mixedness of a quantum state can thus \textit{always} be considered as stemming from ignorance about about additional degrees of freedom. This is in contrast to classical physics, which is not endowed with a purification principle.

To show that such a purification always exists, recall that 
 any mixed state $\rho_\stxt$ is diagonal in its eigenbasis $\{\ket{r}_\stxt\}$, i.e., 
\begin{gather}
 \rho_\stxt = \sum_{r} \lambda_r \ket{r}_\stxt \! \bra{r}, \ \text{with} \ \ \lambda_r\geq 0 \ \ \text{and} \ \ \sum_r \lambda_r = 1.
\end{gather}
This state can, for example, be purified by 
\begin{gather}
 \ket{\Psi}_{\stxt \stxtp} = \sum_r \sqrt{\lambda_r}\ket{r}_S\ket{r}_{\stxtp}.
\end{gather}
More generally, as a consequence of the Schmidt decomposition, \textit{any} state $\ket{\Psi}_{\stxt \stxtp'}$ that purifies $\rho_\stxt$ is of the form $\ket{\Psi}_{\stxt \stxtp'} = \sum_r \sqrt{\lambda_r} \ket{r}_\stxt W (\ket{r}_{\stxtp})$, where $W$ is an isometry from space $\stxtp$ to $\stxtp'$. Importantly, $\ket{\Psi}_{\stxt\stxtp}$ is entangled between $\stxt$ and $\stxtp$ as soon as $\rho_\stxt$ is mixed, i.e., as soon as $\lambda_r<1$ for all $r$. 

As entangled states lie outside of what can be captured by classical theories, classical mixed states do not admit a purification in the above sense -- at least not one that lies within the framework of classical physics. Randomness in classical physics can thus not be considered as stemming from ignorance of parts of a pure state in a higher-dimensional space, but it has to be inserted manually into the theory. On the other hand, \textit{any} quantum state can be purified within quantum mechanics, and thus randomness can always be understood as ignorance about extra degrees of freedom. 

\textbf{Purification example.} To provide an explicit example, consider the purification of a maximally mixed state on a $d$-dimensional system $\rho_\stxt = \tfrac{1}{d}\sum_{i=r}^{d} \ket{r}_\stxt \! \bra{r}$. Following the above reasoning, this state is, for example, purified by $\ket{\Phi^+}_{\stxt\stxtp} := \tfrac{1}{\sqrt{d}} \sum_{i=1}^d \ket{r}_\stxt\ket{r}_{\stxtp}$, which is the maximally entangled state. 

Remarkably the purification principle also holds for dynamics, i.e., \textit{any} quantum channel can be understood as stemming from a unitary interaction with an ancillary system, while this does not hold true for classical dynamics. This former statement can be most easily seen by direct construction. As we have seen, quantum channels can be represented in terms of their Kraus operators as
\begin{gather}
 \Ecal[\rho_\stxt] = \sum_{j} K_j \rho_\stxt K^{\dagger}_j, \quad \text{where} \quad \sum_{j} K_j^{\dagger} K_j = \openone_S.
\end{gather}
The above can be easily rewritten as an isometry in terms of operators $K_j \in \Bcal(\Hcal_\stxt)$ and vectors $\ket{j}_\etxt \in \Hcal_\etxt$:
\begin{gather}
\label{eqn::KrausIso}
V_{S\rightarrow \setxt}:= \sum_j K_j \otimes \ket{j}_\etxt =: V,
\end{gather}
satisfying $V^\dagger V = \openone_{\stxt}$. Consequently, the number of Kraus operators determines the dimension $d_\etxt$ of the environment that is used for the construction.\ftnt{Note that there always exists a minimal number of Kraus operators, and, as such, a minimal environment dimension that allows one to dilate the map $\Ecal$. \textit{Any} map $\Ecal$ can be dilated in a space of dimension $d_\etxt \leq d_\stxt^2$.} With this, we have 
\begin{gather}\label{eq:cpdilation}
 \Ecal[\rho_\stxt] = \tr_\etxt(V \rho_\stxt V^\dagger) = \tr_\etxt(U \rho_\stxt \otimes \ketbra{0}{0}_\etxt U^\dagger).
\end{gather}
The second equality comes from the fact that any isometry $V$ can be completed to a unitary $U_{\setxt \rightarrow \setxt} =:U$ (see, for example, Ref.~\cite{buscemi_physical_2003} for different possible constructions). For completeness, here, we provide a simple way to obtain $U$ from $V$: Let $\{\ket{\ell}_\stxt \}_{\ell=0}^{d_\stxt-1}$ ($\{\ket{\alpha}_\etxt\}_{\alpha=0}^{d_\etxt-1}$) be an orthogonal basis of the system (environment) Hilbert space. By construction (see Eq.~\eqref{eq:cpdilation}), we have $U\ket{\ell 0}_{\setxt} = V \ket{\ell}_\stxt$. Consequently, $U$ can be written as
\begin{gather}
\label{eqn::Unitary_Stine}
 U = \sum_{\ell=0} V \ket{\ell}_\stxt \bra{\ell0}_{\setxt} + \sum_{\substack{\ell=0,\\\alpha=1}} \ket{\vartheta_{\ell,\alpha}}_{\setxt} \bra{\ell\alpha}_{\setxt},
\end{gather}
where ${\ }_{\setxt}\braket{\vartheta_{\ell',\alpha}|V |{\ell}}_\stxt = 0$ for all $\{\ell, \ell'\}$ and $\alpha\geq 1$ and $\braket{\vartheta_{\ell',\alpha'}|\vartheta_{\ell,\alpha}} = \delta_{\ell\ell'}\delta_{\alpha\alpha'}$. Such a set  $\{\ket{\vartheta_{\ell',\alpha}}\}$ of orthogonal vectors can be readily found via a Gram-Schmidt orthogonalization procedure. It is easy to verify that the above matrix $U$ is indeed unitary.

The fact that any quantum channel can be understood as part of a unitary process is often referred to as \textit{Stinespring dilation}~\cite{stinespring1955}. Together with the possibility to purify every quantum state, it implies that all dynamics in quantum mechanics can be seen as reversible processes, and randomness only arises due to lack of knowledge. In particular, we have 
\begin{gather}
\label{eqn::DilatedChannel}
 \Ecal[\rho_\stxt] = \tr_{\etxt\stxtp}[U (\Psi_{\stxt\stxtp}\otimes \ketbra{0}{0}_{\etxt}) U^\dagger],
\end{gather}
where we have omitted the respective identity matrices.

On the other hand, in the classical case, the initial state of the environment is definite, but unknown, in each run of the experiment. Supposing that the system too is initialized in a definite state, then any pure interaction, i.e., a permutation will always yield the system in a definite state. In other words, if any randomness exists in the final state, the randomness must have been present \textit{a priori}, either in the initial state of the environment or in the interaction. A classical dynamics that transforms pure states to mixed states, i.e., one that is described by stochastic maps, can thus not have come from a permutation and pure states only, and stochasticity in classical physics does not stem from ignorance about additional degrees of freedom alone. 

As both of these statements, the purifiability of quantum states and quantum channels ensure the fundamental reversibility of quantum mechanics, purification postulates have been employed as one of the axioms in reconstructing quantum mechanics from first principles~\cite{hardy_quantum_2001, dariano_quantum_2017}.

\textbf{Purification of dephasing dynamics.} Before advancing, it is insightful to provide the dilation of an explicit quantum channel. Here, we choose the so-called dephasing map on a single qubit:
\begin{gather}
\label{eq:dephasing}
\begin{split}
 &\Ecal^{DD}_{(t:0)} [\rho(0)] = \rho(t) \\ 
 &\rho(0) = \left(\begin{matrix} \rho_{00} & \rho_{01} \\
 \rho_{10} & \rho_{11} \end{matrix}\right)
 \mapsto
 \rho(t) = \left(\begin{matrix} \rho_{00} & e^{-\gamma t}\rho_{01} \\
 e^{-\gamma t} \rho_{10} & \rho_{11} \end{matrix}\right).
\end{split} 
\end{gather}
In what follows, whenever a dynamics is such that the off-diagonal elements vanish exponentially in a given basis, we will call it \textit{pure} dephasing. The above channel can be represented with two Kraus matrices 
\begin{gather}
K_0(t) = \sqrt{\f{1+e^{-\gamma t}}{2}}\sigma_0, \quad
K_1(t) = \sqrt{\f{1-e^{-\gamma t}}{2}} \sigma_3.
\end{gather}
Following Eq.~\eqref{eqn::KrausIso}, the corresponding isometry is given by $V(t) = K_0 \otimes \ket{0}_\etxt + K_1 \otimes \ket{1}_\etxt$, which implies 
\begin{gather}
\begin{split}
 &V(t)\ket{0}_\stxt = \sqrt{\f{1+e^{-\gamma t}}{2}} \ket{00}_{\setxt} + \sqrt{\f{1-e^{-\gamma t}}{2}} \ket{01}_{\setxt}, \\
 &V(t)\ket{1}_\stxt = \sqrt{\f{1+e^{-\gamma t}}{2}} \ket{10}_{\setxt} - \sqrt{\f{1-e^{-\gamma t}}{2}} \ket{11}_{\setxt}. \\
\end{split}
\end{gather}
From this, we can construct the two remaining vectors $\ket{\vartheta_{01}}_{\setxt}(t)$ and $\ket{\vartheta_{11}}_{\setxt}(t)$ to complete $V(t)$ to a unitary $U(t)$ by means of Eq.~\eqref{eqn::Unitary_Stine}. For example, we can make the choice 
\begin{gather}
\begin{split}
 &\ket{\vartheta_{01}(t)}_{\setxt} = \sqrt{\f{1-e^{-\gamma t}}{2}} \ket{10}_{\setxt} + \sqrt{\f{1+e^{-\gamma t}}{2}} \ket{11}_{\setxt}, \\
 &\ket{\vartheta_{11}(t)}_{\setxt} = \sqrt{\f{1-e^{-\gamma t}}{2}} \ket{00}_{\setxt} - \sqrt{\f{1+e^{-\gamma t}}{2}} \ket{01}_{\setxt}. \\
\end{split}
\end{gather}
It is easy to check that these vectors indeed satisfy ${\ }_{\setxt}\braket{\vartheta_{\ell',\alpha}|V |{\ell}}_\stxt = 0$ for all $\{\ell, \ell'\}$ and $\alpha\geq 1$, as well as $\braket{\vartheta_{\ell',\alpha'}|\vartheta_{\ell,\alpha}} = \delta_{\ell\ell'}\delta_{\alpha\alpha'}$. This, then, provides a unitary matrix $U(t)$ that leads to the above dephasing dynamics: 
\begin{gather}
\begin{split}
 U(t) =& V(t)\ket{0}_\stxt\bra{00}_{\setxt} + V(t)\ket{1}_\stxt\bra{10}_{\setxt} \\
 &+ \ketbra{\vartheta_{01}(t)}{01}_{\setxt} + \ketbra{\vartheta_{11}(t)}{11}_{\setxt}.
\end{split}
\end{gather}
Insertion into Eq.~\eqref{eq:cpdilation} then shows that the thusly defined unitary evolution indeed leads to dephasing dynamics on the system. 

Finally, while we chose to introduce quantum channels in terms of the natural properties one would demand from them, we could have chosen the converse route, starting from the assumption that every dynamics can be understood as a unitary one in a bigger spaces, thus positing an equation along the lines of Eq.~\eqref{eqn::DilatedChannel} as the starting point. Unsurprisingly, this, too, would have yielded CPTP maps, since we have 
\begin{gather}
     \Ecal[\rho_\stxt] = \tr_{\etxt\stxtp}[U (\Psi_{\stxt\stxtp}\otimes \ketbra{0}{0}_{\etxt}) U^\dagger] = \sum_{\alpha} K_\alpha \rho_\stxt K_\alpha^\dagger\, ,
\end{gather}
where $K_\alpha:= \braket{\alpha|U|0}$, $\{\alpha\}$ is a basis of $\Hcal_{\etxt}$ and $\rho_\stxt = \tr_{\stxtp}(\Psi_{\stxt \stxtp})$. Since this is a Kraus decomposition that satisfies $\sum_\alpha K_\alpha^\dagger K_\alpha = \ident_\stxt$, the dynamics of an initial system state that is initially uncorrelated with the environment (here in the state $\ketbra{0}{0}_\etxt$) and evolves unitarily on system plus environment, is always given by a CPTP map on the level of the system alone. We will use this fact -- amongst others -- later on when we lay out how to detect non-Markovianity based on quantum master equation approaches.
\subsection{Quantum Master Equations}
\label{sec::QuasterEq}

\begin{figure}[t]
 \centering
 \includegraphics[width=0.95\linewidth]{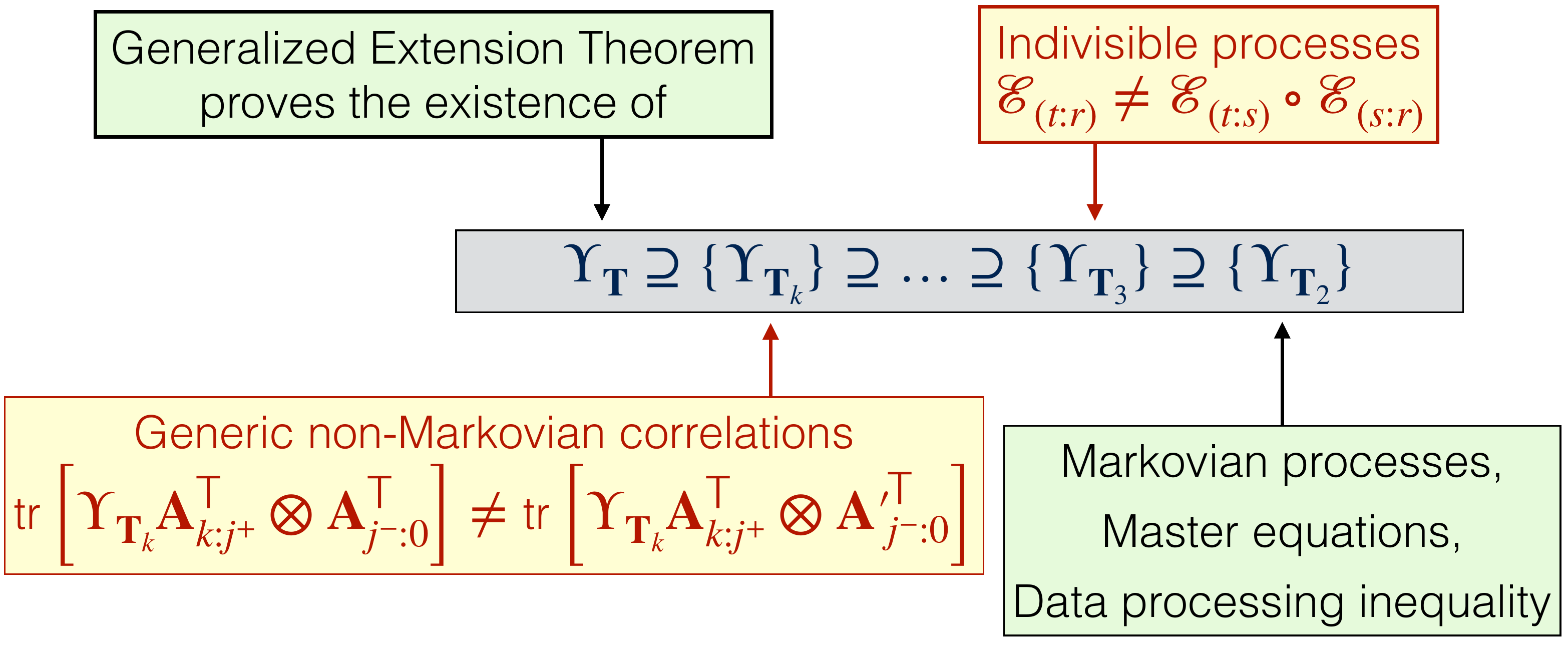}
 \caption{\textit{Hierarchy of multi-time quantum processes}. A quantum stochastic process is the process tensor over all times. Of course, in practice one looks only at finite time statistics. However, the generalized extension theorem tells us that the set of all $k$-time process tensors $\{\Upsk\}$ contain, as marginals, all $j$-time probability distributions $\{\Upsilon_{\Tset_{j}}\}$ for $j<k$. Moreover, the set of two and three time processes play a significant roles in the theory of quantum stochastic processes. Here, we only display a small part of the multi-faceted structure of non-Markovian quantum processes. For a much more comprehensive stratification, see Ref.~\cite{Rivas2014, Li2018}.}
 \label{fig:q-hierarchy}
\end{figure}

While we have yet to formalize the theory of quantum stochastic processes (in the sense that we have yet to explore how to obtain multi-time statistics in quantum mechanics), the quantum stochastic matrix formalism is enough to keep us occupied for a long time. In fact, much of the active research in the field of open quantum system dynamics is concerned with the properties of families of quantum channels. It should be already clear that the quantum stochastic matrix, like its classical counterpart, only deals with two-time correlations, see Figures~\ref{fig:hierarchy} and ~\ref{fig:q-hierarchy}, and can thus not provide a complete description of quantum stochastic processes. This analogy goes further; as is the case on the classical side, an important family of stochastic matrices corresponds to quantum master equations.\ftnt{This is a pedagogical statement, not a historical one.} Before fully generalizing the concept of stochastic processes to the quantum realm, let us thus have a quick -- and very superficial -- look at quantum master equations and witnesses of non-Markovianity that are based on them.

Quantum master equations have a long and rich history dating back to the 1920s. Right at the inception of modern quantum theory, Landau derived a master equation for light interacting with charged matter~\cite{landau}. This should not be surprising because master equations play a key role in understanding the real phenomena observed in the lab. For the same reason, they are widely used tools for theoretical physicists and beyond, including quantum chemistry, condensed matter physics, high-energy physics, material science, and so on. However, the formal derivation of overarching master equations took another thirty years. Nakajima and Zwanzig independently derived exact memory kernel master equations using the so-called projection operator method. Since then there have an enormous number of studies of non-Markovian master and stochastic equations~\cite{diosi_non-markovian_1997, diosi_non-markovian_1998, strunz_open_1999, yu_post-markov_2000, wiseman_complete_2001, breuer_stochastic_2009, breuer_structure_2009, PhysRevB.70.045323, PhysRevA.68.062104,	PhysRevLett.101.140401,	PhysRevA.66.012108,	PhysRevE.76.031115, PhysRevB.94.214308,	PhysRevA.98.052129,	PhysRevLett.104.070406,	PhysRevA.98.022123, PhysRevB.84.075150, PhysRevA.98.042119, PhysRevA.71.020101, PhysRevLett.100.180402, vacchini_quantum_2020, smirne_rate_2020, megier_interplay_2020, megier_evolution_2020}, spanning from exploring their mathematical structure, studying the transition between the Markovian and non-Markovian regime~\cite{liu_experimental_2011, garrido_transition_2016} and applying them to chemistry or condensed matter systems. Here, we will not concern ourselves with these details and limit our discussion to the overarching structure of the master equation, and in particular how to tell Markovian ones apart from non-Markovian ones. We refer the reader to standard textbooks for more details~\cite{Gardiner, Wiseman, BreuerPetruccione} on these aspects as well as proper derivations, which we will not provide in this section.

The most general quantum master has a form already familiar to us. We simply replace the probability distribution in Eq.~\eqref{eq:master} with a density matrix to obtain the Nakajima-Zwanzig master equation\ftnt{Some authors would not call this a master equation and refer to it as a memory-kernel equation. We are being a bit liberal here.}
\begin{gather}\label{eq:qmaster}
 \frac{d}{d t} \rho(t) = \int_{s}^t \mathcal{K}(t,\tau)[\rho(\tau)] \ d\tau.
\end{gather}
Above, $\mathcal{K}(t,\tau)$ is a super-operator\ftnt{In the classical case it was an operator acting on a vector. Here, it is called a super-operator because it acts on $\rho$, which is an operator on the Hilbert space.} that is called the memory kernel. Often, this equation is written in two parts
\begin{gather}\label{eq:qmaster1}
 \frac{d}{d t} \rho(t) = -i[H,\rho(t)] + \Dcal[\rho(t)] + \int_{s}^t \mathcal{K}(t,\tau)[\rho(\tau)] \ d\tau,
\end{gather}
where $\Dcal$ is called the dissipator with the form
\begin{gather}\label{eq:qgen}
 \Dcal[\rho(t)] = \sum_{n,m=1}^{d^2} \gamma_j \left(L_j \rho(t) L_j^\dag - \frac{1}{2}\left\{L^\dag_j L_j, \rho(t)\right\}\right).
\end{gather}
Above, the first term on the RHS corresponds to a unitary dynamics, the second term is the dissipative part of the process, and the third terms carries the memory (which can also be dissipative). We note in passing that we have yet to define what Markovian and non-Markovian actually mean in the quantum realm, and how the ensuing definitions relate to their classical counterparts. We will provide a rigorous definition of these concepts in Sec.~\ref{sec::PropQuant}, while here, for the moment, we shall content ourselves with the vague `definition' that non-Markovian processes are those where memory effects play a non-negligible role; Markovian processes are those where memory effects are absent.

While the Nakajima-Zwanzig equation is the most general quantum master equation, the rage in the 1960s and 1970s was to derive the most general Markovian master equation. It took well over a decade to get there, after many attempts, see Ref.~\cite{gksl} for more on this history and  Ref.~\cite{lindped} for a pedagogical treatment. Those who failed in this endeavor were missing a key ingredient, complete positivity. In 1976, this feat was finally achieved by Gorini-Kossakowski-Sudarshan~\cite{sudarshangorini} and Lindblad~\cite{lindblad75} independently.\ftnt{Around the same time Franke proposed a very similar equation, though he seems to be unaware of complete positivity at that time~\cite{franke_general_1976}.} A quantum Markov process can be described by this master equation, now known as the GKSL master equation. Eq.~\eqref{eq:qmaster1} already contains the GKSL master equation in the sense that the final term vanishes for the Markov process
\begin{gather}\label{eq:qmarkovmaster}
\frac{d}{d t} \rho_t = \mathcal{L}[\rho_t] \quad \mbox{with} \quad
\mathcal{L}[\sbt \,] = -i[H,\sbt \,]+\Dcal[\sbt \,],
\end{gather}
and, here $\mathcal{L}$ stands for Louivillian, but often called Lindbladian. Intuitively, the above master equation is considered memoryless, since it does not contain the integral over past states that is present in Eq.~\eqref{eq:qmaster1} (we will see in Sec.~\ref{sec::ExDivMark} that this intuition is somewhat too simplistic, since there are processes that carry memory effects but can nonetheless be modeled by a GKSL equation.) If $\mathcal{L}$ is time independent, then the above Master equation has the formal solution 
\begin{gather}
    \rho_t = \Ecal_{t:r}[\rho_r] = e^{\mathcal{L}(t-r)}[\rho_r]\, ,
\end{gather}
where $\rho_r$ is the system state at time $r$. From this, we see that the respective dynamics between two arbitrary times $r$ and $t$ only depends on $t-r$, but not the absolute times $r$ and $t$ (or any earlier times). Using $ e^{\mathcal{L}(t-r)} = e^{\mathcal{L}(t-s)} e^{\mathcal{L}(s-r)}$, this implies the often used semi-group property of Markovian dynamics 
\begin{gather}
\label{eqn::semigroup}
     \Ecal_{t:r} = \Ecal_{t:s} \circ \Ecal_{s:r}
\end{gather}
for $t\geq s \geq r$. 

Some remarks are in order. The decomposition in Eq.~\eqref{eq:qmaster1} is not always unique. Often, a term dubbed as the \textit{inhomogeneous} term is present and it is due to the initial system-environment correlations. As we will outline below, describing dynamics with initial correlations in terms of quantum channels (and thus, master equations), is operationally dubious and the interpretation of an inhomogeneous term as stemming from initial correlations thus somewhat misdirected. 

In the Markovian case, the super-operators in Eq.~\eqref{eq:qmarkovmaster} should be time-independent. In fact, it is possible to derive master equations for non-Markovian processes that look just like Eq.~\eqref{eq:qmarkovmaster} but then the super-operators will be time-dependent and the rates $\{\gamma_j\}$ may be negative~\cite{breuer_stochastic_1999,andersson_finding_2007, PhysRevLett.104.070406, bassano1, laine_local--time_2012, hall_canonical_2014} (while they are always positive in the Markovian case). For a Markovian master equation, the operators $\{L_j\}$ are directly related to the Kraus operators of the resulting quantum channels~\cite{rodriguez_theory_2008}. Since Eq.~\eqref{eq:qmaster1} is the most general form of a  quantum master equation it contains equations due to Redfield, Landau, Pauli, and others. To reach this equation one usually expands and approximates the memory kernel. This is a field of its own and we cannot do justice to these methods or the reasoning behind the approximations here (for a comparison of the validity of the different employed approximations, see, for example, Ref.~\cite{rivas_markovian_2010, hartmann_accuracy_2020}).

As with the classical case, the above master equations express the statistical quantum state continuously in time.\ftnt{This fact notwithstanding, many master equations can be derived as limits of discrete time collision models~\cite{scarani_thermalizing_2002, ziman_diluting_2002, giovannetti_dynamical_2005, vacchini_non-markovian_2013, vacchini_general_2014, caruso_quantum_2014, vacchini_generalized_2016, ciccarello_collision_2017}, both in the Markovian~\cite{ziman_description_2005, ziman_all_2005, attal_repeated_2006} and the non-Markovian~\cite{pellegrini_non-markovian_2009, Giovannetti2012, filippovjphysb, giovannetti_master_2012JPB, ciccarello_collision-model-based_2013, PhysRevA.89.052120, lorenzo_heat_2015, kretschmer_collision_2016, Lorenzo2017, PhysRevA.96.022109, arXiv:2008.00765} case.} They can be either derived from first principles by making the right approximations, or as ad hoc phenomenological dynamical equations that model the pertinent peroperties of a process at hand (see, for example, Ref.~\cite{BreuerPetruccione} for a wide array of different derivations of quantum master equations).

As before, it may be tempting to think that the master equation is equivalent to a stochastic process as defined above. However, just as in the classical case, the quantum master equation only accounts for two-point correlations. This can be seen intuitively by realizing that the solution of a master equation is a family of quantum channels, each correpsonding to two-time correlations, or, more directly, by employing the transfer tensor method~\cite{CerrilloCao2014, Rosenbach2016, Pollock-quantum}, which shows that the RHS of Eq.~\eqref{eq:qmaster1} can be expressed as a linear combination of product of quantum maps $\Ecal_{(c:b)} \circ \Ecal_{(b:a)}$, with $c$ being either $t$ or $t-dt$, $b=s$, and $a$ being either the initial time. A quantum map $\Ecal_{(b:a)}$ is a mapping of a preparation at time $a$ to a density matrix at time $b$. Thus, it only contains correlations between two times $a$ and $b$. The LHS can be computed by setting $b=t$ and $a=t-dt$. Another formal method for showing that the RHS can be expressed as a product of two stochastic matrices can be done by means of the Laplace transform~\cite{bassano1, vacchini}.

This also puts into question our somewhat lax use of the term `Markovian' in the above discussion. As we discussed in the classical case, Markovianity is a statement about conditional independence for multi-time probability distributions. How, then, can a master equation that is concerned with two-point correlations only, be Markovian? Indeed, as we shall see below it is possible to have physical non-Markovian processes (i.e., processes that do not display the correct conditional independence) that can be described by what we dubbed a Markovian master equation in Eq.~\eqref{eq:qmarkovmaster}. That is, the implication only goes one way; a Markov process always leads to a master equation of Eq.~\eqref{eq:qmaster1} form with the final term vanishing. The converse does not hold. We detail an example below, but to fully appreciate it we must have a better understanding of multi-time quantum correlations. Nonetheless, while it is not possible to unambiguously deduce the Markovianity of a process from limited statistical data only, one can, just like in the classical case, already use it to detect the presence of memory effects.

\subsection{Witnessing non-Markovianity}
\label{sec::Witness}

As mentioned above, Markovian processes will lead to master equations of the form of Eq.~\eqref{eq:qmarkovmaster} and, in turn, can be fully described by the resulting family of CPTP maps. Thus, having access to the stochastic matrix and master equation is already sufficient to witness departures from Markovianity. That is, there are certain features and properties that must belong to any Markovian quantum processes, which then allows for discriminating between Markov and non-Markov processes.

\subsubsection{Initial correlations}
\label{sec::InitCorr}

Consider the dynamics of a system from an initial time to some final time. When the system interacts with an environment the process on the system can be described by a map $\Ecal_{(t:0)}$. As we showed in Eq.~\eqref{eq:cpdilation}, such a map can be thought to come from unitary system-environment dynamics, with the caveat that the initial system-environment has no correlations (in Eq.~\eqref{eq:cpdilation}, it was of the form $\rho_S\otimes \ketbra{0}{0}$). Already in the 1980s and 1990s, researchers began to wonder what happens if the initial system-environment state has correlations~\cite{park, pechukas, Alicki95}. Though this may -- at first glance -- seem unrelated to the issue of non-Markovianity, the detectable presence of initial correlation is already a non-Markovian effect. This is because initial correlations indicate past interactions and if the initial correlations affect the future dynamics then the future dynamics are a function of the state of the system at $t=0$, as well as further back in the past. As this is in line with an intuitive definition of non-Markovianity (a thorough one will be provided later), the observable presence of initial correlations constitutes an experimentally accessible witness for memory~\cite{IC-breuer, PhysRevLett.107.180402, smirne_experimental_2011, gessner_local_2013, smirne_quantum_2013, Gessner:2014kl}. 

We emphasize, that the presence of initial correlations does not make the resulting process non-Markovian per se; suppose there are initial correlations whose presence \textit{cannot} be detected on the level of the system, then these initial system-environment correlations do not lead to non-Markovianity. If, however, it is possible to detect an influence of such correlations on the behavior of the system (for example, by observing a breakdown of complete positivity~\cite{pechukas, jordan:052110, StelmachovicBuzek01} or by means of a local dephasing map~\cite{PhysRevLett.107.180402,gessner_local_2013}), then the corresponding process is non-Markovian. With this in mind, in what follows, by `presence' of correlations, we will always mean `detectable presence'.

A pioneering result on initial correlations and open system dynamics was due to Pechukas in 1995~\cite{pechukas}. He argued that either there are no initial correlations or we must give up either the complete positivity or the linearity of the dynamics. Around the same time, many experiments began to experimentally reconstruct quantum maps~\cite{Wein:121.13, myrskog:013615, orien:080502}. Surprisingly, many of these experiments revealed not completely positive maps. This began a flurry of theoretical research either arguing for not-completely-positive (NCP) dynamics or reinterpreting the experimental results~\cite{jordan:052110, shaji_whos_2005, ctz, ziman, Ziman06, rodriguez2008completely, ctz, PhysRevLett.102.100402, modi_role_2010, PhysRevA.87.042301}. However, this does not add to the physical legitimacy of NCP processes~\cite{vacchini_reduced_2016}. Nevertheless, NCP dynamics remains as a witness for non-Markovianity. We will show below that all dynamics, including non-Markovian ones, must be completely positive. We do this by getting around the Pechukas theorem by paying attention to what it means to have a state in quantum mechanics. For the moment though, the take-home message is that, one way to detect memory is to devise experiments that can detect initial correlations between the system and the environment.

\textbf{Two illustrative examples.}
Let us conclude this discussion of initial correlations with two examples that highlight the problems encountered in the presence of initial correlations. To this end, we consider a single qubit system interacting with a single qubit environment. We let the initial correlated state be 
\begin{gather}\label{eq:initstateex}
    \rho_{\setxt}(0)  = \frac{1}{4} \left(\openone_\stxt \otimes \openone_\etxt + \vec{a}\cdot \vec{\sigma}_\stxt \otimes \openone_\etxt + g \ \sigma^y_\stxt \otimes \sigma^z_\etxt \right).
\end{gather}
The system-environment interaction is chosen to be 
\begin{gather}\label{eq:unitaryex}
    U_{\setxt} = \prod_{j=x,y,z}\left\{\cos(\omega t) \openone_\stxt \otimes \openone_\etxt -\sin(\omega t) \sigma^j_\stxt \otimes \sigma^j_\etxt\right\}.
\end{gather}

We are of course interested in only the reduced dynamics of the system. The initial state of the system is $\rho_\stxt(0) = \frac{1}{2} \left(\openone_\stxt + \vec{a}\cdot \vec{\sigma}_\stxt \right)$ and under the above unitary, it evolves to 
\begin{gather}
\begin{split}
    \rho_\stxt(t) =& \frac{1}{2} \left[\openone_\stxt + c_\omega^2 \vec{a}\cdot \vec{\sigma}_\stxt -g \ c_\omega s_\omega\sigma^x_\stxt \right].
\end{split}
\end{gather}
where where $c_\omega := \cos(2\omega t)$ and $s_\omega := \sin(2\omega t)$.

\emph{Example 1.} In the first instance, in order to provide a full basis of input states, we will fix the correlation term (i.e., the third term in Eq.~\eqref{eq:initstateex}) and vary the system state alone. By choosing $\vec{a}$ to be $\kappa(\pm1,0,0)^\trps$, $\kappa(0,1,0)^\trps$, and $\kappa(0,0,1)^\trps$ gives us a linearly independent set of input states given in Eq.~\eqref{eq:statebasis}. Here, $\kappa$ is a number less than 1 to ensure that the total $\setxt$ state is positive. The quantum stochastic matrix is straightforwardly constructed plugging the output states along with the dual basis in Eq.~\eqref{eq:dualbasis} in Eq.~\eqref{eqn::TomoVec}. 

This process is easily shown to be not completely-positivity. To do so, we compute the Choi state using Eq.~\eqref{eq:tomrepmap} to get
\begin{gather}
\Upsilon_\Ecal = \frac{1}{2} \begin{pmatrix}
1+c_\omega^2 & 0 & -g\ c_\omega s_\omega & 2c_\omega^2\\
0 & 1-c_\omega^2 & 0 & -g\ c_\omega s_\omega\\
-g\ c_\omega s_\omega & 0 & 1-c_\omega^2 & 0\\
2c_\omega^2 & -g\ c_\omega s_\omega & 0 & 1+c_\omega^2 \\
\end{pmatrix}.
\end{gather}
Two of the four eigenvalues of this Choi matrix turn out to be $\frac{1}{2}\left(1- \cos^2(2\omega t) \pm g \cos(2 \omega t) \sin(2\omega t) \right)$, which are not always positive, i.e., the process is not completely-positive. Seemingly then, initial correlations lead to dynamics that are not CP. 

\emph{Example 2.} One argument against the above procedure is that there is no operational mechanism for varying the state of the system, while keeping the correlation term fixed. This is indeed true and a major flaw with the programme of NCP dynamics. However, we could envisage the case where the initial state of the system is \textit{prepared}, starting from the correlated state of Eq.~\eqref{eq:initstateex}, by means of projective operations along directions of the vectors $\vec{a} \in \{\kappa(\pm1,0,0)^\trps$, $\kappa(0,1,0)^\trps,\kappa(0,0,1)^\trps\}$. The Choi state of the resulting map, in this case, turns out to be
\begin{gather}
\label{eqn::NCPnonLin}
\Upsilon_\Ecal = \frac{1}{2} \begin{pmatrix}
1+c_\omega^2 & 0 & 0 & f \\
0 & s_\omega^2 & -\frac{i g}{2} s_{2\omega} & 0\\
0 & \frac{i g}{2} s_{2\omega} & s_\omega^2 & 0\\
f^* & 0 & 0 & 1+c_\omega^2 \\
\end{pmatrix},
\end{gather}
where $f := -1 -c_{2\omega} +\frac{i g}{2} s_{2\omega}$. This map is perfectly operational (in the sense that there is a clear procedure of how to obtain it experimentally) and still we find that the dynamics are NCP, since the above matrix is not positive semidefinite..

One of the key assumptions in the construction of CP maps (and linear maps in general) is that the state of the environment is the same for all possible inputs of the system, and thus does not depend on how the system of interest was prepared. This is clearly not the case here, as the correlation term vanishes for three of these projections, but not for the projection along eigenvectors of $\sigma^y_\stxt$. This is the source of NCP dynamics and this is the reason why NCP dynamics are an indicator for non-Markovianity, since system-environtment correlations constitute a memory of past interactions. 

However, there is a bigger question looming over us: how do we decide whether the first or the second map is the valid one? Worse yet, we can also construct a third map; let us assume that the initial state has the Bloch vector $\vec{a} = \kappa(1,0,0)$, and we prepare the  basis of input states by applying local unitary operations to this initial state. In this case, the correlation term will be different for each preparation and we will get another map that will also be NCP. It turns out that there are infinite ways of preparing the input states~\cite{modiosid} and each will lead to a different dynamical map, see Ref.~\cite{modi_role_2010} several examples related to the two presented here. This is not tenable as we do not have a unique description for the process (since the maps we obtain depend on how we created the basis of input states) and it violates the CP condition. Moreover, all of these maps will further violate the linearity condition. That is, a map should have value in predicting the future state of the system, given an arbitrary input state. The above map
in example 2, has little to no predictive power; when preparing an initial system state that is not one of the original basis states, the action of the map of Eq.~\eqref{eqn::NCPnonLin} will not yield the correct output state. The map in example 1 does have predictive power, but it is not an operationally feasible object because, as mentioned before, there is no way to manipulate $\vec{a}$ without changing the correlation term. In Sec.~\ref{subsec::IC} we will show how these problems are avoided in a systematic, and operationally well-defined manner manner.

\subsubsection{Completely positive and divisible processes}
\label{sec::CPDiv}

Going beyond this rather static marker of non-Markovianity in terms of initial system-environment correlations, we can extend the concept of divisibility that we first discussed in Sec.~\ref{sec:cl-div} for the classical case to the quantum case.
A quantum process is called divisible if
\begin{gather}
 \Ecal_{(t:r)} = \Ecal_{(t:s)}\circ \Ecal_{(s:r)}, \quad \forall r,s, t.
\end{gather}
Here, $\circ$ stands for the composition of two quantum maps. Since they are not necessarily matrices, the composition may -- depending on the chosen representation -- not be a simple matrix product. Moreover, in the quantum case we now further require that each map here is completely positive and thus such a class of processes is referred to as \textit{CP divisible} processes. 

Understanding the divisibility of quantum maps and giving it an operational interpretation is a highly active area of research~\cite{divcomp, buscemi_complete_2014, PhysRevA.93.012101, PhysRevA.96.032111, chruscinski2016, PhysRevLett.121.080407, PhysRevA.95.012112, PhysRevA.99.042105, PhysRevA.93.042120, filippovjphysb}, and we will only scratch the surface. Importantly, as we have seen above, processes that satisfy a GKSL equation are divisible (see Eq.~\eqref{eqn::semigroup}), making the breakdown of divisibility a witness of non-Markovianity.

Now, if $r=0$ then we can certainly run a set of experiments to determine the quantum maps $\{\Ecal_{(t:0)}\}$ for all $t$ by means of quantum process tomography outlined above. These maps will be CP as long as there are no initial correlations. But how do we determine the intermediate dynamics $\Ecal_{(t:s)}$ for $s > 0$? One possible way is to infer an intermediate process from the family of maps $\{\Ecal_{(t:0)}\}$ by inversion
\begin{gather}\label{eq:defindiv}
 \zeta_{(t:s)} := \Ecal_{(t:0)} \circ \Ecal_{(s:0)}^{-1},
\end{gather}
provided the maps $\{\Ecal_{(t:0)}\}$ are invertible. We deliberately label this map with a different letter $\zeta$, as it may not actually represent a physical process~\cite{Milz2019}. Now, if the process is Markovian then $\zeta_{(t:s)} = \Ecal_{(t:s)}$, i.e., it corresponds indeed to the physical evolution between $s$ and $t$, and it will be completely positive. Conversely, if we find that $\zeta_{(t:s)}$ is not CP then we know that process is non-Markovian.

\textbf{Example of an indivisible process.} To provide some more concrete insight, let us provide an example of a process that is \textit{not} divisible. To this end, we consider the initial state in Eq.~\eqref{eq:initstateex}, along with the interaction in Eq.~\eqref{eq:unitaryex}. Here, we will take the limit of $g \to 0$, thus dropping the correlation term and rendering the initial system-environment state uncorrelated. The Choi matrix for the dynamics of the system is then given by 
\begin{gather}\label{eq:ecalxz}
    \Upsilon_\Ecal =\frac{1}{2} \begin{pmatrix}
    1+c_\omega^2 & 0 & 0 & -2c_\omega^2\\
    0 & 1-c_\omega^2 & 0 & 0\\
    0 & 0 & 1-c_\omega^2 & 0\\
    -2c_\omega^2 & 0 & 0 & 1+c_\omega^2 \\
    \end{pmatrix}.
\end{gather}
While this process is CP when considered from the initial time (since system and environment are initially uncorrelated), it is not divisible, simply because it is not possible to `divide' $\cos(2\omega t)$ into a product of two function readily. More concretely, due to the oscillatory nature of the process, many of the possible inferred maps  $\zeta_{(t:s)}$ of Eq.~\eqref{eq:defindiv} would be NCP. On the other hand, for a process where $c_\omega$ is replaced by something like $\exp(-\omega t)$, the process would become divisible, see Eq.~\eqref{eq:dephasingqstochmat}.

Working with divisible processes has several advantages. Two that we have already discussed in the classical case are the straightforward connection to the master equation and data processing inequality (which also holds in the quantum case). We can use these to construct further witnesses for non-Markovianity, such as those based on the trace distance measure~\cite{PhysRevLett.103.210401}. The amplitude damping channel in Eq.~\eqref{eq:ADchannel} and the dephasing channel in Eq.~\eqref{eq:dephasing} are both divisible as long as they relax exponentially. Otherwise, they are indivisible processes, which is easily checked numerically. As in the classical case, the logic surrounding CP divisibility and its relationship to Markovianity is as follows: If a process is Markovian, it is also CP divisible (the converse does not necessarily hold). A breakdown of CP divisibiltiy thus signals non-Markovian effects, without the need to investigate multi-time statistics. Pursuing this line of reasoning further, there are properties that hold for CP divisible processes, like, for example, quantum data processing inequalities (see below). Instead of checking for the breakdown of CP divisibility, one can thus check the breakdown of other properties as a proxy. This, however, will lead to succesively weaker (but potentially more easily accessible) witnesses of non-Markovianity.

We mention briefly that CP divisibility is not the only type of divisibility of open quantum system dynamics. There exists a vast body of research on different types of divisibility for quantum processes, their stratification and  interconnectedness~\cite{wolf_dividing_2008, chruscinski_generalized_2016, davalos_divisibility_2019,chruscinski_dissipative_2019}, as well as the closely related question of simulatability of quantum and classical channels adn dynamical maps~\cite{PhysRevLett.101.150402, puchala_pauli_2019, korzekwa_quantum_2020, shahbeigi_log-convex_2020}. Here,we will not dive into these fields in detail.

\subsubsection{Snapshot}
\label{sec::Snap}

As in the classical case, when a process is divisible it will be governed by Markovian master equation of GKSL type of Eq.~\eqref{eq:qmarkovmaster}. Following the classical case, in Eq.~\eqref{eq:markov-master}, the Liouvillian for the quantum process can be obtained via
\begin{gather}\label{eq:divME}
\frac{d}{dt}\rho(t) = \lim_{\Delta t \to 0}\frac{\Ecal_{t+\Delta t: t } - \mathcal{I}}{\Delta t} \rho(t) = \mathcal{L}[\rho_t].
\end{gather}
This, in turn, means that 
\begin{gather}
 \rho(t) = \Ecal_{(t:0)}[\rho(0)] \quad \mbox{with} \quad \Ecal_{(t:0)}= e^{\mathcal{L} t}
\end{gather}

We can now reverse the implication to see if a process is Markovian by considering the process $\Ecal_{(t:0)}$ for some $t$. We can take the $\log$ of this map, which has to be done carefully, to obtain $\mathcal{L}$. If the process is Markovian then $\exp(\mathcal{L} s)$ will be CP for all values of $s$. If this fails then the process must be non-Markovian, provided it is also symmetric under time translation. That is, we may have a Markovian process that slowly varies in time, and may fail this test. This witness was one of the first proposed for quantum processes~\cite{wolf_dividing_2008, PhysRevLett.101.150402}. Once again, note that here only two-time correlations are accounted for and, again, we use the term `Markovian master equation' in a rather lax sense; most importantly, besides reasoning by intuition, we have not yet defined what a Markovian quantum process actually is). Unsurprisingly then, this witness will miss processes that are non-Markovian at higher orders.

\textbf{Dephasing dynamics.} Let us clarify these concepts by means of a concrete example. The dephasing process introduced in the previous subsection is divisible and thus can be described by a Markovian master equations. To obtain this, we simply differentiate the state at time $t$ to get
\begin{gather}\label{eq:dephasingmaster}
 \f{d}{dt} \rho(t) = \frac{\gamma e^{-\gamma t}}{2} (\sigma_3 \rho \sigma_3 -\rho).
\end{gather}
The quantum stochastic matrix for this process is
\begin{gather}\label{eq:dephasingqstochmat}
 \breve\Ecal_{(t:0)} = 
 \left( \begin{matrix}
 1 & 0 & 0 & 0 \\
 0 & e^{-\gamma t} & 0 & 0 \\
 0 & 0 & e^{-\gamma t} & 0 \\
 0 & 0 & 0 & 1 \\
 \end{matrix}\right).
\end{gather}
Since the matrix is diagonal, it can be trivially seen to be divisible by the fact that $\exp{(-\gamma t)} = \exp{(-\gamma (t-s))} \exp{(-\gamma s)}$. Consequently, the underlying process could -- in principle -- be Markovian. In Sec.~\ref{sec::ExDivMark} we will revisit this example and show that there are non-Markovian processes, where the two point correlations have this exact form.

\subsubsection{Quantum data processing inequalities}
\label{sec::QDPI}

As mentioned, just like in the classical case, in quantum mechanics, data processing inequalities hold, with the distinction that here, they do not apply to stochastic matrices, but to quantum channels, i.e., CPTP maps. Specifically, there are several distance measure that are proven to be contractive under CPTP dynamics~\cite{lindblad_expectations_1974, Lindblad1975}:
\begin{gather}
 f[\rho,\sigma] \ge f[\Ecal(\rho),\Ecal(\sigma)].
\end{gather}
Three prominent examples are the quantum trace distance
\begin{gather}
 \|\rho-\sigma\|_1 := \tr|\rho-\sigma|,
\end{gather}
the quantum mutual information, 
\begin{gather}
 S(A:B) = S(\rho_A)+S(\rho_B)- S(\rho_{AB}),
\end{gather}
and the quantum relative entropy
\begin{gather}
 S(\rho\|\sigma) = -\tr[\rho \log(\sigma) - \log (\rho)].
\end{gather}
All of these are defined as in the classical case, with the sole difference that for the latter two we replace Shannon entropy with von Neumann entropy, $S(\rho) := -\tr[\rho \log(\rho)]$. Since divisible processes can be composed of independent CPTP maps, they have to satisfy data processing inequalities between any two points in time. Violation of the DPI thus implies a break-down of CP divisibility, and thus heralds the presence of memory effects.

Two of the most popular witnesses of non-Markovianity~\cite{PhysRevLett.103.210401, laine_measure_2010, PhysRevLett.105.050403}, derived using the first two data processing inequalities, were introduced about a decade ago. In particular, Ref.~\cite{PhysRevLett.105.050403} proposed to prepare a maximally entangled state of a system and an ancilla. The system is then subjected to a quantum process. Under this process, if the quantum mutual information (or any other correlation measure) between the system and ancilla behaves non-monotonically then the process must be non-Markovian. A similar argument was proposed by Ref.~\cite{PhysRevLett.103.210401} using the trace distance measure. It can be shown that the former is a stronger witness than the latter~\cite{PhysRevA.83.052128}. Nonetheless, even this latter witness of non-Markovianity is generally not equivalent to the break-down of CP divisibility, as there are processes that behave monotonically under the above distance measures, but are \textit{not} CP divisible~\cite{Addis2014, bylicka_constructive_2017, megier_eternal_2017, PhysRevLett.121.080407}. We will not delve into the details of these measures here since there are excellent reviews on these topics readily available~\cite{RevModPhys.88.021002, Rivas2014} (for an in-depth \textit{quantitative} study of the sensitivity to memory effects of correlation-based measures, see, for example, Ref.~\cite{de_santis_witnessing_2020}).

With this, we come to the end of our very cursory discussion of quantum stochastic processes in terms of master equations and two-point correlations. We emphasize that the above is meant less as a pedagogical introduction to the field, but rather as a brief (incomplete) overview of the machinery that exists to model processes and detect memory by means of master equations and their properties. While very powerful and widely applicable in experimental settings, the reader should also have noted the natural shortcomings of this approach. On the one hand, it cannot account for multi-time statistics, thus not providing a complete framework for the definition and treatment of quantum stochastic processes. On the other hand, as a direct consequence of these shortcommings, we somehow had to awkwardly beat around the bush when it came to a proper definition of Markovianity in the quantum case. The remainder of this tutorial will be focused on working out the explicit origin of the difficulties with defining quantum stochastic processes, and how to overcome them.
\subsection{Troubles with quantum stochastic processes}
\label{sec::Troubles}

Do we need more (sophisticated) machinery than families of quantum stochastic maps and quantum master equations\ftnt{We have already argued that the two are equivalent, so only one will suffice.} to describe stochastic quantum phenomena? Perhaps for a long time, the machinery introduced above was sufficient. However, as quantum technologies gain sophistication and as we uncover unexpected natural phenomena with quantum underpinnings, the above tools do not suffice~\cite{PhysRevLett.94.200403, PhysRevA.73.022102, smirne_coherence_2017}. Take for example the pioneering experiments that have argued for the persistence of quantum effects on time scales relevant for photosynthetic processes~\cite{Lambert:2013xi,fmo1,plenio_dephasing-assisted_2008,fmo2,fmo3,caruso_entanglement_2010}, and, in particular, that these processes might exploit complex quantum memory effects arising from the interplay of the electronic degrees of freedom -- constituting the system of interest -- and the vibrational degrees of freedom -- playing the role of the environment. In these experiments, three ultra-short laser pulses are fired at the sample and then a signal from the sample is measured. The time between each pulse, as well as the final measurement, are varied. This system itself is mesoscopic and therefore certainly an open system. The conclusion from these experiments is based on the wave-like feature in the signal, see the video in the supplementary materials of Ref.~\cite{fmo1}. This experiment is fundamentally making use of four-time correlations and thus requiring more sophistication for its description than the above machinery affords us. 

Another important example is the mitigation of non-Markovian noise in quantum computers and other quantum technologies~\cite{brown, robin, flammia, White2020, crosstalk} which can display non-trivial multi-time statistics. Finally, as we already mentioned in our discussion of classical Master equations, in order to make assertions about multi-time statistics, it is inevitable to account for intermediate measurements, which cannot be done within approaches to quantum stochastic processes based on master equations. It seems reasonable then, to aim for a direct generalization of the description of classical stochastic processes in terms of multi-time statistics to the quantum realm. However, as we unveil next, there are fundamental problems that we must overcome first before we can describe multi-time quantum correlations as a stochastic process.

\subsubsection{Break down of KET in quantum mechanics}
\label{subsec::KET_QM}

\begin{figure}
 \centering
 \includegraphics[width=0.85\linewidth]{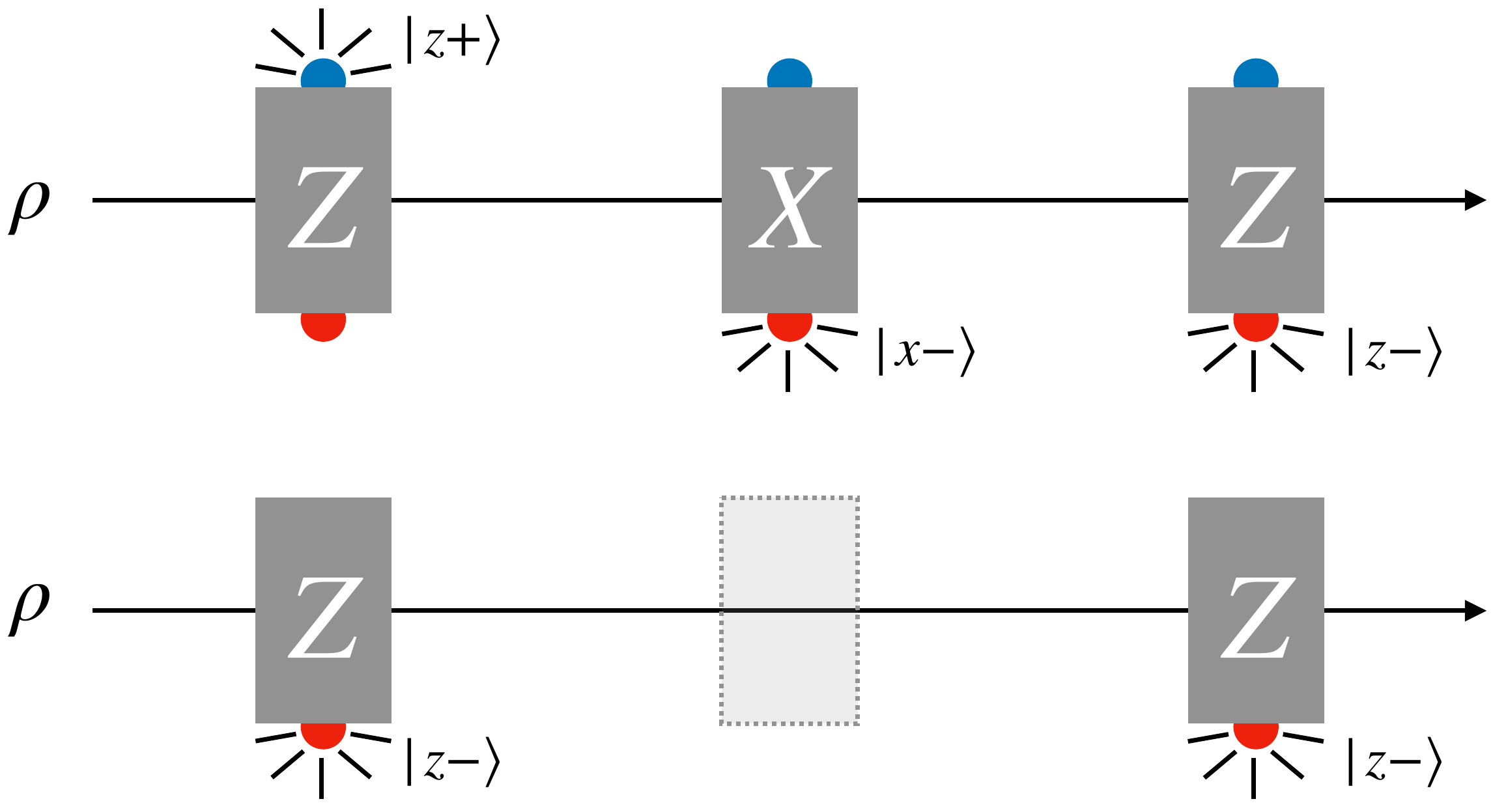}
 \caption{\textbf{Simple quantum process that violates the assumptions of the KET.} Successive measurements of the spin of a spin-$\tfrac{1}{2}$ particle do not allow one to predict the statistics if the intermediate measurement is not conducted. Here, measuring in the $x-$basis is invasive, and thus summing over the respective outcomes is not the same as not having done the measurement at all.}\label{fig:mz}
\end{figure}

As we have mentioned in Sec~\ref{sec::KET}, one of the fundamental theorems for the theory of classical stochastic processes, and the starting point of most books on them, is the Kolmogorov extension theorem (\textbf{KET}). It hinges on the fact that joint probability distributions of a random variable $S$ pertaining to a classical stochastic process satisfy consistency conditions amongst each other, like, for example, $\sum_{s_2} \Pprob(S_3, S_2=s_2,S_1) = \Pprob(S_3, S_1)$; a joint distribution on a set of times can always be obtained by marginalization from one on a larger set of times. Fundamentally, this is a requirement of non-invasiveness, as it implies that not performing a measurement at a time is the same as performing a measurement but forgetting the outcomes. 

While seemingly innocuous, this requirement is not fulfilled in quantum mechanics, leading to a breakdown of the KET~\cite{accardi_topics_1981}. To see this, consider the following concatenated Stern-Gerlach experiment~\cite{Milz2017KET} (depicted in Figure~\ref{fig:mz}): Let a qubit initially be in the state $\ket{x+} = \tfrac{1}{\sqrt{2}}(\ket{z+} + \ket{z-})$, where $\{\ket{z+}, \ket{z-}\}$ are the pure states corresponding to projective measurements in the $z$-basis yielding outcomes $z+,z-$. Now, the state is measured sequentially (with no intermediate dynamics happening) in the $z$-, $x$- and $z$-direction at times $t_1, t_2$ and $t_3$ (see Figure~\ref{fig:mz}). These measurements have the possible outcomes $\{z+,z+\}$ and $\{x+,x-\}$ for the measurement in $z$- and $x$-direction, respectively. It is easy to see that the probability for any possible sequence of outcomes is equal to $1/8$. For example, we have
\begin{gather}
\Pprob(z+,x+,z+) = \frac{1}{8}.
\end{gather}

Now, summing outcomes at time $t_2$, we obtain the marginal probability $\sum_{s_2=x\pm}\Pprob(z+,s_2,z+) = 1/4$. However, by considering the case where the measurement is \textit{not} made at $t_2$, it is easy to see that $\Pprob(S_3=z+,S_1=z+) = 1/2$. The intermediate measurement changes the state of the system, and the corresponding probability distributions for different sets of times are not compatible anymore~\cite{RevModPhys.88.021002, shrapnel_causation_2018}. Here, for example, when summing over the outcomes at $t_2$, the corresponding transformation of the state of the system corresponds to 
\begin{gather}
\begin{split}
    \rho_{t_2^-} \mapsto \rho_{t_2^+} &= \braket{x+|\rho_{t_2^-}|x+}\ketbra{x+}{x+} \\
    &\phantom{=}+ \braket{x-|\rho_{t_2^-}|x-}\ketbra{x-}{x-}\, ,
    \end{split}
\end{gather}
which, in general, does not coincide with the state $\rho_{t_2^-}$ right before $t_2$. Does this then mean that there is no singular object that can describe the joint probability for a sequence of quantum events? Alternatively, what object would describe a quantum stochastic process if it cannot be a joint probability distribution?

Seemingly, the breakdown of consistency conditions prevents one from properly reconciling the idea of an underlying process with its manifestation on finite sets of times, as we did in classical theory by means of the KET. However, somewhat unsurprisingly, this obstacle is one of formalism, and not a fundamental one, in the sense that marginalization is more subtle for quantum processes than it is for classical ones; introducing a proper framework for the description of quantum stochastic processes -- as we shall do below in Sec.~\ref{sec:QStochProc} -- brings with it a natural way of marginalization in quantum mechanics, that contains the classical version as a special case, and alleviates the aforementioned problems. 

\begin{figure}
\centering
\includegraphics[width=0.7\linewidth]{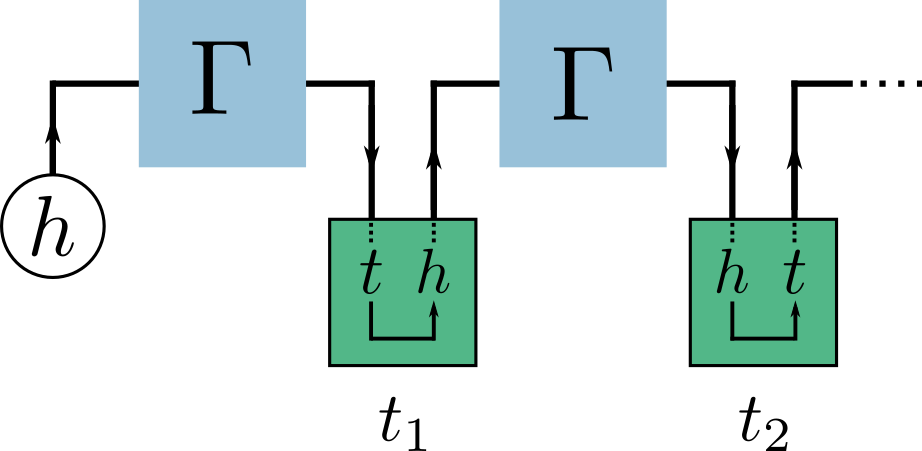}
\caption{\textbf{Perturbed coin with interventions.} Between measurements, the coin -- which initially shows heads -- is perturbed and stays on its side with probability $p$ and flips with probability $1-p$, leading to a stochastic matrix $\Gamma$ between measurements. Using their instrument, upon measuring an outcome, the experimenter flips the coin. Here, this is shown for the outcome of $hh$. For most values of the probability $p$, this process -- despite being fully classical -- does not satisfy the requirement of the KET.}
 \label{fig::Output_Input}
\end{figure}

\subsubsection{Input / output processes}
\label{sec::Input/Output}

As outlined above, the breakdown of the KET comes from the fact that in general, quantum measurements are invasive. Analogously, our understanding of classical stochastic processes, and with it the consistency between different observed joint probability distributions are built upon the idea that classical measurements are \textit{non}-invasive. However, depending on the `instrument' $\Jcal$ an experimenter uses to probe a system, this assumption of non-invasiveness might not be fulfilled, even in classical physics. 

To see this, consider the example of a perturbed coin, that flips with probability $p$ and stays on the same side with probability $1-p$ (see Figure~\ref{fig::Output_Input}). Instead of merely observing outcomes, an experimenter could actively interfere with the process. As there are many different ways, how the experimenter could interfere at each point in time, we have to specify the way in which they probe, or, in anticipation of later matters, what \textit{instrument} they use, which we will denote by $\Jcal$. 

For example, upon observing heads or tails, they could always flip the coin to tails and continue perturbing it. Or, upon observing an outcome, they could flip the coin, i.e., $h\mapsto t$ and $t\mapsto h$. Finally, they could just leave it on the side they found it in and let the perturbation process continue. Let us refer to the latter two instruments $\Jcal_\Fcal$ and $\Jcal_\Ical$, respectively.

Now, let us assume, that, before the first perturbation, the coin shows heads. Then, if at $t_1$ we choose the instruments $\Jcal_1 = \Jcal_\Fcal$ that, upon observing an outcome, flips the coin, we obtain, e.g.,
\begin{gather}
\label{eqn::SumProbClass}
 \begin{split}
 &\Pprob(F_2=h,F_1=t|\Jcal_1 =\Jcal_\Fcal) = p (1-p) \\
 &\Pprob(F_2=h,F_1=h|\Jcal_1 = \Jcal_\Fcal) = p (1-p).
 \end{split}
\end{gather}
This means that $\Pprob(F_2=h)=2p(1-p)$ when $\Jcal_1 = \Jcal_\Fcal$. On the other hand, if the experimenter does not actively change the state of the coin at the first time, i.e., $\Jcal_1 = \Jcal_\Ical$, upon perturbation, the coin will show $h$ with probability $1-p$ and $t$ with probability $p$ at time $t_1$. Then, the probability to observe $h$ at time $t_2$ is given by 
\begin{gather}
 \Pprob(F_2 = h) = (1-p)^2 + p^2, 
\end{gather} 
which does not coincide with $2p(1-p)$ except if $p=\tfrac{1}{2}$. Thus the two cases do generally not coincide and the requirements of the KET are not fulfilled. 

As, here, the experimenter can observe the output of the process, and freely choose what they input into the process, these processes are often called input-output processes and are subject of investigation in the field of computational mechanics~\cite{barnett_computational_2015}. A priori, it might seem arbitrary to allow for active interventions in classical physics. However, such operations naturally occur in the field of causal modeling~\cite{pearl_causality_2009}, where they can be used to deduce causal relations between events; indeed, the only way to see whether two events are causally connected is to change a parameter at one of them and see if this change affects the outcome statistics at the other one. 

On the other hand, while in classical physics, it is a choice (up to experimental accuracy that is) to actively interfere with the process at hand, in quantum mechanics, such an active intervention due to measurements -- even projective ones -- can generally not be avoided. Considering classical processes with interventions thus points us in the right direction as to how quantum stochastic processes should be modeled. 

Concretely, when active interventions are employed, the outcome statistics are conditional on the choices of instruments the experimenter made to probe a process at hand. Consequently, such a setup would not be described by a single joint probability distribution $\Pprob(F_n, \dots, F_1)$, but rather by conditional probabilities of the form $\Pprob(F_n, \dots, F_1|\Jcal_n, \dots, \Jcal_1)$. It is exactly this dependence of observed probability distributions on the employed instruments that we will encounter again when describing quantum stochastic processes. 

Given that the breakdown of the KET can even occur in classical physics, one might again pause and wonder if there actually exists such a thing as a classical stochastic process with interventions. Put differently, is there an underlying statistical object that is independent of the made interventions and can thus be considered the underlying process. While we will discuss in detail that this is indeed the case, recalling the above example of interventions that are used to unveil causal relations between events already tells us why the answer will be affirmative. Indeed, causal relations between events, and the strength with which different events can potentially influence each other are independent of what experimental interventions are employed to probe them.

Interestingly, the breakdown of the requirements of the KET is closely related to the violation of Leggett-Garg inequalities in quantum mechanics~\cite{leggett_quantum_1985, emary_leggettgarg_2014}, which, in brief, were derived to distinguish between the statistics of classical and non-classical processes. These inequalities are derived on the assumption of realism \textit{per se} and non-invasive measurability. While realism \textit{per se} implies that joint probability distributions for a set of times can be expressed as marginals of a respective joint probability distribution for more times, non-invasiveness means that all finite distributions are marginals of the \textit{same} distribution. Naturally then, as soon as one of these conditions does not hold, the KET can fail and Leggett-Garg inequalities can be violated. More precisely, if one allows for active interventions in the classical setting, without any additional restrictions, then classical processes can exhibit exactly the same joint probability distributions as quantum mechanics~\cite{budroni_memory_2019} (this equivalence changes once one imposes, for example, dimensional restrictions).

\subsubsection{KET and spatial quantum states}
\label{sec::SpatStat}

\begin{figure}
 \centering
 \includegraphics[width=0.9\linewidth]{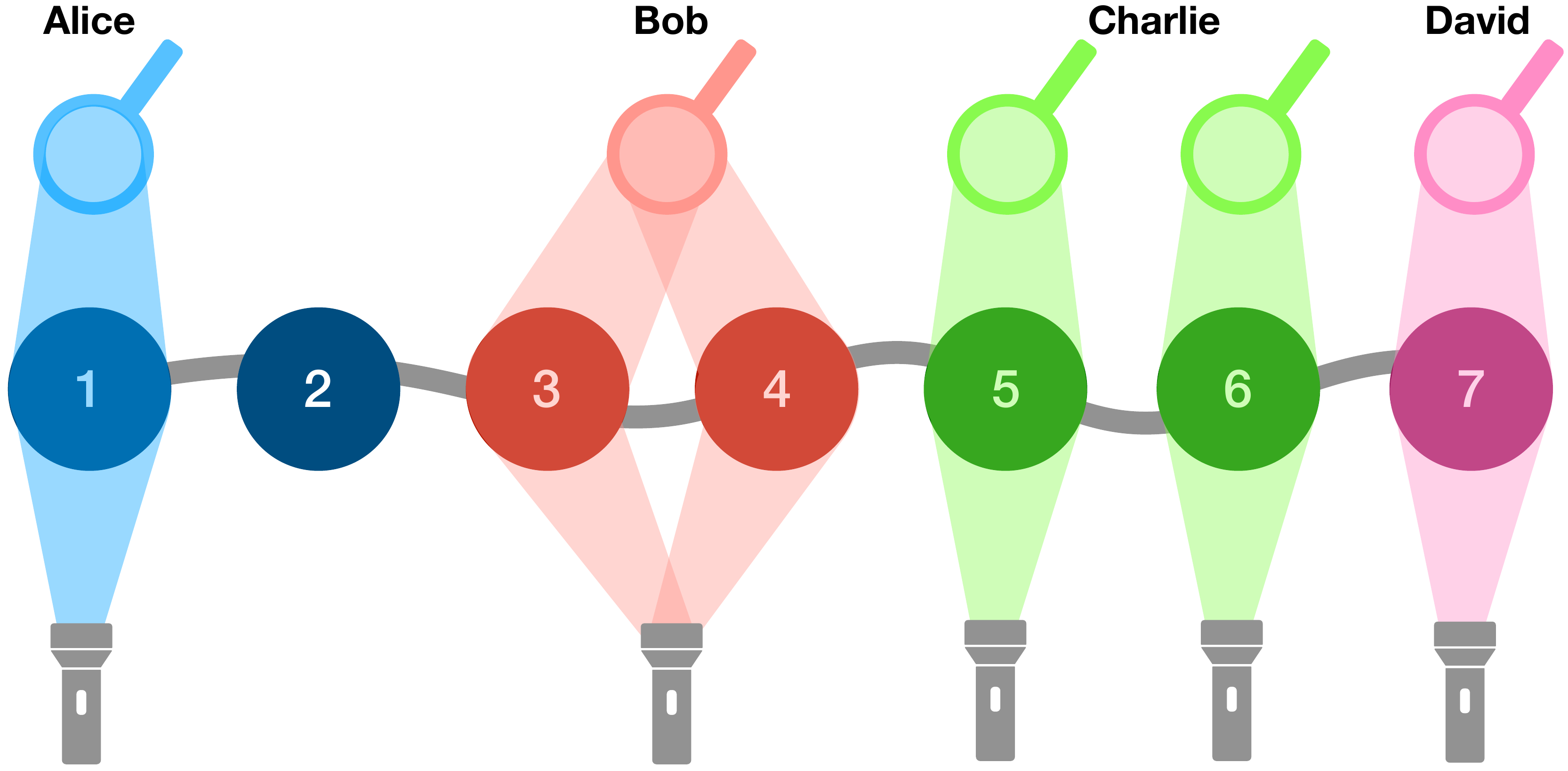}
 \caption{\textbf{Spatial Measurements.} Alice, Bob, Charlie, and David perform measurements on a seven-partite quantum state $\rho$. Both Bob and Charlie have access to two parts of said state, respectively, but while Bob can perform correlated measurements on said systems, Charlie can only access them independently. The probabilities corresponding to the respective outcomes are computed via the Born rule (see Eq.~\eqref{eqn::Multi_Born}).}\label{fig:spacemeas}
\end{figure}

Before finally advancing to quantum stochastic processes, it is instructive -- as a preparation -- to reconsider the concept of states in quantum mechanics, in the context of measurements. To this end, consider the situation depicted in Figure~\ref{fig:spacemeas}, where four parties (Alice, Bob, Charlie, and David) measure separate parts of a multipartite quantum state. In the general case, their measurements are given by Positive operator-valued measures (POVMs) denoted by $\Jcal_X$, where each outcome $j$ corresponds to a positive matrix $X_j$, and we have $\sum_j X_j = \openone$. Then, according to the Born rule, probabilities for the measurements depicted in Figure~\ref{fig:spacemeas} are computed via
\begin{gather}
\begin{split}
 &\Pprob(a,b,c,d|\Jcal_A,\Jcal_B,\Jcal_C,\Jcal_D) \\
 & = \tr[\rho (A_{a_1} \otimes \openone_{2} \otimes B_{b_{34}} \otimes C_{c_5} \otimes C_{c_6} \otimes D_{d_7})],
 \end{split}
 \label{eqn::Multi_Born}
\end{gather}
where $\rho:=\rho_{1234567}$ is the probed multipartite state, and $\{X_{a_m}\}$ is the POVM operator for party $X$ with outcome $a$ when measuring system $m$. We use the double subscript notation to label the operator index and the system at once. The above probability depends crucially on the respective POVMs the parties use to probe their part of the state $\rho$. This dependence is denoted by making the probability contingent on the instruments $\Jcal_X$. As soon as $\rho$ is known, all joint probabilities for all possible choices of instruments can be computed via the above Born rule. In this sense, a quantum state represents the maximal statistical information that can be inferred about spatially separated measurements.

While, pictographically, Figure~\ref{fig:spacemeas} appears to be a direct quantum version of the classical stochastic processes we encountered previously, there is a fundamental difference between spatially and temporally separated measurements: In the spatial setting, none of the parties can signal to the others. For example, we have 
\begin{gather}
\begin{split}
&\sum_c \Pprob(a,b,c,d|\Jcal_A,\Jcal_B,\Jcal_C,\Jcal_D) = \\
& \quad \sum_{c'} \Pprob(a,b,c',d|\Jcal_A,\Jcal_B,\Jcal_C',\Jcal_D)
\end{split}
\end{gather}
for all instruments. Put differently, the quantum state a subset of parties sees is independent of the choice of instruments of the remaining parties. This is also mirrored by the fact that we model the respective measurement outcomes by POVM elements, which make no assertion about how the state at hand transforms upon measurement. 

On the other hand, the possible breakdown of the KET in quantum mechanics and classical processes with interventions shows that, in temporal processes, an instrument choice at an earlier time can influence the statistics at later times. To accommodate for this kind of signaling between different measurements, we will have to employ a more general description of measurements, that accounts for the transformations a quantum state undergoes upon measurement. However, the general idea of how to describe temporal processes can be directly lifted from the spatial case: as soon as we know how to compute the statistics for all sequences of measurements and all choices of (generalized) instruments, there is nothing more that can be learned about the process at hand. Unsurprisingly then, we will recover a temporal version of the Born rule~\cite{chiribella_memory_2008, Shrapnel_2018}, where the POVM elements are replaced by more general completely positive (CP) maps, and the spatial quantum state is replaced by a more general quantum comb that contains all detectable spatial \textit{and} temporal correlations.

\section{Quantum Stochastic Processes}
\label{sec:QStochProc}

In the last section, we saw various methods to look at two-time quantum correlations. While indispensable tools for the description of many experimental scenarios, these methods are not well-suited to describe multi-time statistics, and as such do not allow one to extend the notion of Markovianity -- or absence thereof -- to the quantum case in a way that boils down to the classical one in the correct limit. We now introduce tools that will allow us to consistently describe multi-time quantum correlations, independently of the choice of measurement. Before doing this, it is worth elaborating on the source of the troubles in way of a theory of quantum stochastic processes.

\subsection{Subtleties of the quantum state and quantum measurement}
\label{subsec::Qmeas}

Let us use the initial correlation problem in quantum mechanics as an example. This problem has been fraught with controversies for decades now~\cite{shaji_whos_2005} as some researchers have argued that, in presence of initial correlations, a dynamical map is not well defined~\cite{pechukas2}, while others have argued to give up complete positivity or linearity~\cite{Alicki95, shaji_whos_2005}. What is the underlying reason for these disagreements? And does the same problem exist in classical mechanics? 

The answer to this latter question is no. The crucial difference being that it is possible to observe classical states without disturbing the system, while the same cannot be said for quantum states. Consider a classical experiment that starts with an initial system-environment state that is correlated between the system of interest and some environment. The overall process takes the initial state $\Lambda_{(t:0)}: \Pprob(\stxt_0\etxt_0) \mapsto \Pprob(\stxt_t\etxt_t)$. Of course, we can simply observe the system (without disturbing it) and measure the frequencies for $\stxt_0 = s_j \mapsto \stxt_t = s_k$. This is already enough to construct joint distribution $\Pprob(\stxt_t,\stxt_0)$ and from it, we can construct a stochastic matrix $\Gamma_{(t:0)}$ that takes the initial system state to the final state. In other words, the initial correlations pose no obstacles at all here. This should not be surprising, after all, a multi-time classical process will have system-environment correlations at some point. And we have already argued that it is always possible to construct a stochastic matrix between any two points.

If we try to repeat the same reconstruction process for the quantum case, we quickly run into trouble. Again, without any controversy, we can imagine that an initial system-environment quantum state is being transformed into a final one, $\rho_{\setxt}(0) \mapsto \rho_{\setxt}(t)$. It may be then tempting to say that we can also have a transformation on the reduced state of the system $\rho_{\stxt}(0) \mapsto \rho_{\stxt}(t)$. However, we run into trouble as soon as we try to determine the process $\rho_{\stxt}(0) \mapsto \rho_{\stxt}(t)$. In order to do this, we need to relate different initial states and the corresponding final states. Do we then mean that there is an initial set of states $\{\rho_{\setxt}(0)\}$ and for each element of the set here we have a different initial state for the system, i.e., $\{\rho_{\stxt}(0) = \tr_E [\rho_{\setxt}(0)]\}$? This is possible of course but requires knowledge about the environment, which goes against the spirit of the theory of open systems where we assume that the experimenter does not have control over the environmental degrees of freedom.

Our problem is still more profound when we focus solely on $\stxt$. Suppose that the above setup is true and the initial system states $\{\rho_{\stxt}(0)\}$ are linearly independent, constituting an input basis. But then in a given run of the experiment, how do we know which element of this set we have at our disposal? Quantum mechanics fundamentally forbids us to unambiguously discriminate a set of non-orthogonal states; if the set $\{\rho_{\stxt}(0)\}$ contains $d^2$ linearly independent basis elements, then at most $d$ of them can be orthogonal. Therefore, quantum mechanics fundamentally forbids us from experimentally deducing the dynamical map when there are initial correlations!

This contextual nature of quantum states is the key subtlety that forces a fundamentally different structure for quantum stochastic processes than classical ones. Perhaps, a theorist may be tempted to say that never mind the experiments and let us construct the map with a theoretical calculation, i.e., first, properly define the $\setxt$ dynamics then infer the process on $\stxt$ alone. This is in fact what was done by many theorists in the past two decades. They asked what happened if we fixed the correlations in the initial state $\rho_{\setxt}(0)$ and consider the family of $\rho_{\stxt}(0)$ states that are compatible with the former, see Example 1 in Sec.~\ref{sec::InitCorr}. Can we construct a map? These types of constructions are precisely what led to not completely positive maps. However, do such calculations have a correspondence with reality then~\cite{modi_role_2010}? 
The real source of the problem (in the technical sense) is that -- as we have seen when we discussed the experimental reconstruction of quantum channels -- we need an informationally complete set of initial states and corresponding final states to have a well-defined map. For an experimenter, there is an `easy' solution. You simply go ahead and prepare the initial state as desired, which can even be noisy~\cite{modiosid}. Then let this initial state evolve and measure the corresponding final state. In fact, this is the only way, in quantum mechanics, to ensure that we have a linearly independent set of input states, whose output states are also accessible. Without preparation at the initial time, we only have a single point in the state space, i.e., the reduced state $\rho_\stxt = \tr_\etxt(\rho_{\stxt\etxt})$, and a map is only defined on a dense domain. However, in general, when there are system-environment correlations, the preparation of input states will affect the state of the environment, which, in turn, will influence the subsequent dynamics, seemingly making it non-linear in the sense that the dynamics itself depends on the input state. See example two in Sec.~\ref{sec::InitCorr} and subsequent discussion.

This, in turn, raises the question whether finite set of such experiments have enough information to construct a well-defined dynamical map? And will this mapping be linear (and, in a sense to be defined below, CP and trace preserving)? Somewhat surprisingly, despite all the apparent roadblocks we sketched above, the answer is yes! However, to achieve this goal, it is necessary to switch our understanding of what a dynamical map actually is when there are initial correlations. Giving away the punchline of the following sections, here, it is not meaningful to define a mapping from initial to final states, but rather from initial \textit{preparations} to final states. It is easy to show that there is only a finite number of preparations that are linearly independent (for finite-dimensional systems). And therefore, any other preparations can be expressed as a linear combination of a fixed set of preparations. Since for each initial preparation it is possible to determine the final output state, very much in the same way as we already saw for quantum channel tomography, one can unambiguously reconstruct a map that correctly maps all inputs (here: the initial preparations) to the final output states. We will flesh out and extend these ideas in this and the following sections.

First, we lay out the mathematical foundations for the notion of preparation, which is historically known as an instrument and which generalizes POVMs. With these tools, we will show that the solution to the initial correlation problem is well-defined, completely positive, and linear all at once~\cite{modiscirep}. Moreover, this is then a pathway to laying down the foundations for quantum stochastic processes, since it will be directly generalizable to multi-time scenarios.

\subsection{Quantum measurement and instrument}
\label{sec::Queasurement}
As mentioned in Sec.~\ref{sec::Troubles}, unlike in the case of spatially separate measurements, in the temporal case, it is important to keep track of how the state of the system of interest changes upon being measured, as this change will influence the statistics of subsequent measurements. In order to take this into account, we work with the concept of generalized instruments introduced by Davies and Lewis~\cite{Davies:1970ez}. This will both allow us to overcome the problems with initial correlations lined out above, as well as provide a fully fledged theory of quantum stochastic processes that can account for intermediate measurements, and, as such, multi-time correlations.

To this end, first, recall the definition of a POVM provided in Sec.~\ref{subsec::POVM}. A POVM is a collection of positive matrices $\Jcal=\{E_j\}_{j=1}^n$ with the property $\sum_j E_j=\ident$. Each element of $\Jcal$ corresponds to a possible outcome of the measurement. Intuitively, a POVM allocates an operator to each measurement outcome of the measurement device that allows one to compute outcome statistics for arbitrary quantum states that are being probed. However, it does not enable one to deduce how the state changes upon observation of one of the outcomes. 

To account for state changes, we have to modify the concept of a POVM; this generalization is known as a generalized instrument~\cite{davies76a, lindblad_non-markovian_1979}. As POVMs turn out to be a special case of (generalized) instruments, we will denote them by $\Jcal$, too. An instrument corresponding to a measurement with outcomes $j=\{1,\dots, n\}$ is a collection of CP maps $\Jcal = \{\Acal_j\}$ that add up to a CPTP map, i.e., $\Acal = \sum_{j=1}^n \Acal_j$. Each of the CP maps corresponds to one of the possible outcomes, while their sum corresponds to the overall transformation of the state at hand due to the application of the respective instrument (it is exactly the invasiveness of said map that leads to a breakdown of the KET in quantum mechanics). For example, returning to the case of a measurement of a qubit in the computational basis, the corresponding instrument is given by 
\begin{gather}
\{\Acal_0[\!\sbt] := \ketbra{0}{0} \sbt\,\ketbra{0}{0}, \ \Acal_1[\!\sbt] := \ketbra{1}{1}\sbt\, \ketbra{1}{1}\},
\end{gather}
assuming, that after projecting the state onto the computational basis, it is sent forward unchanged.

Importantly, an instrument allows one to compute \textit{both} the probability to obtain different outcomes \textit{and} the state change upon measurement. The latter is given by 
\begin{gather}
\rho'_j = \Acal_j[\rho]
\end{gather}
when the system in state $\rho$ is interrogated by the instrument $\Jcal$, yielding outcome $j$. The state after said interrogation, given the outcome, is obtained via the action of the corresponding element of the instrument. Importantly, this state is in general \textit{not} normalized. Its trace provides the probability to observe a given outcome. Concretely, we have 
\begin{gather}
\label{eqn::ProbCP}
 \Pprob(j|\Jcal) = \tr(\Acal_j[\rho]) = \tr\Big(\sum_{\alpha_j} K_{\alpha_j} \rho K_{\alpha_j}^{\dagger}\Big), 
\end{gather}
where the sum runs over all Kraus operators that pertain to the CP map $\Acal_j$, and we have $\sum_{\alpha_j} K_{\alpha_j}^\dagger K_{\alpha_j} < \ident$ if $\Acal_j$ is not trace preserving.
The requirement that all CP maps of an instrument add up to a CPTP map ensures -- just like in the analogous case for POVMs -- the normalization of probabilities: 
\begin{gather}
 \sum_{j=1}^n \Pprob(j|\Jcal) 
 = \tr\left(\sum_{j=1}^n \sum_{\alpha_j} K_{\alpha_j} \rho K_{\alpha_j}^\dagger\right)
 =\tr(\Acal[\rho]),
 \end{gather}
which is 1 for all $\rho$. Naturally, the concept of generalized instruments contains POVMs as a special case, namely as those generalized instruments, where the output space of the respective CP maps is trivial. Put differently, if one simply wants to compute the probabilities of measurements on a quantum state, generalized instruments are not necessary. Concretely, we have 
\begin{gather}
 \Pprob(j|\Jcal) 
 = \tr(\rho E_j) \quad \mbox{with} \quad E_j = \sum_{\alpha_j} K_{\alpha_j}^\dagger K_{\alpha_j}.
\end{gather}
This is because for a single measurement, the state transformation is not of interest. However, as we will see in the next section, this situation drastically changes, as soon as sequential measurements are considered. There, POVMs are not sufficient anymore to correctly compute statistics. 

Before advancing, it is insightful to make the connection between the CP maps of an instrument and the elements of its corresponding POVM explicit. This is most easily done via the Choi states we introduced in Sec.~\ref{subsec::Representation}. There, we discussed that the action of a map $\Acal_j$ on a state $\rho$ can be expressed as 
\begin{gather}
 \rho'_j = \Acal_j[\rho] = \tr_\inp[\Asf^\trps_j(\rho \otimes \openone^\out)], 
\end{gather}
where $\Asf_j \in \Bcal(\Hcal^\inp \otimes \Hcal^\out)$ is the Choi state of the map $\Acal_j$ and $\rho \in \Bcal(\Hcal^\inp)$, and we have moved the transposition onto $\Asf_j$ instead of $\rho$. Using this expression to compute probabilities, we obtain 
\begin{gather}
 \Pprob(j|\Jcal) = \tr_{\inp\out}[\Asf^\trps_j (\rho \otimes \openone^\out)] = \tr(E_j \rho). 
\end{gather}
Comparing this last expression with the Born rule, we see that the POVM element $E_j$ corresponding to $\Acal_j$ is given by $E_j = \tr_\out(\Asf_j^\trps)$, where the additional transpose stems from our definition of the CJI. This definition indeed yields a POVM, as the partial trace of a positive matrix is also positive, and we have $\sum_{j=1}^n E_j = \sum_{j=1}^n \tr_\out (\Asf_j^\trps) = \openone$, where we have used that the Choi state $\Asf$ of $\Acal$ satisfies $\tr_\out(\Asf) = \openone$. Discarding the outputs of an instruments thus yields a POVM. This implies that different instruments can have the same corresponding POVM. For example, the instrument that measures in the computational basis and feeds forward the resulting state, has the same corresponding POVM as the instrument that measures in the computational basis, but feeds forward a maximally mixed state, indiscriminate of the outcome. While both of these instruments lead to the same POVM, their influence on future statistics is very different.

\subsubsection{POVMs, Instruments, and probability spaces}
\label{sec::POVMInst}
Before advancing to the description of multi-time quantum processes, let us quickly connect POVMs and instruments to the more formal discussion of stochastic processes we conducted earlier. The benefit of making this connection transparent is two-fold; on the one hand, it recalls the original ideas, stemming from the theory of probability measures, that led to their introduction in quantum mechanics. On the other hand, it renders the following discussions of quantum stochastic processes a natural extension of both of the concepts of instruments, as well as the theory of classical stochastic processes. 

In the classical case, we described a probability space as a $\sigma$-algebra (where each observable outcome corresponds to an element of the $\sigma$-algebra) and a probability measure $\omega$ that allocates a probability to each element of said $\sigma$-algebra. Without dwelling on the technical details (see, e.g., Refs.~\cite{Davies1969, lindblad_non-markovian_1979} for more thorough discussions), this definition can be straightforwardly extended to POVMs and instruments. However, instead of directly mapping observable outcomes to probabilities, in quantum mechanics, we have to specify \textit{how} we probe the system at hand. Mathematically, this means that instead of mapping the elements of our $\sigma$-algebra to probabilities, we map them to positive operators via a function $\xi$ that satisfies the properties of a probability measure (hence the name positive operator-valued \textit{measure}). For example, the POVM element corresponding to the union of two disjoint elements of the $\sigma$-algebras is the sum of the two individual POVM elements, and so on. Together with the Born rule, each POVM then leads to a distinct probability measure on the respective $\sigma$-algebra. Concretely, denoting the Born rule corresponding to a state $\rho$ by 
$\chi_\rho[E] = \tr(\rho E)$, then $\omega_\rho = \chi_\rho \circ \xi$ is a probability measure on the considered $\sigma$-algebra. 

For instruments, the above construction is analogous, but with POVM elements replaced by CP maps. It is then a natural step to assume that, in order to obtain probabilities, a \textit{generalized} Born rule~\cite{chiribella_memory_2008, Shrapnel_2018}, that maps CP maps to the corresponding probabilities. More generally yet, \textit{sequences} of measurement outcomes correspond to sequences of CP maps, and a full description of the process at hand would be given by a mapping of such sequences to probabilities. In the next section, we will see that this reasoning indeed leads to a consistent description of quantum stochastic processes that -- additionally -- resolves the aforementioned problems, like, e.g., the breakdown of the Kolmogorov extension theorem.

\subsection{Initial correlations and complete positivity}\label{subsec::IC}

With the introduction of the instrument, we are now in a position to operationally resolve the initial correlation problem alluded to above. Importantly, the resolution of this special case will directly point us in the right direction of how to generalize stochastic processes to the quantum realm, which is why we consider it first.

We begin with an initial system-environment quantum state that is correlated. Now, in a meaningful experiment, that aims to characterize the dynamics of the system from the initial time to the final time, one will apply an instrument $\Jcal=\{\Acal_j\}$ on the system alone at the initial time to prepare it into a known (desired) state. To be concrete, this instrument could, for example, be a measurement in the computational basis, such that each $\Acal_j$ is a trace non-increasing CP map with action $\Acal_j[\rho] = \braket{j|\rho|j}\ketbra{j}{j}$. Importantly, though, we impose no limitation on the set of admissible instruments an experimenter could use. Next, the total $\setxt$ state propagates in time via a map 
\begin{gather}\label{eq:ucal}
\Ucalt{(t:0)} [\sbt \,]:= U_{(t:0)} (\sbt \,) U_{(t:0)}^\dag.
\end{gather}
Note that, due to the dilation theorem in Sec.~\ref{sec:higherorderdilation} we can always take the system-environment propagator to be unitary. Taking the propagator to be a CPTP map would make no difference at the system level. The full process can written down as
\begin{gather}\label{eq:superchannel}
 \rho_j(t) = \tr_\etxt \{\Ucalt{(t:0)} \circ (\Acal_j \otimes \Ical)[ \rho_{\setxt}(0)]\}.
\end{gather}
Above $\Ical$ is the identity map on the $\etxt$ as the instrument acts only on $\stxt$. While perfectly correct, whenever $\rho_{\setxt}(0)$ is not of product form, the above does not allow one to obtain a (physically meaningful) mapping that takes input states of the system and maps them to the corresponding output states at a later time.

Now, let us recall that a map (in quantum, classical, and beyond physics) is nothing more than a relationship between experimentally controllable inputs and measurabble outputs. Here, the inputs are the choice of the instrument $\Jcal = \{\Acal_j\}$ -- which can be freely chosen by the experimenter -- and the corresponding outcome is $\rho_\stxt(t)$ -- which can be determined by means of quantum state tomography. Then, right away, by combining everything that is unknown to the experimenter in Eq.~\eqref{eq:superchannel} into one object, we have the map
\begin{gather}
\label{eq:superchannelIntro}
\rho_j(t) = \Tcal_{(t:0)}[\Acal_j].
\end{gather}
The map $\Tcal_{(t:0)}$ was introduced in Ref.~\cite{modiscirep} and was referred to as the superchannel in Ref.~\cite{PhysRevLett.114.090402}, where it was first realized experimentally. By comparing Eqs.~\eqref{eq:superchannel} and~\eqref{eq:superchannelIntro}, we see that the action of the superchannel is given by 
\begin{gather}
\label{eqn::DefSupChann}
    \Tcal_{(t:0)}[\sbt] = \tr_\etxt \{\Ucalt{(t:0)} \circ (\sbt \otimes \Ical)[ \rho_{\setxt}(0)]\}\, ,
\end{gather}
which is a linear map on the operations $\Acal_i$ it is defined on. While, in contrast to the case of quantum channels, $\Tcal_{(t:0)}$ does not act on states but on \textit{operations}, we emphasize the operational similarities between quantum channels and superchannels. On the one hand, they are both `made up' of all the parts of the evolution that are not directly accessible to the experimenter; the initial state of the environment and the unitary system-environment evolution in the case of quantum channels, and the initial system-environment state and the unitary system-environment evolution in the case of superchannels. Additionally, they both constitute a mapping of what can be freely chosen by the experimenter to a later state of the system at time $t$. Unsurprisingly then, as we shall see below, quantum channels can be considered to be just a special case of superchannels.

Ref.~\cite{modiscirep} proved that -- besides being linear -- the map $\Tcal_{(t:0)}$ is completely positive and trace-preserving (in a well-defined sense); and clearly, it is well-defined for \textit{any} initial preparation $\Acal_j$. The trace-preservation property means that if $\Acal$ is CPTP then the output will be unit-trace. See Ref.~\cite{PhysRevA.100.042120} for further discussion and theoretical development with respect to open system dynamics of initially correlated systems.

\begin{figure}[t]
 \centering
 \includegraphics[width=0.7\linewidth]{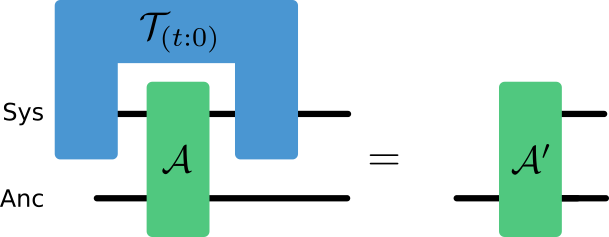}
 \caption{\textbf{Complete Positivity and Trace Preservation for Superchannels.} A superchannel is said to be CP if it maps CP maps to CP maps (even when acting on only a part of them), and we call it CPTP if it maps CPTP maps to CPTP maps. Here, $\Tcal_{(t:0)}$ is CP (CPTP) if for all CP (CPTP) maps $\Acal$ and all possible ancilla sizes, the resulting map $\Acal'$ is also CP (CPTP). Note that for the TP part, it is already sufficient that $\Tcal_{(t:0)}$ maps all CPTP maps on the system to a unit trace object.}
 \label{fig::CPPTTP}
\end{figure}

The meaning of complete positivity for this map is operationally clear and very analogous to the case of quantum channels; suppose the instrument $\Jcal$ acts not only on the system $\stxt$, but also on an ancilla. Then the superchannel's complete positivity guarantees that the result of its action on \textit{any} CP map -- which could be  acting on the system and an additional ancilla -- is again a CP map (see Figure~\ref{fig::CPPTTP} for a graphical depiction). We will not provide a direct proof of this statement here. However, as we will discuss below, it is easy to see that $\Tcal_{(t:0)}$ has a positive Choi state, which implies complete positivity in the above sense. Since the explicit computation of $\Tcal_{(t:0)}$ from $\Ucal_{(t:0)}$ and $\rho_{\setxt}(0)$ requires -- as for the case of quantum channels -- the choice of an explicit representation, we will also relegate it to Sec.~\ref{sec::ManyChoi}, where we discuss Choi states of higher order quantum maps in more detail.

The `TP' property of superchannels means that $\Tcal_{(t:0)}$ maps any trace preserving map $\Acal$ to a unit trace object. Indeed, with the Definition~\eqref{eqn::DefSupChann} of superchannels in mind, we see that, since $\tr_\etxt$ and $\Ucal_{(t:0)}$ are trace preserving and $\rho_{\setxt}(0)$ is a unit trace state, $\Tcal_{(t:0)}[\Acal]$ amounts to a concatenation of trace preserving maps acting on a unit trace state, thus yielding a unit trace object. This, then, also implies that, whenever $\Acal_j$ is trace non-increasing, $\rho_j(t) = \Tcal_{(t:0)}[\Acal_j]$ is subnormalized and its trace amounts to the probability of the map $\Acal_j$ occurring. This is a simple consequence of the fact that $\Tcal_{(t:0)}$ is made up of trace preserving elements only, and, as we have discussed around Eq.~\eqref{eqn::ProbCP}, the trace of the output of a CP map yields its implementation probability.

Importantly, as mentioned, the superchannel is a higher-order map as its domain is the set of CP maps and the image is density operators. Clearly, this is different from the quantum stochastic matrix. In fact, the superchannel is the first step beyond two point quantum correlations. This is most easily seen from its Choi state, which is a bounded operator on three Hilbert spaces: $\Upsilon_{(t:0)} \in \Bcal(\Hcal_{0}^\inp \otimes \Hcal_{0}^\out \otimes \Hcal_{1}^\inp)$ (details for constructing the Choi state of higher order maps can be found below and in Sec.~\ref{sec::ManyChoi}). Moreover, the superchannel contains `normal' quantum channels as a limiting case: when there are no initial correlations, i.e., $\rho_{\setxt}(0) = \rho_\stxt(0) \otimes \rho_\etxt(0)$, then the superchannel reduces to the usual CPTP map:
\begin{gather}
\begin{split}
&\Tcal_{(t:0)}[\Acal_j] = (\Ecal_{(t:0)} \circ \Acal_j) [\rho_\stxt(0)], \quad \mbox{where} \\
&\Ecal_{(t:0)}[\!\sbt] = \tr_{\etxt}\{\Ucalt{(t:0)} [\sbt \otimes \rho_\etxt]\},
\end{split}
\end{gather}
which can be seen by direct insertion of the product state assumption into Eq.~\eqref{eqn::DefSupChann}.

The superchannel is a primitive for constructing the descriptor of quantum stochastic processes. As such, it should be operationally accessible via a set of experiments, in the same vein as quantum channels are experimentally reconstructable. Somewhat unsurprisingly, the reconstruction procedure for superchannels works in a similar way as that for quantum channels; the input of the superchannel, CP maps, span a vector space that has a basis consisting of CP maps. This means that the superchannel is fully determined by its action on the CP maps $\{\hat \Acal_j\}$ that form a linear basis. Concretely, let the output states corresponding to this basis of input operations be
\begin{gather}
\Tcal_{(t:0)} [\hat\Acal_j] = \rho_j(t). 
\end{gather}
Now, this informationally complete input-output relation can be used to represent the superchannel $\Tcal_{(t:0)}$. As was already the case for channels, this can be done in terms of duals, but this time not in terms of duals for a set of input \textit{states}, but a set of input \textit{operations} $\{\hat \Acal_j\}$. While there is no conceptual problem with duals of maps, let us avoid this additional level of abstraction. Rather, here, we opt to directly choose the Choi state representation of the superchannel, in the same spirit as Eq.~\eqref{eq:tomrepmap}. To this end, let $\{\hat{\mathsf{A}}_j\}$ be the Choi states of the maps $\{\hat \Acal_j\}$, and let $\{\hat D_j\}$ be the corresponding set of dual matrices, i.e., $\tr({\hat D_j}^\dagger {\hat{\mathsf{A}}}_k) = \delta_{jk}$. Then, the Choi state of $\Tcal_{(t:0)}$ can be written as
\begin{gather}
\label{eqn::supChoiConstr}
\Upsilon_{\Tcal_{(t:0)}} = \sum_j \rho_j(t) \otimes \hat{D}_j^*, 
\end{gather}
and its action on an arbitrary map $\Acal$ is given by 
\begin{gather}
\label{eqn::ActionSupChoi}
    \Tcal_{(t:0)} [\Acal] = \tr_\inp[(\openone^\out \otimes \mathsf{A}^\trps) \Upsilon_{\Tcal_{(t:0)}}]
\end{gather}
where $\mathsf{A}$ is the Choi state of the map $\Acal$. While somewhat more complex than in the case of quantum channels, this form should not come as a surprise; indeed, it simply expresses a linear input-output relation. Eq.~\eqref{eqn::supChoiConstr} `attaches' the correct output state to each dual of a basis map, and Eq.~\eqref{eqn::ActionSupChoi} guarantees that the action of $\Tcal_{(t:0)}$ is properly defined on all basis elements, and thus on all maps $\Acal$. The fact that we went via the Choi representation of the maps is then rather a mathematical convenience than a conceptual leap. Below, we will discuss Choi states of higher-order maps in more detail, and also argue why the above object $\Upsilon_{\Tcal_{(t:0)}}$ can rightfully be called a Choi state of $\Tcal_{(t:0)}$. For the moment, let us emphasize once again that $\Upsilon_{\Tcal_{(t:0)}}$ together with the action given by Eq.~\eqref{eqn::ActionSupChoi} yields the correct output state for \textit{any} input operation $\Acal$; any CP map can be cast as a linear sum of the basis maps as $\Acal = \sum_{j} \alpha_{j} \hat\Acal_j $. The action of $\Tcal_{(t:0)}$ defined above yields the correct output state for a basis $\{\hat \Acal_j\}$, since 
\begin{gather}
\begin{split}
    \Tcal_{(t:0)} [\hat{\Acal}_j] &=  \sum_{k=1}^{d^2} \tr_\inp[(\openone^\out \otimes {\hat{\mathsf{A}}}_j^\trps) (\rho_k(t) \otimes \hat{D}_k^*)] \\
    &=  \sum_{k=1}^{d^2} \rho_k(t)  \tr(\hat{D}_k^\dagger \hat{\mathsf{A}}_j) = \rho_j(t), 
\end{split}
 \end{gather}
where we have used the duality of $\hat{D}_k^\dagger$ and $\hat{\mathsf{A}}_j$. Consequently, the superchannel defined in this way indeed provides the correct mapping on on all conceivable CP maps $\Acal$.

\emph{Example.} Armed with this new operational understanding of open system dynamics, we now revisit the example from Sec.~\ref{sec::InitCorr}, where we discussed open system dynamics in the presence of initial correlations. Previously, we saw that the usual CPTP map fails to describe this process. Now, as promised, we will see that the superchannel can describe this process adequately. To do so, let us first write down linear basis of CP maps on a qubit as
\begin{gather}
    \hat\Acal_{ij}[\sbt \ ] := \ketbra{\pi_i}{\pi_j} (\sbt\ ) \ketbra{\pi_j}{\pi_i}
\end{gather}
where $\ket{\pi_i},\ket{\pi_j} \in \{\ket{x+},\ket{y+},\ket{z+},\ket{x-}\}$. Intuitively, each of the maps $\Acal_{ij}$ performs a projective measurement and, depending on the outcome, feeds forward a different state. Since the corresponding pure states form a linear basis on the matrix space $\Bcal(\Hcal)$, their cross-combination forms a basis on the instrument space (we will discuss the necessary number of basis elements in more detail below). The corresponding Choi states are given by $\hat{\mathsf{A}}_ij = \ketbra{\pi_i}{\pi_i} \otimes \ketbra{\pi_j}{\pi_j}$, with corresponding duals $D_{k\ell} = D_k \otimes D_\ell$ (given in Eq.~\eqref{eq:dualbasis}). Using these instruments we can tomographically construct the superchannel according to Eq.~\eqref{eqn::supChoiConstr} by computing the corresponding output states $\rho_{ij}(t)$ for this scenario. Doing this, we obtain:
\begin{widetext}
\begin{align}\label{mmapex}
\Upsilon_{\Tcal_{t:0}}=
\begin{pmatrix}
\frac{a_3^+C^2_+}{2} &
0 & 0 &
a_3^+c_\omega^2  & 
 \frac{-i  g  s_\omega^2 +a^- C^2_+}{2} & 
 0 & 0 &
 -g c_\omega s_\omega +a^- c_\omega^2 \\
 0 & 
 \frac{a_3^+C^2_-}{2} & 
 0 & 0 & 0 & 
 \frac{i  g + a^-}{2}s_\omega^2 & 
 0 & 0 \\
 0 & 0 & 
 \frac{a_3^+C^2_-}{2} & 
 0 & 0 & 0 &
 \frac{-i  g + a^-}{2}s_\omega^2 & 
 0 \\
a_3^+c_\omega^2 & 
0 & 0 & 
\frac{a_3^+C^2_+}{2} & 
g  c_\omega s_\omega+a^-c_\omega^2 & 
0 & 0 & 
\frac{i  g s_\omega^2+a^-C^2_+}{2} \\
\frac{i g s_\omega^2 +a^+C^2_+}{2} & 
0 & 0 & 
g c_\omega s_\omega+a^+ c_\omega^2 & 
\frac{a_3^- C^2_+}{2}
& 0 & 0 & 
a_3^- c_\omega^2  \\
0 & 
\frac{-i g +a^+}{2}s_\omega^2 & 
0 & 0 & 0 & 
\frac{a_3^-C^2_-}{2} & 
0 & 0 \\
0 & 0 & 
\frac{i  g + a^+}{2} s_\omega^2 & 
0 & 0 & 0 & 
\frac{a_3^-C^2_-}{2} &
0 \\
- g c_\omega s_\omega+a^+ c_\omega^2 & 
0 &  0 &
\frac{-i g s_\omega^2+a^+ C^2_+}{2} &
a_3^- c_\omega^2
& 0 & 0 &
\frac{a_3^-C^2_+}{2}
\end{pmatrix},
\end{align}
\end{widetext}
where $c_\omega = \cos(2\omega t)$, $s_\omega = \sin(2\omega t)$, $C^2_\pm=1\pm c_\omega^2$, $a^\pm_3=1\pm a_3$, $a^+=a_1+ia_2$, $a^-=a_1-ia_2$, and $g$ is the correlation coefficient. Importantly, the above matrix is positive semidefinite, making -- as we shall see when we discuss higher order quantum maps in more generality -- the corresponding superchannel a completely positive map. Additionally, the above procedure is fully operational; the resulting $\Tcal_{(t:0)}$ is independent of the respective maps $\hat{\Acal})_{ij}$, and, once reconstructed, can be applied to \textit{any} preparation $\Acal$ to yield the correct output state. 
This also implies that the superchannel is constructed with finite number of experiments. Consequently, the examples in Sec.~\ref{sec::InitCorr} are all contained here as limiting cases. Finally, while a CPTP map acting on a $d$-dimensional system would be represented by $d^2\times d^2$ matrix. The superchannel, on the other hand, is a $d^3 \times d^3$ matrix that contracts with CP map to yield an output that is $d\times d$ matrix.

In fact, the superchannel has been observed in the laboratory~\cite{PhysRevLett.114.090402} and proven to be effective at dealing with initial correlations without giving up either linearity or complete positivity. One then might wonder how does this get around Pechukas' theorem? To retain both linearity and complete positivity we have given up the notion of the initial state. In fact, as we argued in Sec.~\ref{subsec::Qmeas}, in presence of correlation, quantum mechanics does not allow for a well-defined local state beyond a singular point in the Hilbert space. Therefore a map on this singular point alone is not very much meaningful, hence there is no big loss in giving up the notion of the initial state as a relevant concept for the dynamics.\ftnt{The singular point state is contained in $\Tcal_{(t:0)}$ and can be obtained by tracing over spaces $\Hcal_0^{\out} \otimes \Hcal_1^\inp$.} Finally, it should be said that this line of reasoning is very close to that of Pearl~\cite{pearl_causality_2009} in classical causal modeling, which goes beyond the framework of classical stochastic processes and allows for interventions.

At this point, it is insightful to quickly take stock of what we have achieved in this section, what we have implicitly assumed, and how to generalize these ideas to finally obtain a fully-fledged description of quantum stochastic processes. Firstly, in order to deal with initial correlations, we have switched perspective and described the dynamics in terms of a mapping from initial operations (instead of initial states) to final states at time $t$. While seemingly odd from a mathematical perspective, from the operational perspective it is only reasonable: experimentally meaningful maps should map from initial objects that can independently be controlled/prepared by the experimenter to objects that can be measured. In the presence of initial system-environment correlations, the experimenter does not have control over the initial state -- at least not without potentially influencing the remaining parameters of the dynamics, i.e., the correlations with the environment). However, they have control over what operation they implemented, and the dynamics from those operations to the final states can be defined and reconstructed, and is fully independent of what the experimenter does (in the sense that $\Tcal_{(t:0)}$ is independent of the experimenter's action. Consequently, switching perspective as we have done is simply natural from an operational point of view. 

Since the superchannel already deals with `intermediate' CP maps performed by the experimenter, it also directly points out how to go beyond experimental scenarios where the experimenter only acts at two times; in principle, nothing keeps us from also performing CP maps at intermediate times, and then reconstructing the final state for \textit{sequences} of CP maps, instead of only one CP map, as we have done here. It should not come as a surprise that this is exactly what we are going to do in the next section. 

It remains to quickly comment on the mathematical details that we deliberately brushed over in this section. Naturally, to make things simpler, we have chosen the most insightful representation of the superchannel in terms of a $d^3\times d^3$. Unsurprisingly, there is also a vectorized version of the superchannel~\cite{arXiv:1708.00769}, or we could have kept things entirely abstract and phrase everything in terms of maps acting on maps. Again, we emphasize that representation has no bearing on physical properties, but employing the represenation we chose will prove very advantageous; for example, it allows us to easily derive the dimension of the spaces we work with, as well as express the properties of higher-order quantum maps in a concise way. Additionally, as was the case for quantum channels, we will see that this representation indeed has an interpretation in terms of a (multipartite) quantum state, which is why we already called it the Choi representation throughout this section.

\subsection{Multi-time statistics in quantum processes}
\label{sec::Multi-time}

\begin{figure}
 \centering
 \includegraphics[width=0.95\linewidth]{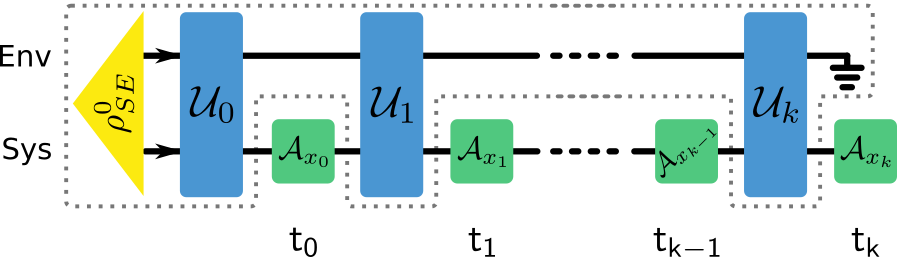}
 \caption{\textbf{General quantum stochastic process.} System of interest is coupled to an unknown environment and probed at times $\Tsetk = \{t_0,t_1, \dots,t_k\}$ with corresponding CP maps $\Asetck = \{\mc{A}_{x_0},\mc{A}_{x_1},\dots,\mc{A}_{x_k}\}$. In between measurements, the system and the environment together undergo closed, i.e., unitary dynamics. The corresponding multi-time joint probabilities can be computed by means of the process tensor corresponding to the process at hand (depicted by the grey dotted outline).}
 \label{fig::ProcTens}
\end{figure}

Following the above resolution for the initial correlation problem in quantum mechanics, we are now in a position to provide a fully-fledged framework for the description of multi-time quantum processes. Here, we predominantly focus on the case of finitely many times at which the process of interest is interrogated (for an in-depth discussion of \textit{continuous} measurements, see, for example, Refs.~\cite{ holevo_statistical_2001, barchielli_quantum_2009}). Note that here, we can but scratch the surface of the different approaches that exist to the theory of multi-time quantum processes. For a much more in-depth investigation of the relation between different concepts of memory in quantum physics, see Ref.~\cite{Li2018}.

In principle, there are two ways to motivate this framework. On the one hand, by generalizing joint probabilities, the descriptor of classical stochastic processes, to the quantum realm, and taking into consideration that, in quantum mechanics, we have to specify the instruments that were used to interrogate the system. This approach would then yield a temporal Born rule~\cite{chiribella_memory_2008, Shrapnel_2018}, and provide a natural descriptor of quantum stochastic processes in terms of a `quantum state over time'. We will circle back to this approach below. Here, we shall take the second possible route to the description of multi-time open quantum processes, which -- just like in the case of initial correlations -- is motivated by considering the underlying dynamics of a quantum stochastic process. As we shall see, though, both approaches are equivalent and lead to the same descriptor of quantum stochastic processes.

As we have seen, the initial correlation problem was solved by taking the preparation procedure into account, and to construct a consistent mapping of the preparation operations to final states. To obtain a consistent description of a \textit{multi}-time process, consider -- as before -- a system of interest $\stxt$ coupled to an environment $\etxt$. Initially, the joint system-environment ($\setxt$) is in state $\rho_{\setxt}(0)$ which might be correlated. Together, we consider $\setxt$ to be closed, such that, between any two times, the system-environment state evolves unitarily -- described by the unitary map
\begin{gather}
\rho_{\setxt}(t_{j+1}) = \Ucalt{(j+1:j)} [\rho_{\setxt}(t_j)] =: \Ucalt{j} [\rho_{\setxt}(t_j)].
\end{gather}
For brevity we have contracted the subscript on $\Ucal$. Next, in order to minimize notational clutter we define several sets
\begin{align}
 &\Tsetk := \{t_0, t_1, \dots, t_{k-1}, t_k \} \label{eq:Tset}\\ 
&\Jsetk := \{\Jcal_0, \Jcal_1, \dots, \Jcal_{k-1}, \Jcal_k\} \label{eq:Jset}\\ 
&\xsetk := \{x_0, x_1, \dots, x_{k-1}, x_k\} \label{eq:xset}\\
&\Asetck := \{\Acal_{x_0}, \Acal_{x_1}, \dots, \Acal_{x_{k-1}}, \Acal_{x_k}\}. \label{eq:Asetc}
\end{align}
The first set, $\Tsetk$ is the set of times on which the process is defined. At these times the system $\stxt$ is interrogated with a set of instruments $\Jsetk$ yielding a set of outcomes $\xsetk$. The set of outcomes corresponds to a set of CP map $\Asetck$. Note, that while we have let the instruments at each time be independent of each other, we can also allow for correlated instruments, also known as testers, see Sec~\ref{sec::Testers}. 

Now, in clear analogy to both the classical case, as well as the quantum case with initial correlations, we envision an experimenter that probes the system of interest at times $\Tsetk$ by means of instruments $\Jsetk$ and we are interested in a consistent descriptor of this experimental situation. For example, they could perform measurements in the computational basis, such that each outcome $x_j$ at a time $t_j$ would correspond to the (trace non-increasing) transformation $\rho \mapsto \braket{x_j|\rho|x_j} \ketbra{x_j}{x_j}$. However, importantly, we do \textit{not} limit the set of allowed operations in any way, shape, or form (besides them being trace non-increasing CP maps). The overall system-environment dynamics is thus a sequence of unitary maps on the system \textit{and} the environment, interspersed by CP maps that act on the system alone, each of them corresponding to a measurement outcome; this is shown in Figure~\ref{fig::ProcTens}. This continues until a final intervention at $t_k$, and then the environmental degrees of freedom are discarded. We emphasize that, as we do not limit or specify the size of the environment $\etxt$, this setup is fully general; as we outlined above, due to the Stinespring dilation, any quantum evolution between two points in time can be understood as a unitary evolution on a larger space. As such, our envisioned setup is the most general description of the evolution of an open quantum system that is probed at times $\Tset_k$. We will see below that this statement even holds in more generality: there are no conceivable quantum stochastic processes that cannot be represented in the above way, as sequences of unitaries on a system-environment space, interspersed by CP maps that act on the system alone. 

The probability to observe a sequence of quantum events, i.e., the outcomes $\xsetk$ corresponding CP to maps $\Asetck$, can then be straightforwardly computed via 
\begin{gather}
\label{eqn::ProcTensDef}
\Pprob(\xsetk|\Jsetk) = \tr\{ \Acal_{x_k} \bigcirc_{j=0}^{k-1} \Ucalt{j} \circ \Acal_{x_j} [\rho_{SE}(0)] \}.
\end{gather}
 Above, $\bigcirc$ denotes the composition of maps, the maps $\Acal$ act on $\stxt$ alone, while the maps $\Ucal$ act on $\setxt$, but we have omitted $\Ical$ on $\etxt$ for brevity. This last equation is just quantum mechanics, as well as simply a multi-time version of Eq.~\eqref{eq:superchannel}, which defines the superchannel. Of course, the challenge is to now turn this equation into a clear descriptor for a multi-time quantum process.

This can be done by noting that the above expression is a multi-linear map with respect to the maps $\Asetc_{\xset_k}$~\cite{PhysRevA.97.012127}. This is similar to the superchannel case we discussed in the previous section, which was linear with respect to the preparation maps $\Acal_j$. It is then possible to write Eq.~\eqref{eqn::ProcTensDef} as a multi-linear functional $\Tcalk$, which we call the \textit{process tensor}:
\begin{gather}\label{eq:tcalmap}
 \Pprob(\xsetk|\Jsetk) =: \Tcalk[\Asetck].
\end{gather}
While seemingly a mere mathematical rearrangement, the above description of an open system dynamics in terms of the process tensor~\cite{PhysRevLett.120.040405, PhysRevA.97.012127, 1367-2630-18-6-063032} $\Tcalk$ is of conceptual relevance; it allows one to separate the parts of the dynamics that are controlled by the experimenter, i.e., the maps $\Asetck$ from the unknown and inaccessible parts of the dynamics, i.e., the initial system-environment state and the system-environment interactions. This clean separation means that when we speak of a quantum stochastic process we only need to refer to $\Tcalk$, and then for any choice of instrument, we can compute the probability for the sequence of outcomes by means of Eq.~\eqref{eq:tcalmap}. As already mentioned in the discussion of superchannels, this is akin to the well-known case of quantum channels, where we separate the part of the process that cannot be controlled -- that is, the quantum channel $\Ecal$ from the parts of the process that are controlled by the experimenter -- that is, the initial system. Here, while the respective objects are somewhat more involved, the underlying idea is exactly the same. Consequently, $\Tcalk$ is the clear generalization of the superchannel $\Tcal_{(t:0)}$, which is, in turn, is a generalization of CPTP maps as discussed in Sec~\ref{subsec::IC}.

Moreover, this separation will later help us to resolve the aforementioned issues with the KET in quantum mechanics, where, apparently, the possible invasiveness of measurements prevented a consistent description of quantum stochastic processes. This will be possible because $\Tcalk$ does not depend on the maps $\Asetc_{\xsetk}$, and as such provides a description of open quantum system dynamics that is independent of the way in which the process at hand is probed. 

We now discuss several key properties of the process tensor. To remain close the classical case in spirit, we will focus on probabilities, i.e, understand $\Tcalk$ as a mapping that allocates the correct probability to any sequence of measurement outcomes for a given given choice of instruments. At first glance, this is not in line with the superchannel, which constituted a mapping from CP maps to final states. However, we could also understand the process tensor as a mapping from operations to a final \textit{state} at time $t_k$; as it can act on all sequences of CP maps, one can choose to not apply an instrument at the last time $t_k$. Consequently, $\Tcalk$ allows for the construction of a related map \begin{gather}
 \widetilde \Tcal_{\Tsetk} [\Asetc_{\xset_{\Tset_{k-1}}}] = \rho(t_k|\xset_{\Tset_{k-1}},\Jset_{\Tset_{k-1}})
\end{gather}
whose output is a quantum state at $t_k$ conditioned on the sequence of CP maps $\Asetck$ at times $\Tsetk$. Often, in what follows, we will not explicitly distinguish between process tensors that return probabilities and those that return states, and the respective case will either be clear from context, or irrelevant for the point we aim to make. With this somewhat technical point out of the way, let us now become more concrete and discuss both the experimental reconstruction as well as the representation of process tensors. Unsurprisingly, we can directly generalize the ideas we developed above to the multi-time case.

\subsubsection{Linearity and tomography}
\label{sec::LinTomMulti}

As mentioned above, $\Tcalk$ is a multi-linear functional on sequences of CP maps $\Asetck$. Consequently, once all the probabilities for the occurrence of a basis of such sequences are known, the full process tensor is determined. This is analogous to the classical case, where the full process at hand was completely characterized once all joint probabilities for all possible combinations of different outcomes was known. Here, the only difference is that different measurements can be applied at each point in time, making the reconstruction a little bit more cumbersome. As the space of sequences of CP maps is finite-dimensional (for $d<\infty$),  $\Tcalk$ can be reconstructed in a finite number of experiments, in a similar vein to the reconstruction of quantum channels and superchannels discussed above. The instrument $\Jcal_{t_j}$ at any time $t_j$ is a set of CP maps 
\begin{gather}
\{\Acal_{x_j}: \Bcal(\Hcal_j^\inp) \rightarrow \Bcal(\Hcal_j^\out)\}.
\end{gather}
The space spanned by such CP maps, i.e., the space that contains all maps of the form $\sum_{x_j} c_{x_j} \Acal_{x_j}$ is $(d_{j^\inp} d_{j^\out})^2$-dimensional since it is -- as we have seen in our discussion of the CJI in Sec.~\ref{subsec::Representation} -- isomorphic to the matrix space $\Bcal(\Hcal_j^\out \otimes \Hcal_j^\inp)\}$ (in what follows, we will assume $d_{j^\inp} = d_{j^\out} = d$). Since we can choose an instrument at each time independently of other times, we can form a multi-time basis consisting of basis elements at each time, which forms a linear basis on all times $\Tsetk$ (in the same way as the basis of a multipartite quantum system can be constructed from combinations of local basis elements):
\begin{gather}
\hat\Jset_{\Tsetk} = \left\{\hat\Asetc_{\xsetk} := \{\hat \Acal_{x_k}, \cdots, \hat \Acal_{x_0}\}\right\}_{x_j=1}^{d^4}.
\end{gather}
Importantly, \textit{any} other sequence of operations (and also temporally correlated ones, see below) can be written as a linear combination of such a complete set of basis operations.

The action of the process tensor on the multi-time basis gives us the probability to observe the sequence $\xsetk$ as
\begin{gather}
\Pprob(\xsetk|\hat\Jset_{\Tsetk}) := \Tcalk[\hat\Asetc_{\xsetk}].
\end{gather}
From our discussion of the reconstruction of the superchannel, reconstructing the multi-time object $\Tcalk$ is now a straightforward endeavour, which, again, we will carry out using Choi states. To this end, we note that, since all operations performed at different times are uncorrelated, their overall Choi state $\Asetk$ is simply a tensor product of the Choi states of the individual operations, i.e., 
\begin{gather}
\label{eqn::ProcTensChoi}
    \hat{\mathbf{A}}_{\xsetk} = \hat{\Asf}_{x_k} \otimes \hat{\Asf}_{x_{k-1}} \otimes \dots \otimes \hat{\Asf}_{x_{1}} \otimes {\Asf}_{x_0},
\end{gather}
The Choi state of the process tensor can then be written -- in the same spirit as Eq.~\eqref{eq:tomrepmap} -- as
\begin{gather}
\label{eq:tcalChoi}
 \Upsilon_{\Tsetk} = \sum_{\xsetk} \Pprob(\xsetk|\hat\Jset_{\Tsetk}) \ \hat{D}_{x_k}^\ast \otimes \cdots \otimes \hat{D}^\ast_{x_0}
 \end{gather}
 with the action of the process tensor given by 
 \begin{gather}
 \label{eq:tcalaction}
      \Tcal_{\Tsetk} [\Asetc] = \tr[
 \Asetk^\trps
\Upsk].
 \end{gather}
Here, again, $\{\hat{D}_{x_k}\}$ forms the dual basis to the Choi states of the basis operations $\{\hat{\mathcal{A}}_{x_k}\}$ at each time $t_k$, i.e., $\tr[\hat{D}_{x_i}^\dag \hat{\mathsf{A}}_{x_j}] = \delta_{ij}$.

Again, by construction, the process tensor above yields the correct probabilities for any of the basis sequences in $\hat\Jset_{\Tsetk}$ (which can be seen by direct insertion of Eq.~\eqref{eq:tcalChoi} into Eq.~\eqref{eq:tcalaction}):
\begin{gather}
\begin{split}
\Tcal_{\Tsetk}  [ \hat{\mathbf{A}}_{\xsetk}] 
   &=  \sum_{\xsetk'} \tr[( \hat{\Asf}_{x_k}^\mathrm{T} \otimes  \dots \otimes  \hat{\Asf}_{x_0}^\mathrm{T}) \\
   &\phantom{\sum_{\xsetk'}}\cdot(\Pprob(\xsetk'|\hat\Jset_{\Tsetk}) \ \hat{D}_{x'_k}^\ast \otimes \cdots \otimes \hat{D}^\ast_{x'_0})] \\
   &= \sum_{\xsetk'}\Pprob(\xsetk'|\hat\Jset_{\Tsetk}) \tr( \hat{D}_{x'_k}^\dagger \hat{\Asf}_{x_k})\cdots \tr(\hat{D}_{x'_0}^\dagger \hat{\Asf}_{x_0}) \\
   &= \Pprob(\xsetk|\hat\Jset_{\Tsetk}),
 \end{split}
\end{gather} 
thus yielding the correct probability for any basis sequence of measurements, implying that it yields the correct probability for \textit{any} conceivable operation on the set of times $\Tsetk$. In order to reconstruct a process tensor on times $\Tsetk$, an experimenter would hence have to probe the process using informationally complete instruments $\{\hat\Jcal_j\}$ -- in the sense that its elements span the whole space of CP maps. More concretely, the duals $D_{x_j}$ can be computed, the joint probabilities $\Pprob(\xsetk|\hat\Jset_{\Tsetk})$ can be measured, and Eq.~\eqref{eq:tcalChoi} tells us how to combine them to yield the correct matrix $\Upsk$ (see below and Refs.~\cite{PhysRevA.97.012127, Milz2018A} for details on the reconstruction of $\Upsk$).  This reconstruction also applies in case that the experimenter does not have access to informationally complete instruments, yielding a `restricted' process tensor~\cite{Milz2018A, White2020}, that only meaningfully applies to operations that lie in the span of those that can be implemented.

While the number of necessary sequences for the reconstruction of a process tensor scales exponentially with the number of times (if there are $N$ times, then there are $d^{4N}$ different sequences, for which the probabilities would have to be determined), the number is still finite, and thus, in principle, feasible. We note that classical processes are scaled by a similar exponential scaling problem. If there are $d$ different outcomes at each time, then the joint probabilities for $d^N$ different sequences of $N$ outcomes. Let us now discuss some concrete properties and interpretations of these above considerations.

\subsubsection{Spatiotemporal Born rule and the link product}
\label{sec:bornrule}

As before, let
\begin{gather}\label{eq:Aset}
\Asetk = \Asf_{x_k} \otimes \Asf_{x_{k-1}} \otimes \dots \otimes \Asf_{x_{1}} \otimes \Asf_{x_0},
\end{gather}
be a set of Choi states corresponding to a sequence of independent CP maps. Then, as we have seen, the probability to obtain this sequence is given by
\begin{gather}
\Pprob(\xsetk|\Jsetk)= \tr[\Upsk \Aset_{\xsetk}^{\trps}],
\label{eq:process}
\end{gather}
where $\Upsk$ is the Choi state of $\Tcalk$ (see below for a discussion as to why it actually constitutes a Choi state). The advantage of representing the process tensor by its action in this way is two-fold. On the one hand, all objects are now rather concrete (and not abstract maps), and we can easily talk about their properties (see below).  On the other hand, the fact that $\Upsk$ is a matrix and not an abstract map will allow us to freely talk about temporal correlations in quantum mechanics in the same way that we do in the spatial setting.

Additionally, the above Eq.~\eqref{eq:process} constitutes a multi-time generalization of the Born rule~\cite{chiribella_memory_2008, Shrapnel_2018}, where $\Upsk$ plays the role of a quantum state over time, and the Choi states $\Asetk$ play a role that is analogous to that of POVM elements in the spatial setting. In principle, $\Upsk$ can be computed from the underlying dynamics by means of the \textit{link product} $\star$ defined in Ref.~\cite{chiribella_theoretical_2009} as
\begin{gather}
\label{eqn::ChoiLink}
\Upsk = \tr_E[\texttt{U}_{k} \star \dots \star \texttt{U}_{0} \star \rho_{\setxt}(0)].
\end{gather}
Here, $\texttt{U}_{j}$ is the Choi state of the map $\Ucal_{j}$ and the link product acts like a matrix product on the space $E$ and a tensor product on space $S$. Basically, the link product translates concatenation of maps onto their corresponding Choi matrices, i.e., if $\mathsf{A}$ and $\mathsf{C}$ are the Choi states $\Acal$ and $\Ccal$, respectively, then $\mathsf{D} = \mathsf{C} \star \mathsf{A}$ is the Choi state of $\Dcal = \Ccal \circ \Acal$. We will not employ the link product frequently in this Tutorial, but will quickly provide its definition and motivation here (see, for example, Ref.~\cite{chiribella_theoretical_2009} for more details). Concretely, as an exemplary case, let $\Acal: \Bcal(\Hcal_1) \rightarrow \Bcal(\Hcal_3 \otimes \Hcal_4)$ and $\Ccal: \Bcal(\Hcal_4 \otimes \Hcal_5) \rightarrow \Bcal(\Hcal_6)$, then $\Dcal = \Ccal \circ \Acal: \Bcal(\Hcal_1 \otimes \Hcal_5) \rightarrow \Bcal(\Hcal_3 \otimes \Hcal_6)$. Correspondingly, for the respective Choi states we have $\mathsf{C} \in \Bcal(\Hcal_4\otimes \Hcal_5 \otimes \Hcal_6)$, $\mathsf{A} \in \Bcal(\Hcal_1\otimes \Hcal_3 \otimes \Hcal_4)$, and $\mathsf{D} \in \Bcal(\Hcal_1\otimes \Hcal_5 \otimes \Hcal_3 \otimes \Hcal_6)$. Using Eq.~\eqref{eqn::ChoiAction}, one can rewrite the action of the resulting map $\Dcal$ on an arbitrary matrix $\rho \in \Bcal(\Hcal_1 \otimes \Hcal_5))$ in terms of its Choi state, which yields
\begin{gather}
    \Dcal[\rho] = \tr_{15}[\mathsf{D}(\rho^{\mathrm{T}} \otimes \ident^{36})] =: \tr_{15}[(\mathsf{C} \star \mathsf{A})(\rho^{\mathrm{T}} \otimes \ident^{36})]\,,
\end{gather}
where $\ident^{36}$ is the identity matrix on $\Hcal_3 \otimes \Hcal_6$. Now, using the above equation, one can directly read off the form of $\mathsf{C} \star \mathsf{A}$ as 
\begin{gather}
\label{eqn::LinkDef}
\mathsf{C} \star \mathsf{A} = \tr_{4}[(\mathsf{C} \otimes \ident^{13})(\mathsf{A}^{\mathrm{T}_4} \otimes \ident^{56})]\, .
\end{gather}
The derivation of the above relation from Eq.~\eqref{eqn::ChoiLink} is straightforward but somewhat lenghty and left as an exercise to the reader. Intuitively, the above tells us that the link product between two matrices consists of i) tensoring both matrices with identity matrices so that they live on the same space, ii) partially transposing one of the matrices with respect to the spaces both of the matrices share, and iii) taking the trace of the product of the obtained objects with respect to the spaces both of the matrices share. This recipe holds for all conceivable situations where the Choi matrix of the concatenation of maps is to be computed.  As mentioned, we will not make much use of the link product here (with the exception of Sec.~\ref{sec::MemBond}), but it can be very convenient when working out the Choi states of higher order quantum maps like the process tensor. Let us mention in passing that the link product has many appealing properties, like, for example, commutativity (for all intents and purposes~\cite{chiribella_theoretical_2009}) and associativity, which allows us to write the Choi state $\Upsk$ in Eq.~\eqref{eqn::ChoiLink} as a multi-link product without having to care about in what order we carry out the `multiplication' $\star$. 

As it will turn out, $\Upsk$ is a many-body density matrix (up to a normalization), therefore constituting a very natural generalization for a classical stochastic process which is a joint probability distribution over many random variables. Since it allows the compact phrasing of many of the subsequent results, we will often opt for a representation of $\Tcal_{\Tsetk}$ in terms of its Choi matrix $\Upsk$ in what follows (there, we will also see why it is justified to dub it a Choi matrix), and we will, for simplicity, often call both of them the process tensor. Nonetheless, for better accessibility, we will also express our results in terms of maps whenever appropriate.

Before advancing, let us recapitulate what has been achieved by introducing the process tensor for the description of general quantum processes. First, the effects on the system due to interaction with the environment have been isolated in the process tensor $\Upsk$. All of the details of the instruments and their outcomes are encapsulated in $\Asetk$, while all inaccessible effects and influences are contained in the process tensor. In this way, $\Upsk$ is a complete representation of the stochastic quantum process, containing all accessible multi-time correlations~\cite{PhysRevLett.122.140401, PhysRevA.99.042108, Phil_MemStr, arXiv:1811.03722}.\ftnt{This tensor, in general, is also a quantum comb, where the bond represents information fed forward through an ancillary system.} The process tensor can be formally shown to be the quantum generalization of a classical stochastic process~\cite{Milz2017KET}, and it reduces to classical stochastic process in the correct limit~\cite{Milz2017KET, PhysRevA.100.022120, milz_when_2020} (we will get back to this point below).

We emphasize that open quantum system dynamics is not the only field of physics where an object like the process tensor (or variants thereof) crop up naturally. See, for example Refs.~\cite{supermaps, chiribella_quantum_2008, chiribella_theoretical_2009,kretschmann_quantum_2005, caruso_quantum_2014,portmann_causal_2015, hardy_operational_2016, hardy_operator_2012, cotler_superdensity_2017, OreshkovETAL2012, 1367-2630-18-6-063032, oreshkov_causal_2016, PhysRevX.7.031021, gutoski2007toward, Gutoski2018fidelityofquantum} for an incomplete collection of works where similar mathematical objects have been used for the study of higher-order quantum maps, causal automata/non-anticipatory channels, quantum networks with modular elements, quantum information in general relativistic space-time, quantum causal modeling, and quantum games (see also Table~\ref{tab::Fields}). In open quantum system dynamics, they have been used in the disguise of so-called correlation kernels already in early works on multi-time quantum processes~\cite{Davies1969,lindblad_non-markovian_1979, accardi}.
\begingroup
\setlength{\tabcolsep}{5.5pt}
\begin{table}
    \centering
    \begin{tabular}{Sl||Sc|Sc}
          &  \textbf{Name} & \textbf{Application}\\
         \hline
         \makecell[l]{Quantum \\ Information} & \makecell{Quantum comb/ \\ Causal box} & \makecell{Quantum circuit \\ architecture} \\
         \hline
         \makecell[l]{Open Quantum \\ system dynamics} & \makecell{Correlation kernel/ \\Process tensor} & \makecell{Study of \\ temporal \\ correlations} \\
         \hline
         \makecell[l]{Quantum \\ Games} & Strategy & \makecell{Computation of \\ winning probabilities} \\
         \hline
         \makecell[l]{Quantum \\ Causality} & Process matrix & \makecell{Processes without \\ definitive causal \\ order} \\
         \hline
         \makecell[l]{Quantum causal \\ modelling} & Process matrix & \makecell{Causal relations \\ in quantum \\ processes} \\
         \hline
         \makecell[l]{Quantum \\ Shannon Theory} & \makecell{Causal \\ automaton/ \\ non-anticipatory \\ channel} & \makecell{Quantum channels \\ with memory}
    \end{tabular}
    \caption{\textbf{`Process tensors' in different fields of quantum mechanics.} Mathematical objects that are similar in spirit to the process tensor crop up frequently in quantum mechanics. The above table is an incomplete list of the respective fields and commonly used names. Note that, even within these fields, the respective names and concrete applications differ. Additionally, some of the objects that occur on the above list might have slightly different properties than the process tensor (for example, process matrices do not have to display a global causal order), and might look very different than the process tensor (for example, it is not obvious that the correlation kernels used in open quantum system dynamics are indeed variants of process tensors in disguise). These disparities notwithstanding, the objects in the above table are close both in spirit, as well as the related mathematical framework.}
    \label{tab::Fields}
\end{table}
\endgroup

\subsubsection{Many-body Choi state}
\label{sec::ManyChoi}
\begin{figure}
 \centering
 \includegraphics[width=0.95\linewidth]{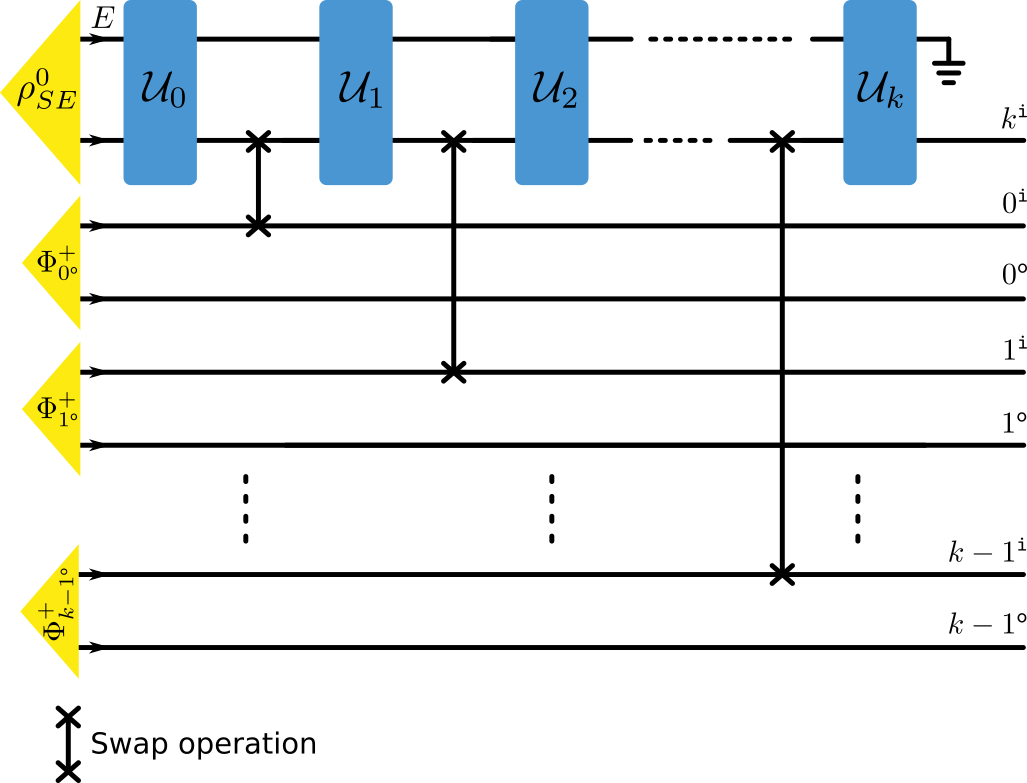}
 \caption{\textbf{Choi state of a process tensor.} At each time, half of an unnormalized maximally entangled state is fed into the process. For better book-keeping, all spaces are labeled by their respective time. The resulting many-body state $\Upsk$ contains all spatio-temporal correlations of the corresponding process as spatial correlations.}
 \label{fig::Choi_Process}
\end{figure}

While we now know how to experimentally reconstruct it, it remains to provide a physical interpretation for $\Upsk$, and discuss its properties (and justify why we called it the Choi matrix of $\Tcal_{\Tsetk}$ above). We start with the former. For the case of quantum channels, the interpretation of the Choi state $\Upsilon_\Ecal$ is clear; it is the state that results from letting $\Ecal$ act on half of an unnormalized maximally entangled state. $\Upsilon_\Ecal$ then contains exactly the same information as the original map $\Ecal$. Somewhat unsurprisingly, in the multi-time case, the CJI is similar to the two-time scenario of quantum channels. Here, however, instead of feeding one half of a (unnormalized) maximally entangled state into the process once, we have to do so at each time in ${\Tsetk}$ (see Figure~\ref{fig::Choi_Process} for a graphical representation). From Eq.~\eqref{eq:process}, we see that $\Upsilon_{\Tsetk}$ must be an element of $\Bcal(\Hcal_k^{\inp} \otimes \Hcal_{k-1}^{\out} \otimes \cdots \otimes \Hcal_0^\out \otimes \Hcal_0^\inp)$. Labeling the maximally entangled states in Figure~\ref{fig::Choi_Process} diligently, and distinguishing between input and output spaces, we see that the resulting state $\Upsk$ lives on exactly the right space. Checking that the matrix $\Upsk$ constructed in this way indeed yields the correct process tensor can be seen by direct insertion. Indeed, by using the Choi state of Figure~\ref{fig::Choi_Process} and the definition of the Choi state $\{\Asf_{x_j}\}$, one sees that Eq.~\eqref{eq:process} holds. While straightforward, this derivation is somewhat arduous and left as an exercise to the reader. 

We thus see that $\Upsk$ is proportional to a many-body quantum state, and the spatio-temporal correlations of the underlying process are mapped onto spatial correlations of $\Upsk$ via the CJI and each time corresponds to two Hilbert spaces (one for the input space, and one for the output space). Specifically, statements  like `correlations between different time' now translate to statements about correlations between different Hilbert spaces the state $\Upsk$ is defined on. These properties lend themselves to convenient methods for treating a multi-time process as a many-body state with applications for efficient simulations and learning the process~\cite{Luchnikov2018, costa2019, Luchnikov2019L, Luchnikov2019, guochu2020}. 

Additionally, the CJI for quantum channels as well as superchannels are simply special cases of the more general CJI presented here. We emphasize that, with this, expressing the action of a process tensor in terms of matrices has become more than just a convenient trick. Knowing that $\Upsk$ is (proportional to) a quantum state tells us straight away that it is positive, and all spatio-temporal correlations present in the process can now be conveniently be understood as spatial correlations in the state $\Upsk$. This convenience is the main reason why most of our results will be phrased in terms of Choi matrices in what follows.

\subsubsection{Complete positivity and trace preservation}
\label{sec::CPTPMulti}
Just like for the case of quantum channels, the properties of a multi-time process can be most easily read off its Choi state. First, as we have seen above, $\Upsk$ is positive. Like in the case of channels, and superchannels, this property implies complete positivity of the process at hand. As was the case for superchannels, complete positivity here has a particular meaning: let the process act on any sequence of CP maps 
\begin{gather}
\begin{split}
&\mathscr{B}_{\Tset_k} =    \{\Bcal_{x_0},\Bcal_{x_1},\dots,\Bcal_{x_{k-1}},\Bcal_{x_k}\}\\
&\Bsetk := \Bsf_{x_k} \otimes \Bsf_{x_{k-1}} \cdots \otimes \Bsf_{x_1} \otimes \Bsf_{x_0},
\end{split}
\end{gather}
where $\Bsf_x$ is the Choi state of $\Bcal_{x}$. These superoperators act both on the $\stxt$ of interest, as well as some external ancillas, which we collectively denote by $B$, which do not interact with the environment $\etxt$ that is part of the process tensor. We can see the complete positivity of the process tensor directly in terms of the positivity of the process' Choi state
\begin{gather}
 \tr[(\Upsk \otimes \ident) \ \Bsetk^\trps] \geq 0.
\end{gather}
Above $\Upsk$ acts on $\stxt$ at times $\Tsetk$ and $\ident_B$ is the identity matrix on the ancillary degrees of freedom $B$. As the positivity of the Choi state implies complete positivity of the underlying map, \textit{any} sequence $\Bsetk$ of CP maps is mapped to a CP map by $\Tcalk$. Analogously, we could have expressed the above equation in terms of maps, yielding
\begin{gather}
    (\Tcal_{\Tsetk} \otimes \Ical_B)[\Bcal_{x_0},\Bcal_{x_1},\dots,\Bcal_{x_{k-1}},\Bcal_{x_k}] \ \text{is CP}.
\end{gather}
However, as mentioned, the properties of process tensors are much more easily represented in terms of their Choi matrices.

\begin{figure}
 \centering
 \includegraphics[width=0.8\linewidth]{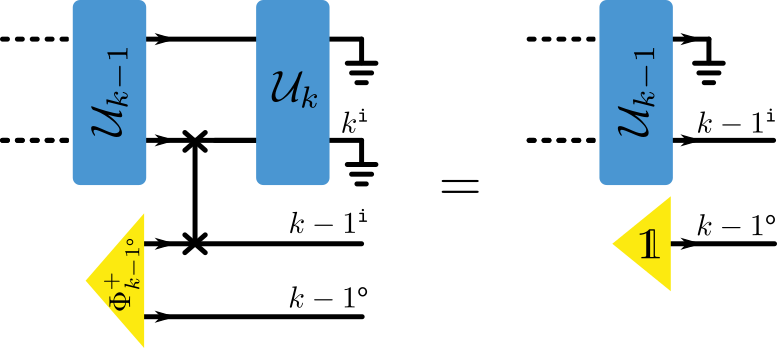}
 \caption{\textbf{Trace conditions on process tensors.} Displayed is the pertinent part of Figure~\ref{fig::Choi_Process}. As $\tr \circ \Ucal = \tr$ for all CPTP maps $\Ucal$, tracing out the final degree of freedom of $\Upsilon_{\Tset_k}$, denoted by $k^\inp$, amounts to a partial trace of $\Phi^+_{k-1^\out}$. This, in turn, yields a tensor product between $\ident_{{k-1}^\inp}$ and a process tensor on one step less. As in Figure~\ref{fig::Choi_Process}, the swap operation is represented by a vertical line with crosses at its ends.}
 \label{fig::RelevantPart}
\end{figure}

In clear analogy to the case of quantum channels, process tensors should also satisfy a property akin to trace preservation. At its core, trace preservation is a statement about normalization of probabilities. As CPTP maps can be implemented with unit probability, at first glance, the natural generalization of trace preservation thus appears to be 
\begin{gather}
 \tr[\Upsk^\trps (\Asf_k \otimes \cdots \otimes \Asf_0)] = 1 
\end{gather}
for all CPTP maps $\Asf_0, \dots, \Asf_k$. However, this requirement on its own is too weak, as it does not encapsulate the temporal ordering of the process at hand~\cite{OreshkovETAL2012}. If only the above requirement was fulfilled, then, actions at a time $t_j$ could in principle influence the statistics at an earlier time $t_j' < t_j$. This should be forbidden by causality, though. Fortunately, $\Upsk$ already encapsulates the causal ordering of the underlying process by construction. Specifically, tracing over the degrees of freedom of $\Upsk$ that correspond to the last time (i.e., the degrees of freedom labeled by $k^\inp$ in Figure~\ref{fig::Choi_Process}) yields 
\begin{gather}
\label{eqn::hierarchyFirst}
 \tr_{k^\inp} \Upsk = \ident_{{k-1}^\out} \otimes \Upsilon_{\Tset_{k-1}},
\end{gather}
where $\Upsilon_{\Tset_{k-1}}$ is the process tensor on times $\Tset_{k-1}$ with a final output degree of freedom denoted by $k-1^\inp$. The above property trickles down, in the sense that 
\begin{gather}
\label{eqn::Hierarchy}
\begin{split}
 \tr_{k-1^\inp} \Upsilon_{\Tset_{k-1}} &= \ident_{{k-2}^\out} \otimes \Upsilon_{\Tset_{k-2}},\\
 \tr_{k-2^\inp} \Upsilon_{\Tset_{k-2}} &= \ident_{{k-3}^\out} \otimes \Upsilon_{\Tset_{k-3}}, \\
 &\vdots \\
 \tr_{1^\inp} \Upsilon_{\Tset_{1}} &= \ident_{{0}^\out} \otimes \Upsilon_{\Tset_{0}},\\
 \tr_{0^\inp} \Upsilon_{\Tset_{0}}&=1.
 \end{split}
\end{gather}
Before elucidating why these properties indeed ensure causal ordering, let us quickly lay out why they hold. To this end, it is actually sufficient to only prove the first condition~\eqref{eqn::hierarchyFirst}, as the others follow in the same vein. A rigorous version of this proof can, for example, be found in~\cite{chiribella_theoretical_2009, PhysRevA.97.012127}. Here, we will prove it by means of Figure~\ref{fig::Choi_Process}. Consider tracing out the degrees of freedom denoted by $k^\inp$ in said figure. This, then, amounts to tracing out all output degrees of freedom of the map $\Ucalt{k}$. As $\Ucalt{k}$ is CPTP, tracing out all outputs after applying $\Ucalt{k}$ is the same as simply tracing out the outputs without having applied $\Ucalt{k}$, i.e., $\tr\circ \Ucalt{k} = \tr$. This, then, implies a partial trace of the unnormalized maximally entangled state $\Phi^+_{k-1^\out}$, yielding $\ident_{{k-1}^\out}$, as well as a trace over the environmental output degrees of freedom of $\Ucalt{k-1}$ (see Figure~\ref{fig::RelevantPart} for a detailed graphical representation). The remaining part, i.e., the part besides $\ident_{{k-1}^\out}$ is then a process tensor on the times $\Tset_{k-1} = \{t_0,\dots,t_{k-1}\}$. Iterating these arguments then leads to the hierarchy of trace conditions in Eq.~\eqref{eqn::Hierarchy}. While a little bit tedious algebraically, these relations can very easily be read off from the graphical representation provided in Figure~\ref{fig::RelevantPart}.

Showing that the above trace conditions indeed imply correct causal ordering of the process tensor now amounts to showing that a CPTP map at a later time does not have an influence on the remaining process tensor at earlier times. We start with a CPTP map at $t_k$. This map does not have an output space. The only CPTP map with trivial output space is the trace operation, which has a Choi state $\ident_{k^\inp}$. Thus, letting $\Upsk$ act on it amounts to a partial trace $\tr_{k^\inp} \Upsk$, which is equal to $\ident_{{k-1}^\out} \otimes \Upsilon_{\Tset_{k-1}}$. Letting this remaining process tensor act on a CPTP map $\Asf_{k-1}$ at time $t_{k-1}$ yields 
\begin{gather}
 \tr_{k-1}[(\ident_{{k-1}^\out} \otimes \Upsilon_{\Tset_{k-1}})\Asf^{\trps}_{k-1}] = \ident_{{k-2}^\out} \otimes \Upsilon_{\Tset_{k-2}},
\end{gather}
where $\tr_{k-1}$ denotes the trace over $k-1^\inp$ and $k-1^\out$, and we have used the property of CPTP maps that $\tr_{k-1^\out} (\Asf_{k-1}) = \ident_{{k-1}^\inp}$. As the LHS of the above equation does not depend on the specific choice of $\Asf_{k-1}$, no statistics before $t_{k-1}$ will depend on the choice of $\Asf_{k-1}$ either. Again, iterating this argument then shows that the above hierarchy of trace conditions implies proper causal ordering. 

\subsubsection{`Reduced' process tensors}
\label{subsec::Red}
Importantly, this independence on earlier CPTP maps implies that we can uniquely define certain `reduced' process tensors. Say, we have a process tensor $\Upsk$ that is defined on times $t_1 < \cdots < t_k$ and we want to obtain the correct process tensor only on the first couple $\{t_0,\dots,t_j\}$ with $t_j < t_k$. To this end, at first glance, it seems like we would have to specify what instruments $\Jcal_i$ we aim to apply at times $t_i > t_j$. However, since $\Upsk$ satisfies the above causality constraints, earlier statistics, and with them, the corresponding process tensors are independent of later CPTP maps. This is in contrast to later statistics, that can, due to causal influences, absolutely depend on earlier CPTP maps. Long story short, while `tracing out' later times is a unique operation on combs, `tracing out' earlier ones is not, and the corresponding resulting process tensor would depend on the CPTP maps that were used for the tracing out operations. This, unsurprisingly, is in contrast to the spatial case, where local CPTP maps never influence the statistics of other parties, for the simple reason that in the spatial case, there is no signalling between different parties happening. To be more concrete, in order to obtain a process tensor $\Upsilon_{\Tset_j}$ from $\Upsk$, where $t_j<t_k$, we could `contract' $\Upsk$ with \textit{any} sequence of CPTP maps $\Asf_{j+1},\dots,\Asf_k$: 
\begin{gather}
    \Upsilon_{\Tset_j} = \tr_{k:j+1}[\Upsk(\Asf_{j+1}^\trps \otimes \cdots \otimes \Asf_{k}^\trps)).
\end{gather}
Since $\Upsk$ satisfies the causality constraints of Eqs.~\eqref{eqn::hierarchyFirst} and~\eqref{eqn::Hierarchy} the above $\Upsilon_{\Tset_j}$ is independent of the choice of CPTP maps $\Asf_{j+1}, \dots, \Asf_{k}$ and correctly reproduces all statistics on $\Tset_j$. Since the choice of CPTP maps is arbitrary, we can take the simple choice $\Asf_{i} =\tfrac{1}{d_{i^\out}} \ident_i^\inp \otimes \ident_{i^\out}$, which yields 
\begin{gather}
    \Upsilon_{\Tset_j} = \tfrac{1}{\Pi_{i = j+1}^{k}d_{i^\out}}\tr_{k:j+1}(\Upsk)\, .
\end{gather}
Again, we emphasize that the causality constraints on $\Upsk$ only apply in a fixed order -- that is, the order that is given by causal ordering of the times in $\Tset_k$, such that a `reduced' process tensor on later times is a meaningful concept, but would in general depend on the CPTP maps that were applied at earlier times. For example, we would generally have 
\begin{gather}
    \Upsilon_{t_k,\dots,t_1}^{(\Asf_0)} := \tr_0(\Upsk\Asf_0^\trps) \neq \tr_0(\Upsk\Asf_0^{\prime \mathrm{T}}) =:\Upsilon_{t_k,\dots,t_1}^{(\Asf'_0)}
\end{gather}
 
As before, the above results can, equivalently, be stated in terms of maps. However, the corresponding equations would not be very enlightening. To summarize, process tensors, just like channels and superchannels satisfy complete positivity and trace preservation, albeit with slightly different interpretations than was the case for channels.

At this point, it is insightful to return to the two different ways of motivating the discussion of quantum stochastic processes we alluded to at the beginning of Sec.~\ref{sec::Multi-time}. Naturally, based on a reasoning by analogy, we could have introduced the process tensor as a positive linear functional that maps sequences of CP maps to probabilities and respects the causal order of the process. After all, coming from classical stochastic processes and knowing about how measurements are described in quantum mechanics, this would have been a very natural route to take. This, then, might in principle have led to a larger set of process tensors than the ones we obtained from underlying circuits. However, this is not the case; as we shall see in the next section, any object that is positive and satisfies the trace hierarchy above actually corresponds to a quantum circuit with only pure states and unitary intermediate maps. Consequently, and somewhat reassuringly, both the axiomatic perspective, as well as the operational one we took here, lead to the same resulting descriptors of quantum stochastic processes.

Finally, one might wonder, why we never discussed the question of causality in the case of classical stochastic processes. There, however, causality does not play a role per se if only non-invasive measurements are considered. It is only through the invasiveness of measurements/interrogations that influences between different events, and, as such, causal relations can be discerned. A joint probability distribution obtained from making non-invasive measurements does thus not contain information about causal relations. This, naturally, changes drastically, as soon as active interventions are taken into considerations, as is done actively in the field of classical causal modeling~\cite{pearl_causality_2009}, and as cannot be avoided in quantum mechanics~\cite{Milz2017KET}.

\subsubsection{Testers: Temporally correlated `instruments'}
\label{sec::Testers}

\begin{figure}
 \centering
 \includegraphics[width=0.95\linewidth]{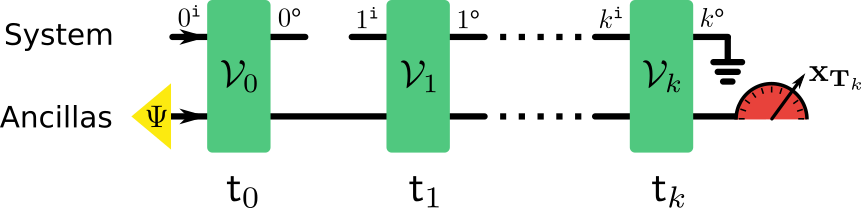}
 \caption{\textbf{Tester Element.} In the most general case, an experimenter can correlate the system of interest with an ancilla (here, initially in state $\ket{\Psi}$), use said ancilla again at the next time, etc., and make a final measurement with outcome $\mathbf{x}$ in the end. As the unitaries $\Vcal_j$ can also act trivially on parts of the ancilla, this scenario includes all conceivable measurements an experimenter can perform. Summing over the outcomes $\xsetk$ amounts to tracing out the ancillas, thus yielding a proper comb (compare with Figure~\ref{fig::ProcTens}. Note that the inputs (outputs) of the resulting tester elements correspond to the outputs (inputs) of the process tensor, and the system of interest corresponds to the top line, not the bottom line.}
 \label{fig::tester_element}
\end{figure}

So far, we have only considered the application of independent instruments, which have the form given in Eq.~\eqref{eq:Aset}. However, these are not the only operations a process tensor can meaningfully act on. In principle, an experimenter could, for example, condition their choice of instrument at time $t_{j'}$ on all outcomes they recorded at times $t_j < t_{j'}$. This would lead to a (classically) temporally correlated `instrument', which is commonly practiced in quantum optics experiments~\cite{slh}. More generally, at times $\Tsetk$, the experimenter could correlate the system of interest with  external ancillas, which are re-used, and measure said ancillas at time $t_k$ (see Figure~\ref{fig::tester_element}). This, then, would result in a generalized instrument that has temporal quantum correlations. 

We can always express such correlated operations using a local linear basis as
\begin{gather}\label{eq:Asetcorr}
 \Asetk = \sum_{\xsetk} \alpha_{\xsetk} \hat\Asf_{x_k} \otimes \hat\Asf_{x_{k-1}} \otimes \dots \otimes \hat\Asf_{x_{1}} \otimes \hat\Asf_{x_0}.
\end{gather}
The LHS side of this equation is labeled the same as Eq.~\eqref{eq:Aset}. This is because the above equation contains Eq.~\eqref{eq:Aset} as a special case. Here, $\{\hat\Asf_{x_{j}}\}$ form a linear basis for operations at time $t_j$ and $\alpha_{\xsetk}$ are generic coefficient that can be non-positive. In other words, the above correlated operation can carry `entanglement in time' since not only convex combinations of product operations are possible. Temporally correlated operations can be performed as part of a temporally correlated instrument. Such generalizations of instruments have been called `testers' in the literature~\cite{chiribella_memory_2008, chiribella_theoretical_2009, chiribella_optimal_2016}. 

In the case of `normal' instruments, the respective elements CP maps have to add up to a CPTP map. Here, in clear analogy, the elements of a tester have to add up to a proper process tensor. In terms of Choi states, this means that the elements $\{\Asetk\}$ of a tester have to be positive, and add up to a matrix $\Aset = \sum_k \Asetk$ that satisfies the hierarchy of trace conditions of Eqs.~\eqref{eqn::hierarchyFirst} and~\eqref{eqn::Hierarchy}. We emphasize that the possible outcomes $\xsetk$ that label the tester elements do \textit{not} have to correspond to sequences $x_0, \dots x_k$ of individual outcomes at times $t_0,\dots t_k$. As outlined above, for correlated tester elements, all measurements could happen at the last time, only, or at any subset of times. Consequently, in what follows, unless explicitly stated otherwise, $\xsetk$ will label `collective' measurement outcomes and not necessarily sequences of individual outcomes. Interestingly, since tester elements add up to a proper process tensor, this discussion of testers already points us to the interpretation of correlations between different times in $\Upsk$; each type of them corresponds to a different type of information that is transmitted between different points in time by the environment -- just like in the tester case classical correlations correspond to classical information that is fed forward, while entanglement between different times relates to quantum information being processed. We will make these points clearer below, but already want to emphasize the dual role that testers and process tensors play.

However, for a tester, the roles of input and output are reversed with respect to the process tensors that act on them; an output of the process tensor is an input for the tester and vice versa. Consequently, keeping the labeling of spaces consistent with the above, and assuming that testers end on the last output space $k^\inp$, the trace hierarchy for testers starts with $\tr_{k^\out}(\Asetk) = \ident_{k^\inp} \otimes \Aset_{\Tset_{k-1}}$, $\tr_{k-1^\out} (\Aset_{\Tset_{k-1}}) = \ident_{{k-1}^\inp} \otimes \Aset_{\Tset_{k-2}}$, etc., implying that, with respect to Eqs.~\eqref{eqn::hierarchyFirst} and~\eqref{eqn::Hierarchy}, the roles of $\inp$ and $\out$ in the trace hierarchy are simply exchanged. Naturally, testers generalize both POVMs and instruments to the multi-time case.

\begin{figure}[t]
 \centering
 \includegraphics[width=0.95\linewidth]{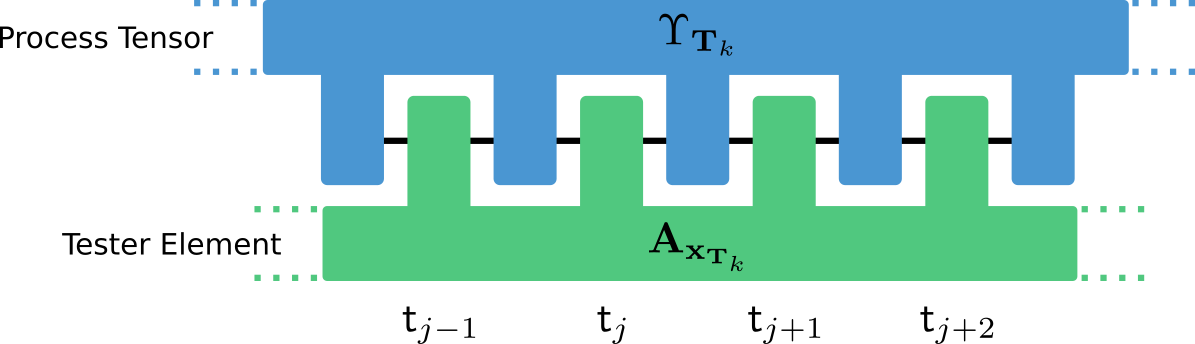}
 \caption{\textbf{Action of a process tensor on a tester element.} `Contracting' a process tensor (depicted in blue) with a temporally correlated measurement, i.e., a tester element (depicted in green), yields the probability for the occurrence of said tester element.}
 \label{fig::TensTest}
\end{figure}

Importantly, for any element $\Asetk$ of a tester that is ordered in the same way as the underlying process tensor, we have 
\begin{gather}
 0\leq \tr(\Upsk \Asetk^\trps) \leq 1, \quad \text{and} \quad \tr(\Upsk \Aset_{\Tset_k}^\trps) = 1,
\end{gather}
which can be seen by employing the hierarchy of trace conditions that hold for process tensors and testers. Similarly to the case of POVMs and instruments, letting a process tensor act on a tester element yields the probability to observe the outcome $\xsetk$ that corresponds to $\Asetk$ (see Figure~\ref{fig::TensTest} for a graphical representation). Below, we will encounter temporally correlated tester elements when discussing Markovianity and Markov order in the quantum setting.

As before, one might wonder why temporally correlated measurements -- unlike temporally correlated joint probability distributions -- made no occurrence in our discussion of classical stochastic processes. As before, the answer is rather simple; in our discussion of classical stochastic processes, there was no notion of different instruments that could be used at different times, so that there was also no means of temporally correlating them. Had we allowed for different classical instruments, i.e., had we allowed for different kinds of active interventions, then classically correlated instruments would have played a role as well. However, these instrument would have only displayed classical correlations between different points in time, since in no quantum information can be sent between classical instruments.

\subsubsection{Causality and dilation}

\label{sec:higherorderdilation}
In Sec.~\ref{sec:get}, we will see that, besides being a handy mathematical tool, process tensors allow for the derivation of a \textit{generalized} extension theorem, thus appearing to be the natural extension of stochastic processes to the quantum realm on a fundamental level. Here, we will, in a first step, connect process tensors to underlying dynamics. In classical physics, it is clear that every conceivable joint probability distribution can be realized by some -- potentially highly exotic -- classical dynamics. On the other hand, so far, it is unclear if the same holds for process tensors. By this, we mean, that, we have not shown the claim made above, that every process tensor, i.e., every positive matrix that satisfies the trace hierarchy of Eqs.~\eqref{eqn::hierarchyFirst} and~\eqref{eqn::Hierarchy} can actually be realized in quantum mechanics. We will provide a short `proof' by example here; more rigorous treatments can, for instance, be found in Refs.~\cite{eggeling_semicausal_2002, kretschmann_quantum_2005, chiribella_theoretical_2009, PhysRevA.97.012127}. 

Concretely, showing that any process tensor can be realized in quantum mechanics amounts to showing that they admit a quantum circuit that is only composed of pure states and unitary dynamics. This is akin to the Stinespring dilation we discussed in Sec.~\ref{sec::PurDil}, which allowed us to represent any quantum channel in terms of pure objects only. In this sense, the following dilation theorem will even be more general than the analogous statement in the classical case, where randomness has to be inserted `by hand'. 
\begin{figure}[t]
\centering
\includegraphics[width = 0.45\linewidth]{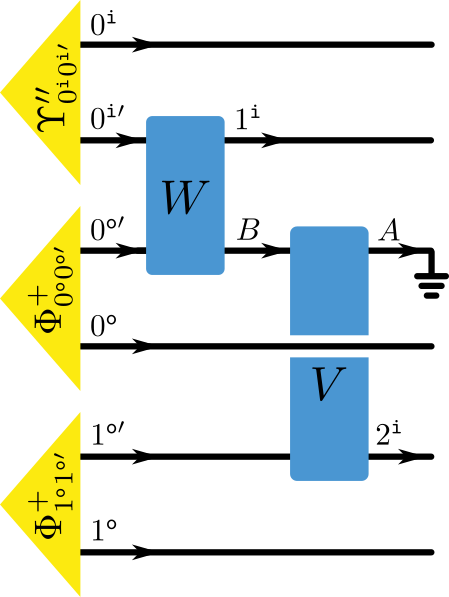}
\caption{\textbf{Dilation of the Choi state of a process tensor.} Up to normalization, Eqs.~\eqref{eqn::HighDil1} and~\eqref{eqn::HighDil2} together yield a quantum circuit for the implementation of the Choi state of a two-step process tensor that only consists of pure states and isometries (which could be further dilated to unitaries).}
 \label{fig::ChoiCircuit}
\end{figure}
\begin{figure}[t]
 \includegraphics[width=0.7\linewidth]{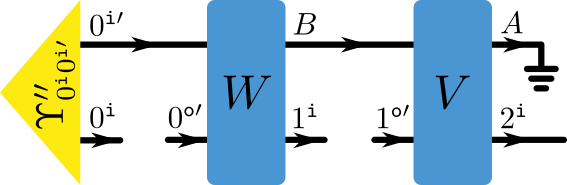}
 \caption{\textbf{Process tensor corresponding to Figure~\ref{fig::ChoiCircuit}}. Rearranging the wires of the circuit of Figure~\ref{fig::ChoiCircuit} and maximally entangled states (i.e., undoing the CJI), yields the representation of a process tensor we have already encountered in the previous section. As above, note that $\Hcal_{0}^{\out} \cong \Hcal_{0'}^{\out}$ and $\Hcal_{1}^{\out} \cong \Hcal_{1^\prime}^{\out}$, such that this is indeed the correct dilation of $\Upsilon_{0^\inp 0^\out 1^\inp 1^\out 2^\inp}$.}
 \label{fig::ProcTensRep}
\end{figure}

We use the property that all quantum states are purifiable to obtain a representation for general process tensors. For concreteness, let us consider a three-step process tensor $\Upsilon_{0^\inp 0^\out 1^\inp 1^\out 2^\inp}$, defined on three times $\{t_0,t_1,t_2\}$. Now, due to the causality constraints of Eqs.~\eqref{eqn::hierarchyFirst} and~\eqref{eqn::Hierarchy}, we have $\tr_{2^\inp}(\Upsilon_{0^\inp 0^\out 1^\inp 1^\out 2^\inp}) = \ident_{{1}^\out} \otimes \Upsilon'_{0^\inp 0^\out 1^\inp}$, where the prime is added for clearer notation in what follows. Since each of its components is proportional to a quantum state, this latter term can be dilated in at least two different ways: 
\begin{gather}
\begin{split}
 \ident_{{1}^\out} \otimes \Upsilon'_{0^\inp 0^\out 1^\inp} &= \tr_{2^\inp A}(\ketbra{\Upsilon}{\Upsilon}_{0^\inp 0^\out 1^\inp 1^\out 2^\inp A}) \\&= d_{1^\out} \tr_{1^{\out\prime}B}(\Phi_{1^\out1^{\out \prime}}^+ \otimes \ketbra{\Upsilon'}{\Upsilon'}_{0^\inp 0^\out 1^\inp B}),
 \end{split}
\end{gather}
where $\ket{\Upsilon}_{0^\inp 0^\out 1^\inp 1^\out 2^\inp A}$ and $\ket{\Upsilon'}_{0^\inp 0^\out 1^\inp B}$ are purifications of $\Upsilon_{0^\inp 0^\out 1^\inp 1^\out 2^\inp}$ and $\Upsilon'_{0^\inp 0^\out 1^\inp }$, respectively (with corresponding ancillary purification spaces $A$ and $\{1^{\out \prime},B\}$), and the additional pre-factor $d_{1^\out} = \tr(\ident_{1^\out})$ is required for proper normalization. These two different dilations of the same object are related by an isometry $V_{1^{\out\prime} B \rightarrow 2^\inp A} =: V$ that only acts on the dilation spaces, i.e.,
\begin{gather}
\label{eqn::HighDil1}
\begin{split}
 &\ketbra{\Upsilon}{\Upsilon}_{0^\inp 0^\out 1^\inp 1^\out 2^\inp A} \\
 &= d_{1^\out} V(\Phi_{1^\out1^{\out \prime}}^+ \otimes \ketbra{\Upsilon'}{\Upsilon'}_{0^\inp 0^\out 1^\inp B}) V^\dagger
\end{split}
\end{gather}
In the same vein, due to the causality constraints of $\Upsilon'_{0^\inp 0^\out 1^\inp}$, we can show that there exists an isometry $W_{0^{\out \prime} 0^{\inp \prime} \rightarrow 1^\inp B}=:W$, such that 
\begin{gather}
\label{eqn::HighDil2}
\begin{split}
&\ketbra{\Upsilon'}{\Upsilon'}_{0^\inp 0^\out 1^\inp B} \\
&=d_{0^\out} W (\Phi_{0^\out 0^{\out \prime}}^+ \otimes \ketbra{\Upsilon''}{\Upsilon''}_{0^\inp 0^{\inp \prime}}) W^\dagger,
\end{split}
\end{gather}
where $\ket{\Upsilon''}_{0^\inp 0^{\inp \prime}}$ is a pure quantum state. Inserting this into Eq.~\eqref{eqn::HighDil1} and using that $\Upsilon_{0^\inp 0^\out 1^\inp 1^\out 2^\inp} = \tr_{A}(\ketbra{\Upsilon}{\Upsilon}_{0^\inp 0^\out 1^\inp 1^\out 2^\inp A})$ yields -- up to normalization -- a representation of $\Upsilon_{0^\inp 0^\out 1^\inp 1^\out 2^\inp}$ in terms of a pure initial state $\ketbra{\Upsilon''}{\Upsilon''}_{0^\inp 0^{\inp \prime}} \otimes \Phi_{0^\out 0^{\out \prime}}^+ \otimes \Phi_{1^\out1^{\out \prime}}^+$ and subsequent isometries $W_{0^{\out \prime} 0^{\inp \prime} \rightarrow 1^\inp B}$ and $V_{1^{\out\prime} B \rightarrow 2^\inp A} $ (see Figure~\ref{fig::ChoiCircuit}). As any isometry can be completed to a unitary, this implies that $\Upsilon_{0^\inp 0^\out 1^\inp 1^\out 2^\inp}$ can indeed be understood as stemming from a quantum circuit consisting only of pure states and unitaries. This circuit simply provides the CJI of the corresponding process tensor, as can easily be seen by `removing' the maximally entangled states, and rearranging the wires in a more insightful way (see Figure~\ref{fig::ProcTensRep}). Naturally, these arguments can be extended to any number of times. Here, we sacrificed some of the mathematical rigor for brevity and clarity of the exposition; as mentioned, for a more rigorous derivation, see Refs.~\cite{eggeling_semicausal_2002, kretschmann_quantum_2005, chiribella_theoretical_2009, PhysRevA.97.012127}. Importantly, with this dilation propewrty at hand, we can be sure that \textit{every} process tensor actually has a physical representation, i.e., it describes a conceivable physical situation. This is akin to the case of channels, where the Stinespring dilation guaranteed that every CPTP map could be implemented in the real world. With these loose ends wrapped up, it is now time to discuss process tensors and quantum stochastic processes on a more axiomatic level.

\subsection{\texorpdfstring{Some mathematical rigor \\ Generalized extension theorem (GET)}{}}
\label{sec:get}

Above, we provided a consistent way to describe quantum stochastic processes. Importantly, this description given by process tensors can deal with the inherent invasiveness of quantum measurements, as it separates the measurements made by the experimenter from the underlying process they probe. Unsurprisingly then, employing this approach to quantum stochastic processes, the previously mentioned breakdown of the KET in quantum mechanics can be resolved in a satisfactory manner~\cite{accardi, Milz2017KET}. 

\begin{figure}[t]
 \centering
 \includegraphics[width=0.98\linewidth]{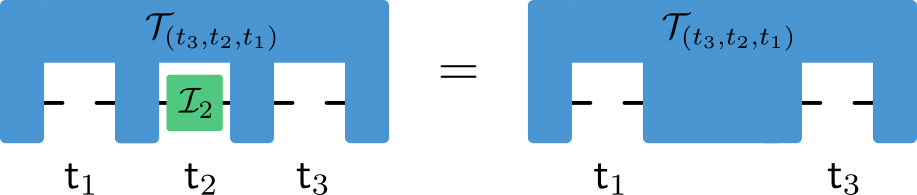}
 \caption{\textbf{Consistency condition for Quantum Stochastic Processes}. Letting a process tensor act on an identity (here at time $t_2$) yields the correct process tensor on the remaining times.}
 \label{fig::ConsQu}
\end{figure}

Recall that one of the ingredients of the Kolmogorov extension theorem -- which does not hold in quantum mechanics -- was the fact that a multi-time joint probability distribution contains all joint probability distributions for fewer times. In quantum mechanics on the other hand, a joint probability distribution, say, at times $\{t_1, t_2, t_3\}$ for instruments $\{\Jcal_1, \Jcal_2, \Jcal_3\}$ does not contain the information of what statistics one would have recorded, had one not measured at $t_2$, but only at times $\{t_1,t_3\}$. More generally, $\Pprob(x_3,x_2,x_1|\Jcal_3,\Jcal_2,\Jcal_1)$ does not allow one to predict probabilities for different instruments $\{\Jcal_1',\Jcal_2',\Jcal_3'\}$. On the other hand, the process tensor allows one to -- on the set of times it is defined on -- compute \textit{all} joint probabilities for \textit{all} employed instruments, in particular, for the case where one of more of the instruments are the `do-nothing' instrument. Consequently, it is easy to see that for a given process on, say, times $\{t_1,t_2,t_3\}$, the corresponding process tensor $\Tcal_{\{t_1,t_2,t_3\}}$ -- where for concreteness, here we use $\Tcal_{\{t_1,t_2,t_3\}}$ instead of $\Tcal_{\Tset_3}$ -- contains the correct process tensors for any subset of $\{t_1,t_2,t_3\}$. For example, we have 
\begin{gather}
 \Tcal_{\{t_1,t_3\}}[\sbt \,,\sbt \,] = \Tcal_{\{t_1,t_3\}}[\sbt \,,\Ical_2,\sbt \,],
\end{gather}
where $\Ical_2$ is the identity map $\Ical[\rho] = \rho$ at time $t_2$ (see Figure~\ref{fig::ConsQu} for a graphical representation). 

This, in turn, implies that process tensors satisfy a \textit{generalized} consistency condition. Importantly, as $\Ical$ is a unitary operation, letting $\Tcal$ act on an identity does generally \textit{not} coincide with the summation over measurement outcomes. Concretely, for any instrument with more than one outcome, we have $\sum_{x}\Acal_{x} \neq \Ical$, and thus summation over outcomes is not the correct way to `marginalize' process tensors. We will discuss below, why it works nonetheless for classical processes. To make this concept of compatibility for process tensors more manifest, let us revisit the concatenated Stern-Gerlach experiment we presented in Sec.~\ref{subsec::KET_QM} when we discussed the breakdown of the Kolmogorov extension theorem in quantum mechanics. There, the system of interest underwent trivial dynamics (given by the identity channel $\Ical$), interspersed by measurements in the $z$-, $x$-, and $z$- direction (see Figure~\ref{fig:mz}). Choosing the initial state of the system to be fixed and equal to $\ket{+}$ (as we did in Sec.~\ref{subsec::KET_QM}) then yields a corresponding process tensor that actso on CP map $\{\Acal_{x_1},\Acal_{z_2},\Acal_{x_3}\}$ at times $\{t_1,t_2,t_3\}$ as 
\begin{gather}
\begin{split}
    &\Tcal_{\{t_1,t_2,t_3\}}[\Acal_{x_1},\Acal_{z_2},\Acal_{x_3}] \\
    &= \tr[\Acal_{x_3}\circ \Ical_{2\rightarrow 3} \circ \Acal_{z_2}\circ \Ical_{1\rightarrow 2} \circ \Acal_{x_1})[\ketbra{+}{+}]]. 
\end{split}
\end{gather} 
Now, replacing $\Acal_{z_2}$ in the above by $\Ical_2$, since $\Ical_{2\rightarrow 3} \circ \Ical_2 \circ \Ical_{1\rightarrow 2} = \Ical_{1\rightarrow 3}$ we see that we exactly obtain the process tensor for trivial dynamics between $t_1$ and $t_3$, i.e., 
\begin{gather}
\begin{split}
   \Tcal_{\{t_1,t_2,t_3\}}[\Acal_{x_1},\Ical_2,\Acal_{x_3}] \!&=\! \tr[\Acal_{x_3}\circ \Ical_{1\rightarrow 3} \circ \Acal_{x_1})[\ketbra{+}{+}]] \\
   &= \Tcal_{\{t_1,t_3\}}[\Acal_{x_1},\Acal_{x_3}]. 
\end{split}
\end{gather}
On the other hand, summing over the outcomes at $t_2$ (as one would do in the classical case), we would \textit{not} obtain the correct process tensor in absence of a measurement at $t_2$. Specifically, setting $\Acal_z = \sum_{z_2}\Acal_{z_2}$, we obtain 
\begin{gather}
\begin{split}
    &\sum_{z_2} \Tcal_{\{t_1,t_2,t_3\}}[\Acal_{x_1},\Acal_{z_2},\Acal_{x_3}] \\
    &= \tr[\Acal_{x_3}\circ \Acal_z \circ \Acal_{x_1})[\ketbra{+}{+}]] \neq \Tcal_{\{t_1,t_3\}}[\Acal_{x_1},\Acal_{x_3}]. 
\end{split}
\end{gather}
The quantum process we discussed in Sec.~\ref{subsec::KET_QM}, and more generally, \textit{all} quantum processes thus satisfy consistency properties, however, not in exactly the same sense as classical processes do.

With this generalized consistency condition at hand, a \textit{generalized} extension theorem (GET) in the spirit of the KET can be proven for quantum processes~\cite{accardi, Milz2017KET}; any underlying quantum process on a set of times $\Tset$ leads to a family of process tensors $\{\Tcalk\}_{\Tsetk \subset \Tset}$ that are compatible with each other, while any family of compatible process tensors implies the existence of a process tensor that has all of them as marginals in the above sense. More precisely, setting 
\begin{gather}
 \Tcall^{|\Tsetk}[\sbt\,]:= \Tcall \bigg[ \bigotimes_{\alpha\in \Tsetl \setminus \Tsetk} \Ical_\alpha,\sbt\,\bigg],
\end{gather}
where we employ the shorthand notation $\bigotimes_{\alpha\in \Tsetl \setminus \Tsetk} \Ical_\alpha$ to denote that the identity map is `implemented' at each time $t_\alpha \in \Tsetl \setminus \Tsetk$, we have the following Theorem~\cite{accardi, Milz2017KET}:

\vspace{5pt}
\textbf{Theorem. (GET)} \textit{Let $\Tset$ be a set of times. For each finite $\Tsetk \subset \Tset$, let $\Tcalk$ be a process tensor. There exists a process tensor $\Tcal_\Tset$ that has all finite ones as `marginals', i.e., $\Tcalk = \Tcal_\Tset^{|\Tsetk}$ iff all finite process tensors satisfy the consistency condition, i.e., $\Tcalk = \Tcall^{|\Tsetk}$ for all finite $\Tsetk \subset \Tsetl \subset \Tset$.}
\vspace{5pt}

As the proof of the GET is somewhat technical, we will not provide it here and refer the interested reader to Refs.~\cite{accardi, Milz2017KET}. We emphasize though, that, since the basic idea of the GET is -- just like the KET -- based on compatibility of descriptors on different sets of times, it can be proven in a way that is rather similar to the proof of the KET~\cite{Milz2017KET}.

\begin{figure}
 \centering
 \includegraphics[width=0.8\linewidth]{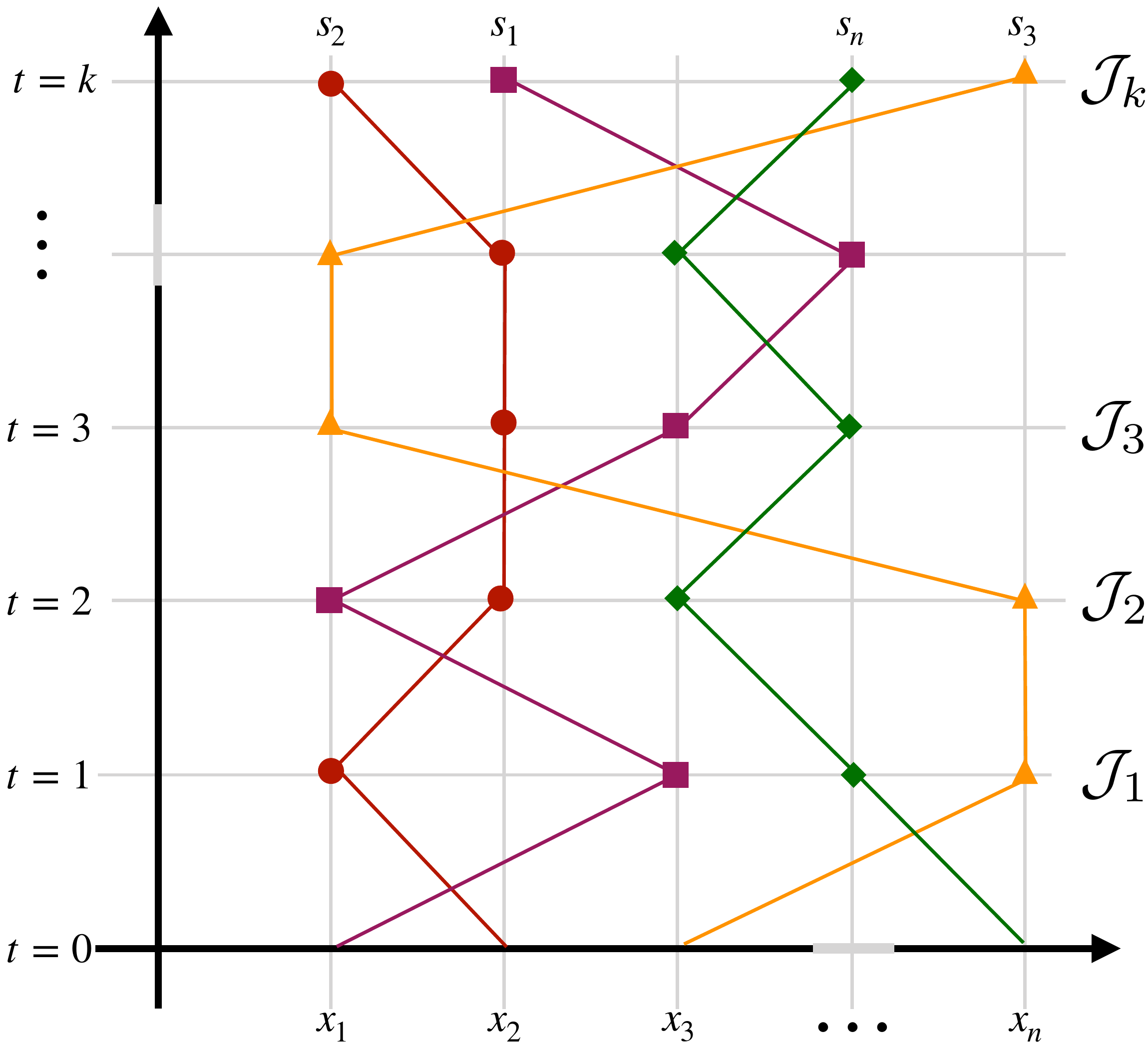}
 \caption{\textbf{`Trajectories' of a quantum stochastic process}. An open quantum process is fully described once all joint probabilities for sequences of outcomes are known for all possible instruments an experimenter can employ to probe the process. Like in the classical case, each sequence of outcomes can be considered a trajectory, but unlike in the classical case, there is no ontology attached to such trajectories. Additionally, each sequence of outcomes in the quantum case corresponds to a sequence of measurement operators, not just labels. If both the process and the allowed (non-invasive) measurements are diagonal in the same fixed basis, then the above figure coincides with Figure~\ref{fig:traj}, where trajectories of classical stochastic processes were considered. Importantly, while in classical physics, only probabilistic mixtures of different trajectories are possible, quantum mechanics allows for the coherent superposition of `trajectories'~\cite{Sakuldee2018}.}
 \label{fig::Quantum_Traj}
\end{figure}

\begin{figure}[t]
\centering
\includegraphics[width=0.75\linewidth]{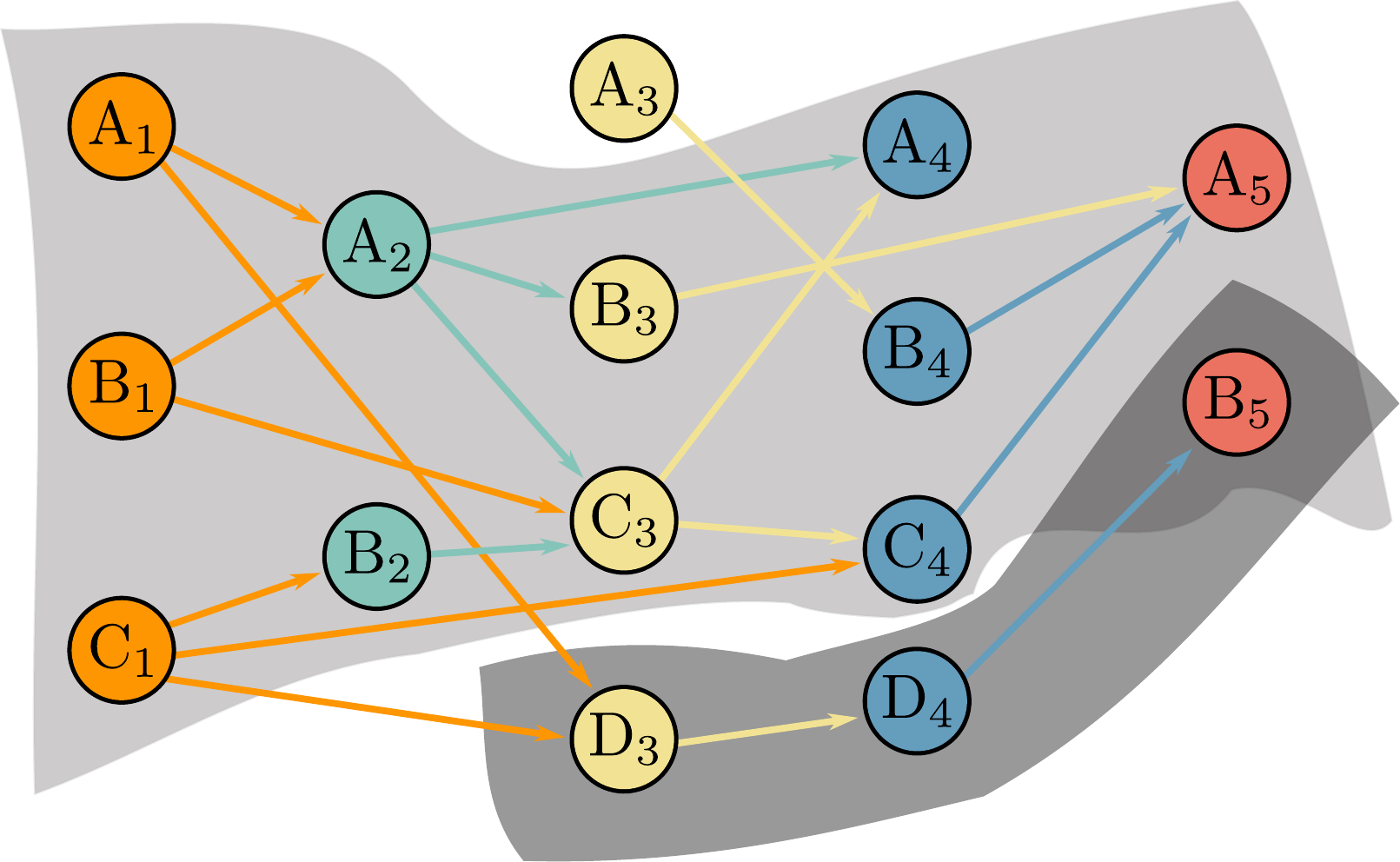}
\caption{\textbf{(Quantum) Causal network.} Performing different interventions allows for the causal relations between different events (denoted by $X_j$) to be probed. For example, in the figure the event $B_1$ directly influences the events $C_3$ and $A_2$, while $A_3$ influences only $B_4$. As not all pertinent degrees of freedom are necessarily in the control of the experimenter, such scenarios can equivalently be understood as an open system dynamics. Any such scenario can be described by a process tensor~\cite{chiribella_theoretical_2009}, and the GET applies, even though active interventions must be performed to discern causal relations. For example, the events $D_3, D_4, B_5$ could be successive (\textit{e.g.}, at times $t_3,t_4$ and $t_5$) spin measurements in $z$-, $x$- and $z$-direction, respectively. Summing over the results of the spin measurement in $x$-direction at $t_4$ would not yield the correct probability distribution for two measurements in $z$-direction at $t_3$ and $t_5$ only, but consistency still holds on the level of process tensors (see also Sec.~\ref{subsec::KET_QM}).}
\label{fig::image3}
\end{figure}

Importantly, this theorem contains the KET as a special case, namely the one where all involved process tensors and operations are classical. Consequently, introducing process tensors for the description of quantum stochastic processes closes the apparent conceptual gaps we discussed earlier, and provides a direct connection to their classical counterpart; while quantum stochastic processes can still be considered as mappings from sequences of outcomes to joint probabilities, in quantum mechanics, a full description requires that these probabilities are known for \textit{all} instruments an experimenter could employ (see Figure~\ref{fig::Quantum_Traj}). Additionally, the GET provides satisfactory mathematical underpinnings for physical situations, where active interventions are purposefully employed, for example, to discern different causal relations and mechanisms. This is for instance the case in classical and quantum causal modeling~\cite{pearl_causality_2009, 1367-2630-18-6-063032, PhysRevX.7.031021, barrett_quantum_2020} (see Figure~\ref{fig::image3} for a graphical representation).

In light of the fact that, mathematically, summing over outcomes of measurements does not amount to an identity map -- even in the classical case -- it is worth reiterating from a mathematical point of view, why the KET holds in classical physics. For a classical stochastic process, we always implicitly assume that measurements are made in a fixed basis (the computational basis), and no active interventions are implemented. Mathematically, this implies that the considered CP maps are of the form $\Acal_{x_j}[\rho] = \braket{x_j|\rho|x_j}\ketbra{x_j}{x_j}$. Summing over these CP maps yields the completely dephasing CPTP map $\Delta_j[\rho] := \sum_{x_j} \braket{x_j|\rho|x_j}\ketbra{x_j}{x_j}$, which does not coincide with the identity map. However, on the set of states that are diagonal in the computational basis, the action of both maps coincides, i.e., $\Delta_j[\rho] = \Ical_j[\rho]$ for all $\rho = \sum_{x_j} \lambda_{x_j} \ketbra{x_j}{x_j}$. More generally, their action coincides with the set of all combs that describe classical processes~\cite{milz_when_2020}. In a sense then, mathematically speaking, the KET works because, in classical physics, only particular operations, as well as particular process tensors are considered. Going beyond either of these sets requires -- already in classical physics -- a more general way of `marginalization', leading to an extension theorem that naturally contains the classical one as a special case. 

With the GET, which, here, we only looked at in a very cursory manner, we have answered the final foundational question about process tensors and have established them as the natural generalization of the descriptors of classical stochastic processes to the quantum realm, both from an operational as well as an axiomatic perspective. While rather obvious in hindsight, it required the introduction of some machinery to be able to properly describe measurements in quantum mechanics. For the remainder of this Tutorial, we will employ the developed machinery to discuss questions properties of quantum stochastic processes, in particular that of Markovianity and Markov order.

\section{Properties of quantum stochastic processes}
\label{sec::PropQuant}

\subsection{Quantum Markov conditions and Causal break}
\label{sec::QuarkovCausalBr}

Now, armed with a clear description for quantum stochastic processes, i.e., the process tensor, we are in the position to ask when is a quantum process Markovian. We will formulate a quantum Markov condition~\cite{PhysRevLett.120.040405, 1367-2630-18-6-063032} by employing the notion of \textit{causal breaks}. As we have seen in the classical case, Markovianity is a statement about conditional independence of the past and the future. Intuitively speaking, information of the past can be transmitted to the future in two different ways: via the system itself and via the inaccessible environment. In a Markovian process, the environment does not transmit any past system information to the future process on the system. This condition is encapsulated in the classical Markov condition 
\begin{gather}
\label{eqn::MarkovClassRev}
    \Pprob(x_k| x_{k-1}, \dots, x_0) = \Pprob(x_k| x_{k-1}) \quad \forall k.
\end{gather}
Conditioning on a given outcome blocks the information flow from the past to the future through the system (since it is set to a fixed outcome), and conditional independence from the past then tells us that there is no information that travels through the environment.

A causal break allows one to extend this classical intuition to the quantum case. It is designed to block the information transmitted by the system itself and at the same time look for the dependence of the future dynamics of the system conditioned on the past control operations on the system. If the future depends on the past controls, then we must conclude that the past information is transmitted to the future by the environment, which is exactly the non-Markovian memory.

\begin{figure}[t]
 \centering
 \includegraphics[width=0.95\linewidth]{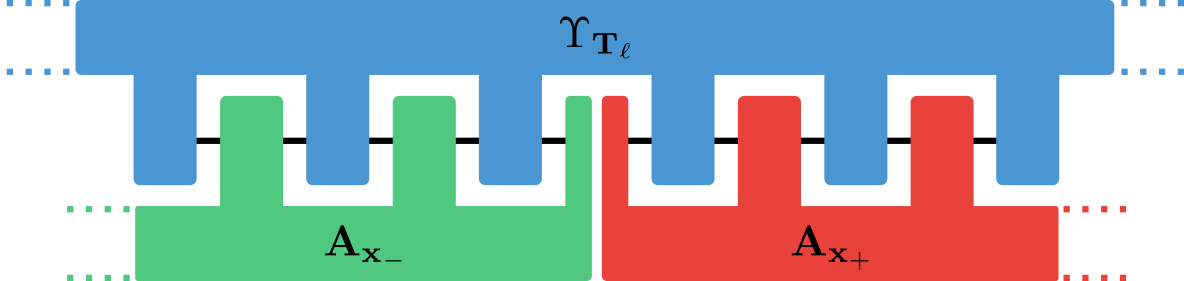}
\caption{\textbf{Determining whether a quantum process is Markovian.} Generalized testers (multi-time instruments) $\Aset_{\xset_-}$ and $\Aset_{\xset_+}$ are applied to the system during a quantum process, where the subscripts represent the outcomes. The testers are chosen to implement a causal break at a timestep $t_k$, which ensures that the only way the future outcomes depend on the past if the process is non-Markovian. Thus by checking if the future depends on the past for a basis of instruments we can certify the process to be Markovian or not. \label{image-cloud}}
 \label{fig::Quarkov_condition}
\end{figure}

Let us begin by explicitly denoting the process $\Upsilon_{\Tsetl}$ on a set of time $\Tsetl = \{t_0, \dots, t_k, \dots, t_\ell\}$. We break this set into two subsets at an intermediate time step $k < l$ as $\Tset_- = \{t_0, \dots, t_{k^-}\}$ and
$\Tset_+ = \{t_{k^+}, \dots, t_\ell\}$ where $t_{k^-}$ and $t_{k^+}$ respectively are the times corresponding to the spaces in the Choi state of the process denoted by $k^\inp$ and $k^\out$. In the first segment, we implement a tester element $\Aset_{\xset_{-}}$ belonging to instrument $\Jcal_-$ with outcomes $\{\xset_{-}\}$. In the next time segment, as the system evolves to time step $\ell$, we implement a tester element $\Aset_{\xset_{+}}$ belonging to instrument $\Jcal_+$ with outcomes $\{\xset_{+}\}$ (see Figure~\ref{fig::Quarkov_condition}). Together, we have applied two independent tester elements $\Aset_{\xset_+} \otimes \Aset_{\xset_-}$, where the simple tensor product between the two testers implies their independence. In detail, the two testers split the timestep $k$: the first instrument ends with a measurement on the output of the process at time $t_k$ (labeled as $t_{k^-}$). The second instrument begins with preparing a fresh state at the same time (labeled as $t_{k^+}$). Importantly, it implies a \textit{causal break} that prevents any information transmission between the past $\Tset_{-}$ and the future $\Tset_{+}$ via the system which is similar to our reasoning in the classical case. Thus, detecting an influence of past operations on future statistics when implementing a causal break implies the presence of memory effects mediated by the environment. Additionally, it is easy to see that causal breaks can span a basis of the space of all testers on $\Tset$, a property we will make use of below.

For future convenience, let us define the process tensor $\Upsilon_{\Tset_-} := \tfrac{1}{d_{O^+}} \tr_+(\Upsilon_{\Tset_\ell})$ that is defined on $\Tset_-$ only. As we have seen in our discussion of causal ordering of process tensor, $\Upsilon_{\Tset_-}$ is well-defined and reproduces the statistics on $\Tset_-$ correctly. We now focus on the conditional outcome statistics of the future process, which are given by Eq.~\eqref{eq:process}
\begin{gather}
\label{eqn::CondProbTester}
\Pprob(\xset_{+}| \Jcal_+, \xset_- ) = \tr[\Upsilon_{\Tset_+}^{(\Aset_{\xset_-})} \ \Aset_{\xset_+}^\trps].
\end{gather}
Note that we have added a second condition $\xset_-$ on the LHS as well as an additional superscript on the RHS because the future process, in general, may depend on the outcomes for the past instrument $\Jcal_-$. Importantly, above we have set 
\begin{gather}
\label{eqn::Renormalize}
    \Upsilon_{\Tset_+}^{(\Aset_{\xset_-})} := \tr_-(\Upsilon_{\Tset_\ell} \Aset_{\xset_-}^\trps)/\tr(\Upsilon_{\Tset_-} \Aset_{\xset_-}^\trps)\, ,
\end{gather}
making the $\Pprob(\xset_{+}| \Jcal_+, \xset_- )$ given in Eq.~\eqref{eqn::CondProbTester} a proper conditional probability.
This operationally well-defined conditional probability is fully consistent with the conditional classical probability distributions in Eq.~\eqref{eq:conddist}.

The causal break at timestep $k$ guarantees that the system itself cannot carry any information of the past into the future beyond step $k$. The only way the future process $\Upsilon_{\Tset_+}^{(\Aset_{\xset_-})}$ could depend on the past is if the information of the past is carried across the causal break via the environment. We have depicted this in Figure~\ref{image-cloud}, where the only possible way the past information can go to the future is through the process tensor itself. This immediately results in the following operational criterion for a Markov process:

\vspace{5pt}
\textbf{Quantum Markov Condition.} \textit{A quantum process is Markovian when the future process statistics, after a causal break at time step $k$ (with $l>k$), is independent of the past instrument outcomes $\xset_-$
\begin{gather}\label{eq:qcondind}
\Pprob(\xset_+| \Jcal_+, \xset_-) =
\Pprob(\xset_+| \Jcal_+),
\end{gather}
$\forall \, \Jcal_+, \Jcal_-$ and $\forall \, k \in \Tset$.}
\vspace{5pt}

Alternatively, the above Markov condition says that a quantum process is non-Markovian iff there exist two past testers outcomes, $\xset_-$ and $\xset'_-$, such that after a causal break at time step $k$, the conditional future process statistics are different for a some future instrument $\Jcal_+$:
\begin{gather}\label{CLMarkovCond1}
\Pprob(\xset_+| \Jcal_+, \xset_- ) \ne
\Pprob(\xset_+| \Jcal_+, \xset'_-).
\end{gather}
Conversely, if the statistics remain unchanged for all possible past controls, then the process is Markovian. Naturally, the Markov condition of Eq.~\eqref{eq:qcondind} should ring a bell and remind the reader of the exactly analogous Markov condition in the classical case. 

The above quantum Markov condition is fully operational (since it is phrased in terms of conditional probabilities, which can, in principle, be determined experimentally) and thus it is testable with a finite number of experiments\ftnt{It can even be tested for if the ordering of events is not given a priori~\cite{giarmatzi_quantum_2018}}. Suppose the conditional independence in Eq.~\eqref{eq:qcondind} holds for a complete basis of past and future testers $\{\hat{\Aset}_{\xset_-} \otimes \hat{\Aset}_{\xset_+}\}$, then, by linearity it holds for any instruments, and the future is always conditionally independent of the past. It is worth noting that this definition is the quantum generalization of the \emph{causal Markov condition} for classical stochastic evolutions where interventions are allowed~\cite{Pearl}.

Additionally, in spirit, the above definition is similar to the satisfaction of the \textit{Quantum regression formula (QRF)}~\cite{Gardiner, carmichael_open_1993, BreuerPetruccione}. Indeed, its equivalence to a generalized QRF has been shown in~\cite{Li2018}, while satisfaction of the generalized QRF has been used in~\cite{lindblad_non-markovian_1979, accardi} as a basis for the definition of quantum Markovian processes. On the other hand, the relation of the QRF and the witnesses of non-Markovianity we discussed in Sec.~\ref{sec::Witness} has been investigated~\cite{PhysRevLett.94.200403, Guarnieri2014}. Here, we opt for the understanding of Markovianity in terms of conditional future-past independence, an approach fully equivalent to the one taken in the aforementioned works~\cite{lindblad_non-markovian_1979, accardi, Li2018}.

\subsubsection{Quantum Markov processes}
\label{sec:Markovprocess}

Intuitively, the quantum Markov condition implies that any segment of the process  is uncorrelated with the remainder of the process. Put differently, at \textit{every} time in $\Tsetl$, a Markovian process is conditionally independent of its past. This right away means that a Markov process must have a rather remarkably simple structure. Translating the idea of conditional independence to the correlations that can persist in $\Upsilon_{\Tsetl}$, one would expect that $\Upsilon_{\Tsetl}$ cannot contain any correlations between distant times if the underlying process is Markovian. And indeed, the Choi state of the process tensor for a Markov process can be shown to be simply a product state
\begin{gather}\label{eq:qmarkov}
 \Upsilon_{\Tsetl}^\markov = \rho_0 \otimes \bigotimes_{j=0}^{\ell-1} \Upsilon_{\Ecal_{({j+1}^-:j^+)}},
\end{gather}
where each $\Upsilon_{\Ecal_{({j+1}^-:j^+)}}$ is the Choi matrix of a CPTP map from $t_{j^+}$ to $t_{{j+1}^-}$ and $\rho_0$ is the initial state of the system. Before commenting on the origin of Eq.~\eqref{eq:qmarkov}, let us first comment on the meaning of its structure. The above equation simply says that there are no temporal correlations in the process, other than those between neighboring time steps facilitated by the channel on the system itself, i.e., there is no memory transmitted via the environment. An obvious example of such a process is a closed process, i.e., a unitary process. Here each $\Upsilon_{\Ecal_{({j+1}^+:j^-)}}$ will be maximally entangled (since quantum information is transmitted perfectly by unitary maps) and corresponds to a unitary evolution, respectively. However, there are no other memory effects between distant times present, since in a closed process there is no environment that could transport such memory.

By inserting Eq.~\eqref{eq:qmarkov} into Eq.{eq:process}, where the action of the process tensor in terms of its Choi state was defined, we see that for a sequence of CP maps $\Asf_{x_0} \otimes \cdots \Asf_{x_\ell}$ performed by the experimenter, we see that, after rearrangement, we have 
\begin{gather}
\label{eqn::ActionMarkow}
    \Pprob(\xset_{\Tsetl}|\Jsetl) = \tr[\Asf_{x_\ell}^\trps \Upsilon_{\Ecal_{(\ell^-:\ell-1^+)}}\Asf_{x_{\ell-1}}^\trps \cdots \Upsilon_{\Ecal_{(1^-:0^+)}} \Asf_{x_0}^\trps \rho_0],
\end{gather}
which simply looks like a concatenation of mutually independent maps that act on the system alone, as one would expect from a Markovian process. This becomes even clearer when we represent the above equation in terms of quantum maps. Then, the action of the corresponding process tensor $\Tcal^\markov_{\Tsetk}$ CP maps $\{\Acal_{x_i}\}$ can be expressed equivalently to Eq.~\eqref{eqn::ActionMarkow} as
\begin{gather}
\begin{split}
    &\Tcal^\markov_{\Tsetk}[\rho,\Acal_{x_1},\dots,\Acal_k] = \tr\{(\Acal_k\circ\Ecal_{(k^-:{k-1}^+)}\circ 
    \cdots \\
    &\phantom{\Tcal^\markov_{\Tsetk}[\rho,\Acal_{x_1},\dots}\cdots \circ\Ecal_{({2}^-:1^+)} \circ \Acal_{x_1} \circ \Ecal_{({1}^-:0^+)})[\rho_0]\}, 
\end{split}
\end{gather}
where all $\Ecal_{({j+1}^-:j^+)}$ (corresponding to $\Upsilon_{\Ecal_{({j+1}^-:j^+)}}$) are mutually independent CPTP maps that act on the system alone, and $\rho_0$ is the initial system state. While this property of independent CPTP maps -- at first sight -- seems equivalent to CP divisibility, we emphasize that it is strictly stronger, as the mutual independence of the respective maps has to hold for arbitrary interventions at all times in $\Tset_k$~\cite{PhysRevLett.120.040405,PhysRevA.97.012127, Milz2019}  and is thus -- unlike CP divisibility -- a genuine multi-time statement.

As an aside, the above form for Markov processes does not mean that we need to do experiments with causal breaks in order to decide Markovianity of a process. We simply need to determine if the process tensor has any correlations in time which can also be done using noisy or temporally correlated instruments that do not correspond to causal breaks. We can infer the correlations in a process once we have reconstructed the propcess tensor. This can be done -- as outlined above -- by tomography, which only requires applying a linear basis of instruments, causal breaks or not. Causal breaks, however, have the conceptual upside that they make the relation to the classical Markov condition transparent. Additionally, deviation from Markovianity can already be witnessed -- and assertions about the size of the memory can be made -- even if not a full basis of operations is available to the experimenter~\cite{Milz2018A, costa2019, Morris2019, White2020}, but we will not delve into these details here. 

Finally, let us comment on how we actually arrived at the product structure of Markovian process tensors, starting from the requirement of conditional independence of Eq.~\eqref{eq:qcondind}. Slightly rewritten in terms of process tensors, conditional independence of the future and the past implies that 
\begin{gather}
    \tr_{-}(\Upsilon_{{\Tsetl}} \Asf_{{\xset_+}}^\trps) \propto \tr_-(\Upsilon_{\Tsetl} \Asf_{\xset_+}^{\prime\, \mathrm{T}}) \quad \forall  \Asf_{\xset_+}, \Asf_{\xset_+}^{\prime},
\end{gather}
where the proportionality sign $\propto$ is used instead of an equality, since $\Asf_{\xset_+}$ and $\Asf_{\xset_+}^{\prime}$ generally occur with different probabilities (this has no bearing on the \textit{conditional} future probabilities though, since they are renormalized by the respective past probabilities (see Eqs.~\eqref{eqn::CondProbTester} and~\eqref{eqn::Renormalize}). Since the above equation has to hold for \textit{all} conceivable tester elements $\Asf_{\xset_+}$ and $\Asf_{\xset_+}^{\prime})$ (and thus, by linearity, for all matrices), it is easy to see that the corresponding $\Upsilon_{\Tsetl}$ has to be of product form, i.e., $\Upsilon_{\Tsetl} = \Upsilon_+ \otimes \Upsilon_-$. Demanding that conditional independence holds for \textit{all} times in $\Tsetl$ then implies that $\Upsilon_{\Tsetl}$ is indeed of the product form postulated in Eq.~\eqref{eq:qmarkov}. More detailed proofs of this statement can, for example, be found in~\cite{PhysRevA.97.012127, PhysRevLett.120.040405, 1367-2630-18-6-063032}. 

It is important to emphasize that we did not start out by postulating this product structure of Markovian processes. While tempting -- and eventually correct -- it would not have been an a priori operationally motivated postulate, but rather one guided by purely mathematical consideration. Here, we rather started from a fully operational quantum definition of Markovianity, phrased entirely in terms of experimentally accessible quantities, and in line with its classical counterpart.

Besides following the same logic as the classical definition of Markovianity, that is, conditional independence of the future and the past, the above notion of Markovianity also explicitly boils down to the classical one in the correct limit: Choosing \textit{fixed} instruments at each time in $\Tset_k$ yields a probability distribution $\Pprob(x_k,\dots,x_1)$ for the possible combinations of outcomes. Now, if each of the instruments only consists of causal breaks -- which is the case in the study of classical processes -- then a (quantum) Markovian process yields a joint probability distribution for those instruments that satisfies the classical Markov condition of Eq.~\eqref{eqn::MarkovClassRev}. The quantum notion of Markovianity thus contains the classical one as a special case. One might go further in the restriction to the classical case, by demanding that the resulting statistics also satisfy the Kolmogorov consistency conditions we discussed earlier. However, on the one hand, there are quantum processes that do not satisfy Kolmogorov consistency conditions, independent of the choice of probing instruments~\cite{milz_when_2020}. On the other hand, Markovianity is also a meaningful concept for classical processes with interventions~\cite{pearl_causality_2009}, where Kolmogorov conditions are generally not satisfied. Independent of how one aims to restrict to the classical case, the notion of Markovianity we introduced here for the quantum case would simplify to the respective notion of Markovianity in the classical case.

Below we will use the structure of Markovian processes to construct operationally meaningful measures for non-Markovianity that go beyond the witnesses of non-Markovianity based on two-time correlations we presented in Sec.~\ref{sec::Witness}. We then discuss the concept of quantum Markov order to close this section.

Before doing this we will shortly discuss how the quantum Markov condition we introduced above relates to the aforementioned witnesses of non-Markovianity we have already encountered. In contrast to the non-Markovianity witnesses discussed in Sec.~\ref{sec::Witness}, the above condition is necessary \textit{and} sufficient for memorylessness. That is, if it holds, then there is no physical experiment that will see conditional dependence between the past and future processes. If it does not hold then there exists some experiment that will be able to measure some conditional dependence between the past and the future processes. In fact, a large list of non-Markovian witnesses, defined in Refs.~\cite{PhysRevLett.103.210401, PhysRevLett.105.050403, hou2011, PhysRevLett.105.050403, PhysRevA.82.042103, mazzola2012dynamical, PhysRevA.82.042107, rodriguez2008completely, modiscirep, IC-breuer, rodriguez2012unification, PhysRevLett.107.180402, Gessner:2014kl, rodriguez2012unification, PhysRevA.86.044101, zhihe2017, fanchini2014, bylicka2014, pineda2016, bylicka_constructive_2017, PhysRevA.83.022109, PhysRevLett.101.150402, rajagopal, sabri}, herald the breaking of the quantum Markov condition. However, there are always non-Markovian processes that will not be identified by most of these witnesses. This happens because these witnesses above usually only account for (at most) three-time correlations. Many of them are based on the divisibility of the process. A Markov process, which has the form of Eq.~\eqref{eq:qmarkov}, will always be CP-divisible, while the converse does not hold true in general~\cite{accardi, Milz2019}. To see this latter point, it suffices to show an example of a CP-divisible process that is non-Markovian.

\subsubsection{Examples of divisible non-Markovian processes}
\label{sec::ExDivMark}

\begin{figure}[t]
\begin{center}
\includegraphics[width=.90\linewidth]
{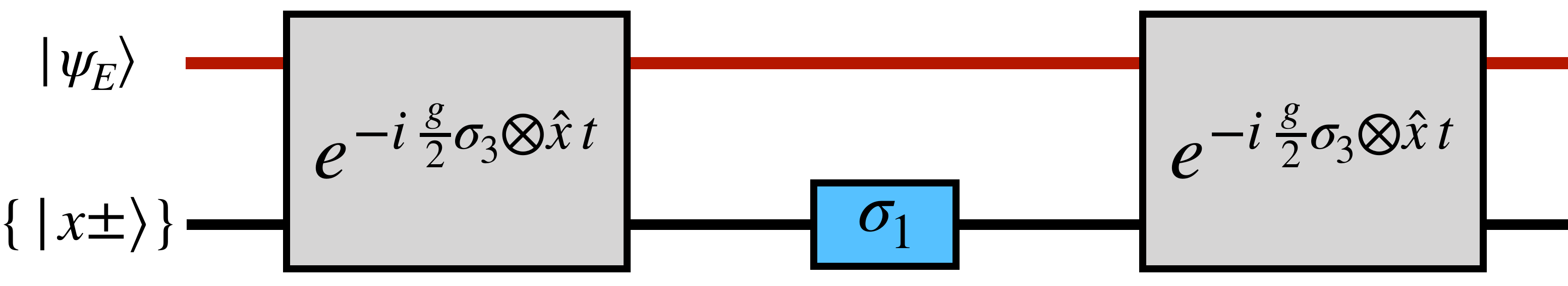}
\caption{\textbf{A CP-divisible but non-Markovian process.} A qubit system is prepared in states $\ket{x\pm}$ and evolves along with an environment. The uninterrupted dynamics of the system is pure dephasing, which will be certified as Markovian by two-point witnesses. However, when an instrument $X$ is applied at time $t$, the system dynamics reverse and the system returns to its original state, which is only possible in the presence of non-Markovian memory. \label{image-cpdivisibility}}
\end{center}
\end{figure}

A completely positive and divisible process on a single qubit system can be acquired by following the prescription in Refs.~\cite{Lindblad1980, accardi, PhysRevA.92.022102}, where the so-called shallow pocket model was discussed. We begin with the system in an arbitrary state $\rho(0)$ that interacts with an environment whose initial state is a Lorentzian wavefunction 
\begin{gather}\label{eq:lorentzian}
\braket{x|\psi_\etxt} = \psi_\etxt(x) = \sqrt{\frac{\mathscr{G}}{\pi}} \frac{1}{x+i\mathscr{G}}.
\end{gather}
We assume the initial state to be uncorrelated, i.e. of the form $\rho(0) \otimes \ketbra{\psi_\etxt}{\psi_{\etxt}}$. The two evolve together according to the Hamiltonian $H_{\setxt}= \frac{g}{2}\sigma_3 \otimes \hat{x}$, where $\hat{x}$ is the environmental position degree of freedom. The total $\setxt$ dynamics are then due to the unitary operator
\begin{gather}\label{eq:shallowU}
 U_t = e^{-i H_{\setxt} t}.
\end{gather}

It is easy to show, by partial tracing of the environment $\etxt$, that the reduced dynamics of the system $\stxt$ is pure dephasing in the $z$-basis (see Eq.~\eqref{eq:dephasingmaster} in Sec.~\ref{sec::Snap}), and can be written exactly in GKSL form, i.e., if the system is not interfered with, the evolution between any two points is a CPTP map of the following form:
\begin{gather}
\begin{split}
&\rho(t_j) = \Ecal_{(t_j-t_i)}[\rho(t_i)] \quad \mbox{with} \\
&\Ecal_{(t_j-t_i)}=\exp\{\mathcal{L} (t_j-t_i)\}.
\end{split}
\end{gather}
As we argued above, such a process is both completely positive, fully divisible~\cite{sabri, Rivas2014, RevModPhys.88.021002}, and also has a `Markovian' generator as required by the snapshot method~\cite{PhysRevLett.101.150402}. However, as we will see now, this process is not Markovian, since there are instruments that can detect memory in the process.

Now suppose we start the system in initial states $\rho_\pm(0) := \{\ket{x\pm}\!\bra{x\pm} \}$. After some time $t$, these states will have the form
\begin{gather}\label{eq:statesatt}
 \rho_\pm(t) := 
 \frac{1}{2}
 \begin{pmatrix}
 1 & \pm e^{-\gamma t}\\
 \pm e^{-\gamma t} & 1
 \end{pmatrix} \quad \mbox{with} \quad \gamma = g \ \mathscr{G}.
\end{gather}

It is then easy to see that the trace distance between the two states will monotonically decrease:
\begin{gather}
 \Dcal[\rho_+(t),\rho_-(t)] := \tfrac{1}{2}\|\rho_+(t) -\rho_-(t)\| = e^{-\gamma t}.
\end{gather}
This means that the non-Markovianity witness based on non-monotonicity of the trace-distance measure, given in Ref.~\cite{PhysRevLett.103.210401}, would call this a Markovian process. This is not surprising as the process is divisible, which a stronger witness for non-Markovianity than the trace distance~\cite{Rivas2014, PhysRevA.83.052128}. This process will also be labeled as Markovian by the snap-shot approach, as the generator of the dynamics of the system alone will always lead to CP maps. In fact, we have already shown in Sec.~\ref{sec::Witness} that divisibility-based witnesses will not see any non-Markovianity in a pure dephasing process. However, as we have discussed, Markovianity is a multi-time phenomenon that should be decided on conditional independence of events at different points in time.

To take this argument further, let us split the process from $0 \to 2t$ into two segments: $0 \to t$ and $t \to 2t$. If the process is indeed Markovian then we can treat it identically in each segment, i.e., the dynamical map for both segments will be the same.  This fact should be independent of whether or not an instrument was performed at time $t$; observing a change of the dynamics from $t$ to $2t$ would thus constitute a memory effect. Now, using this intuition, we show that this process, while divisible, is indeed non-Markovian. However, the usual witnesses fail to detect temporal correlations as the process only reveals non-Markovianity when an instrument at an intermediate time is applied, see Figure~\ref{image-cpdivisibility}. 

Suppose, we apply a single element (unitary) instrument $\Jcal_1 = X[\sbt \, ] := \sigma_1 (\sbt\, ) \, \sigma_1$ at time $t$. Doing so should not break the Markovianity of the process. Moreover, the process should not change at all because the states in Eq.~\eqref{eq:statesatt} commute with $\sigma_1$. Thus, continuing the process to time $2t$ should continue to decrease the trace distance monotonically, 
\begin{gather}\label{eq:moredephasing}
 \Dcal[\rho_+(2t),\rho_-(2t)] \to \exp(-\gamma 2t).
\end{gather}
Indeed this is what happens if the instrument $X$ is not applied. However, when the instrument $X$ is applied, the dynamics in the second segment reverses the dephasing. This is most easily seen by the fact the total system-environment unitary is mapped to its adjoint by $X$ as $U^\dag = \sigma_1 U \sigma_1$. Concretely we have
\begin{gather}
\label{eqn::unitaryChannelContr}
\begin{split}
 \rho(t) =& \tre[U_t \sigma_1 U_t \rho(0) \otimes \rho_\etxt U^\dag_t \sigma_1 U^\dag_t]\\
 =&\tre[\sigma_1 U^\dag_t U_t \rho(0) \otimes \rho_\etxt U^\dag_t U_t \sigma_1] \\
 =& \sigma_1 \rho(0) \sigma_1.\\
\end{split}
\end{gather}
Above $\rho_\etxt = \ket{\psi_\etxt} \! \bra{\psi_\etxt}$. This calculations shows that the state at time $2t$ is unitarily equivalent to the initial state of the system, which is in contrast to Eq.~\eqref{eq:moredephasing}.

There are a few take away messages here. First, the initial states, $\rho_\pm(0)$, which were monotonically moving closer to each other during the first time segment, will begin to move apart monotonically if the CPTP map $X$ is applied on the system at time $t$. In other words, during the second segment, they are becoming more and more distinguishable. This means that the trace distance monotonically grows for a time greater than $t$ (until $2t$ that is). Therefore, with the addition of an intermediate instrument, the process is no longer seen to be Markovian. Indeed, if the process were Markovian then an addition of an intermediate instrument would not break the monotonicity of the trace distance. In other words, this is breaking a data processing inequality, and therefore the process was non-Markovian from the beginning.

Second, the dynamics in the second segment are restoring the initial state of the system, which means that the dynamical map in the second segment depends on the initial condition. If the process was Markovian, then the total dynamics would have to have the following form 
\begin{gather}
\Ecal_{(2t:0)} = \Ecal_{(2t:t)} \circ X_{t} \circ \Ecal_{(t:0)}.
\end{gather}
However, as we saw above, this is not the same as the total dynamics , which is  simply a unitary transformation. Therefore, the process is not divisible anymore when an intermediate instrument is applied. Again, if the process were Markovian, adding an intermediate instrument will not break the divisibility of the process, and therefore the process was non-Markovian from the beginning. 

Third, the snapshot witness~\cite{PhysRevLett.101.150402} would not be able to attribute CP dynamics to the second segment if the map $X$ is applied at $t$) and thus it too would conclude that the process is non-Markovian. In fact, it is possible to construct dynamics that look Markovian for arbitrary times and then reveal themselves to be non-Markovian~\cite{arXiv:2009.10605}.

\begin{figure}[t]
\begin{center}
\includegraphics[width=0.65\linewidth]{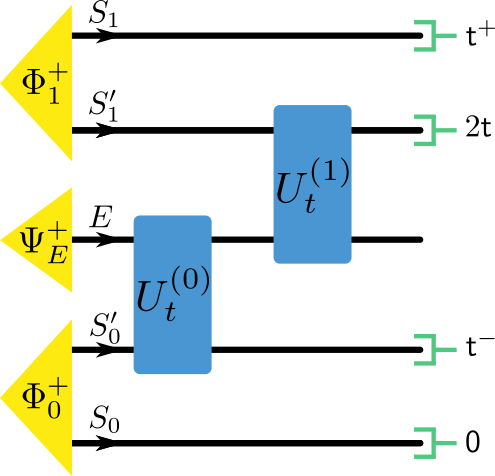}
\caption{\textbf{Choi state for the shallow pocket model.} Each wire of the Choi state of the shallow pocket model (defined in Eq.~\eqref{eq:SEchoi} corresponds to a different time (where $t_1^-$ and $t_1^+$ are the input and output wire at time $t_1$, respectively). The intermediate system-environment unitaries $U_t$ are given by Eq.~\eqref{eq:controlu}. Note that, in contrast to previously depicted Choi states, here, the environmental degree of freedom $E$ is not traced out yet.}
\label{fig::Choi_shallowpocket}
\end{center}
\end{figure}

To be clear, unlike the snapshot method, the process tensor for the whole process will always be completely positive. Let us then write down the process tensor $\Upsilon_{\{2t,t,0\}}$ for this process for three time $\{2t,t,0\}$~\cite{milz_when_2020}. To do so, we first notice that the action of the system-environment unitary has the following simple form
\begin{gather}\label{eq:controlu}
\begin{split}
 U_t \ket{0\psi_\etxt} &= \ket{0} e^{-i \tfrac{g t}{2} \hat{x}}\ket{\psi_\etxt} =\ket{0} u_t\ket{\psi_\etxt}\\
 U_t \ket{1\psi_\etxt} &= \ket{1} e^{i \tfrac{g t}{2} \hat{x}} \ket{\psi_\etxt} = 
 \ket{1} u^\dag_t \ket{\psi_\etxt},
\end{split}
\end{gather}
where $u_t := \exp(-i \tfrac{g t}{2} \hat{x})$ is a unitary operator on $\etxt$ alone. Next, to construct the Choi state for this process we will `feed' half of two maximally entangled states into the process. That is, we prepare two maximally entangled states for the system: $\ket{\Phi^+}_{0} \in \Hcal_{\stxt_0} \otimes \Hcal_{\stxtp_0} $ and $\ket{\Phi^+}_{1} \in \Hcal_{\stxt_1} \otimes \Hcal_{\stxtp_1}$ and let the part $\stxtp_0$ interact with the environment in time segment one and then $\stxtp_1$ in time segment two. Namely, let $U_t^{(0)} \in \Bcal(\Hcal_{\stxtp_0} \otimes \Hcal_{E})$ and $U_t^{(1)} \in \Bcal(\Hcal_{\stxtp_1} \otimes \Hcal_{E})$, where these are the interaction unitary matrices for the two segments. We first write down the process tensor for the whole $\setxt$, i.e., without the final trace on the environment (see Figure~\ref{fig::Choi_shallowpocket}):
\begin{align}\label{eq:SEchoi}
\ket{\Upsilon_{\{2t,t,0\}}^{\setxt}} \!=& U_{t}^{(1)} U_{t}^{(0)} \ket{\Phi^+}_{1} \!\otimes\! \ket{\Phi^+}_{0} \!\otimes \!\ket{\psi_\etxt}\\ \notag
=& \ket{0000} u_t^2\ket{\psi_\etxt}+ \ket{0011} u_t u_t^\dag\ket{\psi_\etxt}\\ \notag
 &+ \ket{1100}u_t^\dag u_t \ket{\psi_\etxt}+\ket{1111} (u_t^\dag)^2 \ket{\psi_\etxt}\\ \notag
=& \ket{\mathbf{00} u_{2t}\psi_\etxt} + \ket{\mathbf{01} \psi_\etxt} + \ket{\mathbf{10}\psi_\etxt} + \ket{\mathbf{11} u_{2t}^\dag \psi_\etxt},
\end{align}
where we have defined $\ket{\mathbf{0}}\!:=\!\ket{00} \in \Hcal_{S_i} \otimes \Hcal_{S_i'}$ and $\ket{\mathbf{1}}\!:=\!\ket{11} \in \Hcal_{S_i} \otimes \Hcal_{S_i'}$ for brevity, where $i\in \{0,1\}$. 

Combining Eq.~\eqref{eq:SEchoi} with Eq.~\eqref{eq:controlu} and tracing over the environment (i.e., $\Upsilon_{\{2t,t,0\}} = \tr_\etxt(\ketbra{\Upsilon_{\{2t,t,0\}}^{\setxt}}{\Upsilon_{\{2t,t,0\}}^{\setxt}})$), we get the Choi state of the process in the compressed basis $\ket{\mathbf{0}}\!:=\!\ket{00}$ and $\ket{\mathbf{1}}\!:=\!\ket{11}$:
\begin{gather}
\label{eqn::ShallowPocketProc}
 \Upsilon_{\{2t,t,0\}} = 
 \begin{pmatrix} 
 1 & e^{-\gamma t} & e^{-\gamma t} & e^{-2\gamma t}\\
 e^{-\gamma t} & 1 & 1 & e^{-2\gamma t}\\
 e^{-\gamma t} & 1 & 1 & e^{-2\gamma t}\\
 e^{-2\gamma t} & e^{-\gamma t} & e^{-\gamma t} & 1\\
 \end{pmatrix}.
\end{gather}
We have used the fact that
\begin{gather}
\tr[u_t \ket{\psi_\etxt} \!\! \bra{\psi_\etxt}] = \tr[u^\dag_t \ket{\psi_\etxt} \!\! \bra{\psi_\etxt}] = e^{-\gamma t},
\end{gather}
where again $\gamma = g \ \mathscr{G}$ and we have employed the explicit form of $\ket{\Psi_\etxt}$ provided in Eq.~\eqref{eq:lorentzian}. Note that the process tensor is really a $16 \times 16$ matrix, but we have expressed it in the compressed basis. In other words, all elements of the process that are not of the form $\ket{jjll}\bra{mmnn}$ are vanishing.

Looking at the Choi state it is clear that there are correlations between time steps $0$ and $2$. This is most easily seen by computing the mutual information. We can think of the process tensor in Eq.~\eqref{eqn::ShallowPocketProc} as a two-qubit state, where the first qubit represents spaces $\stxt_0\stxtp_0$ and the second qubit represents $\stxt_1\stxtp_1$ (see Figure~\ref{fig::Choi_shallowpocket}). Moreover, the $S'_0$ and $S'_1$ spaces are the output of the process at times $t$ and $2t$ respectively. Computing the mutual information information between these spaces thus gives us an idea of whether the process correlates the initial and the final time. If it does, it cannot be of product form, and thus it is not Markovian. For our chosen example, the mutual information between the respective spaces of interest is about 0.35 for large values of $\gamma t$. Therefore it does not have the form of Eq.~\eqref{eq:qmarkov} and the process is non-Markovian. This non-Markovianity will also be detectable if causal breaks are applied at $t$. However, it is not detectable by witnesses of non-Markovianity that are based on CP divisibility only.

For completeness, let us detail how to obtain quantum channels from the more general object $\Upsilon_{\{2t,t,0\}}$. The quantum stochastic matrix from $0 \to 2t$ can be obtained by contracting the process tensor with the instrument at time $t$:
\begin{gather}
\label{eqn::ContractionProcTen}
\Upsilon_{\Ecal_{(2t:0)}^{(\Asf_t)}} := \tr_{t^\inp t^\out}[\Upsilon_{\{2t,t,0\}} \Asf_t^\trps]
\end{gather}
Note that this cannot be done in the compressed basis as the instruments live on the $S'_0S'_1$ spaces, i.e., one would have to fully write out the process tensor of Eq.~\eqref{eqn::ShallowPocketProc}, which we leave as an exercise to the reader. Naturally, the channel resulting from Eq.~\eqref{eqn::ContractionProcTen} depends on the operation $\Asf_t$ (even in the Markovian case). Applying the identity, i.e., contracting with $\Phi^+_t$, which is the Choi state of the identity channel, will give us -- as expected -- exactly Choi state of the dephasing channel 
\begin{gather}
 \Upsilon_{\Ecal_{(2t:0)}^{(\Ical_t)}} =
 \begin{pmatrix} 
 1 & 0 & 0 & e^{-2\gamma t}\\
 0 & 0 & 0 & 0\\
 0 & 0 & 0 & 0\\
 e^{-2\gamma t} & 0 & 0 & 1\\
 \end{pmatrix}.
\end{gather}
On the other hand, applying the $X_t$ instrument will give us a unitary channel, as we already know from Eq.~\eqref{eqn::unitaryChannelContr}: $\Ecal_{(2t:0)}^{(X_t)}[\rho] = \sigma_1 \rho \sigma_1$. 

This example shows that there are non-Markovian effects that can only be detected by interventions. This is not a purely quantum phenomenon; the same can be done in the classical setting and this is the key distinction between stochastic processes and causal modeling, that is, between theories with and without interventions. Naturally, a similar comparison between other traditional witnesses for non-Markovianity in the quantum case, and the results obtained by means of the process tensor can be conducted, too; we point the interested reader to Ref.~\cite{PhysRevA.100.012120} for a detailed analysis.

\subsection{Measures of non-Markovianity for Multi-time processes}
\label{sec::MeasuresNonMark}

Having encountered the shortcomings of traditional witnesses and measures of non-Markovianity in quantum mechanics, it is natural to construct new, more sensitive ones, based on the process tensor approach. Importantly, we already know what Markovian processes `look like' making the quantification of the deviation of a process from the set of Markovian ones a relatively straight forward endeavour. Concretely, the Choi state $\Upsilon$ of a quantum process translates the correlations between timesteps into spatial correlations. A multi-time process is then described by a many-body density operator. This general description then affords the freedom to use any method for quantifying many-body correlations to quantify non-Markovianity. However, there are some natural candidates, which we discuss below. We do warn the reader that there will be infinite ways of quantifying non-Markovianity, as there are infinite ways of quantifying entanglement and other correlations. However, there are metrics that are natural for certain operational tasks. We emphasize that, here, we will only provide general memory measures, and will not make a distinction between classical and quantum memory, the latter corresponding to entanglement in the respective splittings of the corresponding Choi matrix~\cite{arXiv:1811.03722, milz_genuine_2020}. Also, we will only provide a cursory overview of possible ways to quantify non-Markovian effects, and will have to refer the reader to the references mentioned in this section for further information. Finally, in what follows, we omit the subscripts on the process tensors, as they will all be understood to be many-time processes.

\subsubsection{Memory bond}
\label{sec::MemBond}

\begin{figure}
\begin{center}
\includegraphics[width=0.75\linewidth]{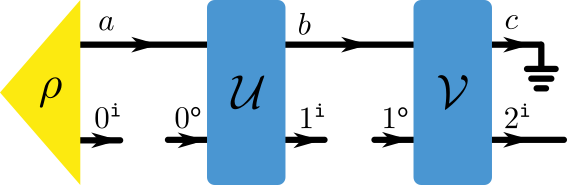}
\caption{\textbf{Three-step process.} Graphic provided as reference for the considerations of Sec.~\ref{sec::MemBond}}
\label{fig::MPOreference}
\end{center}
\end{figure}

We begin by discussing the natural structure of quantum processes. One important feature of the process tensor is that it naturally has the structure of a matrix product operator (MPO)~\cite{VerstraeteCirac2004,zwolak_mixed-state_2004}, i.e., it can be written as a product of matrices that are `contracted' on certain spaces and contain open indices. While this vague notion of an MPO already reminds us of the action of the link product that we introduced in Sec.~\ref{sec:bornrule} let us be more concrete and provide such a matrix product operator for a simple three-step process with an initial system-environment state $\rho$ and two system-environment unitary maps $\Ucal$ and $\Vcal$. Before we do so, we emphasize that we are not attempting to provide a general introduction to MPOs, but rather motivate why there usage in the context of process tensors can be very fruitful. Let us denote the involved system spaces from $0^{\inp}$ to $2^\inp$ and the involved environment spaces by $\{abc\}$ (see Figure~\ref{fig::MPOreference} for reference). As we have seen in our discussion of the link product, we can write the resulting process tensor $\Upsilon$ in terms of the Choi matrices $\mathsf{U}$ and $\mathsf{V}$ as
\begin{gather}
    \Upsilon = \rho^{0^{\inp}a} \star \mathsf{U}^{0^\out 1^\inp ab} \star \tr_c\mathsf{V}^{1^\out 2^\inp bc}\, ,
\end{gather}
where we have added the respective spaces each of the matrices is defined on as subscripts. Using the definition of the link product provided below Eq.~\eqref{eqn::LinkDef}, and recalling that it amounts to a partial transpose and trace over shared spaces, the above can be written as 
\begin{gather}
\label{eqn::MPOEx}
   \Upsilon = \tr_b\{\tr_a[ \rho^{0^{\inp}a} ( \mathsf{U}^{0^\out 1^\inp ab} )^{\mathrm{T}_a}] \tr_c(\mathsf{V}^{1^\out 2^\inp bc})^{\mathrm{T}_b}\}\, , 
\end{gather}
where we have omitted the respective identity matrices. Note that, in the above, the spaces with labels $\{a,b,c\}$ are `contracted', while the remaining spaces are untouched, such that $\Upsilon \in \Bcal(\Hcal_0^\inp \otimes \Hcal_0^\out \otimes \Hcal_1^\inp \otimes \Hcal_1^\out \otimes \Hcal_2^\inp$). This can be made more concrete by rewriting Eq.~\eqref{eqn::MPOEx} as a product of three matrices (without any trace operations). To this end, let us set 
\begin{gather}
\label{eqn::DefMPOmatrix}
\begin{split}
    &\breve{\rho}^{\,0^{\inp}a} = \sum_{i_a} \bra{i_a i_a}\rho^{0^{\inp}a}, \\
    &\breve{\mathsf{U}}^{0^\out 1^\inp ab} = \sum_{i_b,i_a} \braket{i_bi_b|(\mathsf{U}^{0^\out 1^\inp ab} )^{\mathrm{T}_a}|i_ai_a}, \\
    \text{and} \quad & \breve{\mathsf{V}}^{1^\out 2^\inp bc} = \sum_{i_b} \tr_c(\mathsf{V}^{1^\out 2^\inp bc})^{\mathrm{T}_b}\ket{i_bi_b}, 
\end{split}
\end{gather}
where $\{\ket{i_x}\}$ is an orthogonal basis of $\Hcal_x$. Note that each of the objects above now corresponds to a matrix with different input and output spaces, i.e., we have 
\begin{gather}
\begin{split}
    &\breve{\rho}^{\,0^{\inp}a}: \Hcal_0^\inp \rightarrow \Hcal_0^\inp \otimes \Hcal_a^{\otimes 2}, \\
    &\breve{\mathsf{U}}^{0^\out 1^\inp ab}: \Hcal_0^\out \otimes \Hcal_1^\inp \otimes \Hcal_a^{\otimes 2} \rightarrow \Hcal_0^\out \otimes \Hcal_1^\inp \otimes \Hcal_b^{\otimes 2}, \\
    \text{and} \quad & \breve{\mathsf{V}}^{1^\out 2^\inp bc}: \Hcal_1^\out \otimes \Hcal_2^\inp \rightarrow \Hcal_1^\out \otimes \Hcal_2^\inp \otimes \Hcal_b^{\otimes 2}
\end{split}
\end{gather}
Basically, the reshapings in Eq.~\eqref{eqn::DefMPOmatrix} are required such that the trace operations that occur in Eq.~\eqref{eqn::MPOEx} are moved into matrices and Eq.~\eqref{eqn::MPOEx} can be expressed as a simple matrix product. Indeed, we have 
\begin{gather}
\label{eqn::MPOversionUps}
    \Upsilon = \breve{\rho}^{\,0^{\inp}a}  \cdot \breve{\mathsf{U}}^{0^\out 1^\inp ab} \cdot \breve{\mathsf{V}}^{1^\out 2^\inp bc}, 
\end{gather} which can be seen by direct insertion of the expressions of Eq.~\eqref{eqn::DefMPOmatrix} into the above equation: 
\begin{gather}
\begin{split}
    &\breve{\rho}^{\,0^{\inp}a}  \cdot \breve{\mathsf{U}}^{0^\out 1^\inp ab} \cdot \breve{\mathsf{V}}^{1^\out 2^\inp bc} \\
    &= \sum_{i_a, i_b, j_a, j_b} {\color{red}\bra{ {i_a}}} \bra{i_a}\rho^{0^\inp a} {\color{darkgreen}{\bra{i_b}}}\braket{i_b|(\mathsf{U}^{0^\out 1^\inp ab})^{\mathrm{T}_a}|j_a} \color{red}{\ket{j_a}} \\ &\phantom{= \sum_{i_a, i_b, j_a, j_b} }\cdot\tr_c(\mathsf{V}^{1^\out 2^\inp 6c})^{\mathrm{T}_b}\ket{j_b} {\color{darkgreen}{\ket{j_b}}} \\
    &= \sum_{i_a, i_b} \braket{i_b|[(\braket{i_a|\rho^{0^\inp a} \mathsf{U}^{0^\out 1^\inp ab})^{\mathrm{T}_a}|i_a} ) \tr_c(\mathsf{V}^{1^\out 2^\inp bc})^{\mathrm{T}_b}]|i_b}\\
    &=\tr_b\{\tr_a[\rho^{0^\inp a} (\mathsf{U}^{0^\out 1^\inp ab})^{\mathrm{T}_a}] \tr_c(\mathsf{V}^{1^\out 2^\inp bc})^{\mathrm{T}_b}\} = \Upsilon, 
\end{split}
\end{gather}
where, for better orientation, we have color-coded the bras and kets that belong together.
\begin{figure}
    \centering
    \includegraphics[width=0.9\linewidth]{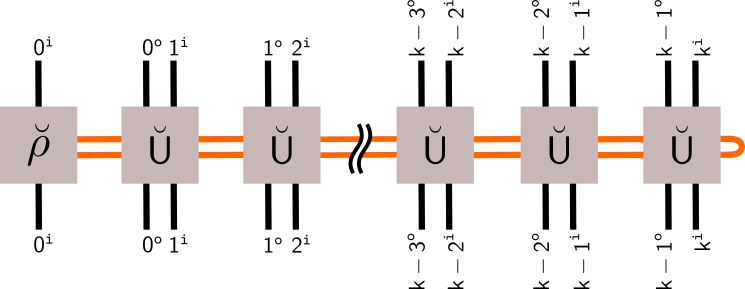}
    \caption{\textbf{MPO representation of a multi-time process $\Upsilon$.} Under the assumptions that all the unitaries in the process are the same -- which, for example, holds true when the times are equidistantly spaced and the generating Hamiltonian is time-independent, then the resulting process tensor can be written as a tensor product of an initial matrix $\breve \rho$ and matrices $\breve{\mathsf{U}}$, together with a final partial trace (bent orange line on the right). The degrees of freedom that they share (orange lines) are the bond dimension of the MPO, the remaining open indeces correspond to the spaces the resulting MPO lives on. The smallest bond dimension of all possible MPO representations of a given $\Upsilon$ can be taken as a measure of non-Markovianity of $\Upsilon$. }
    \label{fig::MPOrep}
\end{figure}
At this point, one might wonder why we went through the ordeal of rewriting our process tensor in terms of a product of matrices, above all in light of the fact that \textit{any} many-body operator can be written as a matrix product operator. To see why this representation is meaningful, let us take a closer look at Eq.~\eqref{eqn::MPOversionUps}; each of the matrices that occur has some `open indices', i.e., spaces that only one of the matrices is defined on, while each of them also shares spaces with their respective neighbors ($a$ is shared between $\breve\rho$ and $\breve{\mathsf{U}}$, $b$ is shared between $\breve{\mathsf{U}}$ and $\breve{\mathsf{V}}$) that are `contracted' over and do not appear in the resulting $\Upsilon$. These dimension of these latter degrees of freedom is the bond dimension of the MPO (in our case, it would be $\max(d_a^2,d_b^2)$). Comparing this to the circuit of Figure~\ref{fig::MPOreference}, we see that the bond dimension directly corresponds to the dimension of the environment. Why would we go through the hustle of working out an MPO reprsentation of $\Upsilon$ then? The reason is twofold. Firstly, in general, we are not given the circuit representation of $\Upsilon$, but only $\Upsilon$ itself. Then, finding a matrix product operator representation gives us a good gauge for the size of the required environment. On the other hand, even if we were given some dilation of $\Upsilon$, there might be other representations that require a smaller environment, thus providing a representation that only uses the effective environment required for the propagation of memory. Concretely then, the bond dimension corresponds to the smallest environment ancilla transporting memory that is required in order to reproduce the process tensor at hand. For a Markov process, naturally, the bond dimension is one. While it is not necessarily straight forward to find a representation of an MPO with minimal bond dimension, by employing methods from the field of tensor networks we can compress the bond and give the process an efficient description~\cite{yang_matrix_2018}.

Additionally, and more importantly for the numerical analysis of multi-time processes, the theory of MPOs provides a large toolbox for the efficient representation of process tensors. This holds particularly true for processes that are time translationally invariant, i.e., each of the matrices that occur in the product is the same (see Figure~\ref{fig::MPOrep}). Since a proper introduction of these techniques would require a tutorial article of its own right (and excellent tutorials on the matter already exist, see, for example, Refs.~\cite{MPSreview,orus_practical_2014}), we will not delve deeper into the theory of tensor networks and MPOs. Let us emphasize though, that their explicit use both for the conceptual and numerical description of open system dynamics is a very active field of research~\cite{prior_efficient_2010, schroder_simulating_2016, wall_simulating_2016, strathearn_efficient_2018} and the corresponding techniques are particularly well-tailored to tackle multi-time processes within the process tenor framework~\cite{Luchnikov2019L, Luchnikov2018, Jorgensen2019, guochu2020}. As mentioned though, these techniques go beyond the scope of this tutorial, and here, we content ourselves with mentioning that the structure of process tensors allows for a very direct representation in terms of MPOs, which i) allows for the whole machinery developed for MPOs to be used for the efficient description of open system dynamics, and ii) enables the interpretation of the minimal necessary bond dimension as a measure of non-Markovinanity.

\subsubsection{Schatten measures}
\label{sec::Schatten}

Next in our discussion of measures of non-Markovianity, we make use of the form of Markov processes given in Eq.~\eqref{eq:qmarkov}, where we saw that the Choi state of a Markovian process tensor is of tensor product form. We remind the reader that this quantum Markov condition contains the classical Markov condition, and any deviations from it in the structure of $\Upsilon$ imply non-Markovianity. Importantly, this structural property of Markovian processes, and the fact that $\Upsilon$ is -- up to normalization -- a quantum state allow for operationally meaningful measures of non-Markovianity. That is, by sampling from a process we can determine if it has memory or not and then also quantify this memory. For instance, if we want to distinguish a given non-Markovian process from the set of Markov processes, we can measure the distance to the closest Markov process for a choice of metric, e.g., the Schatten $p$-norm, 
\begin{gather}
\label{eqn::SchattenMeas}
\mc{N}_p := \min_{\Upsilon^\markov}\|\Upsilon-\Upsilon^\markov\|_p,
\end{gather}
where
$\|X\|_p^p=\tr(|X|^p)$. Here, we are minimizing the distance for a given quantum process $\Upsilon$ over all Markovian processes $\Upsilon^\markov$, which have the form of Eq.~\eqref{eq:qmarkov}. Naturally, this goes to zero if and only if the given process is Markovian. On the other hand, to maximally differentiate between a given process and its closest Markovian process the natural distance choice is the diamond norm:
\begin{gather}
 \mc{N}_\blacklozenge\equiv\f{1}{2}\min_{\Upsilon^\markov}\|\Upsilon-\Upsilon^\markov\|_\blacklozenge, 
 \label{Eq:defdiamond}
\end{gather}
where $\|\sbt \|_{\blacklozenge}$ is the generalized diamond norm for processes~\cite{Phil_MemStr, chiribella_theoretical_2009} and the somewhat random prefactor of $1/2$ is just added for consistency with the literature.  Eq.~\eqref{Eq:defdiamond} then gives the optimal probability to discriminate a process from the closest Markovian one in a single shot, given any set of measurements together with an ancilla. The difference between the diamond norm and Schatten norm is that in the former, we are allowed to use ancillas in the form of quantum memory. This is known to lead to better distinguishability, in general.

Schatten norms play a central role in quantum information theory. Therefore, the family of non-Markovianity measures given above will naturally arise in many applications. For instance, the diamond norm is very convenient to work with when studying the statistical properties of quantum stochastic processes~\cite{Romero2018, Romero2020}. However, while constituting a natural measure, these quantifiers of non-Markovianity have the draw-back that they require a minimization over the whole set of Markovian processes, which makes them computationally hard to access. This problem can be remedied by choosing a different metric in Eq.~\eqref{eqn::SchattenMeas}.

\subsubsection{Relative entropy}
\label{sec::RelEnt}

We could also use any metric or pseudo-metric $\Dcal$ that is contractive under CP operations
\begin{gather}\label{nonmarkovmes}
\mathcal{N} := \min_{\Upsilon_{k:0}^{{(\rm M)}}} \Dcal \left[ \Upsilon_{k:0} \|\Upsilon_{k:0}^{\rm Markov} \right].
\end{gather}
Here, CP contractive means that $\Dcal [\Phi(X) \| \Phi(Y)] \leq \Dcal[X\|Y]$ for any CP map $\Phi$ on the space of generalized Choi states. A metric or pseudo-metric that is not CP contractive, may not lead to consistent measures for non-Markovianity since, for example, it could be increased by the presence of an independent ancillary Markov process. Here, the requirement that $\Dcal$ is a  pseudo-metric means that it satisfies all the properties of a distance except that it may not be symmetric in its arguments. Different quasi-distance measures will then have different operational interpretations for the memory. In general though, they will still be plagued by the problem of minimization that appears in Eq.~\eqref{nonmarkovmes}.

A very convenient pseudo-metric choice is the quantum relative entropy~\cite{vedral2002role}, which we already encountered in Sec.~\ref{sec::QDPI} when we discussed quantum data processing inequalities. In order to be able to use the relative entropy, let us assume for the remainder of this section that all the process tensors we use are normalized, i.e., $\tr \Upsilon = 1$. Besides being contractive under CP maps, this pseudo-metric is very convenient because for any given process $\Upsilon$, the closest (with respect to the quantum relative entropy) Markovian process is straightforwardly found by discarding the correlations. That is, the process made of the marginals of the given process is the closest Markov process, such that
\begin{gather}
    \mathcal{N}_R = \Dcal \left[ \Upsilon_{k:0} \|\Upsilon_{1^-:0^+} \otimes \cdots \otimes \Upsilon_{k^-:k-1^+}\right],
\end{gather}
where the CPTP maps $\{\Upsilon_{j^-:j=1^+}\}$ are the respective marginals (obtained via partial trace) of $\Upsilon_{k:0}$. This follows from the well known fact that, with respect to the quantum relative entropy, the closest product state of a multi-partite quantum state is the one that is simply a tensor product of its marginals~\cite{arXiv:0911.5417}. Moreover, besides alleviating the minimization problem, this measure has a clear operational interpretation as a probability of confusing the given process for being Markovian~\cite{PhysRevA.97.012127}:
\begin{gather}
\Pprob_{\rm confusion} = \exp\{-n \mc{N}_R\} 
\end{gather}
where $\mc{N}_R$ is the relative entropy between the given process and its marginals. Specifically, this measure quantifies the following: suppose a process in an experiment is non-Markovian. The employed model for the experiment is, however, Markovian. The above measure is related to the probability of confusing the model with the experiment after $n$ samplings. If $\mc{N}_R$ is large, then an experimenter will very quickly realize that the hypothesis is false, and the model needs updating, i.e., the experimenter is quick to learn that the Markovian model poorly approximates the (highly) non-Markovian process.

With this, we conclude our short presentation of measures for non-Markovianity in the multi-time setup. The attentive reader will have noticed that we have not touched on witnesses of non-Markovianity here, unlike in Sec.~\ref{sec::Witness} where we discussed (some of the) witnesses for non-Markovianity in the two-time scenario. In principle, such witnesses can be straightforwardly constructed. We have already done so in this tutorial,  when we discussed the shallow pocket model and its non-Markovian features. Also, experimentally implementing some causal breaks and checking for conditional independence would provide a witness for non-Markovianity. However, to date, no experimentally used witness for non-Markovianity in the multi-time setting has crystallized, and a systematic construction of memory witnesses that are attuned to experimental requirements is subject of ongoing research. 

More generally though, `simply' deciding whether a process is Markovian or not seems somewhat blunt for a multi-time process. After all, it is not just of interest \textit{if} there are memory effects, but \textit{what kinds of} memory effects there are. At its core, this is latter question is a question of Markov order for quantum processes. We will thus spend the remainder of this tutorial to provide a proper definition of Markov order in the quantum case, as well as a non-trivial example to illustrate these considerations.

\subsection{Quantum Markov order}
\label{sec::QMarkOrd}
The process tensor allows one to properly define Markovianity for quantum processes. As we have seen, though, in our discussion of the classical case, Markovian processes are not the only possibility. Rather, they constitute the set of processes of Markov order $1$ (and $0$). It is then natural to ask if Markov order is a concept that transfers neatly to the quantum case as well. As we shall see, Markov order is indeed a meaningful concept for quantum processes but turns out to be a more layered phenomenon than in the classical realm. Here, we will only focus on a couple of basic aspects of quantum Markov order. For a more in-depth discussion, see, for example, Refs.~\cite{PhysRevA.99.042108, TarantoThesis}. Additionally, while it is possible to phrase results on quantum Markov order in terms of maps, it proves rather cumbersome which is why the following results will be presented exclusively in terms of Choi states.

Before turning to the quantum case, let us quickly recall (see Sec.~\ref{sec::CondMut}) that for classical processes of Markov order $|M|=\ell$, we had 
\begin{gather}
\label{eqn::MarkovOrder2}
 \Pprob(F|M,H) = \Pprob(F|M), 
\end{gather}
which implied that the conditional mutual information $H(F:H|M)$ between the future $F$ and the history $H$ given the memory $M$ vanished. Additionally, for classical processes of finite Markov order, there exists a recovery map $\Rcal_{M\rightarrow FM}$ that acts only on the memory block and allows one to recover $\Pprob(F,M,H)$ from $\Pprob(M,H)$. 

In the quantum case, an equation like Eq.~\eqref{eqn::MarkovOrder2} is ill-defined on its own, as the respective probabilities depend on the instruments $\{\Jcal_F, \Jcal_M, \Jcal_H\}$ that were used at the respective times to probe the process. With this in mind, we obtain an \textit{instrument-dependent} definition of finite Markov order in the quantum case~\cite{PhysRevLett.122.140401, PhysRevA.99.042108}: 

\vspace{5pt}
\textbf{Quantum Markov order.} \textit{A process is said to be of quantum Markov $|M| = \ell$ with respect to an instrument $\Jcal_M$, if for all possible instruments $\{\Jcal_F,\Jcal_H\}$ the relation
\begin{gather}
\label{eqn::QuarkovOrder}
\Pprob(\xset_F|\mathcal{J}_F;\xset_M,\mathcal{J}_M ; \xset_H , \mathcal{J}_H ) \!=\! \Pprob(\xset_F | \Jcal_F; \xset_M,\mathcal{J}_M)
\end{gather}
is satisfied at all times in $\Tset$}.
\vspace{5pt}

Intuitively, this definition of Markov order is the same as the classical one; once the outcomes on the memory block $M$ are known, the future $F$ and the history $H$ are independent of each other. However, here, we have to specify, what instrument $\Jcal_M$ is used to interrogate the process on $M$. Importantly, demanding that a process has finite Markov order at all times with respect to \textit{all} instruments $\Jcal_M$ is much too strong a requirement, as it can only be satisfied by processes of quantum Markov order $0$, i.e., processes where future statistics do not even depend on the previous outcome~\cite{PhysRevLett.122.140401, PhysRevA.99.042108,capela_monogamy_2020}. 

While seemingly a quantum trait, this instrument dependence of memory length is already implicitly present in the classical case; there, we generally only consider joint probability distributions that stem from sharp, non-invasive measurements. However, as mentioned above, even in classical physics, active interventions, and, as such, different probing instruments, are possible. This, in turn, makes the standard definition of Markov order for classical processes inherently instrument-dependent, albeit without being mentioned explicitly. Indeed, there are classical processes that change their Markov order when the employed instruments are changed (see, e.g., Sec. VI of Ref.~\cite{PhysRevA.99.042108} for a more detailed discussion). 

In the quantum case, there is no `standard' instrument, and the corresponding instrument-dependence of memory effects is dragged into the limelight. Even the definition of Markovianity, i.e., Markov order $1$, that we provided in Sec.~\ref{sec::QuarkovCausalBr} is an inherently instrument-dependent one; quantum processes are Markovian if and only if they do not display memory effects with respect to causal breaks. However, this does not exclude memory effects to appear as soon as other instruments are employed (as these memory effects would be introduced by the instruments and not by the process itself, the instrument-dependent definition of Markovianity still captures all memory that is contained in the process at hand). Just like for the definition of Markovianity, once all process tensors are classical, and all instruments consist of classical measurements only, the above definition of Markov order coincides with the classical one~\cite{PhysRevLett.122.140401}

For generality, in what follows, the instruments on $M$ can be temporally correlated, i.e., they can be testers (however, for conciseness, we will call $\Jcal_F, \Jcal_M$, and $\Jcal_H$ instruments in what follows). While in our above definition of quantum Markov order we fix the instrument $\Jcal_M$ on the memory block, we do not fix the instruments on the future and the history, but require Eq.~\eqref{eqn::QuarkovOrder} to hold for all $\Jcal_F$ and $\Jcal_H$. This, then, ensures, that, if there are any conditional memory effects between future and history for the given instrument on the memory, they would be picked up.

\begin{figure}
 \centering
 \includegraphics[width=0.9\linewidth]{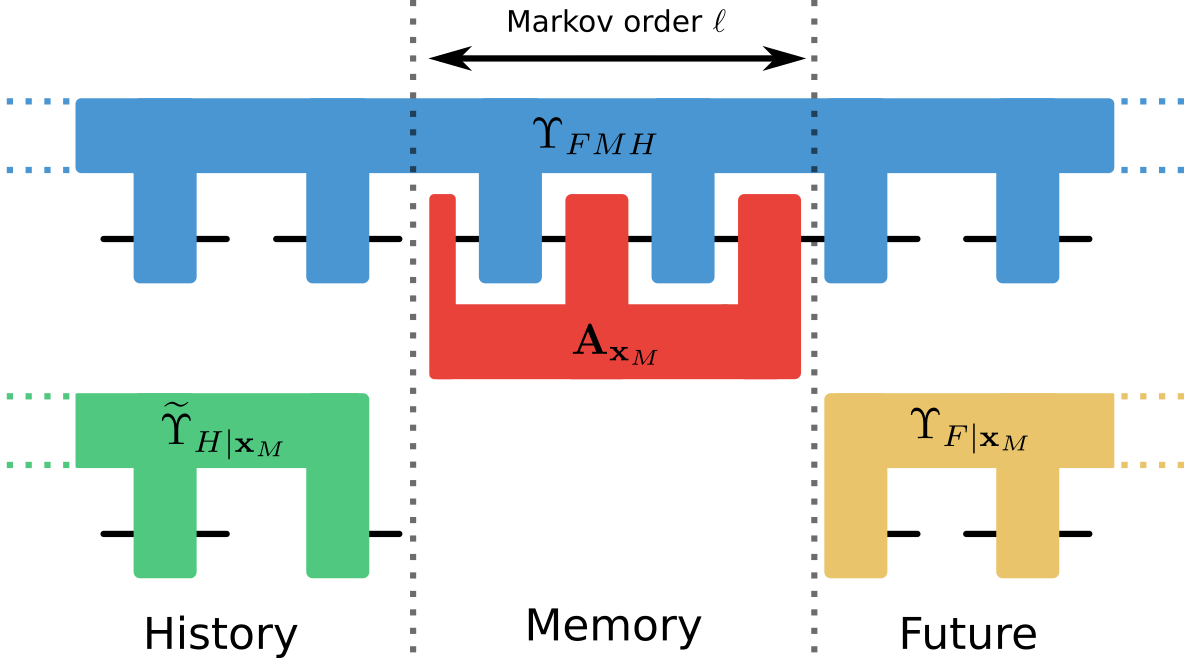}
 \caption{\textbf{Quantum Markov order.} If a process $\Upsilon_{FMH}$ has finite Markov order with respect to an instrument/tester $\Jcal_M = \{\Aset_{\xset_M}\}$ on the memory block, then the application of each of the elements of $\Jcal_M$ leaves the process in a tensor product between future $F$ and history $H$.}
 \label{fig::QuarkovOrder} 
\end{figure}

As all possible temporal correlations are contained in the process tensor $\Upsilon_{FMH}$ that describes the process at hand, vanishing instrument-dependent quantum Markov order has structural consequences for $\Upsilon_{FMH}$. In particular, let $\Jcal_M = \{\Aset_{\xset_M}\}$ be the instrument for which Eq.~\eqref{eqn::QuarkovOrder} is satisfied, and let $\Jcal_F = \{\Aset_{\xset_F}\}$ and $\Jcal_H = \{\Aset_{\xset_H}\}$ be two arbitrary instruments on the future and history. With this, Eq.~\eqref{eqn::QuarkovOrder} implies
\begin{gather}
\label{eqn::Quarkov1}
 \begin{split}
&\frac{\tr[\Upsilon_{FMH}^\trps (\Aset_{\xset_F} \otimes \Aset_{\xset_M} \otimes \Aset_{\xset_H})]}{\tr[\Upsilon_{MH}^\trps (\Aset_{\xset_M} \otimes \Aset_{\xset_H})]} \\
&= \frac{\sum_{\xset_H} \tr[\Upsilon_{FMH}^\trps (\Aset_{\xset_F} \otimes \Aset_{\xset_M} \otimes \Aset_{\xset_H})]}{\sum_{\xset_H} \tr[\Upsilon_{MH}^\trps (\Aset_{\xset_M} \otimes \Aset_{\xset_H})]}, 
 \end{split}
\end{gather}
where the process tensor on $MH$ is $\Upsilon_{MH} = \tfrac{1}{d_{F^\out}} \tr_F(\Upsilon_{FMH})$ (which, due to the causality constraints is independent of $\Jcal_F$) and $d_{F^\out}$ is the dimension of all spaces labeled by $\out$ on the future $F$ (we already encountered this definitition of reduced processes in Sec.~\ref{subsec::Red}). As the relation~\eqref{eqn::Quarkov1} has to hold for all conceivable instruments $\Jcal_H$ and $\Jcal_F$, and all elements of the fixed instrument $\Jcal_M$, it implies that each element $\Aset_{\xset_M} \in \Jcal_M$ `splits' the process tensor in two independent parts, i.e., 
\begin{gather}
\label{eqn::QuarkovCondition}
 \tr_M[\Upsilon_{FMH}^{\trps_M} \Aset_{\xset_M}] = \Upsilon_{F|\xset_M} \otimes \widetilde{\Upsilon}_{H|\xset_M}.
\end{gather}
See Figure~\ref{fig::QuarkovOrder} for a graphical representation. While straightforward, proving the above relation is somewhat tedious, and the reader is referred to Refs.~\cite{PhysRevLett.122.140401, TarantoThesis}, where a detailed derivation can be found. Here, we rather focus on its intuitive content and structure. Most importantly, Eq.~\eqref{eqn::QuarkovCondition} implies that, for any element of the fixed instrument $\Jcal_M$, the remaining `process tensor' on future and history does not contain any correlations; put differently, if one knows the outcome on $M$, the future statistics are fully independent of the past. Conversely, by insertion, it can be seen that any process tensor $\Upsilon_{FMH}$ that satisfies Eq.~\eqref{eqn::QuarkovCondition} for some instrument $\Jcal_M$ also satisfies Eq.~\eqref{eqn::QuarkovOrder}. As an aside, we have already seen this `splitting' of the process tensor due to conditional independence when we discussed Markovian processes. Indeed, the resulting structure of Markovian processes is a particular case of the results for Markov order presented below.

On the structural side, it can be directly seen that the terms $\{\Upsilon_{F|\xset_M}\}$ in Eq.~\eqref{eqn::QuarkovCondition} are proper process tensors, i.e., they are positive and satisfy the causality constraints of Eqs.~\eqref{eqn::hierarchyFirst} and~\eqref{eqn::Hierarchy}. Specifically, contracting $\Upsilon_{FMH}$ with a positive element on the memory block $M$ yields positive elements, and does not alter satisfaction of the hierarchy of trace conditions on the block $F$. This fails to generally hold true on the block $H$. While still positive, the terms $\widetilde{\Upsilon}_{H|\xset_M}$ do not necessarily have to satisfy causality constraints. However, the set $\{\widetilde{\Upsilon}_{H|\xset_M}\}$ forms a tester, i.e., $\sum_{\xset_M} \widetilde{\Upsilon}_{H|\xset_M} = \Upsilon_{H}$ is a process tensor. 

Employing Eq.~\eqref{eqn::QuarkovCondition}, we can derive the most general form of a process tensor $\Upsilon_{FMH}$ that has finite Markov order with respect to the instrument $\Jcal_M=\{\Aset_{\xset_M}\}_{\xset_M=1}^n$. To this end, without loss of generality, let us assume that all $n$ elements of $\Jcal_M$ are linearly independent.\ftnt{If an element $\Aset_{\xset_j}$ is linearly dependent on $\Aset_{\xset_k}$ and $\Aset_{\xset_\ell}$, then either the conditional future and past processes are the same for both outcomes $k$ and $\ell$ or outcome $j$ does yield conditional independence.} Then, this set can be completed to a full basis of the space of matrices on the memory block $M$ by means of other tester elements $\{\bar{\Aset}_{\alpha_M}\}_{\alpha_M= n+1}^{d_M}$, where $d_M$ is the dimension of the space spanned by tester elements on the memory block. As these two sets together form a linear basis, there exists a corresponding dual basis, which we denote as 
\begin{gather}
\left\{ \{ \Delta_{\xset_M} \}_{\mathbf{x_M}=1}^n \bigcup \{\bar{\Delta}_{\alpha_M}\}_{\alpha_M = n+1}^{d_M} \right\}. 
\end{gather}
From this, we obtain the general form of a process tensor $\Upsilon_{FMH}$ with finite Markov order with respect to the instrument $\Jcal_M$~\cite{PhysRevA.99.042108}: 
\begin{gather}
\label{eqn::GenDecompCMI0}
\begin{split}
 \Upsilon_{FMH} =& \sum_{\xset_M = 1}^n \Upsilon_{F|\xset_M} \otimes \Delta_{\xset_M}^{*} \otimes \widetilde{\Upsilon}_{H|\xset_M} \\ 
 &+ \sum_{\alpha_M = n+1}^{d_M} \widetilde{\Upsilon}_{FH|\alpha_M} \otimes \bar{\Delta}_{\alpha_M}^{*}.
\end{split}
\end{gather}
It can be seen directly (by insertion into Eq.~\eqref{eqn::QuarkovCondition}) that the above $\Upsilon_{FMH}$ indeed yields the correct term $\Upsilon_{F|\xset_M} \otimes \widetilde{\Upsilon}_{H|\xset_M}$ for every $\Aset_{\xset_M} \in \Jcal_M$. Using other tester elements, like, for example $\bar{\Aset}_{\alpha_M}$, will however not yield uncorrelated elements on $FH$ (as the terms $\widetilde{\Upsilon}_{FH|\alpha_M}$ do not necessarily have to be uncorrelated). This, basically, is just a different way of saying that an informationally incomplete instrument is not sufficient to fully determine the process at hand~\cite{Milz2018A}. Additionally, most elements of the span of $\Jcal_M$ will not yield uncorrelated elements, either, but rather a linear combination of uncorrelated elements, which is generally correlated.

While remaining a meaningful concept in the quantum domain, quantum Markov order is highly dependent on the choice of instrument $\Jcal_M$, and there exists a whole zoo of processes that show peculiar memory properties for different kinds of instruments, like, for example, processes that only have finite Markov order for unitary instruments, or processes, which have finite Markov order with respect to an informationally complete instrument, but the conditional mutual information does not vanishes~\cite{PhysRevLett.122.140401, TarantoThesis}. 

Before providing a detailed example of a process with finite quantum Markov order, let us discuss this aforementioned connection between quantum Markov order and the quantum version of the conditional mutual information. In analogy to the classical case, one can define a quantum CMI (\textbf{QCMI}) for quantum states $\rho_{FMH}$ shared between parties $F,M,$ and $H$ as 
\begin{gather}
 S(F\!:\!H|M) = S(F|M) \!+\! S(H|M) \!-\! S(FH|M) ,
\end{gather}
where $S(A|B):=S(AB)-S(B)$ and $S(A):=- \tr[A \log(A)]$ (see Sec.~\ref{sec::QDPI}) is the von-Neumann entropy. Quantum states with vanishing QCMI have many appealing properties, like, for example, the fact that they admit a block decomposition~\cite{Hayden2004}, as well as a CPTP recovery map $\Wcal_{M\rightarrow FM}[\rho_{MH}] = \rho_{FMH}$ that only acts on the block $M$~\cite{Petz1986, Petz2003}. Unlike in the classical case, the proof of this latter property is far from trivial and a highly celebrated result. States with vanishing QCMI or, equivalently, states that can be recovered by means of a CPTP map $\Wcal_{M\rightarrow FM}$ are called \textit{quantum Markov chains}~\cite{Ruskai2002, Petz2003, Hayden2004, Ibinson2008, Fawzi2015, Wilde2015, Sutter2016, Sutter2017}. Importantly, for states with approximately vanishing QCMI, the recovery error one makes when employing a map $\Wcal_{M\rightarrow FM}$ can be bounded by a function of the QCMI~\cite{Fawzi2015, Wilde2015, Sutter2016, Sutter2017}. 

As process tensors $\Upsilon_{FMH}$ are, up to normalization, quantum states, all of the aforementioned results can be used for the study of quantum processes with finite Markov order. However, the relation of quantum processes with finite Markov order and the QCMI of the corresponding process tensor is -- unsurprisingly -- more layered than in the classical case. We will present some of the peculiar features here without proof to provide a flavor of the fascinating jungle that is memory effects in quantum mechanics (see, for example, Refs.~\cite{PhysRevA.99.042108, PhysRevLett.122.140401, TarantoThesis} for in-depth discussions). 

Let us begin with a positive result. Using the representation of quantum states with vanishing QCMI provided in Ref.~\cite{Hayden2004}, for any process tensor $\Upsilon_{FMH}$ that satisfies $S(F\!:\!H|M)_{\Upsilon_{FMH}} = 0$, one can construct an instrument on the memory block $M$, that blocks the memory between $H$ and $F$. Put differently, vanishing QCMI implies (instrument-dependent) finite quantum order. 

However, the converse does not hold. This can already be seen from Eq.~\eqref{eqn::GenDecompCMI0}, where the general form of a process tensor with finite Markov order is provided. The occurrence of the second set of terms $\widetilde{\Upsilon}_{FH|\alpha_M} \otimes \bar \Delta_{\alpha_M}$ implies the existence of a wide range of correlations between $H$ and $F$ that can still persist (but not be picked up by the fixed instrument chosen on $M$), making it unlikely that the QCMI of such a process tensor actually vanishes. On the other hand, if the instrument $\Jcal_M$ is informationally complete, then there is a representation of $\Upsilon_{FMH}$ that only contains terms of the form $\Upsilon_{F|\xset_M} \otimes \Delta_{\xset_M}$, which looks more promising in terms of vanishing QCMI (in principle, such a decomposition can also exist when the respective tester elements are not informationally complete, which is the case for classical stochastic processes). However, when the tester elements $\Aset_{\xset_M}$ corresponding to the duals $\Delta_{\xset_M}$ do not commute (which, in general they do not), then, again, the QCMI of $\Upsilon_{FMH}$ does not vanish~\cite{PhysRevA.99.042108, PhysRevLett.122.140401, TarantoThesis}. Nonetheless, for any process tensor of the form 
\begin{gather}
 \Upsilon_{FMH} = \sum_{\xset_M} \Upsilon_{F|\xset_M} \otimes \Delta_{\xset_M}^* \otimes \widetilde{\Upsilon}_{H|\xset_M},
\end{gather}
knowing the outcomes on the memory block (for the instrument $\Jcal_B = \{\Aset_{\xset_M}\}$) allows one to reconstruct the full process tensor. Concretely, using 
\begin{gather}
 \begin{split}
 \Upsilon_{MH} &= \frac{\tr_{F} (\Upsilon_{FMH})}{d_{F^\out}} \\
 &:= \frac{1}{d_{F^\out}} \sum_{\xset_M} c_{\xset_M} \Delta_{\xset_M}^* \otimes \widetilde{\Upsilon}_{H|\xset_M} ,
\end{split}
\end{gather}
where we set $c_{\xset_M}= \tr(\Upsilon_{F|\xset_M})$ and $d_{F^\out}$ is the dimension of all output spaces on the future block, we have
\begin{gather}
\label{eqn::FalseRecov}
\begin{split}
 \Upsilon_{FMH} &= d_{F^\out} \sum_{\xset_M} c_{\xset_M}^{-1} \Upsilon_{F|\xset_M} \otimes \Delta_{\xset_M}^* \otimes \tr_M(\Upsilon_{MH} \Aset_{\xset_M}^\trps) \\
 &=: \widetilde{\Wcal}_{M\rightarrow FM}[\Upsilon_{MH}].
\end{split}
\end{gather}
, asHere, the map $\widetilde{\Wcal}_{M\rightarrow FM}$ appears to play the role of a recovery map. However, as the duals $\{\Delta_{\xset_M}\}$ are not necessarily positive,\ftnt{The duals are, for example, all positive, if the blocking instrument consists of elements with Choi that are orthogonal projectors. In this case, $\widetilde{\Wcal}_{M\rightarrow FM}$ is a proper recovery map and the QCMI vanishes.} $\widetilde{\Wcal}_{M\rightarrow FM}$ in the above equation is generally not CPTP. Nonetheless, with this procedure one can then construct an ansatz for a quantum process with approximate Markov order. The crucial point being that the difference between the ansatz process and the actual process can be quantified by relative entropy between the two~\cite{Phil_MemStr}. Such a construction has applications in taming quantum non-Markovian memory; as we stated earlier, the complexity of a process increases exponentially with the size of the memory. Thus contracting the memory, without the loss of precision, is highly desirable. With this, we conclude our discussion of the properties of quantum processes with finite Markov order. We now provide an explicit example of such a process.

\subsubsection{Non-trivial example of quantum Markov order}
\label{sec::NonTrivEx}

\begin{figure}[t]
\centering
\includegraphics[width=0.95\linewidth]{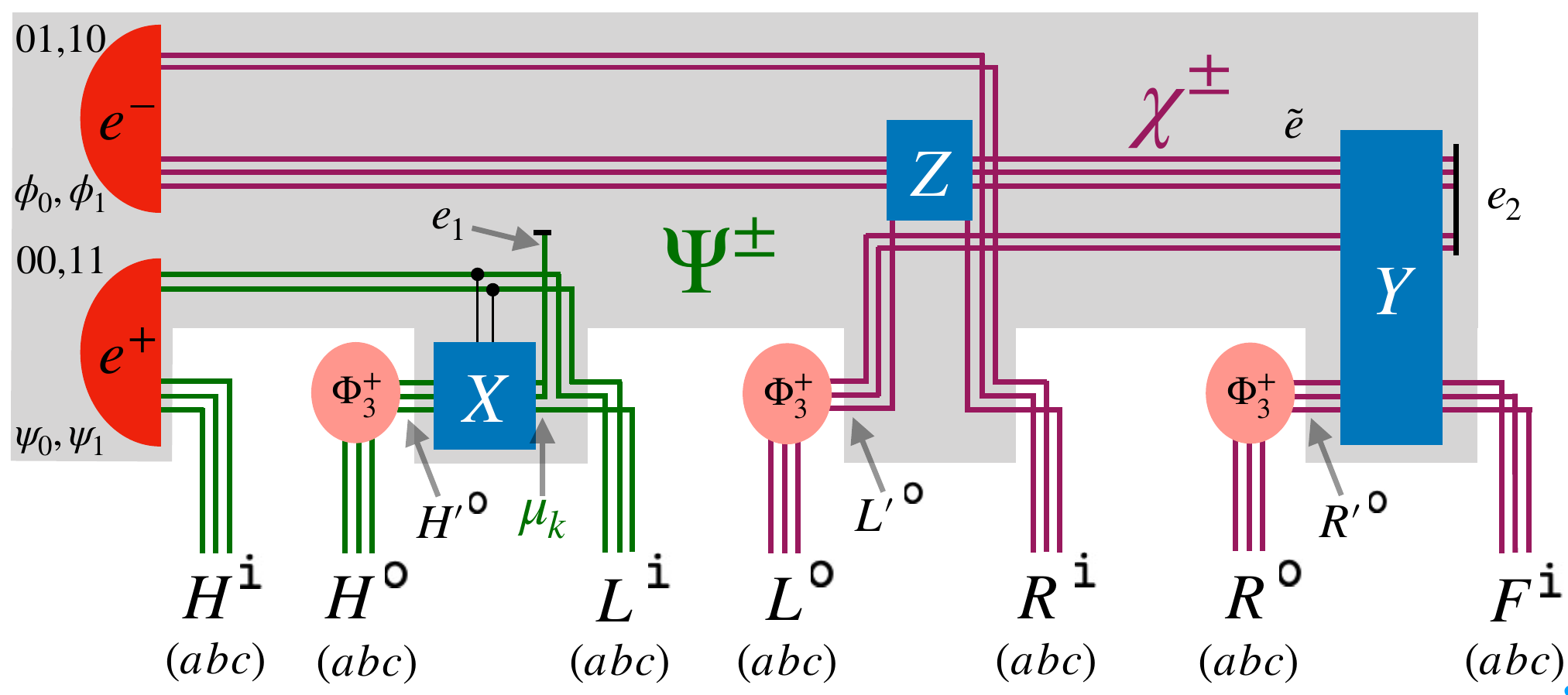}
\includegraphics[width=0.95\linewidth]{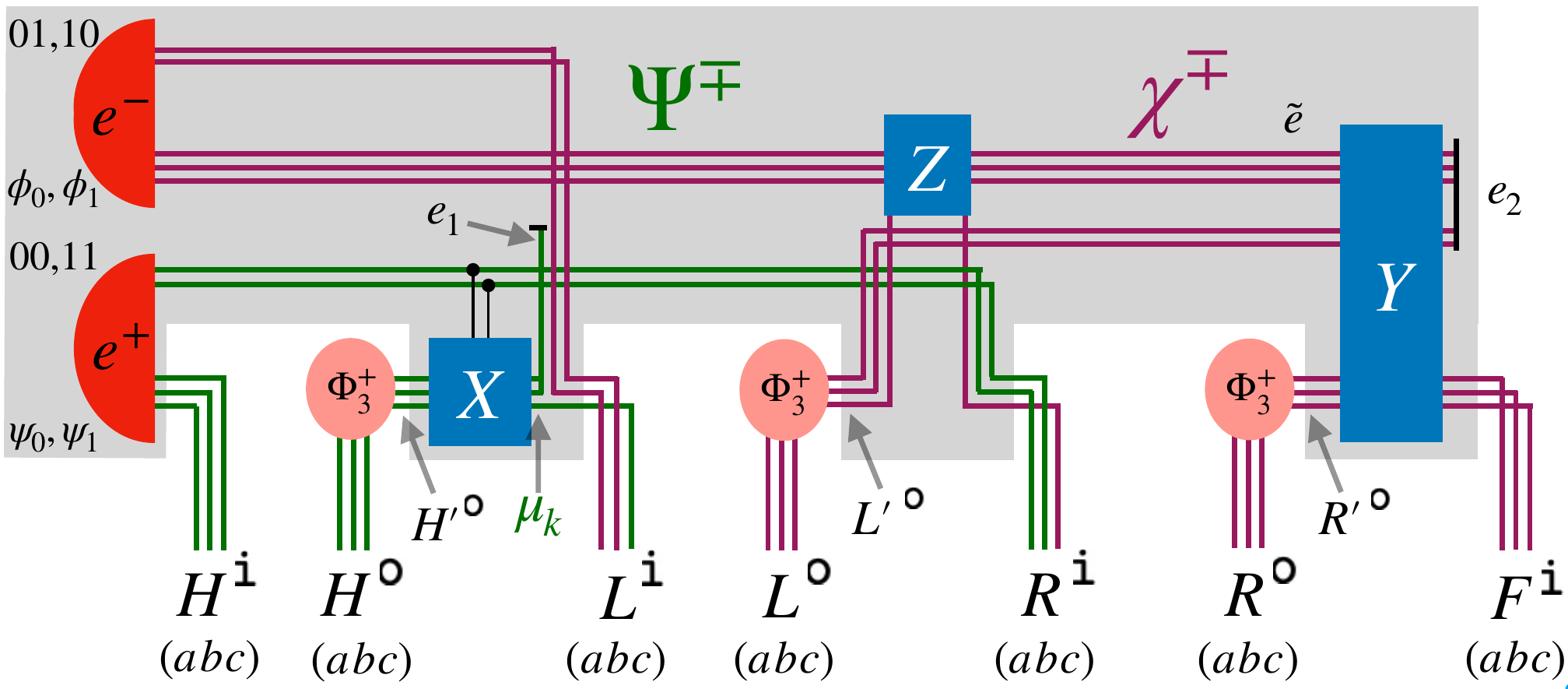}
\caption{\textbf{(Quantum) Markov order network.} A process with finite quantum Markov order with parts of $M$ kept by $H$ and $F$. The top panel shows the first process, in which parts of the common cause state $\ket{e^+}$ is sent to $L^{\inp}$ and $\ket{e^-}$ is sent to $R^{\inp}$. The process in the bottom panel has the recipients flipped. The process tensor is depicted in gray, and entanglement between parties color-coded in green and maroon. The overall process is a probabilistic mixture of both scenarios. Still the process has finite Markov order because it is possible to differentiate between the scenarios by making a parity measurement on $M$.}
\label{fig::markov-order}
\end{figure}

Let us now consider a process, introduced in~\cite{Phil_MemStr} and depicted here in Figure~\ref{fig::markov-order}, which requires leaving parts of $M$ attached to $H$ and $F$ (as we will see shortly). We label the input/output spaces associated to each time of the process as follows $\{H,M,F\} = \{H^{\inp}, H^{\out}, L^{\inp}, L^{\out}, R^{\inp}, R^{\out}, F^{\inp}\}$, where we have subdivided $M$ into left $L$ and right $R$ spaces. At each time, the system of interest comprises three qubits, and so each Hilbert space is of the form $\mathcal{H}_X = \mathcal{H}_{X_a} \otimes \mathcal{H}_{X_b} \otimes \mathcal{H}_{X_c}$, where $X$ takes values for the times and $a,b,c$ are labels for the three qubits; whenever we refer to an individual qubit, we will label the system appropriately, e.g., $L^{\inp}_{a}$ refers to the $a$ qubit of the system $L^{\inp}$; whenever no such label is specified, we are referring to all three qubits. 

The environment first prepares the five-qubit common cause states 
\begin{gather}
\begin{split}
 &\ket{e^+} = \frac{1}{\sqrt{2}}(\alpha \ket{\psi_0,00}+ \beta \ket{\psi_1,11}) 
 \quad \mbox{and}\\
 &\ket{e^-} = \frac{1}{\sqrt{2}}(\gamma \ket{\phi_0,01}+ \delta \ket{\phi_1,10}).
 \end{split}
\end{gather}
Here, we have separated the first register, which is a three qubit state, from the second, which consists of two qubits, with a comma. The first parts of the states $\ket{e^+}$ and $\ket{e^-}$ are respectively sent to $H^{\inp}$ and $F^{\inp}$. The second parts are sent either to $L^{\inp}$ or $R^{\inp}$, according to some probability distribution (see Figure~\ref{fig::markov-order}).

Let the state input at $H^{\out}$ be the first halves of three maximally entangled states $\bigotimes_{x\in \{a,b,c\}} \ket{\Phi^+}_{H_x^{\out} {H'}_x^{\out}}$ with $\ket{\Phi^+} := \tfrac{1}{\sqrt{2}}(\ket{00} +\ket{11})$; here, the prime denotes systems that are fed into the process, whereas the spaces without a prime refer to systems kept outside of it (these maximally entangled states are fed into the process to construct the resulting Choi matrix). The input at $L^{\out}$ and $R^{\out}$ are labeled similarly. In between times $H^{\out}$ and $L^{\inp}$, the process makes use of the second part of the state $\ket{e^+}$ to apply a controlled quantum channel $X$, which acts on all three qubits $a,b,c$. Following this, qubits $a$ and $b$ are discarded. The $ab$ qubits input at $L^{\out}$, as well as all three qubits input at $R^{\out}$, are sent forward into the process, which applies a joint channel $Y$ on all of these systems, as well as the first part of the state $\ket{e^-}$. Three of the output qubits are sent out to $F^{\inp}$, and the rest are discarded. The $c$ qubit input at $L^{\out}$ is sent to $R^{\inp}$, after being subjected to a channel $Z$, which interacts with the first part of the common cause state $\ket{e^-}$, i.e., the $\phi_0,\phi_1$ register.

Consider the process where $\ket{e^+}$ is sent to $H^{\inp}$ and $L^{\inp}$ and $\ket{e^-}$ to $R^{\inp}$ and $F^{\inp}$. The process tensor for this case is
\begin{gather}
\begin{split}
& \Upsilon^{\pm} = \Psi^{\pm}_{H^{\inp}H^{\out}L^{\inp}} \otimes \chi^{\pm}_{L^{\out}R^{\inp}R^{\out}F^{\inp}} \quad \mbox{with}\\
& \Psi^{\pm}_{H^{\inp}H^{\out}L^{\inp}} =\mbox{tr}_{e_1}\left[\ket{G^{\pm}}\bra{G^{\pm}} \right] \quad \mbox{and}\\
& \chi^{\pm}_{R^{\inp}R^{\out}F^{\inp}} =\mbox{tr}_{e_2}\left[\ket{K^{\pm}}\bra{K^{\pm}} \right],
\end{split}
\end{gather}
where
\begin{gather}
\begin{split}
\ket{G^{\pm}} = \frac{1}{\sqrt{2}} & \left(
\alpha \ket{\psi_0}_{H^{\inp}}
 \ket{\mu_{0}}_{H^{\out} e_1 L^{\inp}_c}\ket{00}_{L^{\inp}_{ab}} \right.\\
& \left. +\beta
 \ket{\psi_1}_{H^{\inp}}
 \ket{\mu_{1}}_{H^{\out} e_1 L^{\inp}_c}\ket{11}_{L^{\inp}_{ab}} \right),\\
\end{split}
\end{gather}
with $\ket{\mu_{k}}_{H^{\out} e_1 L_c^{\inp}} = X^{k}_{{H'}^{\out} \to e_1 L_c^{\inp}} \ket{\Phi^+}^{\otimes 3}_{H^{\out} {H'}^{\out}}$
\begin{gather}
\begin{split}
\ket{K^{\pm}} =& \frac{1}{\sqrt{2}} Y_{{L'}_{ab}^{\out} {R'}^{\out} \tilde{e} \to F^{\inp} e_2} 
Z^{k}_{{L'}_c^{\out}\tilde{e} \to R_{c}^{\inp}\tilde{e}} \\
& \times \left(
\gamma \ket{01}_{R_{ab}^{\inp}} \ket{\phi_0}_{\tilde{e}} +\delta \ket{10}_{R_{ab}^{\inp}} \ket{\phi_1}_{\tilde{e}}\right)\\
& \ket{\phi^+}^{\otimes 3}_{L^{\out} {L'}^{\out}} 
\ket{\phi^+}^{\otimes 3}_{R^{\out} {R'}^{\out}}.
\end{split}
\end{gather}

Next, consider the process where $\ket{e^+}$ is sent to $H^{\inp}$ and $M'^{\inp}$ and $\ket{e^-}$ to $M^{\inp}$ and $F^{\inp}$. The process tensor for this scenario is 
\begin{gather}
\begin{split}
&\Upsilon^{\mp} = \Psi^{\mp}_{H^{\inp} H^{\out} L_c^{\inp} R_{ab}^{\inp}} \otimes \chi^{\mp}_{L_{ab}^{\inp} L^{\out} R_c^{\inp} R^{\out} F^{\inp}} \quad \mbox{with} \\
& \Psi^{\mp}_{H^{\inp} H^{\out} L_c^{\inp} R_{ab}^{\inp}} =\mbox{tr}_{e_1}\left[\ket{G^{\mp}}\bra{G^{\mp}} \right] \quad \mbox{and}\\
& \chi^{\mp}_{L_{ab}^{\inp} L^{\out} R_c^{\inp} R^{\out} F^{\inp}} =\mbox{tr}_{e_2}\left[\ket{K^{\mp}}\bra{K^{\mp}} \right]
\end{split}
\end{gather}
where
\begin{gather}
\begin{split}
\ket{G^{\mp}} = \frac{1}{\sqrt{2}} & \left( \alpha \ket{\psi_0}_{H^{\inp}}
 \ket{\mu_{0}}_{H^{\out} e_1 L_c^{\inp}}\ket{00}_{R_{ab}^{\inp}} \right.\\
&\left.+\beta
 \ket{\psi_1}_{H^{\inp}}
 \ket{\mu_{1}}_{H^{\out} e_1 L_c^{\inp} }\ket{11}_{R_{ab}^{\inp}} \right)
\end{split}
\end{gather}
with $\ket{\mu_{k}}_{H^{\out} e_1 L_c^{\inp}} = X^{k}_{{H'}^{\out} \to e_1 L_c^{\inp}} \ket{\Phi^+}^{\otimes 3}_{H^{\out} {H'}^{\out}}$
\begin{gather}
\begin{split}
\ket{K^{\mp}} = & \frac{1}{\sqrt{2}} Y_{{L'}_{ab}^{\out} {R'}^{\out} \tilde{e} \to F^{\inp} e_2} Z^{k}_{{L'}_c^{\out}\tilde{e} \to R_{c}^{\inp}\tilde{e}}\\
& \times \left(
\gamma \ket{01}_{L_{ab}^{\inp}} \ket{\phi_0}_{\tilde{e}}
+\delta \ket{10}_{L_{ab}^{\inp}} \ket{\phi_1}_{\tilde{e}} \right) \\
& \ket{\Phi^+}^{\otimes 3}_{L^{\out} {L'}^{\out}} \ket{\Phi^+}^{\otimes 3}_{R^{\out} {R'}^{\out}}.
\end{split}
\end{gather}

In the first case, there is entanglement between $H^{\inp \out}$ and $L^{\inp}$, as well as between $L^{\out} R^{\inp \out}$ and $F^{\inp}$. In the second case, there is entanglement between $H^{\inp \out}$ and $L_c^{\inp} R_{ab}^{\inp}$, as well as between $L_{ab}^{\inp} L^{\out} R_c^{\inp} R^{\out}$ and $F^{\inp}$. The overall process is the average of these two, which will still have entanglement across the same cuts for generic probability distributions that the common cause states are sent out with. 

This process has a vanishing Markov order because we can make a parity measurement on the $ab$ parts of $L^{\inp}$ and $R^{\inp}$. The parity measurement applies two controlled phases to an ancilla initially prepared in the state $\ket{+}$, with the control registers being qubits $a$ and $b$. If the two control qubits are in states $\ket{00}$ or $\ket{11}$, then $\ket{+} \mapsto \ket{+}$. However, if the control qubits are in states $\ket{01}$ or $\ket{10}$, then $\ket{+} \mapsto \ket{-}$. By measuring the final ancilla, which can be perfectly distinguished since it is in one of two orthogonal states, we can know which process we have in a given run; in either case, there are no $FH$ correlations. Lastly, note that this process also has vanishing QCMI; this agrees with the analysis in Ref.~\cite{PhysRevA.99.042108}, as the instrument that erases the history, comprises only orthogonal projectors.

\section{Conclusions}

We began this tutorial with the basics of classical stochastic processes by means of concrete examples. We then built up to the formal definition of classical stochastic processes. Subsequently, we moved to quantum stochastic processes, covering the early works from half a century ago to modern methods used to differentiate between Markovian and non-Markovian processes in the quantum domain. Our main message throughout has been to show how a formal theory of quantum stochastic processes can be constructed based on ideas akin to those used in the classical domain. The resulting theory is general enough that it contains the theory of classical stochastic processes as a limiting case. On the structural side, we have shown that a quantum stochastic process is described by a many-body density operator (up to a normalization factor). This is a natural generalization for classical processes which are described by joint probability distributions over random variables in time. Along the way, we have attempted to build intuition for the reader by giving several examples.

In particular, the examples in the last section show that, in general, quantum stochastic processes are as complex as many-body quantum states are. However, there is beauty in the simplicity of the framework that encapsulates complex quantum phenomena in an overarching structure. We restrained our discussion to Markov processes and Markov order in the final section, but needless to say, there is much more to explore. Complex processes, in the quantum or classical realm, will have many attributes that are of interest for foundational or technological reasons. We cannot do justice to many (most) of these facets of the theory in this short manuscript. On the plus side, there are a lot of interesting problems left unexplored for current and future researchers. Our tutorial has barely touched the topic of quantum probability, and associated techniques such as quantum trajectories, quantum stochastic calculus, and the SLH\ftnt{S stands for a scattering matrix, L for jump operators, and H for Hamiltonians} framework. This is an extremely active area of research~\cite{accardi_topics_1981, Parthasarathy, qprob2006, qtraj, slh, Nurdin2019} with many overlaps with the ideas presented here; however, a detailed cross-comparison would form a whole tutorial on its own.

We now bring this article to closure by discussing some important open problems and some important applications of the theory of open quantum systems.

The vastness of the theory of classical stochastic processes suggests that there are many open problems in the quantum realm. In this sense, it is a daunting endeavor to even attempt to make a list of interesting problems, and we make no claims of comprehensiveness. On the foundational front, understanding the quantum to classical transition for stochastic processes~\cite{milz_when_2020} should be a far more manageable problem than the elusive connection between pure-state unitary quantum mechanics and phase space classical mechanics. Objectivity of quantum measurements and quantum Darwinism~\cite{darwin} are also enticing topics to reconsider from the process tensor perspective, i.e, as emergent phenomena in time rather than in space. It may also possible to better understand quantum chaos by analyzing multi-time correlations in quantum processes (e.g., out-of-time-ordered correlations already discuss quantum chaos in a similar vein). The list of complex dynamical phenomena includes dynamical phase transitions~\cite{PhysRevLett.110.135704}, dynamical many-body scars~\cite{scar}, measurement induced phase transition~\cite{PhysRevX.9.031009}, and understanding memory in complex quantum processes~\cite{Romero2018}. For all of these areas, higher order quantum maps like the process tensor provide an ideal framework to foster future developments. For practical implementations, quantifying and witnessing entanglement in time (i.e., genuinely non-classical temporal correlations) is of utmost importance complex experimental setups that aim to exploit quantum phenomena in time. On the mathematical side, there are many interesting problems such as embedding coherent dynamics in classical processes, simulating classical processes on quantum devices~\cite{korzekwa_quantum_2020}, or identifying processes that cannot be classical. Finally, there is still much work to be done approximating quantum processes with ansatz type considerations. For instance, what are the best ways (in the sense of minimal error) to truncate quantum memories, how to quantify contextual errors due to finite pulse width for the control operations. i.e., how to deal with experimental operations that cannot be considered to be implemented instantaneously, as we did throughout this tutorial. The process tensor also opens up -- as we aready mentioned -- the whole toolkit of tensor network to characterize, simulate, and manipulate complex quantum processes and multi-time statistics in quantum mechanics. In particular, little attention has been devoted so far to understanding critical quantum processes from a multi-time perspective, i.e., to efficiently describing processes where the correlations decay is slow (powerlaw).

On the application side, the foremost application of the theory of quantum stochastic processes is quantum control, e.g. dynamical decoupling~\cite{PhysRevLett.82.2417, PhysRevLett.83.4888, Arenz2017, addis-dd} (and understanding processes that cannot be decoupled~\cite{arXiv:1904.03627}), decoherence-free subspaces~\cite{zanardi,PhysRevLett.81.2594}, quantum error correction~\cite{lidarQEC}, and quantum Zeno effect~\cite{PhysRevLett.108.080501, PhysRevLett.121.060401, Burgarth2020quantumzenodynamics}. All of these are dynamical phenomena and it remains to see how they fit into the theory described in this tutorial. Moreover, small quantum computers are now readily available, but they suffer from complex noise, i.e., undergo complex non-Markovian stochastic processes. This forms a fertile ground for the process tensor~\cite{White2020} framework to provide new conceptual insights. Additionally, there are natural systems that would also be excellent candidates for an application of the process tensor framework, for instance, control of biological systems~\cite{caruso_coherent_2012}. They are intersting because, it is possible that these systems harness complex noise to achieve efficient and quantum information processing tasks~\cite{plenio_cavity-loss-induced_1999,verstraete_quantum_2009, caruso_highly_2009, caruso_noise-enhanced_2010, kastoryano_precisely_2013}. Already, and even more so in the future, these tools (will) enable quantum technologies in presence of non-Markovian noise~\cite{PhysRevLett.109.233601, weakfieldphase,liu_experimental_2020}. As we attempt to engineer more and more sophisticated quantum devices we will need more sophistication in accounting for the noise due to the environment. These applications will be within reach once we can characterize the noise~\cite{PhysRevLett.108.160402, compen, PhysRevLett.116.150503, PhysRevA.95.022121, PhysRevLett.121.040601, PhysRevLett.118.080404, benedetti_non-markovianity_2014, Schindler} and understand how quantum processes and memory effects can serve as resources~\cite{arXiv:1301.2585, PhysRevX.8.021033, bylicka_thermodynamic_2016, berk, singha_roy_information-theoretic_2019, abiuso_non-markov_2019}.

There are also foundational applications to the frameworks discussed above. For instance to better understand how the theory of thermodynamics fits with the theory of quantum mechanics requires better handling of interventions and memory, and already there is progress on this front~\cite{Strasberg2018, Strasberg2019-1,  strasberg_repeated_2019, strasberg_thermodynamics_2020}. This framework also allows for a method to build a classical-quantum correspondence, i.e., determining quantum stochastic processes that look classical~\cite{milz_when_2020, PhysRevA.100.022120}. Furthermore, it enables one to understand the statistical nature of quantum processes, i.e., when is the memory too complex~\cite{PhysRevA.81.032101, PhysRevA.91.012104, Romero2018, Romero2020}, or when does a system look as if it has equilibrated~\cite{PhysRevX.7.041015, Romero2019}? These latter questions are closely related to ones aiming to derive statistical mechanics from quantum mechanics~\cite{Popescu2006, PhysRevE.79.061103, Masanes2013, DAlessio2016}. In general, non-Markovian effects in many-body systems~\cite{PhysRevLett.99.120504, virmani, PhysRevA.87.012127} and complex single body systems~\cite{PhysRevA.85.060101, PhysRevA.98.053608} will be of keen interest as they will contain rich physics.

Finally, the tools introduced in the article are closely related to those used to examine the role of causal order -- or absence thereof -- in quantum mechanics. As they are tailored to account for active interventions, they are used in the field of quantum causal modeling~\cite{1367-2630-18-6-063032, PhysRevX.7.031021, giarmatzi_quantum_2018, chiribella_quantum_2019, barrett_quantum_2020} to discern causal relations in quantum processes. Beyond such causally ordered situations, the quantum comb and process matrix framework have been employed to explore quantum mechanics in the absence of global causal order~\cite{chiribella_quantum_2013, OreshkovETAL2012}, and it has been shown that such processes would provide advantages in information processing tasks over causally ordered ones~\cite{chiribella_quantum_2013, procopio_experimental_2015,rubino_experimental_2017,ebler_enhanced_2018,goswami_indefinite_2018,goswami_communicating_2020}. The existence of such exotic processes is still under debate and the search for additional principles to limit the set of `allowed' causally disordered processes is an active field of research~\cite{araujo_purification_2017}. Nonetheless, the tools to describe them are -- both mathematically and in spirit -- akin to the process tensors we introduced for the description of open quantum processes, demonstrating the versatility and wide applicability of the ideas and concepts employed in this tutorial. 

\begin{acknowledgments}
We thank Felix A. Pollock and Philip Taranto for valuable discussions, and Heinz-Peter Breuer, Daniel Burgarth, Fabio Costa, Michael Hall, Susana Huelga, Jyrki Piilo, Martin Plenio, {\'A}ngel Rivas,  Andrea Smirne, Phillipp Strasberg, Bassano Vacchini, Howard Wiseman, and Karol {\.Z}yczkowski  for helpful comments on the original manuscript. SM acknowledges funding from the European Union's Horizon 2020 research and innovation programme under the Marie Sk{\l}odowska Curie grant agreement No 801110, and the Austrian Federal Ministry of Education, Science and Research (BMBWF). KM is supported through Australian Research Council Future Fellowship FT160100073.
\end{acknowledgments}
\bibliography{tut-ref.bib}

\begin{thebibliography}{386}%
\makeatletter
\providecommand \@ifxundefined [1]{%
 \@ifx{#1\undefined}
}%
\providecommand \@ifnum [1]{%
 \ifnum #1\expandafter \@firstoftwo
 \else \expandafter \@secondoftwo
 \fi
}%
\providecommand \@ifx [1]{%
 \ifx #1\expandafter \@firstoftwo
 \else \expandafter \@secondoftwo
 \fi
}%
\providecommand \natexlab [1]{#1}%
\providecommand \enquote  [1]{``#1''}%
\providecommand \bibnamefont  [1]{#1}%
\providecommand \bibfnamefont [1]{#1}%
\providecommand \citenamefont [1]{#1}%
\providecommand \href@noop [0]{\@secondoftwo}%
\providecommand \href [0]{\begingroup \@sanitize@url \@href}%
\providecommand \@href[1]{\@@startlink{#1}\@@href}%
\providecommand \@@href[1]{\endgroup#1\@@endlink}%
\providecommand \@sanitize@url [0]{\catcode `\\12\catcode `\$12\catcode
  `\&12\catcode `\#12\catcode `\^12\catcode `\_12\catcode `\%12\relax}%
\providecommand \@@startlink[1]{}%
\providecommand \@@endlink[0]{}%
\providecommand \url  [0]{\begingroup\@sanitize@url \@url }%
\providecommand \@url [1]{\endgroup\@href {#1}{\urlprefix }}%
\providecommand \urlprefix  [0]{URL }%
\providecommand \Eprint [0]{\href }%
\providecommand \doibase [0]{http://dx.doi.org/}%
\providecommand \selectlanguage [0]{\@gobble}%
\providecommand \bibinfo  [0]{\@secondoftwo}%
\providecommand \bibfield  [0]{\@secondoftwo}%
\providecommand \translation [1]{[#1]}%
\providecommand \BibitemOpen [0]{}%
\providecommand \bibitemStop [0]{}%
\providecommand \bibitemNoStop [0]{.\EOS\space}%
\providecommand \EOS [0]{\spacefactor3000\relax}%
\providecommand \BibitemShut  [1]{\csname bibitem#1\endcsname}%
\let\auto@bib@innerbib\@empty
\bibitem [{\citenamefont {Alicki}\ and\ \citenamefont
  {Lendi}(1987)}]{alicki_semi_1987}%
  \BibitemOpen
  \bibfield  {author} {\bibinfo {author} {\bibfnamefont {R.}~\bibnamefont
  {Alicki}}\ and\ \bibinfo {author} {\bibfnamefont {K.}~\bibnamefont {Lendi}},\
  }\href@noop {} {\emph {\bibinfo {title} {Quantum Dynamical Semi-Groups and
  Applications}}}\ (\bibinfo  {publisher} {Springer},\ \bibinfo {address}
  {Berlin},\ \bibinfo {year} {1987})\BibitemShut {NoStop}%
\bibitem [{\citenamefont {Gardiner}\ and\ \citenamefont
  {Zoller}(1991)}]{Gardiner}%
  \BibitemOpen
  \bibfield  {author} {\bibinfo {author} {\bibfnamefont {C.}~\bibnamefont
  {Gardiner}}\ and\ \bibinfo {author} {\bibfnamefont {P.}~\bibnamefont
  {Zoller}},\ }\href@noop {} {\emph {\bibinfo {title} {Quantum Noise}}}\
  (\bibinfo  {publisher} {Springer},\ \bibinfo {year} {1991})\BibitemShut
  {NoStop}%
\bibitem [{\citenamefont {Breuer}\ and\ \citenamefont
  {Petruccione}(2002)}]{BreuerPetruccione}%
  \BibitemOpen
  \bibfield  {author} {\bibinfo {author} {\bibfnamefont {H.-P.}\ \bibnamefont
  {Breuer}}\ and\ \bibinfo {author} {\bibfnamefont {F.}~\bibnamefont
  {Petruccione}},\ }\href@noop {} {\emph {\bibinfo {title} {{The Theory of Open
  Quantum Systems}}}}\ (\bibinfo  {publisher} {Oxford Univ. Press},\ \bibinfo
  {year} {2002})\BibitemShut {NoStop}%
\bibitem [{\citenamefont {Accardi}\ \emph {et~al.}(2002)\citenamefont
  {Accardi}, \citenamefont {Lu},\ and\ \citenamefont
  {Volovich}}]{accardi_quantum_2002}%
  \BibitemOpen
  \bibfield  {author} {\bibinfo {author} {\bibfnamefont {L.}~\bibnamefont
  {Accardi}}, \bibinfo {author} {\bibfnamefont {Y.~G.}\ \bibnamefont {Lu}}, \
  and\ \bibinfo {author} {\bibfnamefont {I.}~\bibnamefont {Volovich}},\ }\href
  {\doibase 10.1007/978-3-662-04929-7} {\emph {\bibinfo {title} {Quantum
  {Theory} and {Its} {Stochastic} {Limit}}}}\ (\bibinfo  {publisher}
  {Springer-Verlag},\ \bibinfo {address} {Berlin Heidelberg},\ \bibinfo {year}
  {2002})\BibitemShut {NoStop}%
\bibitem [{\citenamefont {Wiseman}\ and\ \citenamefont
  {Milburn}(2010)}]{Wiseman}%
  \BibitemOpen
  \bibfield  {author} {\bibinfo {author} {\bibfnamefont {H.~M.}\ \bibnamefont
  {Wiseman}}\ and\ \bibinfo {author} {\bibfnamefont {G.~J.}\ \bibnamefont
  {Milburn}},\ }\href@noop {} {\emph {\bibinfo {title} {{Quantum Measurement
  and Control}}}}\ (\bibinfo  {publisher} {Cambridge Univ. Press},\ \bibinfo
  {year} {2010})\BibitemShut {NoStop}%
\bibitem [{\citenamefont {Bouten}\ \emph {et~al.}(2007)\citenamefont {Bouten},
  \citenamefont {van Handel},\ and\ \citenamefont {James}}]{qprob2006}%
  \BibitemOpen
  \bibfield  {author} {\bibinfo {author} {\bibfnamefont {L.}~\bibnamefont
  {Bouten}}, \bibinfo {author} {\bibfnamefont {R.}~\bibnamefont {van Handel}},
  \ and\ \bibinfo {author} {\bibfnamefont {M.}~\bibnamefont {James}},\ }\href
  {\doibase 10.1137/060651239} {\bibfield  {journal} {\bibinfo  {journal} {SIAM
  J. Control Optim.}\ ,\ \bibinfo {pages} {2199}} (\bibinfo {year}
  {2007})}\BibitemShut {NoStop}%
\bibitem [{\citenamefont {Rivas}\ \emph {et~al.}(2014)\citenamefont {Rivas},
  \citenamefont {Huelga},\ and\ \citenamefont {Plenio}}]{Rivas2014}%
  \BibitemOpen
  \bibfield  {author} {\bibinfo {author} {\bibfnamefont {{\'A}.}~\bibnamefont
  {Rivas}}, \bibinfo {author} {\bibfnamefont {S.~F.}\ \bibnamefont {Huelga}}, \
  and\ \bibinfo {author} {\bibfnamefont {M.~B.}\ \bibnamefont {Plenio}},\
  }\href {\doibase 10.1088/0034-4885/77/9/094001} {\bibfield  {journal}
  {\bibinfo  {journal} {Rep. Prog. Phys.}\ }\textbf {\bibinfo {volume} {77}},\
  \bibinfo {pages} {094001} (\bibinfo {year} {2014})}\BibitemShut {NoStop}%
\bibitem [{\citenamefont {Breuer}\ \emph {et~al.}(2016)\citenamefont {Breuer},
  \citenamefont {Laine}, \citenamefont {Piilo},\ and\ \citenamefont
  {Vacchini}}]{RevModPhys.88.021002}%
  \BibitemOpen
  \bibfield  {author} {\bibinfo {author} {\bibfnamefont {H.-P.}\ \bibnamefont
  {Breuer}}, \bibinfo {author} {\bibfnamefont {E.-M.}\ \bibnamefont {Laine}},
  \bibinfo {author} {\bibfnamefont {J.}~\bibnamefont {Piilo}}, \ and\ \bibinfo
  {author} {\bibfnamefont {B.}~\bibnamefont {Vacchini}},\ }\href {\doibase
  10.1103/RevModPhys.88.021002} {\bibfield  {journal} {\bibinfo  {journal}
  {Rev. Mod. Phys.}\ }\textbf {\bibinfo {volume} {88}},\ \bibinfo {pages}
  {021002} (\bibinfo {year} {2016})}\BibitemShut {NoStop}%
\bibitem [{\citenamefont {de~Vega}\ and\ \citenamefont
  {Alonso}(2017)}]{deVega2017}%
  \BibitemOpen
  \bibfield  {author} {\bibinfo {author} {\bibfnamefont {I.}~\bibnamefont
  {de~Vega}}\ and\ \bibinfo {author} {\bibfnamefont {D.}~\bibnamefont
  {Alonso}},\ }\href {\doibase 10.1103/RevModPhys.89.015001} {\bibfield
  {journal} {\bibinfo  {journal} {Rev. Mod. Phys.}\ }\textbf {\bibinfo {volume}
  {89}},\ \bibinfo {pages} {015001} (\bibinfo {year} {2017})}\BibitemShut
  {NoStop}%
\bibitem [{\citenamefont {Li}\ \emph {et~al.}(2018)\citenamefont {Li},
  \citenamefont {Hall},\ and\ \citenamefont {Wiseman}}]{Li2018}%
  \BibitemOpen
  \bibfield  {author} {\bibinfo {author} {\bibfnamefont {L.}~\bibnamefont
  {Li}}, \bibinfo {author} {\bibfnamefont {M.~J.}\ \bibnamefont {Hall}}, \ and\
  \bibinfo {author} {\bibfnamefont {H.~M.}\ \bibnamefont {Wiseman}},\ }\href
  {http://www.sciencedirect.com/science/article/pii/S0370157318301601}
  {\bibfield  {journal} {\bibinfo  {journal} {Phys. Rep.}\ }\textbf {\bibinfo
  {volume} {759}},\ \bibinfo {pages} {1} (\bibinfo {year} {2018})}\BibitemShut
  {NoStop}%
\bibitem [{\citenamefont {Li}\ \emph {et~al.}(2020)\citenamefont {Li},
  \citenamefont {Guo},\ and\ \citenamefont {Piilo}}]{eplrev2}%
  \BibitemOpen
  \bibfield  {author} {\bibinfo {author} {\bibfnamefont {C.}~\bibnamefont
  {Li}}, \bibinfo {author} {\bibfnamefont {G.}~\bibnamefont {Guo}}, \ and\
  \bibinfo {author} {\bibfnamefont {J.}~\bibnamefont {Piilo}},\ }\href
  {\doibase 10.1209/0295-5075/128/30001} {\bibfield  {journal} {\bibinfo
  {journal} {EPL}\ }\textbf {\bibinfo {volume} {128}},\ \bibinfo {pages}
  {30001} (\bibinfo {year} {2020})}\BibitemShut {NoStop}%
\bibitem [{\citenamefont {Li}\ \emph {et~al.}(2019)\citenamefont {Li},
  \citenamefont {Guo},\ and\ \citenamefont {Piilo}}]{eplrev1}%
  \BibitemOpen
  \bibfield  {author} {\bibinfo {author} {\bibfnamefont {C.}~\bibnamefont
  {Li}}, \bibinfo {author} {\bibfnamefont {G.}~\bibnamefont {Guo}}, \ and\
  \bibinfo {author} {\bibfnamefont {J.}~\bibnamefont {Piilo}},\ }\href
  {\doibase 10.1209/0295-5075/127/50001} {\bibfield  {journal} {\bibinfo
  {journal} {EPL}\ }\textbf {\bibinfo {volume} {127}},\ \bibinfo {pages}
  {50001} (\bibinfo {year} {2019})}\BibitemShut {NoStop}%
\bibitem [{Note1()}]{Note1}%
  \BibitemOpen
  \bibinfo {note} {The sum of the opposite ends of a die always equals
  7.}\BibitemShut {Stop}%
\bibitem [{\citenamefont {{van Kampen}}(1998)}]{vanKampen1998}%
  \BibitemOpen
  \bibfield  {author} {\bibinfo {author} {\bibfnamefont {N.~G.}\ \bibnamefont
  {{van Kampen}}},\ }\href {http://www.sbfisica.org.br/bjp/files/v28_90.pdf}
  {\bibfield  {journal} {\bibinfo  {journal} {Braz. J. Phys.}\ }\textbf
  {\bibinfo {volume} {28}},\ \bibinfo {pages} {90} (\bibinfo {year}
  {1998})}\BibitemShut {NoStop}%
\bibitem [{\citenamefont {Marshall}\ \emph {et~al.}(2011)\citenamefont
  {Marshall}, \citenamefont {Olkin},\ and\ \citenamefont
  {Arnold}}]{marshall_inequalities_2011}%
  \BibitemOpen
  \bibfield  {author} {\bibinfo {author} {\bibfnamefont {A.~W.}\ \bibnamefont
  {Marshall}}, \bibinfo {author} {\bibfnamefont {I.}~\bibnamefont {Olkin}}, \
  and\ \bibinfo {author} {\bibfnamefont {B.~C.}\ \bibnamefont {Arnold}},\
  }\href {\doibase 10.1007/978-0-387-68276-1} {\emph {\bibinfo {title}
  {Inequalities: {Theory} of {Majorization} and {Its} {Applications}}}},\
  \bibinfo {edition} {2nd}\ ed.,\ Springer {Series} in {Statistics}\ (\bibinfo
  {publisher} {Springer-Verlag},\ \bibinfo {address} {New York},\ \bibinfo
  {year} {2011})\BibitemShut {NoStop}%
\bibitem [{\citenamefont {Crutchfield}\ and\ \citenamefont
  {Young}(1989)}]{crutchfield_inferring_1989}%
  \BibitemOpen
  \bibfield  {author} {\bibinfo {author} {\bibfnamefont {J.~P.}\ \bibnamefont
  {Crutchfield}}\ and\ \bibinfo {author} {\bibfnamefont {K.}~\bibnamefont
  {Young}},\ }\href {\doibase 10.1103/PhysRevLett.63.105} {\bibfield  {journal}
  {\bibinfo  {journal} {Phys. Rev. Lett.}\ }\textbf {\bibinfo {volume} {63}},\
  \bibinfo {pages} {105} (\bibinfo {year} {1989})}\BibitemShut {NoStop}%
\bibitem [{\citenamefont {Shalizi}\ and\ \citenamefont
  {Crutchfield}(2001)}]{shalizi_computational_2001}%
  \BibitemOpen
  \bibfield  {author} {\bibinfo {author} {\bibfnamefont {C.~R.}\ \bibnamefont
  {Shalizi}}\ and\ \bibinfo {author} {\bibfnamefont {J.~P.}\ \bibnamefont
  {Crutchfield}},\ }\href {\doibase 10.1023/A:1010388907793} {\bibfield
  {journal} {\bibinfo  {journal} {J. Stat. Phys.}\ }\textbf {\bibinfo {volume}
  {104}},\ \bibinfo {pages} {817} (\bibinfo {year} {2001})}\BibitemShut
  {NoStop}%
\bibitem [{\citenamefont {Crutchfield}(2012)}]{crutchfield_between_2012}%
  \BibitemOpen
  \bibfield  {author} {\bibinfo {author} {\bibfnamefont {J.~P.}\ \bibnamefont
  {Crutchfield}},\ }\href {\doibase 10.1038/nphys2190} {\bibfield  {journal}
  {\bibinfo  {journal} {Nat. Phys.}\ }\textbf {\bibinfo {volume} {8}},\
  \bibinfo {pages} {17} (\bibinfo {year} {2012})}\BibitemShut {NoStop}%
\bibitem [{\citenamefont {Tao}(2011)}]{tao_introduction_2011}%
  \BibitemOpen
  \bibfield  {author} {\bibinfo {author} {\bibfnamefont {T.}~\bibnamefont
  {Tao}},\ }\href@noop {} {\emph {\bibinfo {title} {An Introduction to Measure
  Theory}}}\ (\bibinfo  {publisher} {American Mathematical Society},\ \bibinfo
  {year} {2011})\BibitemShut {NoStop}%
\bibitem [{\citenamefont {Lemons}(2002)}]{lemons_introduction_2002}%
  \BibitemOpen
  \bibfield  {author} {\bibinfo {author} {\bibfnamefont {D.~S.}\ \bibnamefont
  {Lemons}},\ }\href@noop {} {\emph {\bibinfo {title} {An {Introduction} to
  {Stochastic} {Processes} in {Physics}}}}\ (\bibinfo  {publisher} {Johns
  Hopkins University Press},\ \bibinfo {address} {Baltimore},\ \bibinfo {year}
  {2002})\BibitemShut {NoStop}%
\bibitem [{\citenamefont {Kolmogorov}(1933)}]{kolmogorov_foundations_1956}%
  \BibitemOpen
  \bibfield  {author} {\bibinfo {author} {\bibfnamefont {A.~N.}\ \bibnamefont
  {Kolmogorov}},\ }\href@noop {} {\emph {\bibinfo {title} {Grundbegriffe der
  Wahrscheinlichkeitsrechnung}}}\ (\bibinfo  {publisher} {Springer},\ \bibinfo
  {address} {Berlin},\ \bibinfo {year} {1933})\ \bibinfo {note}
  {[\textit{{F}oundations of the {Theory} of {Probability}} (Chelsea, New York,
  1956)]}\BibitemShut {NoStop}%
\bibitem [{\citenamefont {Feller}(1971)}]{feller_introduction_1968}%
  \BibitemOpen
  \bibfield  {author} {\bibinfo {author} {\bibfnamefont {W.}~\bibnamefont
  {Feller}},\ }\href@noop {} {\emph {\bibinfo {title} {An {Introduction} to
  {Probability} {Theory} and {Its} {Applications}}}}\ (\bibinfo  {publisher}
  {Wiley},\ \bibinfo {address} {New York},\ \bibinfo {year} {1971})\BibitemShut
  {NoStop}%
\bibitem [{Note2()}]{Note2}%
  \BibitemOpen
  \bibinfo {note} {Besides consistency, the individual probability
  distributions also have to be inner regular. We will not concern ourselves
  with this technicality, see Ref.~\cite {tao_introduction_2011} for more
  details.}\BibitemShut {Stop}%
\bibitem [{\citenamefont {Wiener}\ \emph {et~al.}(1966)\citenamefont {Wiener},
  \citenamefont {Siegel}, \citenamefont {Rankin},\ and\ \citenamefont
  {Martin}}]{wiener_differential_1966}%
  \BibitemOpen
  \bibinfo {editor} {\bibfnamefont {N.}~\bibnamefont {Wiener}}, \bibinfo
  {editor} {\bibfnamefont {A.}~\bibnamefont {Siegel}}, \bibinfo {editor}
  {\bibfnamefont {B.}~\bibnamefont {Rankin}}, \ and\ \bibinfo {editor}
  {\bibfnamefont {W.~T.}\ \bibnamefont {Martin}},\ eds.,\ \href@noop {} {\emph
  {\bibinfo {title} {Differential {Space}, {Quantum} {Systems}, and
  {Prediction}}}}\ (\bibinfo  {publisher} {The MIT Press},\ \bibinfo {address}
  {Cambridge (MA)},\ \bibinfo {year} {1966})\BibitemShut {NoStop}%
\bibitem [{\citenamefont {L{\'e}vy}(1940)}]{Levy_1940}%
  \BibitemOpen
  \bibfield  {author} {\bibinfo {author} {\bibfnamefont {M.~P.}\ \bibnamefont
  {L{\'e}vy}},\ }\href {https://doi.org/10.2307/2371467} {\bibfield  {journal}
  {\bibinfo  {journal} {Am. J. Math.}\ }\textbf {\bibinfo {volume} {62}},\
  \bibinfo {pages} {487} (\bibinfo {year} {1940})}\BibitemShut {NoStop}%
\bibitem [{\citenamefont {Ciesielski}(1966)}]{ciesielski_lectures_1966}%
  \BibitemOpen
  \bibfield  {author} {\bibinfo {author} {\bibfnamefont {Z.}~\bibnamefont
  {Ciesielski}},\ }\href@noop {} {\emph {\bibinfo {title} {Lectures on
  {Brownian} motion, heat conduction and potential theory}}}\ (\bibinfo
  {publisher} {Aarhus Universitet, Mathematisk Institutt},\ \bibinfo {year}
  {1966})\BibitemShut {NoStop}%
\bibitem [{\citenamefont {Bhattacharya}\ and\ \citenamefont
  {Waymire}(2017)}]{bhattacharya_basic_2017}%
  \BibitemOpen
  \bibfield  {author} {\bibinfo {author} {\bibfnamefont {R.}~\bibnamefont
  {Bhattacharya}}\ and\ \bibinfo {author} {\bibfnamefont {E.~C.}\ \bibnamefont
  {Waymire}},\ }\href@noop {} {\emph {\bibinfo {title} {A {Basic} {Course} in
  {Probability} {Theory}}}}\ (\bibinfo  {publisher} {Springer},\ \bibinfo
  {address} {New York, NY},\ \bibinfo {year} {2017})\BibitemShut {NoStop}%
\bibitem [{\citenamefont {Van~Kampen}(2011)}]{StochProc}%
  \BibitemOpen
  \bibfield  {author} {\bibinfo {author} {\bibfnamefont {N.}~\bibnamefont
  {Van~Kampen}},\ }\href@noop {} {\emph {\bibinfo {title} {{Stochastic
  Processes in Physics and Chemistry}}}}\ (\bibinfo  {publisher} {Elsevier, New
  York},\ \bibinfo {year} {2011})\BibitemShut {NoStop}%
\bibitem [{\citenamefont {H{\"a}nggi}\ and\ \citenamefont
  {Thomas}(1982)}]{hanggi_stochastic_1982}%
  \BibitemOpen
  \bibfield  {author} {\bibinfo {author} {\bibfnamefont {P.}~\bibnamefont
  {H{\"a}nggi}}\ and\ \bibinfo {author} {\bibfnamefont {H.}~\bibnamefont
  {Thomas}},\ }\href {\doibase 10.1016/0370-1573(82)90045-X} {\bibfield
  {journal} {\bibinfo  {journal} {Phys. Rep.}\ }\textbf {\bibinfo {volume}
  {88}},\ \bibinfo {pages} {207} (\bibinfo {year} {1982})}\BibitemShut
  {NoStop}%
\bibitem [{Note3()}]{Note3}%
  \BibitemOpen
  \bibinfo {note} {Some authors will not call this a master equation due to its
  temporal non-locality~\cite {StochProc}.}\BibitemShut {Stop}%
\bibitem [{\citenamefont {Smirne}\ and\ \citenamefont
  {Vacchini}(2010)}]{bassano1}%
  \BibitemOpen
  \bibfield  {author} {\bibinfo {author} {\bibfnamefont {A.}~\bibnamefont
  {Smirne}}\ and\ \bibinfo {author} {\bibfnamefont {B.}~\bibnamefont
  {Vacchini}},\ }\href {\doibase 10.1103/PhysRevA.82.022110} {\bibfield
  {journal} {\bibinfo  {journal} {Phys. Rev. A}\ }\textbf {\bibinfo {volume}
  {82}},\ \bibinfo {pages} {022110} (\bibinfo {year} {2010})}\BibitemShut
  {NoStop}%
\bibitem [{\citenamefont {Vacchini}\ \emph {et~al.}(2011)\citenamefont
  {Vacchini}, \citenamefont {Smirne}, \citenamefont {Laine}, \citenamefont
  {Piilo},\ and\ \citenamefont {Breuer}}]{vacchini}%
  \BibitemOpen
  \bibfield  {author} {\bibinfo {author} {\bibfnamefont {B.}~\bibnamefont
  {Vacchini}}, \bibinfo {author} {\bibfnamefont {A.}~\bibnamefont {Smirne}},
  \bibinfo {author} {\bibfnamefont {E.-M.}\ \bibnamefont {Laine}}, \bibinfo
  {author} {\bibfnamefont {J.}~\bibnamefont {Piilo}}, \ and\ \bibinfo {author}
  {\bibfnamefont {H.-P.}\ \bibnamefont {Breuer}},\ }\href {\doibase
  10.1088/1367-2630/13/9/093004} {\bibfield  {journal} {\bibinfo  {journal}
  {New J. Phys.}\ }\textbf {\bibinfo {volume} {13}},\ \bibinfo {pages} {093004}
  (\bibinfo {year} {2011})}\BibitemShut {NoStop}%
\bibitem [{\citenamefont {Cerrillo}\ and\ \citenamefont
  {Cao}(2014)}]{CerrilloCao2014}%
  \BibitemOpen
  \bibfield  {author} {\bibinfo {author} {\bibfnamefont {J.}~\bibnamefont
  {Cerrillo}}\ and\ \bibinfo {author} {\bibfnamefont {J.}~\bibnamefont {Cao}},\
  }\href {\doibase 10.1103/PhysRevLett.112.110401} {\bibfield  {journal}
  {\bibinfo  {journal} {Phys. Rev. Lett.}\ }\textbf {\bibinfo {volume} {112}},\
  \bibinfo {pages} {110401} (\bibinfo {year} {2014})}\BibitemShut {NoStop}%
\bibitem [{\citenamefont {Rosenbach}\ \emph {et~al.}(2016)\citenamefont
  {Rosenbach}, \citenamefont {Cerrillo}, \citenamefont {Huelga}, \citenamefont
  {Cao},\ and\ \citenamefont {Plenio}}]{Rosenbach2016}%
  \BibitemOpen
  \bibfield  {author} {\bibinfo {author} {\bibfnamefont {R.}~\bibnamefont
  {Rosenbach}}, \bibinfo {author} {\bibfnamefont {J.}~\bibnamefont {Cerrillo}},
  \bibinfo {author} {\bibfnamefont {S.~F.}\ \bibnamefont {Huelga}}, \bibinfo
  {author} {\bibfnamefont {J.}~\bibnamefont {Cao}}, \ and\ \bibinfo {author}
  {\bibfnamefont {M.~B.}\ \bibnamefont {Plenio}},\ }\href
  {http://stacks.iop.org/1367-2630/18/i=2/a=023035} {\bibfield  {journal}
  {\bibinfo  {journal} {New J. Phys.}\ }\textbf {\bibinfo {volume} {18}},\
  \bibinfo {pages} {023035} (\bibinfo {year} {2016})}\BibitemShut {NoStop}%
\bibitem [{\citenamefont {Pollock}\ and\ \citenamefont
  {Modi}(2018)}]{Pollock-quantum}%
  \BibitemOpen
  \bibfield  {author} {\bibinfo {author} {\bibfnamefont {F.~A.}\ \bibnamefont
  {Pollock}}\ and\ \bibinfo {author} {\bibfnamefont {K.}~\bibnamefont {Modi}},\
  }\href {\doibase 10.22331/q-2018-07-11-76} {\bibfield  {journal} {\bibinfo
  {journal} {{Quantum}}\ }\textbf {\bibinfo {volume} {2}},\ \bibinfo {pages}
  {76} (\bibinfo {year} {2018})}\BibitemShut {NoStop}%
\bibitem [{Note4()}]{Note4}%
  \BibitemOpen
  \bibinfo {note} {Due to this inequivalence of divisibility and Markovianity,
  the maps $\Gamma _{t:s}$ in Eq.~\protect \textup {\hbox {\mathsurround \z@
  \protect \normalfont (\ignorespaces \ref {eq:divisible}\unskip \@@italiccorr
  )}} cannot always be considered as matrices containing conditional
  probabilities $\protect \mathbb {P}(R_t|R_s)$ -- as these conditional
  probabilities might depend on prior measurement outcomes -- but rather as
  mapping from a probability distribution at time $s$ to a probability
  distribution at time $t$~\cite {hanggi_time_1977, hanggi_stochastic_1982}.
  This breakdown of interpretation also occurs in quantum mechanics~\cite
  {Milz2019}. In the Markovian case, $\Gamma _{t:s}$ indeed contains
  conditional probabilities}\BibitemShut {NoStop}%
\bibitem [{\citenamefont {Capela}\ \emph {et~al.}(2020)\citenamefont {Capela},
  \citenamefont {C{\'e}leri}, \citenamefont {Modi},\ and\ \citenamefont
  {Chaves}}]{capela_monogamy_2020}%
  \BibitemOpen
  \bibfield  {author} {\bibinfo {author} {\bibfnamefont {M.}~\bibnamefont
  {Capela}}, \bibinfo {author} {\bibfnamefont {L.~C.}\ \bibnamefont
  {C{\'e}leri}}, \bibinfo {author} {\bibfnamefont {K.}~\bibnamefont {Modi}}, \
  and\ \bibinfo {author} {\bibfnamefont {R.}~\bibnamefont {Chaves}},\ }\href
  {\doibase 10.1103/PhysRevResearch.2.013350} {\bibfield  {journal} {\bibinfo
  {journal} {Phys. Rev. Research}\ }\textbf {\bibinfo {volume} {2}},\ \bibinfo
  {pages} {013350} (\bibinfo {year} {2020})}\BibitemShut {NoStop}%
\bibitem [{\citenamefont {Vedral}(2002)}]{vedral2002role}%
  \BibitemOpen
  \bibfield  {author} {\bibinfo {author} {\bibfnamefont {V.}~\bibnamefont
  {Vedral}},\ }\href {http://dx.doi.org/10.1103/RevModPhys.74.197} {\bibfield
  {journal} {\bibinfo  {journal} {Rev. Mod. Phys.}\ }\textbf {\bibinfo {volume}
  {74}},\ \bibinfo {pages} {197} (\bibinfo {year} {2002})}\BibitemShut
  {NoStop}%
\bibitem [{\citenamefont {Strasberg}\ and\ \citenamefont
  {Esposito}(2019)}]{PhysRevE.99.012120}%
  \BibitemOpen
  \bibfield  {author} {\bibinfo {author} {\bibfnamefont {P.}~\bibnamefont
  {Strasberg}}\ and\ \bibinfo {author} {\bibfnamefont {M.}~\bibnamefont
  {Esposito}},\ }\href {\doibase 10.1103/PhysRevE.99.012120} {\bibfield
  {journal} {\bibinfo  {journal} {Phys. Rev. E}\ }\textbf {\bibinfo {volume}
  {99}},\ \bibinfo {pages} {012120} (\bibinfo {year} {2019})}\BibitemShut
  {NoStop}%
\bibitem [{\citenamefont {Polettini}\ and\ \citenamefont
  {Esposito}(2013)}]{PhysRevE.88.012112}%
  \BibitemOpen
  \bibfield  {author} {\bibinfo {author} {\bibfnamefont {M.}~\bibnamefont
  {Polettini}}\ and\ \bibinfo {author} {\bibfnamefont {M.}~\bibnamefont
  {Esposito}},\ }\href {\doibase 10.1103/PhysRevE.88.012112} {\bibfield
  {journal} {\bibinfo  {journal} {Phys. Rev. E}\ }\textbf {\bibinfo {volume}
  {88}},\ \bibinfo {pages} {012112} (\bibinfo {year} {2013})}\BibitemShut
  {NoStop}%
\bibitem [{\citenamefont {Rivas}(2020)}]{rivas_strong_2020}%
  \BibitemOpen
  \bibfield  {author} {\bibinfo {author} {\bibfnamefont {{\'A}.}~\bibnamefont
  {Rivas}},\ }\href {\doibase 10.1103/PhysRevLett.124.160601} {\bibfield
  {journal} {\bibinfo  {journal} {Phys. Rev. Lett.}\ }\textbf {\bibinfo
  {volume} {124}},\ \bibinfo {pages} {160601} (\bibinfo {year} {2020})},\
  \bibinfo {note} {publisher: American Physical Society}\BibitemShut {NoStop}%
\bibitem [{\citenamefont {Yeung}(2002)}]{yeung1}%
  \BibitemOpen
  \bibfield  {author} {\bibinfo {author} {\bibfnamefont {R.~W.}\ \bibnamefont
  {Yeung}},\ }\href@noop {} {\emph {\bibinfo {title} {A First Course in
  Information Theory (Information Technology: Transmission, Processing and
  Storage)}}}\ (\bibinfo  {publisher} {Springer},\ \bibinfo {year}
  {2002})\BibitemShut {NoStop}%
\bibitem [{\citenamefont {Yeung}(2008)}]{yeung2}%
  \BibitemOpen
  \bibfield  {author} {\bibinfo {author} {\bibfnamefont {R.~W.}\ \bibnamefont
  {Yeung}},\ }\href@noop {} {\emph {\bibinfo {title} {Information Theory and
  Network Coding (Information Technology: Transmission, Processing and
  Storage)}}}\ (\bibinfo  {publisher} {Springer},\ \bibinfo {year}
  {2008})\BibitemShut {NoStop}%
\bibitem [{\citenamefont {Janzing}\ \emph {et~al.}(2013)\citenamefont
  {Janzing}, \citenamefont {Balduzzi}, \citenamefont {Grosse-Wentrup},\ and\
  \citenamefont {Sch\"olkopf}}]{Janzing}%
  \BibitemOpen
  \bibfield  {author} {\bibinfo {author} {\bibfnamefont {D.}~\bibnamefont
  {Janzing}}, \bibinfo {author} {\bibfnamefont {D.}~\bibnamefont {Balduzzi}},
  \bibinfo {author} {\bibfnamefont {M.}~\bibnamefont {Grosse-Wentrup}}, \ and\
  \bibinfo {author} {\bibfnamefont {B.}~\bibnamefont {Sch\"olkopf}},\ }\href
  {\doibase 10.1214/13-AOS1145} {\bibfield  {journal} {\bibinfo  {journal}
  {Ann. Stat.}\ }\textbf {\bibinfo {volume} {41}},\ \bibinfo {pages} {2324}
  (\bibinfo {year} {2013})}\BibitemShut {NoStop}%
\bibitem [{\citenamefont {Chaves}\ \emph {et~al.}(2014)\citenamefont {Chaves},
  \citenamefont {Luft},\ and\ \citenamefont {Gross}}]{cone}%
  \BibitemOpen
  \bibfield  {author} {\bibinfo {author} {\bibfnamefont {R.}~\bibnamefont
  {Chaves}}, \bibinfo {author} {\bibfnamefont {L.}~\bibnamefont {Luft}}, \ and\
  \bibinfo {author} {\bibfnamefont {D.}~\bibnamefont {Gross}},\ }\href
  {\doibase 10.1088/1367-2630/16/4/043001} {\bibfield  {journal} {\bibinfo
  {journal} {New J. Phys.}\ }\textbf {\bibinfo {volume} {16}},\ \bibinfo
  {pages} {043001} (\bibinfo {year} {2014})}\BibitemShut {NoStop}%
\bibitem [{\citenamefont {Fawzi}\ and\ \citenamefont
  {Renner}(2015)}]{Fawzi2015}%
  \BibitemOpen
  \bibfield  {author} {\bibinfo {author} {\bibfnamefont {O.}~\bibnamefont
  {Fawzi}}\ and\ \bibinfo {author} {\bibfnamefont {R.}~\bibnamefont {Renner}},\
  }\href {\doibase 10.1007/s00220-015-2466-x} {\bibfield  {journal} {\bibinfo
  {journal} {Commun. Math. Phys.}\ }\textbf {\bibinfo {volume} {340}},\
  \bibinfo {pages} {575} (\bibinfo {year} {2015})}\BibitemShut {NoStop}%
\bibitem [{\citenamefont {Sutter}\ \emph {et~al.}(2016)\citenamefont {Sutter},
  \citenamefont {Fawzi},\ and\ \citenamefont {Renner}}]{Sutter2016}%
  \BibitemOpen
  \bibfield  {author} {\bibinfo {author} {\bibfnamefont {D.}~\bibnamefont
  {Sutter}}, \bibinfo {author} {\bibfnamefont {O.}~\bibnamefont {Fawzi}}, \
  and\ \bibinfo {author} {\bibfnamefont {R.}~\bibnamefont {Renner}},\ }\href
  {\doibase 10.1098/rspa.2015.0623} {\bibfield  {journal} {\bibinfo  {journal}
  {Proc. Royal Soc. A}\ }\textbf {\bibinfo {volume} {472}},\ \bibinfo {pages}
  {2186} (\bibinfo {year} {2016})}\BibitemShut {NoStop}%
\bibitem [{\citenamefont {Nielsen}\ and\ \citenamefont
  {Chuang}(2000)}]{NielsenBook}%
  \BibitemOpen
  \bibfield  {author} {\bibinfo {author} {\bibfnamefont {M.~A.}\ \bibnamefont
  {Nielsen}}\ and\ \bibinfo {author} {\bibfnamefont {I.~L.}\ \bibnamefont
  {Chuang}},\ }\href@noop {} {\emph {\bibinfo {title} {Quantum Computation and
  Quantum Information}}}\ (\bibinfo  {publisher} {Cambridge University Press},\
  \bibinfo {year} {2000})\BibitemShut {NoStop}%
\bibitem [{\citenamefont {Bengtsson}\ and\ \citenamefont
  {Zyczkowski}(2007)}]{bengtsson_geometry_2007}%
  \BibitemOpen
  \bibfield  {author} {\bibinfo {author} {\bibfnamefont {I.}~\bibnamefont
  {Bengtsson}}\ and\ \bibinfo {author} {\bibfnamefont {K.}~\bibnamefont
  {Zyczkowski}},\ }\href@noop {} {\emph {\bibinfo {title} {Geometry of
  {Quantum} {States}: {An} {Introduction} to {Quantum} {Entanglement}}}}\
  (\bibinfo  {publisher} {Cambridge University Press},\ \bibinfo {year}
  {2007})\BibitemShut {NoStop}%
\bibitem [{\citenamefont {Wilde}(2013)}]{wilde2013quantum}%
  \BibitemOpen
  \bibfield  {author} {\bibinfo {author} {\bibfnamefont {M.~M.}\ \bibnamefont
  {Wilde}},\ }\href@noop {} {\emph {\bibinfo {title} {Quantum information
  theory}}}\ (\bibinfo  {publisher} {Cambridge University Press},\ \bibinfo
  {year} {2013})\BibitemShut {NoStop}%
\bibitem [{\citenamefont {Pauli}(1928)}]{Pauli}%
  \BibitemOpen
  \bibfield  {author} {\bibinfo {author} {\bibfnamefont {W.}~\bibnamefont
  {Pauli}},\ }in\ \href@noop {} {\emph {\bibinfo {booktitle} {Festschrift zum
  60. Geburtstage A. Sommerfeld}}}\ (\bibinfo  {publisher} {Hirzel, Leipzig},\
  \bibinfo {year} {1928})\ p.~\bibinfo {pages} {30}\BibitemShut {NoStop}%
\bibitem [{\citenamefont {Jaynes}(1957)}]{jaynes1957}%
  \BibitemOpen
  \bibfield  {author} {\bibinfo {author} {\bibfnamefont {E.~T.}\ \bibnamefont
  {Jaynes}},\ }\href {\doibase 10.1103/PhysRev.106.620} {\bibfield  {journal}
  {\bibinfo  {journal} {Phys. Rev.}\ }\textbf {\bibinfo {volume} {106}},\
  \bibinfo {pages} {620} (\bibinfo {year} {1957})}\BibitemShut {NoStop}%
\bibitem [{\citenamefont {Nakajima}(1958)}]{Nakajima1958}%
  \BibitemOpen
  \bibfield  {author} {\bibinfo {author} {\bibfnamefont {S.}~\bibnamefont
  {Nakajima}},\ }\href {\doibase 10.1143/PTP.20.948} {\bibfield  {journal}
  {\bibinfo  {journal} {Progr. Theo. Phys.}\ }\textbf {\bibinfo {volume}
  {20}},\ \bibinfo {pages} {948} (\bibinfo {year} {1958})}\BibitemShut
  {NoStop}%
\bibitem [{\citenamefont {Zwanzig}(1960)}]{Zwanzig1960}%
  \BibitemOpen
  \bibfield  {author} {\bibinfo {author} {\bibfnamefont {R.}~\bibnamefont
  {Zwanzig}},\ }\href {\doibase 10.1063/1.1731409} {\bibfield  {journal}
  {\bibinfo  {journal} {J. Chem. Phys.}\ }\textbf {\bibinfo {volume} {33}},\
  \bibinfo {pages} {1338} (\bibinfo {year} {1960})}\BibitemShut {NoStop}%
\bibitem [{\citenamefont {Lamb}(1964)}]{lamb_theory_1964}%
  \BibitemOpen
  \bibfield  {author} {\bibinfo {author} {\bibfnamefont {W.~E.}\ \bibnamefont
  {Lamb}},\ }\href {\doibase 10.1103/PhysRev.134.A1429} {\bibfield  {journal}
  {\bibinfo  {journal} {Phys. Rev.}\ }\textbf {\bibinfo {volume} {134}},\
  \bibinfo {pages} {A1429} (\bibinfo {year} {1964})}\BibitemShut {NoStop}%
\bibitem [{\citenamefont {Weidlich}\ and\ \citenamefont
  {Haake}(1965)}]{weidlich_coherence-properties_1965}%
  \BibitemOpen
  \bibfield  {author} {\bibinfo {author} {\bibfnamefont {W.}~\bibnamefont
  {Weidlich}}\ and\ \bibinfo {author} {\bibfnamefont {F.}~\bibnamefont
  {Haake}},\ }\href {\doibase 10.1007/BF01381300} {\bibfield  {journal}
  {\bibinfo  {journal} {Z. Physik}\ }\textbf {\bibinfo {volume} {185}},\
  \bibinfo {pages} {30} (\bibinfo {year} {1965})}\BibitemShut {NoStop}%
\bibitem [{\citenamefont {Weidlich}\ \emph
  {et~al.}(1967{\natexlab{a}})\citenamefont {Weidlich}, \citenamefont
  {Risken},\ and\ \citenamefont {Haken}}]{weidlich_quantummechanical_1967I}%
  \BibitemOpen
  \bibfield  {author} {\bibinfo {author} {\bibfnamefont {W.}~\bibnamefont
  {Weidlich}}, \bibinfo {author} {\bibfnamefont {H.}~\bibnamefont {Risken}}, \
  and\ \bibinfo {author} {\bibfnamefont {H.}~\bibnamefont {Haken}},\ }\href
  {\doibase 10.1007/BF01326573} {\bibfield  {journal} {\bibinfo  {journal} {Z.
  Physik}\ }\textbf {\bibinfo {volume} {201}},\ \bibinfo {pages} {396}
  (\bibinfo {year} {1967}{\natexlab{a}})}\BibitemShut {NoStop}%
\bibitem [{\citenamefont {Weidlich}\ \emph
  {et~al.}(1967{\natexlab{b}})\citenamefont {Weidlich}, \citenamefont
  {Risken},\ and\ \citenamefont {Haken}}]{weidlich_quantummechanical_1967II}%
  \BibitemOpen
  \bibfield  {author} {\bibinfo {author} {\bibfnamefont {W.}~\bibnamefont
  {Weidlich}}, \bibinfo {author} {\bibfnamefont {H.}~\bibnamefont {Risken}}, \
  and\ \bibinfo {author} {\bibfnamefont {H.}~\bibnamefont {Haken}},\ }\href
  {\doibase 10.1007/BF01326197} {\bibfield  {journal} {\bibinfo  {journal} {Z.
  Physik}\ }\textbf {\bibinfo {volume} {204}},\ \bibinfo {pages} {223}
  (\bibinfo {year} {1967}{\natexlab{b}})}\BibitemShut {NoStop}%
\bibitem [{\citenamefont {Gorini}\ \emph {et~al.}(1976)\citenamefont {Gorini},
  \citenamefont {Kossakowski},\ and\ \citenamefont
  {Sudarshan}}]{sudarshangorini}%
  \BibitemOpen
  \bibfield  {author} {\bibinfo {author} {\bibfnamefont {V.}~\bibnamefont
  {Gorini}}, \bibinfo {author} {\bibfnamefont {A.}~\bibnamefont {Kossakowski}},
  \ and\ \bibinfo {author} {\bibfnamefont {E.~C.~G.}\ \bibnamefont
  {Sudarshan}},\ }\href {\doibase 10.1063/1.522979} {\bibfield  {journal}
  {\bibinfo  {journal} {J. Math. Phys.}\ }\textbf {\bibinfo {volume} {17}},\
  \bibinfo {pages} {821} (\bibinfo {year} {1976})}\BibitemShut {NoStop}%
\bibitem [{\citenamefont {Lindblad}(1975{\natexlab{a}})}]{lindblad75}%
  \BibitemOpen
  \bibfield  {author} {\bibinfo {author} {\bibfnamefont {G.}~\bibnamefont
  {Lindblad}},\ }\href {\doibase 10.1007/BF01608499} {\bibfield  {journal}
  {\bibinfo  {journal} {Commun. Math. Phys.}\ }\textbf {\bibinfo {volume}
  {48}},\ \bibinfo {pages} {119} (\bibinfo {year}
  {1975}{\natexlab{a}})}\BibitemShut {NoStop}%
\bibitem [{\citenamefont {Sudarshan}\ \emph {et~al.}(1961)\citenamefont
  {Sudarshan}, \citenamefont {Mathews},\ and\ \citenamefont
  {Rau}}]{SudarshanMatthewsRau61}%
  \BibitemOpen
  \bibfield  {author} {\bibinfo {author} {\bibfnamefont {E.}~\bibnamefont
  {Sudarshan}}, \bibinfo {author} {\bibfnamefont {P.}~\bibnamefont {Mathews}},
  \ and\ \bibinfo {author} {\bibfnamefont {J.}~\bibnamefont {Rau}},\ }\href
  {https://journals.aps.org/pr/abstract/10.1103/PhysRev.121.920} {\bibfield
  {journal} {\bibinfo  {journal} {Phys. Rev.}\ }\textbf {\bibinfo {volume}
  {121}},\ \bibinfo {pages} {920} (\bibinfo {year} {1961})}\BibitemShut
  {NoStop}%
\bibitem [{\citenamefont {Jordan}\ and\ \citenamefont
  {Sudarshan}(1961)}]{jordan_dynamical_1961}%
  \BibitemOpen
  \bibfield  {author} {\bibinfo {author} {\bibfnamefont {T.~F.}\ \bibnamefont
  {Jordan}}\ and\ \bibinfo {author} {\bibfnamefont {E.~C.~G.}\ \bibnamefont
  {Sudarshan}},\ }\href {\doibase 10.1063/1.1724221} {\bibfield  {journal}
  {\bibinfo  {journal} {J. Math. Phys.}\ }\textbf {\bibinfo {volume} {2}},\
  \bibinfo {pages} {772} (\bibinfo {year} {1961})}\BibitemShut {NoStop}%
\bibitem [{\citenamefont {Kraus}(1971)}]{kraus_general_1971}%
  \BibitemOpen
  \bibfield  {author} {\bibinfo {author} {\bibfnamefont {K.}~\bibnamefont
  {Kraus}},\ }\href {\doibase 10.1016/0003-4916(71)90108-4} {\bibfield
  {journal} {\bibinfo  {journal} {Ann. Phys.}\ }\textbf {\bibinfo {volume}
  {64}},\ \bibinfo {pages} {311} (\bibinfo {year} {1971})}\BibitemShut
  {NoStop}%
\bibitem [{\citenamefont
  {Schr{\"o}dinger}(1936)}]{schrodinger_probability_1936}%
  \BibitemOpen
  \bibfield  {author} {\bibinfo {author} {\bibfnamefont {E.}~\bibnamefont
  {Schr{\"o}dinger}},\ }\href {\doibase 10.1017/S0305004100019137} {\bibfield
  {journal} {\bibinfo  {journal} {Math. Proc. Cambridge Philos. Soc.}\ }\textbf
  {\bibinfo {volume} {32}},\ \bibinfo {pages} {446} (\bibinfo {year}
  {1936})}\BibitemShut {NoStop}%
\bibitem [{\citenamefont {Gisin}(1984)}]{gisin_quantum_1984}%
  \BibitemOpen
  \bibfield  {author} {\bibinfo {author} {\bibfnamefont {N.}~\bibnamefont
  {Gisin}},\ }\href {\doibase 10.1103/PhysRevLett.52.1657} {\bibfield
  {journal} {\bibinfo  {journal} {Phys. Rev. Lett.}\ }\textbf {\bibinfo
  {volume} {52}},\ \bibinfo {pages} {1657} (\bibinfo {year}
  {1984})}\BibitemShut {NoStop}%
\bibitem [{\citenamefont {Hughston}\ \emph {et~al.}(1993)\citenamefont
  {Hughston}, \citenamefont {Jozsa},\ and\ \citenamefont
  {Wootters}}]{hughston_complete_1993}%
  \BibitemOpen
  \bibfield  {author} {\bibinfo {author} {\bibfnamefont {L.~P.}\ \bibnamefont
  {Hughston}}, \bibinfo {author} {\bibfnamefont {R.}~\bibnamefont {Jozsa}}, \
  and\ \bibinfo {author} {\bibfnamefont {W.~K.}\ \bibnamefont {Wootters}},\
  }\href {\doibase 10.1016/0375-9601(93)90880-9} {\bibfield  {journal}
  {\bibinfo  {journal} {Phys. Lett. A}\ }\textbf {\bibinfo {volume} {183}},\
  \bibinfo {pages} {14} (\bibinfo {year} {1993})}\BibitemShut {NoStop}%
\bibitem [{\citenamefont {Gleason}(1957)}]{gleason_measures_1975}%
  \BibitemOpen
  \bibfield  {author} {\bibinfo {author} {\bibfnamefont {A.}~\bibnamefont
  {Gleason}},\ }\href
  {https://link.springer.com/chapter/10.1007%2F978-94-010-1795-4_7} {\bibfield
  {journal} {\bibinfo  {journal} {J. Math. Mech.}\ }\textbf {\bibinfo {volume}
  {6}},\ \bibinfo {pages} {885} (\bibinfo {year} {1957})}\BibitemShut {NoStop}%
\bibitem [{\citenamefont {Busch}(2003)}]{busch_quantum_2003}%
  \BibitemOpen
  \bibfield  {author} {\bibinfo {author} {\bibfnamefont {P.}~\bibnamefont
  {Busch}},\ }\href {\doibase 10.1103/PhysRevLett.91.120403} {\bibfield
  {journal} {\bibinfo  {journal} {Phys. Rev. Lett.}\ }\textbf {\bibinfo
  {volume} {91}},\ \bibinfo {pages} {120403} (\bibinfo {year}
  {2003})}\BibitemShut {NoStop}%
\bibitem [{\citenamefont {d'Espagnat}(1966)}]{mixtures}%
  \BibitemOpen
  \bibfield  {author} {\bibinfo {author} {\bibfnamefont {B.}~\bibnamefont
  {d'Espagnat}},\ }\enquote {\bibinfo {title} {In \emph{Preludes in Theoretical
  Physics}},}\ \ (\bibinfo  {publisher} {North Holland Wiley},\ \bibinfo {year}
  {1966})\ Chap.\ \bibinfo {chapter} {An elementary note about
  `mixtures'}\BibitemShut {NoStop}%
\bibitem [{Note5()}]{Note5}%
  \BibitemOpen
  \bibinfo {note} {We denote operator (and later superoperator) basis elements
  with hats.}\BibitemShut {Stop}%
\bibitem [{\citenamefont {Modi}\ \emph {et~al.}(2012)\citenamefont {Modi},
  \citenamefont {Rodr\'{\i}guez-Rosario},\ and\ \citenamefont
  {Aspuru-Guzik}}]{modi_positivity_2012}%
  \BibitemOpen
  \bibfield  {author} {\bibinfo {author} {\bibfnamefont {K.}~\bibnamefont
  {Modi}}, \bibinfo {author} {\bibfnamefont {C.~A.}\ \bibnamefont
  {Rodr\'{\i}guez-Rosario}}, \ and\ \bibinfo {author} {\bibfnamefont
  {A.}~\bibnamefont {Aspuru-Guzik}},\ }\href {\doibase
  10.1103/PhysRevA.86.064102} {\bibfield  {journal} {\bibinfo  {journal} {Phys.
  Rev. A}\ }\textbf {\bibinfo {volume} {86}},\ \bibinfo {pages} {064102}
  (\bibinfo {year} {2012})}\BibitemShut {NoStop}%
\bibitem [{\citenamefont {Milz}\ \emph {et~al.}(2017)\citenamefont {Milz},
  \citenamefont {Pollock},\ and\ \citenamefont {Modi}}]{arXiv:1708.00769}%
  \BibitemOpen
  \bibfield  {author} {\bibinfo {author} {\bibfnamefont {S.}~\bibnamefont
  {Milz}}, \bibinfo {author} {\bibfnamefont {F.~A.}\ \bibnamefont {Pollock}}, \
  and\ \bibinfo {author} {\bibfnamefont {K.}~\bibnamefont {Modi}},\ }\href
  {https://doi.org/10.1142/S1230161217400169} {\bibfield  {journal} {\bibinfo
  {journal} {Open Syst. Inf. Dyn.}\ }\textbf {\bibinfo {volume} {24}},\
  \bibinfo {pages} {1740016} (\bibinfo {year} {2017})}\BibitemShut {NoStop}%
\bibitem [{\citenamefont {Chuang}\ and\ \citenamefont
  {Nielsen}(1997)}]{JModOpt.44.2455}%
  \BibitemOpen
  \bibfield  {author} {\bibinfo {author} {\bibfnamefont {I.~L.}\ \bibnamefont
  {Chuang}}\ and\ \bibinfo {author} {\bibfnamefont {M.~A.}\ \bibnamefont
  {Nielsen}},\ }\href
  {http://www.tandfonline.com/doi/abs/10.1080/09500349708231894} {\bibfield
  {journal} {\bibinfo  {journal} {J. Mod. Opt.}\ }\textbf {\bibinfo {volume}
  {44}},\ \bibinfo {pages} {2455} (\bibinfo {year} {1997})}\BibitemShut
  {NoStop}%
\bibitem [{\citenamefont {Ferrie}(2014)}]{PhysRevLett.113.190404}%
  \BibitemOpen
  \bibfield  {author} {\bibinfo {author} {\bibfnamefont {C.}~\bibnamefont
  {Ferrie}},\ }\href {\doibase 10.1103/PhysRevLett.113.190404} {\bibfield
  {journal} {\bibinfo  {journal} {Phys. Rev. Lett.}\ }\textbf {\bibinfo
  {volume} {113}},\ \bibinfo {pages} {190404} (\bibinfo {year}
  {2014})}\BibitemShut {NoStop}%
\bibitem [{\citenamefont {Kalev}\ \emph {et~al.}(2015)\citenamefont {Kalev},
  \citenamefont {Kosut},\ and\ \citenamefont {Deutsch}}]{kalev}%
  \BibitemOpen
  \bibfield  {author} {\bibinfo {author} {\bibfnamefont {A.}~\bibnamefont
  {Kalev}}, \bibinfo {author} {\bibfnamefont {R.~L.}\ \bibnamefont {Kosut}}, \
  and\ \bibinfo {author} {\bibfnamefont {I.~H.}\ \bibnamefont {Deutsch}},\
  }\href {\doibase 10.1038/npjqi.2015.18} {\bibfield  {journal} {\bibinfo
  {journal} {npj Quantum Inf.}\ }\textbf {\bibinfo {volume} {1}},\ \bibinfo
  {pages} {1} (\bibinfo {year} {2015})}\BibitemShut {NoStop}%
\bibitem [{Note6()}]{Note6}%
  \BibitemOpen
  \bibinfo {note} {It is easy to construct informationally complete POVMs,
  adding symmetric part is hard. For our purposes IC will be
  sufficient.}\BibitemShut {Stop}%
\bibitem [{\citenamefont {Neumark}(1940)}]{neumark}%
  \BibitemOpen
  \bibfield  {author} {\bibinfo {author} {\bibfnamefont {M.}~\bibnamefont
  {Neumark}},\ }\href@noop {} {\bibfield  {journal} {\bibinfo  {journal} {Izv.
  Akad. Nauk SSSR Ser. Mat.}\ }\textbf {\bibinfo {volume} {4}},\ \bibinfo
  {pages} {53} (\bibinfo {year} {1940})}\BibitemShut {NoStop}%
\bibitem [{\citenamefont {Peres}(1995)}]{peresQT}%
  \BibitemOpen
  \bibfield  {author} {\bibinfo {author} {\bibfnamefont {A.}~\bibnamefont
  {Peres}},\ }\href@noop {} {\emph {\bibinfo {title} {Quantum Theory}}}\
  (\bibinfo  {publisher} {Kluwer Academic},\ \bibinfo {year}
  {1995})\BibitemShut {NoStop}%
\bibitem [{\citenamefont {Paulsen}(2003)}]{paulsen_completely_2003}%
  \BibitemOpen
  \bibfield  {author} {\bibinfo {author} {\bibfnamefont {V.}~\bibnamefont
  {Paulsen}},\ }\href@noop {} {\emph {\bibinfo {title} {Completely {Bounded}
  {Maps} and {Operator} {Algebras}}}}\ (\bibinfo  {publisher} {Cambridge
  University Press},\ \bibinfo {year} {2003})\BibitemShut {NoStop}%
\bibitem [{\citenamefont {D{'}Ariano}\ \emph {et~al.}(2000)\citenamefont
  {D{'}Ariano}, \citenamefont {Maccone},\ and\ \citenamefont
  {Paris}}]{dariano_orthogonality_2000}%
  \BibitemOpen
  \bibfield  {author} {\bibinfo {author} {\bibfnamefont {G.}~\bibnamefont
  {D{'}Ariano}}, \bibinfo {author} {\bibfnamefont {L.}~\bibnamefont {Maccone}},
  \ and\ \bibinfo {author} {\bibfnamefont {M.}~\bibnamefont {Paris}},\ }\href
  {\doibase 10.1016/S0375-9601(00)00660-5} {\bibfield  {journal} {\bibinfo
  {journal} {Phys. Lett. A}\ }\textbf {\bibinfo {volume} {276}},\ \bibinfo
  {pages} {25} (\bibinfo {year} {2000})}\BibitemShut {NoStop}%
\bibitem [{\citenamefont {Havel}(2003)}]{havel03}%
  \BibitemOpen
  \bibfield  {author} {\bibinfo {author} {\bibfnamefont {T.~F.}\ \bibnamefont
  {Havel}},\ }\href {\doibase 10.1063/1.1518555} {\bibfield  {journal}
  {\bibinfo  {journal} {J. Math. Phys.}\ }\textbf {\bibinfo {volume} {44}},\
  \bibinfo {pages} {534} (\bibinfo {year} {2003})}\BibitemShut {NoStop}%
\bibitem [{\citenamefont {Gilchrist}\ \emph {et~al.}(2009)\citenamefont
  {Gilchrist}, \citenamefont {Terno},\ and\ \citenamefont
  {Wood}}]{gilchrist_vectorization_2009}%
  \BibitemOpen
  \bibfield  {author} {\bibinfo {author} {\bibfnamefont {A.}~\bibnamefont
  {Gilchrist}}, \bibinfo {author} {\bibfnamefont {D.~R.}\ \bibnamefont
  {Terno}}, \ and\ \bibinfo {author} {\bibfnamefont {C.~J.}\ \bibnamefont
  {Wood}},\ }\href {http://arxiv.org/abs/0911.2539} {\bibfield  {journal}
  {\bibinfo  {journal} {arXiv:0911.2539}\ } (\bibinfo {year}
  {2009})}\BibitemShut {NoStop}%
\bibitem [{Note7()}]{Note7}%
  \BibitemOpen
  \bibinfo {note} {One should think of the map as an abstract object, and
  should be distinguished from its representation, hence we have $\protect
  \mathcal {E}$ versus $\protect \breve {\protect \mathcal {E}}$.}\BibitemShut
  {Stop}%
\bibitem [{\citenamefont {Eisert}\ and\ \citenamefont
  {Wolf}(2007)}]{arXiv:quant-ph/0505151}%
  \BibitemOpen
  \bibfield  {author} {\bibinfo {author} {\bibfnamefont {J.}~\bibnamefont
  {Eisert}}\ and\ \bibinfo {author} {\bibfnamefont {M.~M.}\ \bibnamefont
  {Wolf}},\ }in\ \href {\doibase 10.1142/9781860948169_0002} {\emph {\bibinfo
  {booktitle} {Quantum {Information} with {Continuous} {Variables} of {Atoms}
  and {Light}}}}\ (\bibinfo  {publisher} {Imperial College Press},\ \bibinfo
  {address} {London},\ \bibinfo {year} {2007})\ pp.\ \bibinfo {pages}
  {23--42}\BibitemShut {NoStop}%
\bibitem [{\citenamefont {Weedbrook}\ \emph {et~al.}(2012)\citenamefont
  {Weedbrook}, \citenamefont {Pirandola}, \citenamefont {Garc\'{\i}a-Patr\'on},
  \citenamefont {Cerf}, \citenamefont {Ralph}, \citenamefont {Shapiro},\ and\
  \citenamefont {Lloyd}}]{RevModPhys.84.621}%
  \BibitemOpen
  \bibfield  {author} {\bibinfo {author} {\bibfnamefont {C.}~\bibnamefont
  {Weedbrook}}, \bibinfo {author} {\bibfnamefont {S.}~\bibnamefont
  {Pirandola}}, \bibinfo {author} {\bibfnamefont {R.}~\bibnamefont
  {Garc\'{\i}a-Patr\'on}}, \bibinfo {author} {\bibfnamefont {N.~J.}\
  \bibnamefont {Cerf}}, \bibinfo {author} {\bibfnamefont {T.~C.}\ \bibnamefont
  {Ralph}}, \bibinfo {author} {\bibfnamefont {J.~H.}\ \bibnamefont {Shapiro}},
  \ and\ \bibinfo {author} {\bibfnamefont {S.}~\bibnamefont {Lloyd}},\ }\href
  {\doibase 10.1103/RevModPhys.84.621} {\bibfield  {journal} {\bibinfo
  {journal} {Rev. Mod. Phys.}\ }\textbf {\bibinfo {volume} {84}},\ \bibinfo
  {pages} {621} (\bibinfo {year} {2012})}\BibitemShut {NoStop}%
\bibitem [{\citenamefont {Poyatos}\ \emph {et~al.}(1997)\citenamefont
  {Poyatos}, \citenamefont {Cirac},\ and\ \citenamefont {Zoller}}]{poyatos}%
  \BibitemOpen
  \bibfield  {author} {\bibinfo {author} {\bibfnamefont {J.~F.}\ \bibnamefont
  {Poyatos}}, \bibinfo {author} {\bibfnamefont {J.~I.}\ \bibnamefont {Cirac}},
  \ and\ \bibinfo {author} {\bibfnamefont {P.}~\bibnamefont {Zoller}},\ }\href
  {\doibase 10.1103/PhysRevLett.78.390} {\bibfield  {journal} {\bibinfo
  {journal} {Phys. Rev. Lett.}\ }\textbf {\bibinfo {volume} {78}},\ \bibinfo
  {pages} {390} (\bibinfo {year} {1997})}\BibitemShut {NoStop}%
\bibitem [{\citenamefont {Ringbauer}(2017)}]{ringbauer_quantum_2017}%
  \BibitemOpen
  \bibfield  {author} {\bibinfo {author} {\bibfnamefont {M.}~\bibnamefont
  {Ringbauer}},\ }in\ \href
  {https://link.springer.com/chapter/10.1007/978-3-319-64988-7_2} {\emph
  {\bibinfo {booktitle} {Exploring {Quantum} {Foundations} with {Single}
  {Photons}}}},\ \bibinfo {series and number} {Springer {Theses}}\ (\bibinfo
  {publisher} {Springer, Cham},\ \bibinfo {year} {2017})\BibitemShut {NoStop}%
\bibitem [{\citenamefont {Wood}(2015)}]{wood_thesis}%
  \BibitemOpen
  \bibfield  {author} {\bibinfo {author} {\bibfnamefont {C.~J.}\ \bibnamefont
  {Wood}},\ }\emph {\bibinfo {title} {Initialization and characterization of
  open quantum systems}},\ \href {http://hdl.handle.net/10012/9557} {Ph.D.
  thesis},\ \bibinfo  {school} {UWSpace} (\bibinfo {year} {2015})\BibitemShut
  {NoStop}%
\bibitem [{\citenamefont {G{\"u}hne}\ and\ \citenamefont
  {T{\'o}th}(2009)}]{guhne_entanglement_2009}%
  \BibitemOpen
  \bibfield  {author} {\bibinfo {author} {\bibfnamefont {O.}~\bibnamefont
  {G{\"u}hne}}\ and\ \bibinfo {author} {\bibfnamefont {G.}~\bibnamefont
  {T{\'o}th}},\ }\href {\doibase 10.1016/j.physrep.2009.02.004} {\bibfield
  {journal} {\bibinfo  {journal} {Phys. Rep.}\ }\textbf {\bibinfo {volume}
  {474}},\ \bibinfo {pages} {1} (\bibinfo {year} {2009})}\BibitemShut {NoStop}%
\bibitem [{\citenamefont {Friis}\ \emph {et~al.}(2019)\citenamefont {Friis},
  \citenamefont {Vitagliano}, \citenamefont {Malik},\ and\ \citenamefont
  {Huber}}]{friis_entanglement_2019}%
  \BibitemOpen
  \bibfield  {author} {\bibinfo {author} {\bibfnamefont {N.}~\bibnamefont
  {Friis}}, \bibinfo {author} {\bibfnamefont {G.}~\bibnamefont {Vitagliano}},
  \bibinfo {author} {\bibfnamefont {M.}~\bibnamefont {Malik}}, \ and\ \bibinfo
  {author} {\bibfnamefont {M.}~\bibnamefont {Huber}},\ }\href {\doibase
  10.1038/s42254-018-0003-5} {\bibfield  {journal} {\bibinfo  {journal} {Nat.
  Rev. Phys.}\ }\textbf {\bibinfo {volume} {1}},\ \bibinfo {pages} {72}
  (\bibinfo {year} {2019})}\BibitemShut {NoStop}%
\bibitem [{\citenamefont {Pechukas}(1994)}]{pechukas}%
  \BibitemOpen
  \bibfield  {author} {\bibinfo {author} {\bibfnamefont {P.}~\bibnamefont
  {Pechukas}},\ }\href {\doibase 10.1103/PhysRevLett.73.1060} {\bibfield
  {journal} {\bibinfo  {journal} {Phys. Rev. Lett.}\ }\textbf {\bibinfo
  {volume} {73}},\ \bibinfo {pages} {1060} (\bibinfo {year}
  {1994})}\BibitemShut {NoStop}%
\bibitem [{\citenamefont {Alicki}(1995)}]{Alicki95}%
  \BibitemOpen
  \bibfield  {author} {\bibinfo {author} {\bibfnamefont {R.}~\bibnamefont
  {Alicki}},\ }\href {\doibase 10.1103/PhysRevLett.75.3020} {\bibfield
  {journal} {\bibinfo  {journal} {Phys. Rev. Lett.}\ }\textbf {\bibinfo
  {volume} {75}},\ \bibinfo {pages} {3020} (\bibinfo {year}
  {1995})}\BibitemShut {NoStop}%
\bibitem [{\citenamefont {Pechukas}(1995)}]{pechukas2}%
  \BibitemOpen
  \bibfield  {author} {\bibinfo {author} {\bibfnamefont {P.}~\bibnamefont
  {Pechukas}},\ }\href {\doibase 10.1103/PhysRevLett.75.3021} {\bibfield
  {journal} {\bibinfo  {journal} {Phys. Rev. Lett.}\ }\textbf {\bibinfo
  {volume} {75}},\ \bibinfo {pages} {3021} (\bibinfo {year}
  {1995})}\BibitemShut {NoStop}%
\bibitem [{\citenamefont {Jordan}\ \emph {et~al.}(2004)\citenamefont {Jordan},
  \citenamefont {Shaji},\ and\ \citenamefont {Sudarshan}}]{jordan:052110}%
  \BibitemOpen
  \bibfield  {author} {\bibinfo {author} {\bibfnamefont {T.~F.}\ \bibnamefont
  {Jordan}}, \bibinfo {author} {\bibfnamefont {A.}~\bibnamefont {Shaji}}, \
  and\ \bibinfo {author} {\bibfnamefont {E.~C.~G.}\ \bibnamefont {Sudarshan}},\
  }\href
  {https://journals.aps.org/pra/abstract/10.1103/PhysRevA.70.052110#fulltext}
  {\bibfield  {journal} {\bibinfo  {journal} {Phys. Rev. A}\ }\textbf {\bibinfo
  {volume} {70}},\ \bibinfo {eid} {052110} (\bibinfo {year}
  {2004})}\BibitemShut {NoStop}%
\bibitem [{\citenamefont {\v{S}telmachovi\v{c}}\ and\ \citenamefont
  {Bu\v{z}ek}(2001)}]{StelmachovicBuzek01}%
  \BibitemOpen
  \bibfield  {author} {\bibinfo {author} {\bibfnamefont {P.}~\bibnamefont
  {\v{S}telmachovi\v{c}}}\ and\ \bibinfo {author} {\bibfnamefont
  {V.}~\bibnamefont {Bu\v{z}ek}},\ }\href
  {https://journals.aps.org/pra/abstract/10.1103/PhysRevA.64.062106#fulltext}
  {\bibfield  {journal} {\bibinfo  {journal} {Phys. Rev. A}\ }\textbf {\bibinfo
  {volume} {64}},\ \bibinfo {pages} {062106} (\bibinfo {year}
  {2001})}\BibitemShut {NoStop}%
\bibitem [{\citenamefont {{\.Z}yczkowski}\ and\ \citenamefont
  {Bengtsson}(2004)}]{zyczkowski_duality_2004}%
  \BibitemOpen
  \bibfield  {author} {\bibinfo {author} {\bibfnamefont {K.}~\bibnamefont
  {{\.Z}yczkowski}}\ and\ \bibinfo {author} {\bibfnamefont {I.}~\bibnamefont
  {Bengtsson}},\ }\href {\doibase 10.1023/B:OPSY.0000024753.05661.c2}
  {\bibfield  {journal} {\bibinfo  {journal} {Open Syst. Inf. Dyn.}\ }\textbf
  {\bibinfo {volume} {11}},\ \bibinfo {pages} {3} (\bibinfo {year}
  {2004})}\BibitemShut {NoStop}%
\bibitem [{\citenamefont {Kraus}(1983)}]{kraus_states_1983}%
  \BibitemOpen
  \bibfield  {author} {\bibinfo {author} {\bibfnamefont {K.}~\bibnamefont
  {Kraus}},\ }\href@noop {} {\emph {\bibinfo {title} {States, {Effects}, and
  {Operations}: {Fundamental} {Notions} of {Quantum} {Theory}}}}\ (\bibinfo
  {publisher} {Springer},\ \bibinfo {address} {Berlin, Heidelberg},\ \bibinfo
  {year} {1983})\BibitemShut {NoStop}%
\bibitem [{\citenamefont {Mendl}\ and\ \citenamefont
  {Wolf}(2009)}]{wolfunital}%
  \BibitemOpen
  \bibfield  {author} {\bibinfo {author} {\bibfnamefont {C.~B.}\ \bibnamefont
  {Mendl}}\ and\ \bibinfo {author} {\bibfnamefont {M.~M.}\ \bibnamefont
  {Wolf}},\ }\href {\doibase 10.1007/s00220-009-0824-2} {\bibfield  {journal}
  {\bibinfo  {journal} {Commun. Math. Phys.}\ }\textbf {\bibinfo {volume}
  {289}},\ \bibinfo {pages} {1057} (\bibinfo {year} {2009})}\BibitemShut
  {NoStop}%
\bibitem [{\citenamefont {D{'}Ariano}\ and\ \citenamefont
  {Lo~Presti}(2003)}]{dariano_imprinting_2003}%
  \BibitemOpen
  \bibfield  {author} {\bibinfo {author} {\bibfnamefont {G.~M.}\ \bibnamefont
  {D{'}Ariano}}\ and\ \bibinfo {author} {\bibfnamefont {P.}~\bibnamefont
  {Lo~Presti}},\ }\href {\doibase 10.1103/PhysRevLett.91.047902} {\bibfield
  {journal} {\bibinfo  {journal} {Phys. Rev. Lett.}\ }\textbf {\bibinfo
  {volume} {91}},\ \bibinfo {pages} {047902} (\bibinfo {year}
  {2003})}\BibitemShut {NoStop}%
\bibitem [{\citenamefont {Pillis}(1967)}]{pillis_linear_1967}%
  \BibitemOpen
  \bibfield  {author} {\bibinfo {author} {\bibfnamefont {J.~d.}\ \bibnamefont
  {Pillis}},\ }\href {http://projecteuclid.org/euclid.pjm/1102991990}
  {\bibfield  {journal} {\bibinfo  {journal} {Pacific J. Math.}\ }\textbf
  {\bibinfo {volume} {23}},\ \bibinfo {pages} {129} (\bibinfo {year}
  {1967})}\BibitemShut {NoStop}%
\bibitem [{\citenamefont {Jamio{\l}kowski}(1972)}]{jamiolkowski_linear_1972}%
  \BibitemOpen
  \bibfield  {author} {\bibinfo {author} {\bibfnamefont {A.}~\bibnamefont
  {Jamio{\l}kowski}},\ }\href {\doibase 10.1016/0034-4877(72)90011-0}
  {\bibfield  {journal} {\bibinfo  {journal} {Rep. Math. Phys.}\ }\textbf
  {\bibinfo {volume} {3}},\ \bibinfo {pages} {275} (\bibinfo {year}
  {1972})}\BibitemShut {NoStop}%
\bibitem [{\citenamefont {Choi}(1975)}]{choi75}%
  \BibitemOpen
  \bibfield  {author} {\bibinfo {author} {\bibfnamefont {M.~D.}\ \bibnamefont
  {Choi}},\ }\href
  {http://www.sciencedirect.com/science/article/pii/0024379575900750}
  {\bibfield  {journal} {\bibinfo  {journal} {Linear Algebra Appl.}\ }\textbf
  {\bibinfo {volume} {10}},\ \bibinfo {pages} {285} (\bibinfo {year}
  {1975})}\BibitemShut {NoStop}%
\bibitem [{\citenamefont {Verstraete}\ and\ \citenamefont
  {Verschelde}(2002)}]{verstraete_quantum_2002}%
  \BibitemOpen
  \bibfield  {author} {\bibinfo {author} {\bibfnamefont {F.}~\bibnamefont
  {Verstraete}}\ and\ \bibinfo {author} {\bibfnamefont {H.}~\bibnamefont
  {Verschelde}},\ }\href {http://arxiv.org/abs/quant-ph/0202124} {\bibfield
  {journal} {\bibinfo  {journal} {arXiv:0202124}\ } (\bibinfo {year}
  {2002})}\BibitemShut {NoStop}%
\bibitem [{\citenamefont {Wood}\ \emph {et~al.}(2015)\citenamefont {Wood},
  \citenamefont {Biamonte},\ and\ \citenamefont {Cory}}]{wood_tensor_2011}%
  \BibitemOpen
  \bibfield  {author} {\bibinfo {author} {\bibfnamefont {C.~J.}\ \bibnamefont
  {Wood}}, \bibinfo {author} {\bibfnamefont {J.~D.}\ \bibnamefont {Biamonte}},
  \ and\ \bibinfo {author} {\bibfnamefont {D.~G.}\ \bibnamefont {Cory}},\
  }\href {http://www.rintonpress.com/xxqic15/qic-15-910/0759-0811.pdf}
  {\bibfield  {journal} {\bibinfo  {journal} {Quant. Inf. Comp.}\ }\textbf
  {\bibinfo {volume} {15}},\ \bibinfo {pages} {759} (\bibinfo {year}
  {2015})}\BibitemShut {NoStop}%
\bibitem [{\citenamefont {Altepeter}\ \emph {et~al.}(2003)\citenamefont
  {Altepeter}, \citenamefont {Branning}, \citenamefont {Jeffrey}, \citenamefont
  {Wei}, \citenamefont {Kwiat}, \citenamefont {Thew}, \citenamefont {O'Brien},
  \citenamefont {Nielsen},\ and\ \citenamefont
  {White}}]{PhysRevLett.90.193601}%
  \BibitemOpen
  \bibfield  {author} {\bibinfo {author} {\bibfnamefont {J.~B.}\ \bibnamefont
  {Altepeter}}, \bibinfo {author} {\bibfnamefont {D.}~\bibnamefont {Branning}},
  \bibinfo {author} {\bibfnamefont {E.}~\bibnamefont {Jeffrey}}, \bibinfo
  {author} {\bibfnamefont {T.~C.}\ \bibnamefont {Wei}}, \bibinfo {author}
  {\bibfnamefont {P.~G.}\ \bibnamefont {Kwiat}}, \bibinfo {author}
  {\bibfnamefont {R.~T.}\ \bibnamefont {Thew}}, \bibinfo {author}
  {\bibfnamefont {J.~L.}\ \bibnamefont {O'Brien}}, \bibinfo {author}
  {\bibfnamefont {M.~A.}\ \bibnamefont {Nielsen}}, \ and\ \bibinfo {author}
  {\bibfnamefont {A.~G.}\ \bibnamefont {White}},\ }\href {\doibase
  10.1103/PhysRevLett.90.193601} {\bibfield  {journal} {\bibinfo  {journal}
  {Phys. Rev. Lett.}\ }\textbf {\bibinfo {volume} {90}},\ \bibinfo {pages}
  {193601} (\bibinfo {year} {2003})}\BibitemShut {NoStop}%
\bibitem [{\citenamefont {D{'}Ariano}\ and\ \citenamefont
  {Lo~Presti}(2001)}]{PhysRevLett.86.4195}%
  \BibitemOpen
  \bibfield  {author} {\bibinfo {author} {\bibfnamefont {G.~M.}\ \bibnamefont
  {D{'}Ariano}}\ and\ \bibinfo {author} {\bibfnamefont {P.}~\bibnamefont
  {Lo~Presti}},\ }\href {\doibase 10.1103/PhysRevLett.86.4195} {\bibfield
  {journal} {\bibinfo  {journal} {Phys. Rev. Lett.}\ }\textbf {\bibinfo
  {volume} {86}},\ \bibinfo {pages} {4195} (\bibinfo {year}
  {2001})}\BibitemShut {NoStop}%
\bibitem [{Note8()}]{Note8}%
  \BibitemOpen
  \bibinfo {note} {Note that there always exists a minimal number of Kraus
  operators, and, as such, a minimal environment dimension that allows one to
  dilate the map $\protect \mathcal {E}$. \protect \textit {Any} map $\protect
  \mathcal {E}$ can be dilated in a space of dimension $d_E\leq
  d_S^2$.}\BibitemShut {Stop}%
\bibitem [{\citenamefont {Buscemi}\ \emph {et~al.}(2003)\citenamefont
  {Buscemi}, \citenamefont {D'Ariano},\ and\ \citenamefont
  {Sacchi}}]{buscemi_physical_2003}%
  \BibitemOpen
  \bibfield  {author} {\bibinfo {author} {\bibfnamefont {F.}~\bibnamefont
  {Buscemi}}, \bibinfo {author} {\bibfnamefont {G.~M.}\ \bibnamefont
  {D'Ariano}}, \ and\ \bibinfo {author} {\bibfnamefont {M.~F.}\ \bibnamefont
  {Sacchi}},\ }\href {\doibase 10.1103/PhysRevA.68.042113} {\bibfield
  {journal} {\bibinfo  {journal} {Phys. Rev. A}\ }\textbf {\bibinfo {volume}
  {68}},\ \bibinfo {pages} {042113} (\bibinfo {year} {2003})}\BibitemShut
  {NoStop}%
\bibitem [{\citenamefont {Stinespring}(1955)}]{stinespring1955}%
  \BibitemOpen
  \bibfield  {author} {\bibinfo {author} {\bibfnamefont {W.~F.}\ \bibnamefont
  {Stinespring}},\ }\href {\doibase 10.1090/S0002-9939-1955-0069403-4}
  {\bibfield  {journal} {\bibinfo  {journal} {Proc. Amer. Math. Soc.}\ }\textbf
  {\bibinfo {volume} {6}},\ \bibinfo {pages} {211 } (\bibinfo {year}
  {1955})}\BibitemShut {NoStop}%
\bibitem [{\citenamefont {Hardy}(2001)}]{hardy_quantum_2001}%
  \BibitemOpen
  \bibfield  {author} {\bibinfo {author} {\bibfnamefont {L.}~\bibnamefont
  {Hardy}},\ }\href {http://arxiv.org/abs/quant-ph/0101012} {\bibfield
  {journal} {\bibinfo  {journal} {arXiv:0101012}\ } (\bibinfo {year}
  {2001})}\BibitemShut {NoStop}%
\bibitem [{\citenamefont {D'Ariano}\ \emph {et~al.}(2017)\citenamefont
  {D'Ariano}, \citenamefont {Chiribella},\ and\ \citenamefont
  {Perinotti}}]{dariano_quantum_2017}%
  \BibitemOpen
  \bibfield  {author} {\bibinfo {author} {\bibfnamefont {G.~M.}\ \bibnamefont
  {D'Ariano}}, \bibinfo {author} {\bibfnamefont {G.}~\bibnamefont
  {Chiribella}}, \ and\ \bibinfo {author} {\bibfnamefont {P.}~\bibnamefont
  {Perinotti}},\ }\href@noop {} {\emph {\bibinfo {title} {Quantum {Theory} from
  {First} {Principles}: {An} {Informational} {Approach}}}},\ \bibinfo {edition}
  {1st}\ ed.\ (\bibinfo  {publisher} {Cambridge University Press},\ \bibinfo
  {address} {Cambridge, United Kingdom ; New York, NY},\ \bibinfo {year}
  {2017})\BibitemShut {NoStop}%
\bibitem [{Note9()}]{Note9}%
  \BibitemOpen
  \bibinfo {note} {This is a pedagogical statement, not a historical
  one.}\BibitemShut {Stop}%
\bibitem [{\citenamefont {Landau}(1927)}]{landau}%
  \BibitemOpen
  \bibfield  {author} {\bibinfo {author} {\bibfnamefont {L.~D.}\ \bibnamefont
  {Landau}},\ }\href@noop {} {\bibfield  {journal} {\bibinfo  {journal} {Z.
  Physik}\ }\textbf {\bibinfo {volume} {45}},\ \bibinfo {pages} {430} (\bibinfo
  {year} {1927})}\BibitemShut {NoStop}%
\bibitem [{\citenamefont {Di{\'o}si}\ and\ \citenamefont
  {Strunz}(1997)}]{diosi_non-markovian_1997}%
  \BibitemOpen
  \bibfield  {author} {\bibinfo {author} {\bibfnamefont {L.}~\bibnamefont
  {Di{\'o}si}}\ and\ \bibinfo {author} {\bibfnamefont {W.~T.}\ \bibnamefont
  {Strunz}},\ }\href {\doibase 10.1016/S0375-9601(97)00717-2} {\bibfield
  {journal} {\bibinfo  {journal} {Phys. Lett. A}\ }\textbf {\bibinfo {volume}
  {235}},\ \bibinfo {pages} {569} (\bibinfo {year} {1997})}\BibitemShut
  {NoStop}%
\bibitem [{\citenamefont {Di{\'o}si}\ \emph {et~al.}(1998)\citenamefont
  {Di{\'o}si}, \citenamefont {Gisin},\ and\ \citenamefont
  {Strunz}}]{diosi_non-markovian_1998}%
  \BibitemOpen
  \bibfield  {author} {\bibinfo {author} {\bibfnamefont {L.}~\bibnamefont
  {Di{\'o}si}}, \bibinfo {author} {\bibfnamefont {N.}~\bibnamefont {Gisin}}, \
  and\ \bibinfo {author} {\bibfnamefont {W.~T.}\ \bibnamefont {Strunz}},\
  }\href {\doibase 10.1103/PhysRevA.58.1699} {\bibfield  {journal} {\bibinfo
  {journal} {Phys. Rev. A}\ }\textbf {\bibinfo {volume} {58}},\ \bibinfo
  {pages} {1699} (\bibinfo {year} {1998})}\BibitemShut {NoStop}%
\bibitem [{\citenamefont {Strunz}\ \emph {et~al.}(1999)\citenamefont {Strunz},
  \citenamefont {Di{\'o}si},\ and\ \citenamefont {Gisin}}]{strunz_open_1999}%
  \BibitemOpen
  \bibfield  {author} {\bibinfo {author} {\bibfnamefont {W.~T.}\ \bibnamefont
  {Strunz}}, \bibinfo {author} {\bibfnamefont {L.}~\bibnamefont {Di{\'o}si}}, \
  and\ \bibinfo {author} {\bibfnamefont {N.}~\bibnamefont {Gisin}},\ }\href
  {\doibase 10.1103/PhysRevLett.82.1801} {\bibfield  {journal} {\bibinfo
  {journal} {Phys. Rev. Lett.}\ }\textbf {\bibinfo {volume} {82}},\ \bibinfo
  {pages} {1801} (\bibinfo {year} {1999})}\BibitemShut {NoStop}%
\bibitem [{\citenamefont {Yu}\ \emph {et~al.}(2000)\citenamefont {Yu},
  \citenamefont {Di{\'o}si}, \citenamefont {Gisin},\ and\ \citenamefont
  {Strunz}}]{yu_post-markov_2000}%
  \BibitemOpen
  \bibfield  {author} {\bibinfo {author} {\bibfnamefont {T.}~\bibnamefont
  {Yu}}, \bibinfo {author} {\bibfnamefont {L.}~\bibnamefont {Di{\'o}si}},
  \bibinfo {author} {\bibfnamefont {N.}~\bibnamefont {Gisin}}, \ and\ \bibinfo
  {author} {\bibfnamefont {W.~T.}\ \bibnamefont {Strunz}},\ }\href {\doibase
  10.1016/S0375-9601(00)00014-1} {\bibfield  {journal} {\bibinfo  {journal}
  {Phys. Lett. A}\ }\textbf {\bibinfo {volume} {265}},\ \bibinfo {pages} {331}
  (\bibinfo {year} {2000})}\BibitemShut {NoStop}%
\bibitem [{\citenamefont {Wiseman}\ and\ \citenamefont
  {Di{\'o}si}(2001)}]{wiseman_complete_2001}%
  \BibitemOpen
  \bibfield  {author} {\bibinfo {author} {\bibfnamefont {H.~M.}\ \bibnamefont
  {Wiseman}}\ and\ \bibinfo {author} {\bibfnamefont {L.}~\bibnamefont
  {Di{\'o}si}},\ }\href {\doibase 10.1016/S0301-0104(01)00296-8} {\bibfield
  {journal} {\bibinfo  {journal} {Chem. Phys.}\ }\textbf {\bibinfo {volume}
  {268}},\ \bibinfo {pages} {91} (\bibinfo {year} {2001})}\BibitemShut
  {NoStop}%
\bibitem [{\citenamefont {Breuer}\ and\ \citenamefont
  {Piilo}(2009)}]{breuer_stochastic_2009}%
  \BibitemOpen
  \bibfield  {author} {\bibinfo {author} {\bibfnamefont {H.-P.}\ \bibnamefont
  {Breuer}}\ and\ \bibinfo {author} {\bibfnamefont {J.}~\bibnamefont {Piilo}},\
  }\href {\doibase 10.1209/0295-5075/85/50004} {\bibfield  {journal} {\bibinfo
  {journal} {EPL}\ }\textbf {\bibinfo {volume} {85}},\ \bibinfo {pages} {50004}
  (\bibinfo {year} {2009})}\BibitemShut {NoStop}%
\bibitem [{\citenamefont {Breuer}\ and\ \citenamefont
  {Vacchini}(2009)}]{breuer_structure_2009}%
  \BibitemOpen
  \bibfield  {author} {\bibinfo {author} {\bibfnamefont {H.-P.}\ \bibnamefont
  {Breuer}}\ and\ \bibinfo {author} {\bibfnamefont {B.}~\bibnamefont
  {Vacchini}},\ }\href {\doibase 10.1103/PhysRevE.79.041147} {\bibfield
  {journal} {\bibinfo  {journal} {Phys. Rev. E}\ }\textbf {\bibinfo {volume}
  {79}},\ \bibinfo {pages} {041147} (\bibinfo {year} {2009})}\BibitemShut
  {NoStop}%
\bibitem [{\citenamefont {Breuer}\ \emph {et~al.}(2004)\citenamefont {Breuer},
  \citenamefont {Burgarth},\ and\ \citenamefont
  {Petruccione}}]{PhysRevB.70.045323}%
  \BibitemOpen
  \bibfield  {author} {\bibinfo {author} {\bibfnamefont {H.-P.}\ \bibnamefont
  {Breuer}}, \bibinfo {author} {\bibfnamefont {D.}~\bibnamefont {Burgarth}}, \
  and\ \bibinfo {author} {\bibfnamefont {F.}~\bibnamefont {Petruccione}},\
  }\href {\doibase 10.1103/PhysRevB.70.045323} {\bibfield  {journal} {\bibinfo
  {journal} {Phys. Rev. B}\ }\textbf {\bibinfo {volume} {70}},\ \bibinfo
  {pages} {045323} (\bibinfo {year} {2004})}\BibitemShut {NoStop}%
\bibitem [{\citenamefont {Gambetta}\ and\ \citenamefont
  {Wiseman}(2003)}]{PhysRevA.68.062104}%
  \BibitemOpen
  \bibfield  {author} {\bibinfo {author} {\bibfnamefont {J.}~\bibnamefont
  {Gambetta}}\ and\ \bibinfo {author} {\bibfnamefont {H.~M.}\ \bibnamefont
  {Wiseman}},\ }\href {\doibase 10.1103/PhysRevA.68.062104} {\bibfield
  {journal} {\bibinfo  {journal} {Phys. Rev. A}\ }\textbf {\bibinfo {volume}
  {68}},\ \bibinfo {pages} {062104} (\bibinfo {year} {2003})}\BibitemShut
  {NoStop}%
\bibitem [{\citenamefont {Wiseman}\ and\ \citenamefont
  {Gambetta}(2008)}]{PhysRevLett.101.140401}%
  \BibitemOpen
  \bibfield  {author} {\bibinfo {author} {\bibfnamefont {H.~M.}\ \bibnamefont
  {Wiseman}}\ and\ \bibinfo {author} {\bibfnamefont {J.~M.}\ \bibnamefont
  {Gambetta}},\ }\href {\doibase 10.1103/PhysRevLett.101.140401} {\bibfield
  {journal} {\bibinfo  {journal} {Phys. Rev. Lett.}\ }\textbf {\bibinfo
  {volume} {101}},\ \bibinfo {pages} {140401} (\bibinfo {year}
  {2008})}\BibitemShut {NoStop}%
\bibitem [{\citenamefont {Gambetta}\ and\ \citenamefont
  {Wiseman}(2002)}]{PhysRevA.66.012108}%
  \BibitemOpen
  \bibfield  {author} {\bibinfo {author} {\bibfnamefont {J.}~\bibnamefont
  {Gambetta}}\ and\ \bibinfo {author} {\bibfnamefont {H.~M.}\ \bibnamefont
  {Wiseman}},\ }\href {\doibase 10.1103/PhysRevA.66.012108} {\bibfield
  {journal} {\bibinfo  {journal} {Phys. Rev. A}\ }\textbf {\bibinfo {volume}
  {66}},\ \bibinfo {pages} {012108} (\bibinfo {year} {2002})}\BibitemShut
  {NoStop}%
\bibitem [{\citenamefont {Wichterich}\ \emph {et~al.}(2007)\citenamefont
  {Wichterich}, \citenamefont {Henrich}, \citenamefont {Breuer}, \citenamefont
  {Gemmer},\ and\ \citenamefont {Michel}}]{PhysRevE.76.031115}%
  \BibitemOpen
  \bibfield  {author} {\bibinfo {author} {\bibfnamefont {H.}~\bibnamefont
  {Wichterich}}, \bibinfo {author} {\bibfnamefont {M.~J.}\ \bibnamefont
  {Henrich}}, \bibinfo {author} {\bibfnamefont {H.-P.}\ \bibnamefont {Breuer}},
  \bibinfo {author} {\bibfnamefont {J.}~\bibnamefont {Gemmer}}, \ and\ \bibinfo
  {author} {\bibfnamefont {M.}~\bibnamefont {Michel}},\ }\href {\doibase
  10.1103/PhysRevE.76.031115} {\bibfield  {journal} {\bibinfo  {journal} {Phys.
  Rev. E}\ }\textbf {\bibinfo {volume} {76}},\ \bibinfo {pages} {031115}
  (\bibinfo {year} {2007})}\BibitemShut {NoStop}%
\bibitem [{\citenamefont {Cerrillo}\ \emph {et~al.}(2016)\citenamefont
  {Cerrillo}, \citenamefont {Buser},\ and\ \citenamefont
  {Brandes}}]{PhysRevB.94.214308}%
  \BibitemOpen
  \bibfield  {author} {\bibinfo {author} {\bibfnamefont {J.}~\bibnamefont
  {Cerrillo}}, \bibinfo {author} {\bibfnamefont {M.}~\bibnamefont {Buser}}, \
  and\ \bibinfo {author} {\bibfnamefont {T.}~\bibnamefont {Brandes}},\ }\href
  {\doibase 10.1103/PhysRevB.94.214308} {\bibfield  {journal} {\bibinfo
  {journal} {Phys. Rev. B}\ }\textbf {\bibinfo {volume} {94}},\ \bibinfo
  {pages} {214308} (\bibinfo {year} {2016})}\BibitemShut {NoStop}%
\bibitem [{\citenamefont {Dann}\ \emph {et~al.}(2018)\citenamefont {Dann},
  \citenamefont {Levy},\ and\ \citenamefont {Kosloff}}]{PhysRevA.98.052129}%
  \BibitemOpen
  \bibfield  {author} {\bibinfo {author} {\bibfnamefont {R.}~\bibnamefont
  {Dann}}, \bibinfo {author} {\bibfnamefont {A.}~\bibnamefont {Levy}}, \ and\
  \bibinfo {author} {\bibfnamefont {R.}~\bibnamefont {Kosloff}},\ }\href
  {\doibase 10.1103/PhysRevA.98.052129} {\bibfield  {journal} {\bibinfo
  {journal} {Phys. Rev. A}\ }\textbf {\bibinfo {volume} {98}},\ \bibinfo
  {pages} {052129} (\bibinfo {year} {2018})}\BibitemShut {NoStop}%
\bibitem [{\citenamefont {Chru\ifmmode \acute{s}\else
  \'{s}\fi{}ci\ifmmode~\acute{n}\else \'{n}\fi{}ski}\ and\ \citenamefont
  {Kossakowski}(2010)}]{PhysRevLett.104.070406}%
  \BibitemOpen
  \bibfield  {author} {\bibinfo {author} {\bibfnamefont {D.}~\bibnamefont
  {Chru\ifmmode \acute{s}\else \'{s}\fi{}ci\ifmmode~\acute{n}\else
  \'{n}\fi{}ski}}\ and\ \bibinfo {author} {\bibfnamefont {A.}~\bibnamefont
  {Kossakowski}},\ }\href {\doibase 10.1103/PhysRevLett.104.070406} {\bibfield
  {journal} {\bibinfo  {journal} {Phys. Rev. Lett.}\ }\textbf {\bibinfo
  {volume} {104}},\ \bibinfo {pages} {070406} (\bibinfo {year}
  {2010})}\BibitemShut {NoStop}%
\bibitem [{\citenamefont {Filippov}\ and\ \citenamefont {Chru\ifmmode
  \acute{s}\else \'{s}\fi{}ci\ifmmode~\acute{n}\else
  \'{n}\fi{}ski}(2018)}]{PhysRevA.98.022123}%
  \BibitemOpen
  \bibfield  {author} {\bibinfo {author} {\bibfnamefont {S.~N.}\ \bibnamefont
  {Filippov}}\ and\ \bibinfo {author} {\bibfnamefont {D.}~\bibnamefont
  {Chru\ifmmode \acute{s}\else \'{s}\fi{}ci\ifmmode~\acute{n}\else
  \'{n}\fi{}ski}},\ }\href {\doibase 10.1103/PhysRevA.98.022123} {\bibfield
  {journal} {\bibinfo  {journal} {Phys. Rev. A}\ }\textbf {\bibinfo {volume}
  {98}},\ \bibinfo {pages} {022123} (\bibinfo {year} {2018})}\BibitemShut
  {NoStop}%
\bibitem [{\citenamefont {Cohen}\ and\ \citenamefont
  {Rabani}(2011)}]{PhysRevB.84.075150}%
  \BibitemOpen
  \bibfield  {author} {\bibinfo {author} {\bibfnamefont {G.}~\bibnamefont
  {Cohen}}\ and\ \bibinfo {author} {\bibfnamefont {E.}~\bibnamefont {Rabani}},\
  }\href {\doibase 10.1103/PhysRevB.84.075150} {\bibfield  {journal} {\bibinfo
  {journal} {Phys. Rev. B}\ }\textbf {\bibinfo {volume} {84}},\ \bibinfo
  {pages} {075150} (\bibinfo {year} {2011})}\BibitemShut {NoStop}%
\bibitem [{\citenamefont {Sutherland}\ \emph {et~al.}(2018)\citenamefont
  {Sutherland}, \citenamefont {Brun},\ and\ \citenamefont
  {Lidar}}]{PhysRevA.98.042119}%
  \BibitemOpen
  \bibfield  {author} {\bibinfo {author} {\bibfnamefont {C.}~\bibnamefont
  {Sutherland}}, \bibinfo {author} {\bibfnamefont {T.~A.}\ \bibnamefont
  {Brun}}, \ and\ \bibinfo {author} {\bibfnamefont {D.~A.}\ \bibnamefont
  {Lidar}},\ }\href {\doibase 10.1103/PhysRevA.98.042119} {\bibfield  {journal}
  {\bibinfo  {journal} {Phys. Rev. A}\ }\textbf {\bibinfo {volume} {98}},\
  \bibinfo {pages} {042119} (\bibinfo {year} {2018})}\BibitemShut {NoStop}%
\bibitem [{\citenamefont {Shabani}\ and\ \citenamefont
  {Lidar}(2005)}]{PhysRevA.71.020101}%
  \BibitemOpen
  \bibfield  {author} {\bibinfo {author} {\bibfnamefont {A.}~\bibnamefont
  {Shabani}}\ and\ \bibinfo {author} {\bibfnamefont {D.~A.}\ \bibnamefont
  {Lidar}},\ }\href {\doibase 10.1103/PhysRevA.71.020101} {\bibfield  {journal}
  {\bibinfo  {journal} {Phys. Rev. A}\ }\textbf {\bibinfo {volume} {71}},\
  \bibinfo {pages} {020101} (\bibinfo {year} {2005})}\BibitemShut {NoStop}%
\bibitem [{\citenamefont {Piilo}\ \emph {et~al.}(2008)\citenamefont {Piilo},
  \citenamefont {Maniscalco}, \citenamefont {H\"ark\"onen},\ and\ \citenamefont
  {Suominen}}]{PhysRevLett.100.180402}%
  \BibitemOpen
  \bibfield  {author} {\bibinfo {author} {\bibfnamefont {J.}~\bibnamefont
  {Piilo}}, \bibinfo {author} {\bibfnamefont {S.}~\bibnamefont {Maniscalco}},
  \bibinfo {author} {\bibfnamefont {K.}~\bibnamefont {H\"ark\"onen}}, \ and\
  \bibinfo {author} {\bibfnamefont {K.-A.}\ \bibnamefont {Suominen}},\ }\href
  {\doibase 10.1103/PhysRevLett.100.180402} {\bibfield  {journal} {\bibinfo
  {journal} {Phys. Rev. Lett.}\ }\textbf {\bibinfo {volume} {100}},\ \bibinfo
  {pages} {180402} (\bibinfo {year} {2008})}\BibitemShut {NoStop}%
\bibitem [{\citenamefont {Vacchini}(2020)}]{vacchini_quantum_2020}%
  \BibitemOpen
  \bibfield  {author} {\bibinfo {author} {\bibfnamefont {B.}~\bibnamefont
  {Vacchini}},\ }\href {\doibase 10.1038/s41598-020-62260-z} {\bibfield
  {journal} {\bibinfo  {journal} {Sci. Rep.}\ }\textbf {\bibinfo {volume}
  {10}},\ \bibinfo {pages} {5592} (\bibinfo {year} {2020})}\BibitemShut
  {NoStop}%
\bibitem [{\citenamefont {Smirne}\ \emph {et~al.}(2020)\citenamefont {Smirne},
  \citenamefont {Caiaffa},\ and\ \citenamefont {Piilo}}]{smirne_rate_2020}%
  \BibitemOpen
  \bibfield  {author} {\bibinfo {author} {\bibfnamefont {A.}~\bibnamefont
  {Smirne}}, \bibinfo {author} {\bibfnamefont {M.}~\bibnamefont {Caiaffa}}, \
  and\ \bibinfo {author} {\bibfnamefont {J.}~\bibnamefont {Piilo}},\ }\href
  {\doibase 10.1103/PhysRevLett.124.190402} {\bibfield  {journal} {\bibinfo
  {journal} {Phys. Rev. Lett.}\ }\textbf {\bibinfo {volume} {124}},\ \bibinfo
  {pages} {190402} (\bibinfo {year} {2020})}\BibitemShut {NoStop}%
\bibitem [{\citenamefont {Megier}\ \emph
  {et~al.}(2020{\natexlab{a}})\citenamefont {Megier}, \citenamefont {Smirne},\
  and\ \citenamefont {Vacchini}}]{megier_interplay_2020}%
  \BibitemOpen
  \bibfield  {author} {\bibinfo {author} {\bibfnamefont {N.}~\bibnamefont
  {Megier}}, \bibinfo {author} {\bibfnamefont {A.}~\bibnamefont {Smirne}}, \
  and\ \bibinfo {author} {\bibfnamefont {B.}~\bibnamefont {Vacchini}},\ }\href
  {\doibase 10.1088/1367-2630/ab9f6b} {\bibfield  {journal} {\bibinfo
  {journal} {New J. Phys.}\ }\textbf {\bibinfo {volume} {22}},\ \bibinfo
  {pages} {083011} (\bibinfo {year} {2020}{\natexlab{a}})},\ \bibinfo {note}
  {publisher: IOP Publishing}\BibitemShut {NoStop}%
\bibitem [{\citenamefont {Megier}\ \emph
  {et~al.}(2020{\natexlab{b}})\citenamefont {Megier}, \citenamefont {Smirne},\
  and\ \citenamefont {Vacchini}}]{megier_evolution_2020}%
  \BibitemOpen
  \bibfield  {author} {\bibinfo {author} {\bibfnamefont {N.}~\bibnamefont
  {Megier}}, \bibinfo {author} {\bibfnamefont {A.}~\bibnamefont {Smirne}}, \
  and\ \bibinfo {author} {\bibfnamefont {B.}~\bibnamefont {Vacchini}},\ }\href
  {\doibase 10.3390/e22070796} {\bibfield  {journal} {\bibinfo  {journal}
  {Entropy}\ }\textbf {\bibinfo {volume} {22}},\ \bibinfo {pages} {796}
  (\bibinfo {year} {2020}{\natexlab{b}})}\BibitemShut {NoStop}%
\bibitem [{\citenamefont {Liu}\ \emph {et~al.}(2011)\citenamefont {Liu},
  \citenamefont {Li}, \citenamefont {Huang}, \citenamefont {Li}, \citenamefont
  {Guo}, \citenamefont {Laine}, \citenamefont {Breuer},\ and\ \citenamefont
  {Piilo}}]{liu_experimental_2011}%
  \BibitemOpen
  \bibfield  {author} {\bibinfo {author} {\bibfnamefont {B.-H.}\ \bibnamefont
  {Liu}}, \bibinfo {author} {\bibfnamefont {L.}~\bibnamefont {Li}}, \bibinfo
  {author} {\bibfnamefont {Y.-F.}\ \bibnamefont {Huang}}, \bibinfo {author}
  {\bibfnamefont {C.-F.}\ \bibnamefont {Li}}, \bibinfo {author} {\bibfnamefont
  {G.-C.}\ \bibnamefont {Guo}}, \bibinfo {author} {\bibfnamefont {E.-M.}\
  \bibnamefont {Laine}}, \bibinfo {author} {\bibfnamefont {H.-P.}\ \bibnamefont
  {Breuer}}, \ and\ \bibinfo {author} {\bibfnamefont {J.}~\bibnamefont
  {Piilo}},\ }\href {\doibase 10.1038/nphys2085} {\bibfield  {journal}
  {\bibinfo  {journal} {Nat. Phys.}\ }\textbf {\bibinfo {volume} {7}},\
  \bibinfo {pages} {931} (\bibinfo {year} {2011})}\BibitemShut {NoStop}%
\bibitem [{\citenamefont {Garrido}\ \emph {et~al.}(2016)\citenamefont
  {Garrido}, \citenamefont {Gorin},\ and\ \citenamefont
  {Pineda}}]{garrido_transition_2016}%
  \BibitemOpen
  \bibfield  {author} {\bibinfo {author} {\bibfnamefont {N.}~\bibnamefont
  {Garrido}}, \bibinfo {author} {\bibfnamefont {T.}~\bibnamefont {Gorin}}, \
  and\ \bibinfo {author} {\bibfnamefont {C.}~\bibnamefont {Pineda}},\ }\href
  {\doibase 10.1103/PhysRevA.93.012113} {\bibfield  {journal} {\bibinfo
  {journal} {Phys. Rev. A}\ }\textbf {\bibinfo {volume} {93}},\ \bibinfo
  {pages} {012113} (\bibinfo {year} {2016})}\BibitemShut {NoStop}%
\bibitem [{Note10()}]{Note10}%
  \BibitemOpen
  \bibinfo {note} {Some authors would not call this a master equation and refer
  to it as a memory-kernel equation. We are being a bit liberal
  here.}\BibitemShut {Stop}%
\bibitem [{Note11()}]{Note11}%
  \BibitemOpen
  \bibinfo {note} {In the classical case it was an operator acting on a vector.
  Here, it is called a super-operator because it acts on $\rho $, which is an
  operator on the Hilbert space.}\BibitemShut {Stop}%
\bibitem [{\citenamefont {Chru\`sci\`nski}\ and\ \citenamefont
  {Pascazio}(2017)}]{gksl}%
  \BibitemOpen
  \bibfield  {author} {\bibinfo {author} {\bibfnamefont {D.}~\bibnamefont
  {Chru\`sci\`nski}}\ and\ \bibinfo {author} {\bibfnamefont {S.}~\bibnamefont
  {Pascazio}},\ }\href {\doibase 10.1142/S1230161217400017} {\bibfield
  {journal} {\bibinfo  {journal} {Open Sys. \& Inf. Dyn.}\ }\textbf {\bibinfo
  {volume} {24}},\ \bibinfo {pages} {1740001} (\bibinfo {year}
  {2017})}\BibitemShut {NoStop}%
\bibitem [{\citenamefont {Manzanoa}(2020)}]{lindped}%
  \BibitemOpen
  \bibfield  {author} {\bibinfo {author} {\bibfnamefont {D.}~\bibnamefont
  {Manzanoa}},\ }\href {\doibase 10.1063/1.5115323} {\bibfield  {journal}
  {\bibinfo  {journal} {AIP Advances}\ }\textbf {\bibinfo {volume} {10}},\
  \bibinfo {pages} {025106} (\bibinfo {year} {2020})}\BibitemShut {NoStop}%
\bibitem [{Note12()}]{Note12}%
  \BibitemOpen
  \bibinfo {note} {Around the same time Franke proposed a very similar
  equation, though he seems to be unaware of complete positivity at that
  time~\cite {franke_general_1976}.}\BibitemShut {Stop}%
\bibitem [{\citenamefont {Breuer}\ \emph {et~al.}(1999)\citenamefont {Breuer},
  \citenamefont {Kappler},\ and\ \citenamefont
  {Petruccione}}]{breuer_stochastic_1999}%
  \BibitemOpen
  \bibfield  {author} {\bibinfo {author} {\bibfnamefont {H.-P.}\ \bibnamefont
  {Breuer}}, \bibinfo {author} {\bibfnamefont {B.}~\bibnamefont {Kappler}}, \
  and\ \bibinfo {author} {\bibfnamefont {F.}~\bibnamefont {Petruccione}},\
  }\href {\doibase 10.1103/PhysRevA.59.1633} {\bibfield  {journal} {\bibinfo
  {journal} {Phys. Rev. A}\ }\textbf {\bibinfo {volume} {59}},\ \bibinfo
  {pages} {1633} (\bibinfo {year} {1999})}\BibitemShut {NoStop}%
\bibitem [{\citenamefont {Andersson}\ \emph {et~al.}(2007)\citenamefont
  {Andersson}, \citenamefont {Cresser},\ and\ \citenamefont
  {Hall}}]{andersson_finding_2007}%
  \BibitemOpen
  \bibfield  {author} {\bibinfo {author} {\bibfnamefont {E.}~\bibnamefont
  {Andersson}}, \bibinfo {author} {\bibfnamefont {J.~D.}\ \bibnamefont
  {Cresser}}, \ and\ \bibinfo {author} {\bibfnamefont {M.~J.~W.}\ \bibnamefont
  {Hall}},\ }\href {\doibase 10.1080/09500340701352581} {\bibfield  {journal}
  {\bibinfo  {journal} {J. Mod. Opt.}\ }\textbf {\bibinfo {volume} {54}},\
  \bibinfo {pages} {1695} (\bibinfo {year} {2007})}\BibitemShut {NoStop}%
\bibitem [{\citenamefont {Laine}\ \emph {et~al.}(2012)\citenamefont {Laine},
  \citenamefont {Luoma},\ and\ \citenamefont {Piilo}}]{laine_local--time_2012}%
  \BibitemOpen
  \bibfield  {author} {\bibinfo {author} {\bibfnamefont {E.-M.}\ \bibnamefont
  {Laine}}, \bibinfo {author} {\bibfnamefont {K.}~\bibnamefont {Luoma}}, \ and\
  \bibinfo {author} {\bibfnamefont {J.}~\bibnamefont {Piilo}},\ }\href
  {\doibase 10.1088/0953-4075/45/15/154004} {\bibfield  {journal} {\bibinfo
  {journal} {J. Phys. B}\ }\textbf {\bibinfo {volume} {45}},\ \bibinfo {pages}
  {154004} (\bibinfo {year} {2012})}\BibitemShut {NoStop}%
\bibitem [{\citenamefont {Hall}\ \emph {et~al.}(2014)\citenamefont {Hall},
  \citenamefont {Cresser}, \citenamefont {Li},\ and\ \citenamefont
  {Andersson}}]{hall_canonical_2014}%
  \BibitemOpen
  \bibfield  {author} {\bibinfo {author} {\bibfnamefont {M.~J.~W.}\
  \bibnamefont {Hall}}, \bibinfo {author} {\bibfnamefont {J.~D.}\ \bibnamefont
  {Cresser}}, \bibinfo {author} {\bibfnamefont {L.}~\bibnamefont {Li}}, \ and\
  \bibinfo {author} {\bibfnamefont {E.}~\bibnamefont {Andersson}},\ }\href
  {\doibase 10.1103/PhysRevA.89.042120} {\bibfield  {journal} {\bibinfo
  {journal} {Phys. Rev. A}\ }\textbf {\bibinfo {volume} {89}},\ \bibinfo
  {pages} {042120} (\bibinfo {year} {2014})}\BibitemShut {NoStop}%
\bibitem [{\citenamefont {Rodr{\'\i}guez}(2008)}]{rodriguez_theory_2008}%
  \BibitemOpen
  \bibfield  {author} {\bibinfo {author} {\bibfnamefont {C.~A.}\ \bibnamefont
  {Rodr{\'\i}guez}},\ }\emph {\bibinfo {title} {The theory of non-{Markovian}
  open quantum systems}},\ \href
  {https://repositories.lib.utexas.edu/handle/2152/3929} {Ph.D. thesis},\
  \bibinfo  {school} {The University of Texas, Austin} (\bibinfo {year}
  {2008})\BibitemShut {NoStop}%
\bibitem [{\citenamefont {Rivas}\ \emph
  {et~al.}(2010{\natexlab{a}})\citenamefont {Rivas}, \citenamefont {Plato},
  \citenamefont {Huelga},\ and\ \citenamefont {Plenio}}]{rivas_markovian_2010}%
  \BibitemOpen
  \bibfield  {author} {\bibinfo {author} {\bibfnamefont {{\'A}.}~\bibnamefont
  {Rivas}}, \bibinfo {author} {\bibfnamefont {A.~D.~K.}\ \bibnamefont {Plato}},
  \bibinfo {author} {\bibfnamefont {S.~F.}\ \bibnamefont {Huelga}}, \ and\
  \bibinfo {author} {\bibfnamefont {M.~B.}\ \bibnamefont {Plenio}},\ }\href
  {\doibase 10.1088/1367-2630/12/11/113032} {\bibfield  {journal} {\bibinfo
  {journal} {New J. Phys.}\ }\textbf {\bibinfo {volume} {12}},\ \bibinfo
  {pages} {113032} (\bibinfo {year} {2010}{\natexlab{a}})}\BibitemShut
  {NoStop}%
\bibitem [{\citenamefont {Hartmann}\ and\ \citenamefont
  {Strunz}(2020)}]{hartmann_accuracy_2020}%
  \BibitemOpen
  \bibfield  {author} {\bibinfo {author} {\bibfnamefont {R.}~\bibnamefont
  {Hartmann}}\ and\ \bibinfo {author} {\bibfnamefont {W.~T.}\ \bibnamefont
  {Strunz}},\ }\href {\doibase 10.1103/PhysRevA.101.012103} {\bibfield
  {journal} {\bibinfo  {journal} {Phys. Rev. A}\ }\textbf {\bibinfo {volume}
  {101}},\ \bibinfo {pages} {012103} (\bibinfo {year} {2020})}\BibitemShut
  {NoStop}%
\bibitem [{Note13()}]{Note13}%
  \BibitemOpen
  \bibinfo {note} {This fact notwithstanding, many master equations can be
  derived as limits of discrete time collision models~\cite
  {scarani_thermalizing_2002, ziman_diluting_2002, giovannetti_dynamical_2005,
  vacchini_non-markovian_2013, vacchini_general_2014, caruso_quantum_2014,
  vacchini_generalized_2016, ciccarello_collision_2017}, both in the
  Markovian~\cite {ziman_description_2005, ziman_all_2005, attal_repeated_2006}
  and the non-Markovian~\cite {pellegrini_non-markovian_2009, Giovannetti2012,
  filippovjphysb, giovannetti_master_2012JPB,
  ciccarello_collision-model-based_2013, PhysRevA.89.052120, lorenzo_heat_2015,
  kretschmer_collision_2016, Lorenzo2017, PhysRevA.96.022109, arXiv:2008.00765}
  case.}\BibitemShut {Stop}%
\bibitem [{\citenamefont {Park}\ and\ \citenamefont {Band}(1992)}]{park}%
  \BibitemOpen
  \bibfield  {author} {\bibinfo {author} {\bibfnamefont {J.}~\bibnamefont
  {Park}}\ and\ \bibinfo {author} {\bibfnamefont {W.}~\bibnamefont {Band}},\
  }\href {\doibase 10.1007/BF01889671} {\bibfield  {journal} {\bibinfo
  {journal} {Found. Phys.}\ }\textbf {\bibinfo {volume} {22}},\ \bibinfo
  {pages} {657} (\bibinfo {year} {1992})}\BibitemShut {NoStop}%
\bibitem [{\citenamefont {Laine}\ \emph
  {et~al.}(2010{\natexlab{a}})\citenamefont {Laine}, \citenamefont {Piilo},\
  and\ \citenamefont {Breuer}}]{IC-breuer}%
  \BibitemOpen
  \bibfield  {author} {\bibinfo {author} {\bibfnamefont {E.-M.}\ \bibnamefont
  {Laine}}, \bibinfo {author} {\bibfnamefont {J.}~\bibnamefont {Piilo}}, \ and\
  \bibinfo {author} {\bibfnamefont {H.-P.}\ \bibnamefont {Breuer}},\ }\href
  {http://stacks.iop.org/0295-5075/92/i=6/a=60010} {\bibfield  {journal}
  {\bibinfo  {journal} {EPL}\ }\textbf {\bibinfo {volume} {92}},\ \bibinfo
  {pages} {60010} (\bibinfo {year} {2010}{\natexlab{a}})}\BibitemShut {NoStop}%
\bibitem [{\citenamefont {Gessner}\ and\ \citenamefont
  {Breuer}(2011)}]{PhysRevLett.107.180402}%
  \BibitemOpen
  \bibfield  {author} {\bibinfo {author} {\bibfnamefont {M.}~\bibnamefont
  {Gessner}}\ and\ \bibinfo {author} {\bibfnamefont {H.-P.}\ \bibnamefont
  {Breuer}},\ }\href {\doibase 10.1103/PhysRevLett.107.180402} {\bibfield
  {journal} {\bibinfo  {journal} {Phys. Rev. Lett.}\ }\textbf {\bibinfo
  {volume} {107}},\ \bibinfo {pages} {180402} (\bibinfo {year}
  {2011})}\BibitemShut {NoStop}%
\bibitem [{\citenamefont {Smirne}\ \emph {et~al.}(2011)\citenamefont {Smirne},
  \citenamefont {Brivio}, \citenamefont {Cialdi}, \citenamefont {Vacchini},\
  and\ \citenamefont {Paris}}]{smirne_experimental_2011}%
  \BibitemOpen
  \bibfield  {author} {\bibinfo {author} {\bibfnamefont {A.}~\bibnamefont
  {Smirne}}, \bibinfo {author} {\bibfnamefont {D.}~\bibnamefont {Brivio}},
  \bibinfo {author} {\bibfnamefont {S.}~\bibnamefont {Cialdi}}, \bibinfo
  {author} {\bibfnamefont {B.}~\bibnamefont {Vacchini}}, \ and\ \bibinfo
  {author} {\bibfnamefont {M.~G.~A.}\ \bibnamefont {Paris}},\ }\href {\doibase
  10.1103/PhysRevA.84.032112} {\bibfield  {journal} {\bibinfo  {journal} {Phys.
  Rev. A}\ }\textbf {\bibinfo {volume} {84}},\ \bibinfo {pages} {032112}
  (\bibinfo {year} {2011})}\BibitemShut {NoStop}%
\bibitem [{\citenamefont {Gessner}\ and\ \citenamefont
  {Breuer}(2013)}]{gessner_local_2013}%
  \BibitemOpen
  \bibfield  {author} {\bibinfo {author} {\bibfnamefont {M.}~\bibnamefont
  {Gessner}}\ and\ \bibinfo {author} {\bibfnamefont {H.-P.}\ \bibnamefont
  {Breuer}},\ }\href {\doibase 10.1103/PhysRevA.87.042107} {\bibfield
  {journal} {\bibinfo  {journal} {Phys. Rev. A}\ }\textbf {\bibinfo {volume}
  {87}},\ \bibinfo {pages} {042107} (\bibinfo {year} {2013})}\BibitemShut
  {NoStop}%
\bibitem [{\citenamefont {Smirne}\ \emph {et~al.}(2013)\citenamefont {Smirne},
  \citenamefont {Cialdi}, \citenamefont {Anelli}, \citenamefont {Paris},\ and\
  \citenamefont {Vacchini}}]{smirne_quantum_2013}%
  \BibitemOpen
  \bibfield  {author} {\bibinfo {author} {\bibfnamefont {A.}~\bibnamefont
  {Smirne}}, \bibinfo {author} {\bibfnamefont {S.}~\bibnamefont {Cialdi}},
  \bibinfo {author} {\bibfnamefont {G.}~\bibnamefont {Anelli}}, \bibinfo
  {author} {\bibfnamefont {M.~G.~A.}\ \bibnamefont {Paris}}, \ and\ \bibinfo
  {author} {\bibfnamefont {B.}~\bibnamefont {Vacchini}},\ }\href {\doibase
  10.1103/PhysRevA.88.012108} {\bibfield  {journal} {\bibinfo  {journal} {Phys.
  Rev. A}\ }\textbf {\bibinfo {volume} {88}},\ \bibinfo {pages} {012108}
  (\bibinfo {year} {2013})}\BibitemShut {NoStop}%
\bibitem [{\citenamefont {Gessner}\ \emph {et~al.}(2014)\citenamefont
  {Gessner}, \citenamefont {Ramm}, \citenamefont {Pruttivarasin}, \citenamefont
  {Buchleitner}, \citenamefont {Breuer},\ and\ \citenamefont
  {Haffner}}]{Gessner:2014kl}%
  \BibitemOpen
  \bibfield  {author} {\bibinfo {author} {\bibfnamefont {M.}~\bibnamefont
  {Gessner}}, \bibinfo {author} {\bibfnamefont {M.}~\bibnamefont {Ramm}},
  \bibinfo {author} {\bibfnamefont {T.}~\bibnamefont {Pruttivarasin}}, \bibinfo
  {author} {\bibfnamefont {A.}~\bibnamefont {Buchleitner}}, \bibinfo {author}
  {\bibfnamefont {H.-P.}\ \bibnamefont {Breuer}}, \ and\ \bibinfo {author}
  {\bibfnamefont {H.}~\bibnamefont {Haffner}},\ }\href
  {http://dx.doi.org/10.1038/nphys2829} {\bibfield  {journal} {\bibinfo
  {journal} {Nat. Phys.}\ }\textbf {\bibinfo {volume} {10}},\ \bibinfo {pages}
  {105} (\bibinfo {year} {2014})}\BibitemShut {NoStop}%
\bibitem [{\citenamefont {Weinstein}\ \emph {et~al.}(2004)\citenamefont
  {Weinstein}, \citenamefont {Havel}, \citenamefont {Emerson}, \citenamefont
  {Boulant}, \citenamefont {Saraceno}, \citenamefont {Lloyd},\ and\
  \citenamefont {Cory}}]{Wein:121.13}%
  \BibitemOpen
  \bibfield  {author} {\bibinfo {author} {\bibfnamefont {Y.~S.}\ \bibnamefont
  {Weinstein}}, \bibinfo {author} {\bibfnamefont {T.~F.}\ \bibnamefont
  {Havel}}, \bibinfo {author} {\bibfnamefont {J.}~\bibnamefont {Emerson}},
  \bibinfo {author} {\bibfnamefont {N.}~\bibnamefont {Boulant}}, \bibinfo
  {author} {\bibfnamefont {M.}~\bibnamefont {Saraceno}}, \bibinfo {author}
  {\bibfnamefont {S.}~\bibnamefont {Lloyd}}, \ and\ \bibinfo {author}
  {\bibfnamefont {D.~G.}\ \bibnamefont {Cory}},\ }\href
  {http://aip.scitation.org/doi/abs/10.1063/1.1785151} {\bibfield  {journal}
  {\bibinfo  {journal} {J. Chem. Phys.}\ }\textbf {\bibinfo {volume} {121}},\
  \bibinfo {pages} {6117} (\bibinfo {year} {2004})}\BibitemShut {NoStop}%
\bibitem [{\citenamefont {Myrskog}\ \emph {et~al.}(2005)\citenamefont
  {Myrskog}, \citenamefont {Fox}, \citenamefont {Mitchell},\ and\ \citenamefont
  {Steinberg}}]{myrskog:013615}%
  \BibitemOpen
  \bibfield  {author} {\bibinfo {author} {\bibfnamefont {S.~H.}\ \bibnamefont
  {Myrskog}}, \bibinfo {author} {\bibfnamefont {J.~K.}\ \bibnamefont {Fox}},
  \bibinfo {author} {\bibfnamefont {M.~W.}\ \bibnamefont {Mitchell}}, \ and\
  \bibinfo {author} {\bibfnamefont {A.~M.}\ \bibnamefont {Steinberg}},\ }\href
  {\doibase 10.1103/PhysRevA.72.013615} {\bibfield  {journal} {\bibinfo
  {journal} {Phys. Rev. A}\ }\textbf {\bibinfo {volume} {72}},\ \bibinfo {eid}
  {013615} (\bibinfo {year} {2005})}\BibitemShut {NoStop}%
\bibitem [{\citenamefont {O'Brien}\ \emph {et~al.}(2004)\citenamefont
  {O'Brien}, \citenamefont {Pryde}, \citenamefont {Gilchrist}, \citenamefont
  {James}, \citenamefont {Langford}, \citenamefont {Ralph},\ and\ \citenamefont
  {White}}]{orien:080502}%
  \BibitemOpen
  \bibfield  {author} {\bibinfo {author} {\bibfnamefont {J.~L.}\ \bibnamefont
  {O'Brien}}, \bibinfo {author} {\bibfnamefont {G.~J.}\ \bibnamefont {Pryde}},
  \bibinfo {author} {\bibfnamefont {A.}~\bibnamefont {Gilchrist}}, \bibinfo
  {author} {\bibfnamefont {D.~F.~V.}\ \bibnamefont {James}}, \bibinfo {author}
  {\bibfnamefont {N.~K.}\ \bibnamefont {Langford}}, \bibinfo {author}
  {\bibfnamefont {T.~C.}\ \bibnamefont {Ralph}}, \ and\ \bibinfo {author}
  {\bibfnamefont {A.~G.}\ \bibnamefont {White}},\ }\href
  {https://journals.aps.org/prl/abstract/10.1103/PhysRevLett.93.080502#fulltext}
  {\bibfield  {journal} {\bibinfo  {journal} {Phys. Rev. Lett.}\ }\textbf
  {\bibinfo {volume} {93}},\ \bibinfo {pages} {080502} (\bibinfo {year}
  {2004})}\BibitemShut {NoStop}%
\bibitem [{\citenamefont {Shaji}\ and\ \citenamefont
  {Sudarshan}(2005)}]{shaji_whos_2005}%
  \BibitemOpen
  \bibfield  {author} {\bibinfo {author} {\bibfnamefont {A.}~\bibnamefont
  {Shaji}}\ and\ \bibinfo {author} {\bibfnamefont {E.~C.~G.}\ \bibnamefont
  {Sudarshan}},\ }\href {\doibase 10.1016/j.physleta.2005.04.029} {\bibfield
  {journal} {\bibinfo  {journal} {Phys. Lett. A}\ }\textbf {\bibinfo {volume}
  {341}},\ \bibinfo {pages} {48} (\bibinfo {year} {2005})}\BibitemShut
  {NoStop}%
\bibitem [{\citenamefont {Carteret}\ \emph {et~al.}(2008)\citenamefont
  {Carteret}, \citenamefont {Terno},\ and\ \citenamefont
  {\.{Z}yczkowski}}]{ctz}%
  \BibitemOpen
  \bibfield  {author} {\bibinfo {author} {\bibfnamefont {H.~A.}\ \bibnamefont
  {Carteret}}, \bibinfo {author} {\bibfnamefont {D.~R.}\ \bibnamefont {Terno}},
  \ and\ \bibinfo {author} {\bibfnamefont {K.}~\bibnamefont {\.{Z}yczkowski}},\
  }\href {\doibase 10.1103/PhysRevA.77.042113} {\bibfield  {journal} {\bibinfo
  {journal} {Phys. Rev. A}\ }\textbf {\bibinfo {volume} {77}},\ \bibinfo
  {pages} {042113} (\bibinfo {year} {2008})}\BibitemShut {NoStop}%
\bibitem [{\citenamefont {Ziman}\ \emph {et~al.}(2005)\citenamefont {Ziman},
  \citenamefont {Plesch}, \citenamefont {Bu\v{z}ek},\ and\ \citenamefont
  {\v{S}telmachovi\v{c}}}]{ziman}%
  \BibitemOpen
  \bibfield  {author} {\bibinfo {author} {\bibfnamefont {M.}~\bibnamefont
  {Ziman}}, \bibinfo {author} {\bibfnamefont {M.}~\bibnamefont {Plesch}},
  \bibinfo {author} {\bibfnamefont {V.}~\bibnamefont {Bu\v{z}ek}}, \ and\
  \bibinfo {author} {\bibfnamefont {P.}~\bibnamefont {\v{S}telmachovi\v{c}}},\
  }\href {\doibase 10.1103/PhysRevA.72.022106} {\bibfield  {journal} {\bibinfo
  {journal} {Phys. Rev. A}\ }\textbf {\bibinfo {volume} {72}},\ \bibinfo {eid}
  {022106} (\bibinfo {year} {2005})}\BibitemShut {NoStop}%
\bibitem [{\citenamefont {Ziman}(2006)}]{Ziman06}%
  \BibitemOpen
  \bibfield  {author} {\bibinfo {author} {\bibfnamefont {M.}~\bibnamefont
  {Ziman}},\ }\href {https://arxiv.org/abs/quant-ph/0603166} {\bibfield
  {journal} {\bibinfo  {journal} {arXiv:0603166}\ } (\bibinfo {year}
  {2006})}\BibitemShut {NoStop}%
\bibitem [{\citenamefont {Rodr\'iguez-Rosario}\ \emph
  {et~al.}(2008)\citenamefont {Rodr\'iguez-Rosario}, \citenamefont {Modi},
  \citenamefont {Kuah}, \citenamefont {Shaji},\ and\ \citenamefont
  {Sudarshan}}]{rodriguez2008completely}%
  \BibitemOpen
  \bibfield  {author} {\bibinfo {author} {\bibfnamefont {C.~A.}\ \bibnamefont
  {Rodr\'iguez-Rosario}}, \bibinfo {author} {\bibfnamefont {K.}~\bibnamefont
  {Modi}}, \bibinfo {author} {\bibfnamefont {A.-m.}\ \bibnamefont {Kuah}},
  \bibinfo {author} {\bibfnamefont {A.}~\bibnamefont {Shaji}}, \ and\ \bibinfo
  {author} {\bibfnamefont {E.}~\bibnamefont {Sudarshan}},\ }\href
  {http://iopscience.iop.org/article/10.1088/1751-8113/41/20/205301} {\bibfield
   {journal} {\bibinfo  {journal} {J. Phys. A}\ }\textbf {\bibinfo {volume}
  {41}},\ \bibinfo {pages} {205301} (\bibinfo {year} {2008})}\BibitemShut
  {NoStop}%
\bibitem [{\citenamefont {Shabani}\ and\ \citenamefont
  {Lidar}(2009)}]{PhysRevLett.102.100402}%
  \BibitemOpen
  \bibfield  {author} {\bibinfo {author} {\bibfnamefont {A.}~\bibnamefont
  {Shabani}}\ and\ \bibinfo {author} {\bibfnamefont {D.~A.}\ \bibnamefont
  {Lidar}},\ }\href {\doibase 10.1103/PhysRevLett.102.100402} {\bibfield
  {journal} {\bibinfo  {journal} {Phys. Rev. Lett.}\ }\textbf {\bibinfo
  {volume} {102}},\ \bibinfo {pages} {100402} (\bibinfo {year}
  {2009})}\BibitemShut {NoStop}%
\bibitem [{\citenamefont {Modi}\ and\ \citenamefont
  {Sudarshan}(2010)}]{modi_role_2010}%
  \BibitemOpen
  \bibfield  {author} {\bibinfo {author} {\bibfnamefont {K.}~\bibnamefont
  {Modi}}\ and\ \bibinfo {author} {\bibfnamefont {E.~C.~G.}\ \bibnamefont
  {Sudarshan}},\ }\href {\doibase 10.1103/PhysRevA.81.052119} {\bibfield
  {journal} {\bibinfo  {journal} {Phys. Rev. A}\ }\textbf {\bibinfo {volume}
  {81}},\ \bibinfo {pages} {052119} (\bibinfo {year} {2010})}\BibitemShut
  {NoStop}%
\bibitem [{\citenamefont {Brodutch}\ \emph {et~al.}(2013)\citenamefont
  {Brodutch}, \citenamefont {Datta}, \citenamefont {Modi}, \citenamefont
  {Rivas},\ and\ \citenamefont {Rodr\'{\i}guez-Rosario}}]{PhysRevA.87.042301}%
  \BibitemOpen
  \bibfield  {author} {\bibinfo {author} {\bibfnamefont {A.}~\bibnamefont
  {Brodutch}}, \bibinfo {author} {\bibfnamefont {A.}~\bibnamefont {Datta}},
  \bibinfo {author} {\bibfnamefont {K.}~\bibnamefont {Modi}}, \bibinfo {author}
  {\bibfnamefont {A.}~\bibnamefont {Rivas}}, \ and\ \bibinfo {author}
  {\bibfnamefont {C.~A.}\ \bibnamefont {Rodr\'{\i}guez-Rosario}},\ }\href
  {\doibase 10.1103/PhysRevA.87.042301} {\bibfield  {journal} {\bibinfo
  {journal} {Phys. Rev. A}\ }\textbf {\bibinfo {volume} {87}},\ \bibinfo
  {pages} {042301} (\bibinfo {year} {2013})}\BibitemShut {NoStop}%
\bibitem [{\citenamefont {Vacchini}\ and\ \citenamefont
  {Amato}(2016)}]{vacchini_reduced_2016}%
  \BibitemOpen
  \bibfield  {author} {\bibinfo {author} {\bibfnamefont {B.}~\bibnamefont
  {Vacchini}}\ and\ \bibinfo {author} {\bibfnamefont {G.}~\bibnamefont
  {Amato}},\ }\href {\doibase 10.1038/srep37328} {\bibfield  {journal}
  {\bibinfo  {journal} {Sci. Rep.}\ }\textbf {\bibinfo {volume} {6}},\ \bibinfo
  {pages} {37328} (\bibinfo {year} {2016})}\BibitemShut {NoStop}%
\bibitem [{\citenamefont {Modi}(2011)}]{modiosid}%
  \BibitemOpen
  \bibfield  {author} {\bibinfo {author} {\bibfnamefont {K.}~\bibnamefont
  {Modi}},\ }\href
  {http://www.worldscientific.com/doi/abs/10.1142/S1230161211000170} {\bibfield
   {journal} {\bibinfo  {journal} {Open Sys. \& Info. Dyn.}\ }\textbf {\bibinfo
  {volume} {18}},\ \bibinfo {pages} {253} (\bibinfo {year} {2011})}\BibitemShut
  {NoStop}%
\bibitem [{\citenamefont {Bausch}\ and\ \citenamefont
  {Cubitt}(2016)}]{divcomp}%
  \BibitemOpen
  \bibfield  {author} {\bibinfo {author} {\bibfnamefont {J.}~\bibnamefont
  {Bausch}}\ and\ \bibinfo {author} {\bibfnamefont {T.}~\bibnamefont
  {Cubitt}},\ }\href {\doibase 10.1016/j.laa.2016.03.041} {\bibfield  {journal}
  {\bibinfo  {journal} {Linear Algebra Appl.}\ ,\ \bibinfo {pages} {64}}
  (\bibinfo {year} {2016})}\BibitemShut {NoStop}%
\bibitem [{\citenamefont {Buscemi}(2014)}]{buscemi_complete_2014}%
  \BibitemOpen
  \bibfield  {author} {\bibinfo {author} {\bibfnamefont {F.}~\bibnamefont
  {Buscemi}},\ }\href {\doibase 10.1103/PhysRevLett.113.140502} {\bibfield
  {journal} {\bibinfo  {journal} {Phys. Rev. Lett.}\ }\textbf {\bibinfo
  {volume} {113}},\ \bibinfo {pages} {140502} (\bibinfo {year}
  {2014})}\BibitemShut {NoStop}%
\bibitem [{\citenamefont {Buscemi}\ and\ \citenamefont
  {Datta}(2016)}]{PhysRevA.93.012101}%
  \BibitemOpen
  \bibfield  {author} {\bibinfo {author} {\bibfnamefont {F.}~\bibnamefont
  {Buscemi}}\ and\ \bibinfo {author} {\bibfnamefont {N.}~\bibnamefont
  {Datta}},\ }\href {\doibase 10.1103/PhysRevA.93.012101} {\bibfield  {journal}
  {\bibinfo  {journal} {Phys. Rev. A}\ }\textbf {\bibinfo {volume} {93}},\
  \bibinfo {pages} {012101} (\bibinfo {year} {2016})}\BibitemShut {NoStop}%
\bibitem [{\citenamefont {Filippov}\ \emph {et~al.}(2017)\citenamefont
  {Filippov}, \citenamefont {Piilo}, \citenamefont {Maniscalco},\ and\
  \citenamefont {Ziman}}]{PhysRevA.96.032111}%
  \BibitemOpen
  \bibfield  {author} {\bibinfo {author} {\bibfnamefont {S.~N.}\ \bibnamefont
  {Filippov}}, \bibinfo {author} {\bibfnamefont {J.}~\bibnamefont {Piilo}},
  \bibinfo {author} {\bibfnamefont {S.}~\bibnamefont {Maniscalco}}, \ and\
  \bibinfo {author} {\bibfnamefont {M.}~\bibnamefont {Ziman}},\ }\href
  {\doibase 10.1103/PhysRevA.96.032111} {\bibfield  {journal} {\bibinfo
  {journal} {Phys. Rev. A}\ }\textbf {\bibinfo {volume} {96}},\ \bibinfo
  {pages} {032111} (\bibinfo {year} {2017})}\BibitemShut {NoStop}%
\bibitem [{\citenamefont {Bae}\ and\ \citenamefont
  {Chru\'{s}ci\'{n}ski}(2016)}]{chruscinski2016}%
  \BibitemOpen
  \bibfield  {author} {\bibinfo {author} {\bibfnamefont {J.}~\bibnamefont
  {Bae}}\ and\ \bibinfo {author} {\bibfnamefont {D.}~\bibnamefont
  {Chru\'{s}ci\'{n}ski}},\ }\href {\doibase 10.1103/PhysRevLett.117.050403}
  {\bibfield  {journal} {\bibinfo  {journal} {Phys. Rev. Lett.}\ }\textbf
  {\bibinfo {volume} {117}},\ \bibinfo {pages} {050403} (\bibinfo {year}
  {2016})}\BibitemShut {NoStop}%
\bibitem [{\citenamefont {Chru\ifmmode \acute{s}\else
  \'{s}\fi{}ci\ifmmode~\acute{n}\else \'{n}\fi{}ski}\ \emph
  {et~al.}(2018)\citenamefont {Chru\ifmmode \acute{s}\else
  \'{s}\fi{}ci\ifmmode~\acute{n}\else \'{n}\fi{}ski}, \citenamefont {Rivas},\
  and\ \citenamefont {St\o{}rmer}}]{PhysRevLett.121.080407}%
  \BibitemOpen
  \bibfield  {author} {\bibinfo {author} {\bibfnamefont {D.}~\bibnamefont
  {Chru\ifmmode \acute{s}\else \'{s}\fi{}ci\ifmmode~\acute{n}\else
  \'{n}\fi{}ski}}, \bibinfo {author} {\bibfnamefont {A.}~\bibnamefont {Rivas}},
  \ and\ \bibinfo {author} {\bibfnamefont {E.}~\bibnamefont {St\o{}rmer}},\
  }\href {\doibase 10.1103/PhysRevLett.121.080407} {\bibfield  {journal}
  {\bibinfo  {journal} {Phys. Rev. Lett.}\ }\textbf {\bibinfo {volume} {121}},\
  \bibinfo {pages} {080407} (\bibinfo {year} {2018})}\BibitemShut {NoStop}%
\bibitem [{\citenamefont {Benatti}\ \emph {et~al.}(2017)\citenamefont
  {Benatti}, \citenamefont {Chru\ifmmode \acute{s}\else
  \'{s}\fi{}ci\ifmmode~\acute{n}\else \'{n}\fi{}ski},\ and\ \citenamefont
  {Filippov}}]{PhysRevA.95.012112}%
  \BibitemOpen
  \bibfield  {author} {\bibinfo {author} {\bibfnamefont {F.}~\bibnamefont
  {Benatti}}, \bibinfo {author} {\bibfnamefont {D.}~\bibnamefont {Chru\ifmmode
  \acute{s}\else \'{s}\fi{}ci\ifmmode~\acute{n}\else \'{n}\fi{}ski}}, \ and\
  \bibinfo {author} {\bibfnamefont {S.}~\bibnamefont {Filippov}},\ }\href
  {\doibase 10.1103/PhysRevA.95.012112} {\bibfield  {journal} {\bibinfo
  {journal} {Phys. Rev. A}\ }\textbf {\bibinfo {volume} {95}},\ \bibinfo
  {pages} {012112} (\bibinfo {year} {2017})}\BibitemShut {NoStop}%
\bibitem [{\citenamefont {Chakraborty}\ and\ \citenamefont {Chru\ifmmode
  \acute{s}\else \'{s}\fi{}ci\ifmmode~\acute{n}\else
  \'{n}\fi{}ski}(2019)}]{PhysRevA.99.042105}%
  \BibitemOpen
  \bibfield  {author} {\bibinfo {author} {\bibfnamefont {S.}~\bibnamefont
  {Chakraborty}}\ and\ \bibinfo {author} {\bibfnamefont {D.}~\bibnamefont
  {Chru\ifmmode \acute{s}\else \'{s}\fi{}ci\ifmmode~\acute{n}\else
  \'{n}\fi{}ski}},\ }\href {\doibase 10.1103/PhysRevA.99.042105} {\bibfield
  {journal} {\bibinfo  {journal} {Phys. Rev. A}\ }\textbf {\bibinfo {volume}
  {99}},\ \bibinfo {pages} {042105} (\bibinfo {year} {2019})}\BibitemShut
  {NoStop}%
\bibitem [{\citenamefont {Wudarski}\ and\ \citenamefont {Chru\ifmmode
  \acute{s}\else \'{s}\fi{}ci\ifmmode~\acute{n}\else
  \'{n}\fi{}ski}(2016)}]{PhysRevA.93.042120}%
  \BibitemOpen
  \bibfield  {author} {\bibinfo {author} {\bibfnamefont {F.~A.}\ \bibnamefont
  {Wudarski}}\ and\ \bibinfo {author} {\bibfnamefont {D.}~\bibnamefont
  {Chru\ifmmode \acute{s}\else \'{s}\fi{}ci\ifmmode~\acute{n}\else
  \'{n}\fi{}ski}},\ }\href {\doibase 10.1103/PhysRevA.93.042120} {\bibfield
  {journal} {\bibinfo  {journal} {Phys. Rev. A}\ }\textbf {\bibinfo {volume}
  {93}},\ \bibinfo {pages} {042120} (\bibinfo {year} {2016})}\BibitemShut
  {NoStop}%
\bibitem [{\citenamefont {Ryb\`ar}\ \emph {et~al.}(2012)\citenamefont
  {Ryb\`ar}, \citenamefont {Filippov}, \citenamefont {Ziman},\ and\
  \citenamefont {Bu\v{z}ek}}]{filippovjphysb}%
  \BibitemOpen
  \bibfield  {author} {\bibinfo {author} {\bibfnamefont {T.}~\bibnamefont
  {Ryb\`ar}}, \bibinfo {author} {\bibfnamefont {S.}~\bibnamefont {Filippov}},
  \bibinfo {author} {\bibfnamefont {M.}~\bibnamefont {Ziman}}, \ and\ \bibinfo
  {author} {\bibfnamefont {V.}~\bibnamefont {Bu\v{z}ek}},\ }\href {\doibase
  10.1088/0953-4075/45/15/154006} {\bibfield  {journal} {\bibinfo  {journal}
  {J. Phys. B}\ }\textbf {\bibinfo {volume} {45}},\ \bibinfo {pages} {154006}
  (\bibinfo {year} {2012})}\BibitemShut {NoStop}%
\bibitem [{\citenamefont {Milz}\ \emph {et~al.}(2019)\citenamefont {Milz},
  \citenamefont {Kim}, \citenamefont {Pollock},\ and\ \citenamefont
  {Modi}}]{Milz2019}%
  \BibitemOpen
  \bibfield  {author} {\bibinfo {author} {\bibfnamefont {S.}~\bibnamefont
  {Milz}}, \bibinfo {author} {\bibfnamefont {M.~S.}\ \bibnamefont {Kim}},
  \bibinfo {author} {\bibfnamefont {F.~A.}\ \bibnamefont {Pollock}}, \ and\
  \bibinfo {author} {\bibfnamefont {K.}~\bibnamefont {Modi}},\ }\href {\doibase
  10.1103/PhysRevLett.123.040401} {\bibfield  {journal} {\bibinfo  {journal}
  {Phys. Rev. Lett.}\ }\textbf {\bibinfo {volume} {123}},\ \bibinfo {pages}
  {040401} (\bibinfo {year} {2019})}\BibitemShut {NoStop}%
\bibitem [{\citenamefont {Breuer}\ \emph {et~al.}(2009)\citenamefont {Breuer},
  \citenamefont {Laine},\ and\ \citenamefont {Piilo}}]{PhysRevLett.103.210401}%
  \BibitemOpen
  \bibfield  {author} {\bibinfo {author} {\bibfnamefont {H.-P.}\ \bibnamefont
  {Breuer}}, \bibinfo {author} {\bibfnamefont {E.-M.}\ \bibnamefont {Laine}}, \
  and\ \bibinfo {author} {\bibfnamefont {J.}~\bibnamefont {Piilo}},\ }\href
  {\doibase 10.1103/PhysRevLett.103.210401} {\bibfield  {journal} {\bibinfo
  {journal} {Phys. Rev. Lett.}\ }\textbf {\bibinfo {volume} {103}},\ \bibinfo
  {pages} {210401} (\bibinfo {year} {2009})}\BibitemShut {NoStop}%
\bibitem [{\citenamefont {Wolf}\ and\ \citenamefont
  {Cirac}(2008)}]{wolf_dividing_2008}%
  \BibitemOpen
  \bibfield  {author} {\bibinfo {author} {\bibfnamefont {M.~M.}\ \bibnamefont
  {Wolf}}\ and\ \bibinfo {author} {\bibfnamefont {J.~I.}\ \bibnamefont
  {Cirac}},\ }\href {\doibase 10.1007/s00220-008-0411-y} {\bibfield  {journal}
  {\bibinfo  {journal} {Commun. Math. Phys.}\ }\textbf {\bibinfo {volume}
  {279}},\ \bibinfo {pages} {147} (\bibinfo {year} {2008})}\BibitemShut
  {NoStop}%
\bibitem [{\citenamefont {Chru{\'s}ci{\'n}ski}\ and\ \citenamefont
  {Siudzi{\'n}ska}(2016)}]{chruscinski_generalized_2016}%
  \BibitemOpen
  \bibfield  {author} {\bibinfo {author} {\bibfnamefont {D.}~\bibnamefont
  {Chru{\'s}ci{\'n}ski}}\ and\ \bibinfo {author} {\bibfnamefont
  {K.}~\bibnamefont {Siudzi{\'n}ska}},\ }\href {\doibase
  10.1103/PhysRevA.94.022118} {\bibfield  {journal} {\bibinfo  {journal} {Phys.
  Rev. A}\ }\textbf {\bibinfo {volume} {94}},\ \bibinfo {pages} {022118}
  (\bibinfo {year} {2016})}\BibitemShut {NoStop}%
\bibitem [{\citenamefont {Davalos}\ \emph {et~al.}(2019)\citenamefont
  {Davalos}, \citenamefont {Ziman},\ and\ \citenamefont
  {Pineda}}]{davalos_divisibility_2019}%
  \BibitemOpen
  \bibfield  {author} {\bibinfo {author} {\bibfnamefont {D.}~\bibnamefont
  {Davalos}}, \bibinfo {author} {\bibfnamefont {M.}~\bibnamefont {Ziman}}, \
  and\ \bibinfo {author} {\bibfnamefont {C.}~\bibnamefont {Pineda}},\ }\href
  {\doibase 10.22331/q-2019-05-20-144} {\bibfield  {journal} {\bibinfo
  {journal} {Quantum}\ }\textbf {\bibinfo {volume} {3}},\ \bibinfo {pages}
  {144} (\bibinfo {year} {2019})}\BibitemShut {NoStop}%
\bibitem [{\citenamefont {Chru{\'s}ci{\'n}ski}\ and\ \citenamefont
  {Mukhamedov}(2019)}]{chruscinski_dissipative_2019}%
  \BibitemOpen
  \bibfield  {author} {\bibinfo {author} {\bibfnamefont {D.}~\bibnamefont
  {Chru{\'s}ci{\'n}ski}}\ and\ \bibinfo {author} {\bibfnamefont
  {F.}~\bibnamefont {Mukhamedov}},\ }\href {\doibase
  10.1103/PhysRevA.100.052120} {\bibfield  {journal} {\bibinfo  {journal}
  {Phys. Rev. A}\ }\textbf {\bibinfo {volume} {100}},\ \bibinfo {pages}
  {052120} (\bibinfo {year} {2019})}\BibitemShut {NoStop}%
\bibitem [{\citenamefont {Wolf}\ \emph {et~al.}(2008)\citenamefont {Wolf},
  \citenamefont {Eisert}, \citenamefont {Cubitt},\ and\ \citenamefont
  {Cirac}}]{PhysRevLett.101.150402}%
  \BibitemOpen
  \bibfield  {author} {\bibinfo {author} {\bibfnamefont {M.~M.}\ \bibnamefont
  {Wolf}}, \bibinfo {author} {\bibfnamefont {J.}~\bibnamefont {Eisert}},
  \bibinfo {author} {\bibfnamefont {T.~S.}\ \bibnamefont {Cubitt}}, \ and\
  \bibinfo {author} {\bibfnamefont {J.~I.}\ \bibnamefont {Cirac}},\ }\href
  {\doibase 10.1103/PhysRevLett.101.150402} {\bibfield  {journal} {\bibinfo
  {journal} {Phys. Rev. Lett.}\ }\textbf {\bibinfo {volume} {101}},\ \bibinfo
  {pages} {150402} (\bibinfo {year} {2008})}\BibitemShut {NoStop}%
\bibitem [{\citenamefont {Pucha{\l}a}\ \emph {et~al.}(2019)\citenamefont
  {Pucha{\l}a}, \citenamefont {Rudnicki},\ and\ \citenamefont
  {{\.Z}yczkowski}}]{puchala_pauli_2019}%
  \BibitemOpen
  \bibfield  {author} {\bibinfo {author} {\bibfnamefont {Z.}~\bibnamefont
  {Pucha{\l}a}}, \bibinfo {author} {\bibfnamefont {{\L}.}~\bibnamefont
  {Rudnicki}}, \ and\ \bibinfo {author} {\bibfnamefont {K.}~\bibnamefont
  {{\.Z}yczkowski}},\ }\href {\doibase 10.1016/j.physleta.2019.04.057}
  {\bibfield  {journal} {\bibinfo  {journal} {Phys. Lett. A}\ }\textbf
  {\bibinfo {volume} {383}},\ \bibinfo {pages} {2376} (\bibinfo {year}
  {2019})}\BibitemShut {NoStop}%
\bibitem [{\citenamefont {Korzekwa}\ and\ \citenamefont
  {Lostaglio}(2020)}]{korzekwa_quantum_2020}%
  \BibitemOpen
  \bibfield  {author} {\bibinfo {author} {\bibfnamefont {K.}~\bibnamefont
  {Korzekwa}}\ and\ \bibinfo {author} {\bibfnamefont {M.}~\bibnamefont
  {Lostaglio}},\ }\href {http://arxiv.org/abs/2005.02403} {\bibfield  {journal}
  {\bibinfo  {journal} {arXiv:2005.02403}\ } (\bibinfo {year}
  {2020})}\BibitemShut {NoStop}%
\bibitem [{\citenamefont {Shahbeigi}\ \emph {et~al.}(2020)\citenamefont
  {Shahbeigi}, \citenamefont {Amaro-Alcal{\'a}}, \citenamefont {Pucha{\l}a},\
  and\ \citenamefont {{\.Z}yczkowski}}]{shahbeigi_log-convex_2020}%
  \BibitemOpen
  \bibfield  {author} {\bibinfo {author} {\bibfnamefont {F.}~\bibnamefont
  {Shahbeigi}}, \bibinfo {author} {\bibfnamefont {D.}~\bibnamefont
  {Amaro-Alcal{\'a}}}, \bibinfo {author} {\bibfnamefont {Z.}~\bibnamefont
  {Pucha{\l}a}}, \ and\ \bibinfo {author} {\bibfnamefont {K.}~\bibnamefont
  {{\.Z}yczkowski}},\ }\href {http://arxiv.org/abs/2003.12184} {\bibfield
  {journal} {\bibinfo  {journal} {arXiv:2003.12184}\ } (\bibinfo {year}
  {2020})}\BibitemShut {NoStop}%
\bibitem [{\citenamefont {Lindblad}(1974)}]{lindblad_expectations_1974}%
  \BibitemOpen
  \bibfield  {author} {\bibinfo {author} {\bibfnamefont {G.}~\bibnamefont
  {Lindblad}},\ }\href {\doibase 10.1007/BF01608390} {\bibfield  {journal}
  {\bibinfo  {journal} {Commun. Math. Phys.}\ }\textbf {\bibinfo {volume}
  {39}},\ \bibinfo {pages} {111} (\bibinfo {year} {1974})}\BibitemShut
  {NoStop}%
\bibitem [{\citenamefont {Lindblad}(1975{\natexlab{b}})}]{Lindblad1975}%
  \BibitemOpen
  \bibfield  {author} {\bibinfo {author} {\bibfnamefont {G.}~\bibnamefont
  {Lindblad}},\ }\href {\doibase 10.1007/BF01609396} {\bibfield  {journal}
  {\bibinfo  {journal} {Commun. Math. Phys.}\ }\textbf {\bibinfo {volume}
  {40}},\ \bibinfo {pages} {147} (\bibinfo {year}
  {1975}{\natexlab{b}})}\BibitemShut {NoStop}%
\bibitem [{\citenamefont {Laine}\ \emph
  {et~al.}(2010{\natexlab{b}})\citenamefont {Laine}, \citenamefont {Piilo},\
  and\ \citenamefont {Breuer}}]{laine_measure_2010}%
  \BibitemOpen
  \bibfield  {author} {\bibinfo {author} {\bibfnamefont {E.-M.}\ \bibnamefont
  {Laine}}, \bibinfo {author} {\bibfnamefont {J.}~\bibnamefont {Piilo}}, \ and\
  \bibinfo {author} {\bibfnamefont {H.-P.}\ \bibnamefont {Breuer}},\ }\href
  {\doibase 10.1103/PhysRevA.81.062115} {\bibfield  {journal} {\bibinfo
  {journal} {Phys. Rev. A}\ }\textbf {\bibinfo {volume} {81}},\ \bibinfo
  {pages} {062115} (\bibinfo {year} {2010}{\natexlab{b}})}\BibitemShut
  {NoStop}%
\bibitem [{\citenamefont {Rivas}\ \emph
  {et~al.}(2010{\natexlab{b}})\citenamefont {Rivas}, \citenamefont {Huelga},\
  and\ \citenamefont {Plenio}}]{PhysRevLett.105.050403}%
  \BibitemOpen
  \bibfield  {author} {\bibinfo {author} {\bibfnamefont {A.}~\bibnamefont
  {Rivas}}, \bibinfo {author} {\bibfnamefont {S.~F.}\ \bibnamefont {Huelga}}, \
  and\ \bibinfo {author} {\bibfnamefont {M.~B.}\ \bibnamefont {Plenio}},\
  }\href {\doibase 10.1103/PhysRevLett.105.050403} {\bibfield  {journal}
  {\bibinfo  {journal} {Phys. Rev. Lett.}\ }\textbf {\bibinfo {volume} {105}},\
  \bibinfo {pages} {050403} (\bibinfo {year} {2010}{\natexlab{b}})}\BibitemShut
  {NoStop}%
\bibitem [{\citenamefont {Chru\ifmmode \acute{s}\else
  \'{s}\fi{}ci\ifmmode~\acute{n}\else \'{n}\fi{}ski}\ \emph
  {et~al.}(2011)\citenamefont {Chru\ifmmode \acute{s}\else
  \'{s}\fi{}ci\ifmmode~\acute{n}\else \'{n}\fi{}ski}, \citenamefont
  {Kossakowski},\ and\ \citenamefont {Rivas}}]{PhysRevA.83.052128}%
  \BibitemOpen
  \bibfield  {author} {\bibinfo {author} {\bibfnamefont {D.}~\bibnamefont
  {Chru\ifmmode \acute{s}\else \'{s}\fi{}ci\ifmmode~\acute{n}\else
  \'{n}\fi{}ski}}, \bibinfo {author} {\bibfnamefont {A.}~\bibnamefont
  {Kossakowski}}, \ and\ \bibinfo {author} {\bibfnamefont {A.}~\bibnamefont
  {Rivas}},\ }\href {\doibase 10.1103/PhysRevA.83.052128} {\bibfield  {journal}
  {\bibinfo  {journal} {Phys. Rev. A}\ }\textbf {\bibinfo {volume} {83}},\
  \bibinfo {pages} {052128} (\bibinfo {year} {2011})}\BibitemShut {NoStop}%
\bibitem [{\citenamefont {Addis}\ \emph {et~al.}(2014)\citenamefont {Addis},
  \citenamefont {Bylicka}, \citenamefont {Chru{\'{s}}ci{\'{n}}ski},\ and\
  \citenamefont {Maniscalco}}]{Addis2014}%
  \BibitemOpen
  \bibfield  {author} {\bibinfo {author} {\bibfnamefont {C.}~\bibnamefont
  {Addis}}, \bibinfo {author} {\bibfnamefont {B.}~\bibnamefont {Bylicka}},
  \bibinfo {author} {\bibfnamefont {D.}~\bibnamefont
  {Chru{\'{s}}ci{\'{n}}ski}}, \ and\ \bibinfo {author} {\bibfnamefont
  {S.}~\bibnamefont {Maniscalco}},\ }\href {\doibase
  10.1103/PhysRevA.90.052103} {\bibfield  {journal} {\bibinfo  {journal} {Phys.
  Rev. A}\ }\textbf {\bibinfo {volume} {90}},\ \bibinfo {pages} {052103}
  (\bibinfo {year} {2014})}\BibitemShut {NoStop}%
\bibitem [{\citenamefont {Bylicka}\ \emph {et~al.}(2017)\citenamefont
  {Bylicka}, \citenamefont {Johansson},\ and\ \citenamefont
  {Ac{\'\i}n}}]{bylicka_constructive_2017}%
  \BibitemOpen
  \bibfield  {author} {\bibinfo {author} {\bibfnamefont {B.}~\bibnamefont
  {Bylicka}}, \bibinfo {author} {\bibfnamefont {M.}~\bibnamefont {Johansson}},
  \ and\ \bibinfo {author} {\bibfnamefont {A.}~\bibnamefont {Ac{\'\i}n}},\
  }\href {\doibase 10.1103/PhysRevLett.118.120501} {\bibfield  {journal}
  {\bibinfo  {journal} {Phys. Rev. Lett.}\ }\textbf {\bibinfo {volume} {118}},\
  \bibinfo {pages} {120501} (\bibinfo {year} {2017})}\BibitemShut {NoStop}%
\bibitem [{\citenamefont {Megier}\ \emph {et~al.}(2017)\citenamefont {Megier},
  \citenamefont {Chru{\'s}ci{\'n}ski}, \citenamefont {Piilo},\ and\
  \citenamefont {Strunz}}]{megier_eternal_2017}%
  \BibitemOpen
  \bibfield  {author} {\bibinfo {author} {\bibfnamefont {N.}~\bibnamefont
  {Megier}}, \bibinfo {author} {\bibfnamefont {D.}~\bibnamefont
  {Chru{\'s}ci{\'n}ski}}, \bibinfo {author} {\bibfnamefont {J.}~\bibnamefont
  {Piilo}}, \ and\ \bibinfo {author} {\bibfnamefont {W.~T.}\ \bibnamefont
  {Strunz}},\ }\href {\doibase 10.1038/s41598-017-06059-5} {\bibfield
  {journal} {\bibinfo  {journal} {Sci. Rep.}\ }\textbf {\bibinfo {volume}
  {7}},\ \bibinfo {pages} {6379} (\bibinfo {year} {2017})}\BibitemShut
  {NoStop}%
\bibitem [{\citenamefont {De~Santis}\ \emph {et~al.}(2020)\citenamefont
  {De~Santis}, \citenamefont {Johansson}, \citenamefont {Bylicka},
  \citenamefont {Bernardes},\ and\ \citenamefont
  {Ac{\'\i}n}}]{de_santis_witnessing_2020}%
  \BibitemOpen
  \bibfield  {author} {\bibinfo {author} {\bibfnamefont {D.}~\bibnamefont
  {De~Santis}}, \bibinfo {author} {\bibfnamefont {M.}~\bibnamefont
  {Johansson}}, \bibinfo {author} {\bibfnamefont {B.}~\bibnamefont {Bylicka}},
  \bibinfo {author} {\bibfnamefont {N.~K.}\ \bibnamefont {Bernardes}}, \ and\
  \bibinfo {author} {\bibfnamefont {A.}~\bibnamefont {Ac{\'\i}n}},\ }\href
  {\doibase 10.1103/PhysRevA.102.012214} {\bibfield  {journal} {\bibinfo
  {journal} {Phys. Rev. A}\ }\textbf {\bibinfo {volume} {102}},\ \bibinfo
  {pages} {012214} (\bibinfo {year} {2020})}\BibitemShut {NoStop}%
\bibitem [{Note14()}]{Note14}%
  \BibitemOpen
  \bibinfo {note} {We have already argued that the two are equivalent, so only
  one will suffice.}\BibitemShut {Stop}%
\bibitem [{\citenamefont {Alonso}\ and\ \citenamefont
  {de~Vega}(2005)}]{PhysRevLett.94.200403}%
  \BibitemOpen
  \bibfield  {author} {\bibinfo {author} {\bibfnamefont {D.}~\bibnamefont
  {Alonso}}\ and\ \bibinfo {author} {\bibfnamefont {I.}~\bibnamefont
  {de~Vega}},\ }\href {\doibase 10.1103/PhysRevLett.94.200403} {\bibfield
  {journal} {\bibinfo  {journal} {Phys. Rev. Lett.}\ }\textbf {\bibinfo
  {volume} {94}},\ \bibinfo {pages} {200403} (\bibinfo {year}
  {2005})}\BibitemShut {NoStop}%
\bibitem [{\citenamefont {de~Vega}\ and\ \citenamefont
  {Alonso}(2006)}]{PhysRevA.73.022102}%
  \BibitemOpen
  \bibfield  {author} {\bibinfo {author} {\bibfnamefont {I.}~\bibnamefont
  {de~Vega}}\ and\ \bibinfo {author} {\bibfnamefont {D.}~\bibnamefont
  {Alonso}},\ }\href {\doibase 10.1103/PhysRevA.73.022102} {\bibfield
  {journal} {\bibinfo  {journal} {Phys. Rev. A}\ }\textbf {\bibinfo {volume}
  {73}},\ \bibinfo {pages} {022102} (\bibinfo {year} {2006})}\BibitemShut
  {NoStop}%
\bibitem [{\citenamefont {Smirne}\ \emph {et~al.}(2018)\citenamefont {Smirne},
  \citenamefont {Egloff}, \citenamefont {D{\'\i}az}, \citenamefont {Plenio},\
  and\ \citenamefont {Huelga}}]{smirne_coherence_2017}%
  \BibitemOpen
  \bibfield  {author} {\bibinfo {author} {\bibfnamefont {A.}~\bibnamefont
  {Smirne}}, \bibinfo {author} {\bibfnamefont {D.}~\bibnamefont {Egloff}},
  \bibinfo {author} {\bibfnamefont {M.~G.}\ \bibnamefont {D{\'\i}az}}, \bibinfo
  {author} {\bibfnamefont {M.~B.}\ \bibnamefont {Plenio}}, \ and\ \bibinfo
  {author} {\bibfnamefont {S.~F.}\ \bibnamefont {Huelga}},\ }\href {\doibase
  10.1088/2058-9565/aaebd5} {\bibfield  {journal} {\bibinfo  {journal} {Quantum
  Sci. Technol.}\ }\textbf {\bibinfo {volume} {4}},\ \bibinfo {pages} {01LT01}
  (\bibinfo {year} {2018})}\BibitemShut {NoStop}%
\bibitem [{\citenamefont {Lambert}\ \emph {et~al.}(2013)\citenamefont
  {Lambert}, \citenamefont {Chen}, \citenamefont {Cheng}, \citenamefont {Li},
  \citenamefont {Chen},\ and\ \citenamefont {Nori}}]{Lambert:2013xi}%
  \BibitemOpen
  \bibfield  {author} {\bibinfo {author} {\bibfnamefont {N.}~\bibnamefont
  {Lambert}}, \bibinfo {author} {\bibfnamefont {Y.-N.}\ \bibnamefont {Chen}},
  \bibinfo {author} {\bibfnamefont {Y.-C.}\ \bibnamefont {Cheng}}, \bibinfo
  {author} {\bibfnamefont {C.-M.}\ \bibnamefont {Li}}, \bibinfo {author}
  {\bibfnamefont {G.-Y.}\ \bibnamefont {Chen}}, \ and\ \bibinfo {author}
  {\bibfnamefont {F.}~\bibnamefont {Nori}},\ }\href
  {http://dx.doi.org/10.1038/nphys2474} {\bibfield  {journal} {\bibinfo
  {journal} {Nat. Phys.}\ }\textbf {\bibinfo {volume} {9}},\ \bibinfo {pages}
  {10} (\bibinfo {year} {2013})}\BibitemShut {NoStop}%
\bibitem [{\citenamefont {Engel}\ \emph {et~al.}(2007)\citenamefont {Engel},
  \citenamefont {Calhoun}, \citenamefont {Read}, \citenamefont {Ahn},
  \citenamefont {Man{\v c}al}, \citenamefont {Cheng}, \citenamefont
  {Blankenship},\ and\ \citenamefont {Fleming}}]{fmo1}%
  \BibitemOpen
  \bibfield  {author} {\bibinfo {author} {\bibfnamefont {G.~S.}\ \bibnamefont
  {Engel}}, \bibinfo {author} {\bibfnamefont {T.~R.}\ \bibnamefont {Calhoun}},
  \bibinfo {author} {\bibfnamefont {E.~L.}\ \bibnamefont {Read}}, \bibinfo
  {author} {\bibfnamefont {T.-K.}\ \bibnamefont {Ahn}}, \bibinfo {author}
  {\bibfnamefont {T.}~\bibnamefont {Man{\v c}al}}, \bibinfo {author}
  {\bibfnamefont {Y.-C.}\ \bibnamefont {Cheng}}, \bibinfo {author}
  {\bibfnamefont {R.~E.}\ \bibnamefont {Blankenship}}, \ and\ \bibinfo {author}
  {\bibfnamefont {G.~R.}\ \bibnamefont {Fleming}},\ }\href {\doibase
  10.1038/nature05678} {\bibfield  {journal} {\bibinfo  {journal} {Nature}\
  }\textbf {\bibinfo {volume} {446}},\ \bibinfo {pages} {782} (\bibinfo {year}
  {2007})}\BibitemShut {NoStop}%
\bibitem [{\citenamefont {Plenio}\ and\ \citenamefont
  {Huelga}(2008)}]{plenio_dephasing-assisted_2008}%
  \BibitemOpen
  \bibfield  {author} {\bibinfo {author} {\bibfnamefont {M.~B.}\ \bibnamefont
  {Plenio}}\ and\ \bibinfo {author} {\bibfnamefont {S.~F.}\ \bibnamefont
  {Huelga}},\ }\href {\doibase 10.1088/1367-2630/10/11/113019} {\bibfield
  {journal} {\bibinfo  {journal} {New J. Phys.}\ }\textbf {\bibinfo {volume}
  {10}},\ \bibinfo {pages} {113019} (\bibinfo {year} {2008})}\BibitemShut
  {NoStop}%
\bibitem [{\citenamefont {Chin}\ \emph {et~al.}(2010)\citenamefont {Chin},
  \citenamefont {Datta}, \citenamefont {Caruso}, \citenamefont {Huelga},\ and\
  \citenamefont {Plenio}}]{fmo2}%
  \BibitemOpen
  \bibfield  {author} {\bibinfo {author} {\bibfnamefont {A.~W.}\ \bibnamefont
  {Chin}}, \bibinfo {author} {\bibfnamefont {A.}~\bibnamefont {Datta}},
  \bibinfo {author} {\bibfnamefont {F.}~\bibnamefont {Caruso}}, \bibinfo
  {author} {\bibfnamefont {S.~F.}\ \bibnamefont {Huelga}}, \ and\ \bibinfo
  {author} {\bibfnamefont {M.~B.}\ \bibnamefont {Plenio}},\ }\href {\doibase
  10.1088/1367-2630/12/6/065002} {\bibfield  {journal} {\bibinfo  {journal}
  {New J. Phys.}\ }\textbf {\bibinfo {volume} {12}},\ \bibinfo {pages} {065002}
  (\bibinfo {year} {2010})}\BibitemShut {NoStop}%
\bibitem [{\citenamefont {Mohseni}\ \emph {et~al.}(2008)\citenamefont
  {Mohseni}, \citenamefont {Rebentrost}, \citenamefont {Lloyd},\ and\
  \citenamefont {Aspuru-Guzik}}]{fmo3}%
  \BibitemOpen
  \bibfield  {author} {\bibinfo {author} {\bibfnamefont {M.}~\bibnamefont
  {Mohseni}}, \bibinfo {author} {\bibfnamefont {P.}~\bibnamefont {Rebentrost}},
  \bibinfo {author} {\bibfnamefont {S.}~\bibnamefont {Lloyd}}, \ and\ \bibinfo
  {author} {\bibfnamefont {A.}~\bibnamefont {Aspuru-Guzik}},\ }\href {\doibase
  10.1063/1.3002335} {\bibfield  {journal} {\bibinfo  {journal} {J. Chem.
  Phys.}\ }\textbf {\bibinfo {volume} {129}},\ \bibinfo {pages} {174106}
  (\bibinfo {year} {2008})}\BibitemShut {NoStop}%
\bibitem [{\citenamefont {Caruso}\ \emph
  {et~al.}(2010{\natexlab{a}})\citenamefont {Caruso}, \citenamefont {Chin},
  \citenamefont {Datta}, \citenamefont {Huelga},\ and\ \citenamefont
  {Plenio}}]{caruso_entanglement_2010}%
  \BibitemOpen
  \bibfield  {author} {\bibinfo {author} {\bibfnamefont {F.}~\bibnamefont
  {Caruso}}, \bibinfo {author} {\bibfnamefont {A.~W.}\ \bibnamefont {Chin}},
  \bibinfo {author} {\bibfnamefont {A.}~\bibnamefont {Datta}}, \bibinfo
  {author} {\bibfnamefont {S.~F.}\ \bibnamefont {Huelga}}, \ and\ \bibinfo
  {author} {\bibfnamefont {M.~B.}\ \bibnamefont {Plenio}},\ }\href {\doibase
  10.1103/PhysRevA.81.062346} {\bibfield  {journal} {\bibinfo  {journal} {Phys.
  Rev. A}\ }\textbf {\bibinfo {volume} {81}},\ \bibinfo {pages} {062346}
  (\bibinfo {year} {2010}{\natexlab{a}})}\BibitemShut {NoStop}%
\bibitem [{\citenamefont {Nickerson}\ and\ \citenamefont
  {Brown}(2019)}]{brown}%
  \BibitemOpen
  \bibfield  {author} {\bibinfo {author} {\bibfnamefont {N.~H.}\ \bibnamefont
  {Nickerson}}\ and\ \bibinfo {author} {\bibfnamefont {B.~J.}\ \bibnamefont
  {Brown}},\ }\href {\doibase 10.22331/q-2019-04-08-131} {\bibfield  {journal}
  {\bibinfo  {journal} {{Quantum}}\ }\textbf {\bibinfo {volume} {3}},\ \bibinfo
  {pages} {131} (\bibinfo {year} {2019})}\BibitemShut {NoStop}%
\bibitem [{\citenamefont {Blume-Kohout}\ \emph {et~al.}(2017)\citenamefont
  {Blume-Kohout}, \citenamefont {Gamble}, \citenamefont {Nielsen},
  \citenamefont {Rudinger}, \citenamefont {Mizrahi}, \citenamefont {Fortier},\
  and\ \citenamefont {Maunz}}]{robin}%
  \BibitemOpen
  \bibfield  {author} {\bibinfo {author} {\bibfnamefont {R.}~\bibnamefont
  {Blume-Kohout}}, \bibinfo {author} {\bibfnamefont {J.~K.}\ \bibnamefont
  {Gamble}}, \bibinfo {author} {\bibfnamefont {E.}~\bibnamefont {Nielsen}},
  \bibinfo {author} {\bibfnamefont {K.}~\bibnamefont {Rudinger}}, \bibinfo
  {author} {\bibfnamefont {J.}~\bibnamefont {Mizrahi}}, \bibinfo {author}
  {\bibfnamefont {K.}~\bibnamefont {Fortier}}, \ and\ \bibinfo {author}
  {\bibfnamefont {P.}~\bibnamefont {Maunz}},\ }\href {\doibase
  10.1038/ncomms14485} {\bibfield  {journal} {\bibinfo  {journal} {Nat.
  Commun.}\ }\textbf {\bibinfo {volume} {8}},\ \bibinfo {pages} {1} (\bibinfo
  {year} {2017})}\BibitemShut {NoStop}%
\bibitem [{\citenamefont {Harper}\ \emph {et~al.}(2020)\citenamefont {Harper},
  \citenamefont {Flammia},\ and\ \citenamefont {Wallman}}]{flammia}%
  \BibitemOpen
  \bibfield  {author} {\bibinfo {author} {\bibfnamefont {R.}~\bibnamefont
  {Harper}}, \bibinfo {author} {\bibfnamefont {S.~T.}\ \bibnamefont {Flammia}},
  \ and\ \bibinfo {author} {\bibfnamefont {J.~J.}\ \bibnamefont {Wallman}},\
  }\href {\doibase 10.1038/s41567-020-0992-8} {\bibfield  {journal} {\bibinfo
  {journal} {Nat. Phys.}\ ,\ \bibinfo {pages} {1}} (\bibinfo {year}
  {2020})}\BibitemShut {NoStop}%
\bibitem [{\citenamefont {White}\ \emph {et~al.}(2020)\citenamefont {White},
  \citenamefont {Hill}, \citenamefont {Pollock}, \citenamefont {Hollenberg},\
  and\ \citenamefont {Modi}}]{White2020}%
  \BibitemOpen
  \bibfield  {author} {\bibinfo {author} {\bibfnamefont {G.~A.~L.}\
  \bibnamefont {White}}, \bibinfo {author} {\bibfnamefont {C.~D.}\ \bibnamefont
  {Hill}}, \bibinfo {author} {\bibfnamefont {F.~A.}\ \bibnamefont {Pollock}},
  \bibinfo {author} {\bibfnamefont {L.~C.~L.}\ \bibnamefont {Hollenberg}}, \
  and\ \bibinfo {author} {\bibfnamefont {K.}~\bibnamefont {Modi}},\ }\href@noop
  {} {\bibfield  {journal} {\bibinfo  {journal} {Nat. Comm.}\ ,\ \bibinfo
  {pages} {6301}} (\bibinfo {year} {2020})}\BibitemShut {NoStop}%
\bibitem [{\citenamefont {McKay}\ \emph {et~al.}(2020)\citenamefont {McKay},
  \citenamefont {Cross}, \citenamefont {Wood},\ and\ \citenamefont
  {Gambetta}}]{crosstalk}%
  \BibitemOpen
  \bibfield  {author} {\bibinfo {author} {\bibfnamefont {D.~C.}\ \bibnamefont
  {McKay}}, \bibinfo {author} {\bibfnamefont {A.~W.}\ \bibnamefont {Cross}},
  \bibinfo {author} {\bibfnamefont {C.~J.}\ \bibnamefont {Wood}}, \ and\
  \bibinfo {author} {\bibfnamefont {J.~M.}\ \bibnamefont {Gambetta}},\ }\href
  {https://www.arxiv.org/abs/2003.02354} {\bibfield  {journal} {\bibinfo
  {journal} {arXiv:2003.02354}\ } (\bibinfo {year} {2020})}\BibitemShut
  {NoStop}%
\bibitem [{\citenamefont {Accardi}(1981)}]{accardi_topics_1981}%
  \BibitemOpen
  \bibfield  {author} {\bibinfo {author} {\bibfnamefont {L.}~\bibnamefont
  {Accardi}},\ }\href {\doibase 10.1016/0370-1573(81)90070-3} {\bibfield
  {journal} {\bibinfo  {journal} {Phys. Rep.}\ }\textbf {\bibinfo {volume}
  {77}},\ \bibinfo {pages} {169} (\bibinfo {year} {1981})}\BibitemShut
  {NoStop}%
\bibitem [{\citenamefont {Milz}\ \emph
  {et~al.}(2020{\natexlab{a}})\citenamefont {Milz}, \citenamefont {Sakuldee},
  \citenamefont {Pollock},\ and\ \citenamefont {Modi}}]{Milz2017KET}%
  \BibitemOpen
  \bibfield  {author} {\bibinfo {author} {\bibfnamefont {S.}~\bibnamefont
  {Milz}}, \bibinfo {author} {\bibfnamefont {F.}~\bibnamefont {Sakuldee}},
  \bibinfo {author} {\bibfnamefont {F.~A.}\ \bibnamefont {Pollock}}, \ and\
  \bibinfo {author} {\bibfnamefont {K.}~\bibnamefont {Modi}},\ }\href {\doibase
  10.22331/q-2020-04-20-255} {\bibfield  {journal} {\bibinfo  {journal}
  {{Quantum}}\ }\textbf {\bibinfo {volume} {4}},\ \bibinfo {pages} {255}
  (\bibinfo {year} {2020}{\natexlab{a}})}\BibitemShut {NoStop}%
\bibitem [{\citenamefont {Shrapnel}\ and\ \citenamefont
  {Costa}(2018)}]{shrapnel_causation_2018}%
  \BibitemOpen
  \bibfield  {author} {\bibinfo {author} {\bibfnamefont {S.}~\bibnamefont
  {Shrapnel}}\ and\ \bibinfo {author} {\bibfnamefont {F.}~\bibnamefont
  {Costa}},\ }\href {\doibase 10.22331/q-2018-05-18-63} {\bibfield  {journal}
  {\bibinfo  {journal} {Quantum}\ }\textbf {\bibinfo {volume} {2}},\ \bibinfo
  {pages} {63} (\bibinfo {year} {2018})}\BibitemShut {NoStop}%
\bibitem [{\citenamefont {Barnett}\ and\ \citenamefont
  {Crutchfield}(2015)}]{barnett_computational_2015}%
  \BibitemOpen
  \bibfield  {author} {\bibinfo {author} {\bibfnamefont {N.}~\bibnamefont
  {Barnett}}\ and\ \bibinfo {author} {\bibfnamefont {J.~P.}\ \bibnamefont
  {Crutchfield}},\ }\href {\doibase 10.1007/s10955-015-1327-5} {\bibfield
  {journal} {\bibinfo  {journal} {J. Stat. Phys.}\ }\textbf {\bibinfo {volume}
  {161}},\ \bibinfo {pages} {404} (\bibinfo {year} {2015})}\BibitemShut
  {NoStop}%
\bibitem [{\citenamefont {Pearl}(2009)}]{pearl_causality_2009}%
  \BibitemOpen
  \bibfield  {author} {\bibinfo {author} {\bibfnamefont {J.}~\bibnamefont
  {Pearl}},\ }\href@noop {} {\emph {\bibinfo {title} {Causality: {Models},
  {Reasoning} and {Inference}}}}\ (\bibinfo  {publisher} {Cambridge University
  Press},\ \bibinfo {address} {Cambridge, U.K.; New York},\ \bibinfo {year}
  {2009})\BibitemShut {NoStop}%
\bibitem [{\citenamefont {Leggett}\ and\ \citenamefont
  {Garg}(1985)}]{leggett_quantum_1985}%
  \BibitemOpen
  \bibfield  {author} {\bibinfo {author} {\bibfnamefont {A.~J.}\ \bibnamefont
  {Leggett}}\ and\ \bibinfo {author} {\bibfnamefont {A.}~\bibnamefont {Garg}},\
  }\href {\doibase 10.1103/PhysRevLett.54.857} {\bibfield  {journal} {\bibinfo
  {journal} {Phys. Rev. Lett.}\ }\textbf {\bibinfo {volume} {54}},\ \bibinfo
  {pages} {857} (\bibinfo {year} {1985})}\BibitemShut {NoStop}%
\bibitem [{\citenamefont {Emary}\ \emph {et~al.}(2014)\citenamefont {Emary},
  \citenamefont {Lambert},\ and\ \citenamefont
  {Nori}}]{emary_leggettgarg_2014}%
  \BibitemOpen
  \bibfield  {author} {\bibinfo {author} {\bibfnamefont {C.}~\bibnamefont
  {Emary}}, \bibinfo {author} {\bibfnamefont {N.}~\bibnamefont {Lambert}}, \
  and\ \bibinfo {author} {\bibfnamefont {F.}~\bibnamefont {Nori}},\ }\href
  {\doibase 10.1088/0034-4885/77/1/016001} {\bibfield  {journal} {\bibinfo
  {journal} {Rep. Prog. Phys.}\ }\textbf {\bibinfo {volume} {77}},\ \bibinfo
  {pages} {016001} (\bibinfo {year} {2014})}\BibitemShut {NoStop}%
\bibitem [{\citenamefont {Budroni}\ \emph {et~al.}(2019)\citenamefont
  {Budroni}, \citenamefont {Fagundes},\ and\ \citenamefont
  {Kleinmann}}]{budroni_memory_2019}%
  \BibitemOpen
  \bibfield  {author} {\bibinfo {author} {\bibfnamefont {C.}~\bibnamefont
  {Budroni}}, \bibinfo {author} {\bibfnamefont {G.}~\bibnamefont {Fagundes}}, \
  and\ \bibinfo {author} {\bibfnamefont {M.}~\bibnamefont {Kleinmann}},\ }\href
  {\doibase 10.1088/1367-2630/ab3cb4} {\bibfield  {journal} {\bibinfo
  {journal} {New J. Phys.}\ }\textbf {\bibinfo {volume} {21}},\ \bibinfo
  {pages} {093018} (\bibinfo {year} {2019})},\ \bibinfo {note} {publisher: IOP
  Publishing}\BibitemShut {NoStop}%
\bibitem [{\citenamefont {Chiribella}\ \emph
  {et~al.}(2008{\natexlab{a}})\citenamefont {Chiribella}, \citenamefont
  {D'Ariano},\ and\ \citenamefont {Perinotti}}]{chiribella_memory_2008}%
  \BibitemOpen
  \bibfield  {author} {\bibinfo {author} {\bibfnamefont {G.}~\bibnamefont
  {Chiribella}}, \bibinfo {author} {\bibfnamefont {G.~M.}\ \bibnamefont
  {D'Ariano}}, \ and\ \bibinfo {author} {\bibfnamefont {P.}~\bibnamefont
  {Perinotti}},\ }\href {\doibase 10.1103/PhysRevLett.101.180501} {\bibfield
  {journal} {\bibinfo  {journal} {Phys. Rev. Lett.}\ }\textbf {\bibinfo
  {volume} {101}},\ \bibinfo {pages} {180501} (\bibinfo {year}
  {2008}{\natexlab{a}})}\BibitemShut {NoStop}%
\bibitem [{\citenamefont {Shrapnel}\ \emph
  {et~al.}(2018{\natexlab{a}})\citenamefont {Shrapnel}, \citenamefont {Costa},\
  and\ \citenamefont {Milburn}}]{Shrapnel_2018}%
  \BibitemOpen
  \bibfield  {author} {\bibinfo {author} {\bibfnamefont {S.}~\bibnamefont
  {Shrapnel}}, \bibinfo {author} {\bibfnamefont {F.}~\bibnamefont {Costa}}, \
  and\ \bibinfo {author} {\bibfnamefont {G.}~\bibnamefont {Milburn}},\ }\href
  {\doibase 10.1088/1367-2630/aabe12} {\bibfield  {journal} {\bibinfo
  {journal} {New J. Phys.}\ }\textbf {\bibinfo {volume} {20}},\ \bibinfo
  {pages} {053010} (\bibinfo {year} {2018}{\natexlab{a}})}\BibitemShut
  {NoStop}%
\bibitem [{\citenamefont {Modi}(2012)}]{modiscirep}%
  \BibitemOpen
  \bibfield  {author} {\bibinfo {author} {\bibfnamefont {K.}~\bibnamefont
  {Modi}},\ }\href {http://www.nature.com/articles/srep00581} {\bibfield
  {journal} {\bibinfo  {journal} {Sci. Rep.}\ }\textbf {\bibinfo {volume}
  {2}},\ \bibinfo {pages} {581} (\bibinfo {year} {2012})}\BibitemShut {NoStop}%
\bibitem [{\citenamefont {Davies}\ and\ \citenamefont
  {Lewis}(1970)}]{Davies:1970ez}%
  \BibitemOpen
  \bibfield  {author} {\bibinfo {author} {\bibfnamefont {E.}~\bibnamefont
  {Davies}}\ and\ \bibinfo {author} {\bibfnamefont {J.}~\bibnamefont {Lewis}},\
  }\href {\doibase 10.1007/BF01647093} {\bibfield  {journal} {\bibinfo
  {journal} {Commun. Math. Phys.}\ }\textbf {\bibinfo {volume} {17}},\ \bibinfo
  {pages} {239} (\bibinfo {year} {1970})}\BibitemShut {NoStop}%
\bibitem [{\citenamefont {Davies}(1976)}]{davies76a}%
  \BibitemOpen
  \bibfield  {author} {\bibinfo {author} {\bibfnamefont {E.~B.}\ \bibnamefont
  {Davies}},\ }\href@noop {} {\emph {\bibinfo {title} {Quantum {Theory} of
  {Open} {Systems}}}}\ (\bibinfo  {publisher} {Academic Press Inc},\ \bibinfo
  {address} {London; New York},\ \bibinfo {year} {1976})\BibitemShut {NoStop}%
\bibitem [{\citenamefont {Lindblad}(1979)}]{lindblad_non-markovian_1979}%
  \BibitemOpen
  \bibfield  {author} {\bibinfo {author} {\bibfnamefont {G.}~\bibnamefont
  {Lindblad}},\ }\href {\doibase 10.1007/BF01197883} {\bibfield  {journal}
  {\bibinfo  {journal} {Commun. Math. Phys.}\ }\textbf {\bibinfo {volume}
  {65}},\ \bibinfo {pages} {281} (\bibinfo {year} {1979})}\BibitemShut
  {NoStop}%
\bibitem [{\citenamefont {Davies}(1969)}]{Davies1969}%
  \BibitemOpen
  \bibfield  {author} {\bibinfo {author} {\bibfnamefont {E.~B.}\ \bibnamefont
  {Davies}},\ }\href {\doibase 10.1007/BF01645529} {\bibfield  {journal}
  {\bibinfo  {journal} {Commun. Math. Phys.}\ }\textbf {\bibinfo {volume}
  {15}},\ \bibinfo {pages} {277} (\bibinfo {year} {1969})}\BibitemShut
  {NoStop}%
\bibitem [{\citenamefont {Ringbauer}\ \emph {et~al.}(2015)\citenamefont
  {Ringbauer}, \citenamefont {Wood}, \citenamefont {Modi}, \citenamefont
  {Gilchrist}, \citenamefont {White},\ and\ \citenamefont
  {Fedrizzi}}]{PhysRevLett.114.090402}%
  \BibitemOpen
  \bibfield  {author} {\bibinfo {author} {\bibfnamefont {M.}~\bibnamefont
  {Ringbauer}}, \bibinfo {author} {\bibfnamefont {C.~J.}\ \bibnamefont {Wood}},
  \bibinfo {author} {\bibfnamefont {K.}~\bibnamefont {Modi}}, \bibinfo {author}
  {\bibfnamefont {A.}~\bibnamefont {Gilchrist}}, \bibinfo {author}
  {\bibfnamefont {A.~G.}\ \bibnamefont {White}}, \ and\ \bibinfo {author}
  {\bibfnamefont {A.}~\bibnamefont {Fedrizzi}},\ }\href {\doibase
  10.1103/PhysRevLett.114.090402} {\bibfield  {journal} {\bibinfo  {journal}
  {Phys. Rev. Lett.}\ }\textbf {\bibinfo {volume} {114}},\ \bibinfo {pages}
  {090402} (\bibinfo {year} {2015})}\BibitemShut {NoStop}%
\bibitem [{\citenamefont {Paz-Silva}\ \emph {et~al.}(2019)\citenamefont
  {Paz-Silva}, \citenamefont {Hall},\ and\ \citenamefont
  {Wiseman}}]{PhysRevA.100.042120}%
  \BibitemOpen
  \bibfield  {author} {\bibinfo {author} {\bibfnamefont {G.~A.}\ \bibnamefont
  {Paz-Silva}}, \bibinfo {author} {\bibfnamefont {M.~J.~W.}\ \bibnamefont
  {Hall}}, \ and\ \bibinfo {author} {\bibfnamefont {H.~M.}\ \bibnamefont
  {Wiseman}},\ }\href {\doibase 10.1103/PhysRevA.100.042120} {\bibfield
  {journal} {\bibinfo  {journal} {Phys. Rev. A}\ }\textbf {\bibinfo {volume}
  {100}},\ \bibinfo {pages} {042120} (\bibinfo {year} {2019})}\BibitemShut
  {NoStop}%
\bibitem [{Note15()}]{Note15}%
  \BibitemOpen
  \bibinfo {note} {The singular point state is contained in $\protect \mathcal
  {T}_{(t:0)}$ and can be obtained by tracing over spaces $\protect \mathcal
  {H}_0^{\protect \texttt {o}} \otimes \protect \mathcal {H}_1^\protect \texttt
  {i}$.}\BibitemShut {Stop}%
\bibitem [{\citenamefont {Holevo}(2001)}]{holevo_statistical_2001}%
  \BibitemOpen
  \bibfield  {author} {\bibinfo {author} {\bibfnamefont {A.~S.}\ \bibnamefont
  {Holevo}},\ }\href {\doibase 10.1007/3-540-44998-1} {\emph {\bibinfo {title}
  {Statistical {Structure} of {Quantum} {Theory}}}},\ Lecture {Notes} in
  {Physics} {Monographs}\ (\bibinfo  {publisher} {Springer-Verlag},\ \bibinfo
  {address} {Berlin Heidelberg},\ \bibinfo {year} {2001})\BibitemShut {NoStop}%
\bibitem [{\citenamefont {Barchielli}\ and\ \citenamefont
  {Gregoratti}(2009)}]{barchielli_quantum_2009}%
  \BibitemOpen
  \bibfield  {author} {\bibinfo {author} {\bibfnamefont {A.}~\bibnamefont
  {Barchielli}}\ and\ \bibinfo {author} {\bibfnamefont {M.}~\bibnamefont
  {Gregoratti}},\ }\href {\doibase 10.1007/978-3-642-01298-3} {\emph {\bibinfo
  {title} {Quantum {Trajectories} and {Measurements} in {Continuous} {Time}:
  {The} {Diffusive} {Case}}}},\ Lecture {Notes} in {Physics}\ (\bibinfo
  {publisher} {Springer-Verlag},\ \bibinfo {address} {Berlin Heidelberg},\
  \bibinfo {year} {2009})\BibitemShut {NoStop}%
\bibitem [{\citenamefont {Pollock}\ \emph
  {et~al.}(2018{\natexlab{a}})\citenamefont {Pollock}, \citenamefont
  {Rodr\'{\i}guez-Rosario}, \citenamefont {Frauenheim}, \citenamefont
  {Paternostro},\ and\ \citenamefont {Modi}}]{PhysRevA.97.012127}%
  \BibitemOpen
  \bibfield  {author} {\bibinfo {author} {\bibfnamefont {F.~A.}\ \bibnamefont
  {Pollock}}, \bibinfo {author} {\bibfnamefont {C.}~\bibnamefont
  {Rodr\'{\i}guez-Rosario}}, \bibinfo {author} {\bibfnamefont {T.}~\bibnamefont
  {Frauenheim}}, \bibinfo {author} {\bibfnamefont {M.}~\bibnamefont
  {Paternostro}}, \ and\ \bibinfo {author} {\bibfnamefont {K.}~\bibnamefont
  {Modi}},\ }\href {\doibase 10.1103/PhysRevA.97.012127} {\bibfield  {journal}
  {\bibinfo  {journal} {Phys. Rev. A}\ }\textbf {\bibinfo {volume} {97}},\
  \bibinfo {pages} {012127} (\bibinfo {year} {2018}{\natexlab{a}})}\BibitemShut
  {NoStop}%
\bibitem [{\citenamefont {Pollock}\ \emph
  {et~al.}(2018{\natexlab{b}})\citenamefont {Pollock}, \citenamefont
  {Rodr\'{\i}guez-Rosario}, \citenamefont {Frauenheim}, \citenamefont
  {Paternostro},\ and\ \citenamefont {Modi}}]{PhysRevLett.120.040405}%
  \BibitemOpen
  \bibfield  {author} {\bibinfo {author} {\bibfnamefont {F.~A.}\ \bibnamefont
  {Pollock}}, \bibinfo {author} {\bibfnamefont {C.}~\bibnamefont
  {Rodr\'{\i}guez-Rosario}}, \bibinfo {author} {\bibfnamefont {T.}~\bibnamefont
  {Frauenheim}}, \bibinfo {author} {\bibfnamefont {M.}~\bibnamefont
  {Paternostro}}, \ and\ \bibinfo {author} {\bibfnamefont {K.}~\bibnamefont
  {Modi}},\ }\href {\doibase 10.1103/PhysRevLett.120.040405} {\bibfield
  {journal} {\bibinfo  {journal} {Phys. Rev. Lett.}\ }\textbf {\bibinfo
  {volume} {120}},\ \bibinfo {pages} {040405} (\bibinfo {year}
  {2018}{\natexlab{b}})}\BibitemShut {NoStop}%
\bibitem [{\citenamefont {Costa}\ and\ \citenamefont
  {Shrapnel}(2016)}]{1367-2630-18-6-063032}%
  \BibitemOpen
  \bibfield  {author} {\bibinfo {author} {\bibfnamefont {F.}~\bibnamefont
  {Costa}}\ and\ \bibinfo {author} {\bibfnamefont {S.}~\bibnamefont
  {Shrapnel}},\ }\href {http://stacks.iop.org/1367-2630/18/i=6/a=063032}
  {\bibfield  {journal} {\bibinfo  {journal} {New J. Phys.}\ }\textbf {\bibinfo
  {volume} {18}},\ \bibinfo {pages} {063032} (\bibinfo {year}
  {2016})}\BibitemShut {NoStop}%
\bibitem [{\citenamefont {Milz}\ \emph {et~al.}(2018)\citenamefont {Milz},
  \citenamefont {Pollock},\ and\ \citenamefont {Modi}}]{Milz2018A}%
  \BibitemOpen
  \bibfield  {author} {\bibinfo {author} {\bibfnamefont {S.}~\bibnamefont
  {Milz}}, \bibinfo {author} {\bibfnamefont {F.~A.}\ \bibnamefont {Pollock}}, \
  and\ \bibinfo {author} {\bibfnamefont {K.}~\bibnamefont {Modi}},\ }\href
  {\doibase 10.1103/PhysRevA.98.012108} {\bibfield  {journal} {\bibinfo
  {journal} {Phys. Rev. A}\ }\textbf {\bibinfo {volume} {98}},\ \bibinfo
  {pages} {012108} (\bibinfo {year} {2018})}\BibitemShut {NoStop}%
\bibitem [{\citenamefont {Chiribella}\ \emph {et~al.}(2009)\citenamefont
  {Chiribella}, \citenamefont {D{'}Ariano},\ and\ \citenamefont
  {Perinotti}}]{chiribella_theoretical_2009}%
  \BibitemOpen
  \bibfield  {author} {\bibinfo {author} {\bibfnamefont {G.}~\bibnamefont
  {Chiribella}}, \bibinfo {author} {\bibfnamefont {G.~M.}\ \bibnamefont
  {D{'}Ariano}}, \ and\ \bibinfo {author} {\bibfnamefont {P.}~\bibnamefont
  {Perinotti}},\ }\href {\doibase 10.1103/PhysRevA.80.022339} {\bibfield
  {journal} {\bibinfo  {journal} {Phys. Rev. A}\ }\textbf {\bibinfo {volume}
  {80}},\ \bibinfo {pages} {022339} (\bibinfo {year} {2009})}\BibitemShut
  {NoStop}%
\bibitem [{\citenamefont {Taranto}\ \emph
  {et~al.}(2019{\natexlab{a}})\citenamefont {Taranto}, \citenamefont {Pollock},
  \citenamefont {Milz}, \citenamefont {Tomamichel},\ and\ \citenamefont
  {Modi}}]{PhysRevLett.122.140401}%
  \BibitemOpen
  \bibfield  {author} {\bibinfo {author} {\bibfnamefont {P.}~\bibnamefont
  {Taranto}}, \bibinfo {author} {\bibfnamefont {F.~A.}\ \bibnamefont
  {Pollock}}, \bibinfo {author} {\bibfnamefont {S.}~\bibnamefont {Milz}},
  \bibinfo {author} {\bibfnamefont {M.}~\bibnamefont {Tomamichel}}, \ and\
  \bibinfo {author} {\bibfnamefont {K.}~\bibnamefont {Modi}},\ }\href {\doibase
  10.1103/PhysRevLett.122.140401} {\bibfield  {journal} {\bibinfo  {journal}
  {Phys. Rev. Lett.}\ }\textbf {\bibinfo {volume} {122}},\ \bibinfo {pages}
  {140401} (\bibinfo {year} {2019}{\natexlab{a}})}\BibitemShut {NoStop}%
\bibitem [{\citenamefont {Taranto}\ \emph
  {et~al.}(2019{\natexlab{b}})\citenamefont {Taranto}, \citenamefont {Milz},
  \citenamefont {Pollock},\ and\ \citenamefont {Modi}}]{PhysRevA.99.042108}%
  \BibitemOpen
  \bibfield  {author} {\bibinfo {author} {\bibfnamefont {P.}~\bibnamefont
  {Taranto}}, \bibinfo {author} {\bibfnamefont {S.}~\bibnamefont {Milz}},
  \bibinfo {author} {\bibfnamefont {F.~A.}\ \bibnamefont {Pollock}}, \ and\
  \bibinfo {author} {\bibfnamefont {K.}~\bibnamefont {Modi}},\ }\href {\doibase
  10.1103/PhysRevA.99.042108} {\bibfield  {journal} {\bibinfo  {journal} {Phys.
  Rev. A}\ }\textbf {\bibinfo {volume} {99}},\ \bibinfo {pages} {042108}
  (\bibinfo {year} {2019}{\natexlab{b}})}\BibitemShut {NoStop}%
\bibitem [{\citenamefont {Taranto}\ \emph
  {et~al.}(2019{\natexlab{c}})\citenamefont {Taranto}, \citenamefont
  {Pollock},\ and\ \citenamefont {Modi}}]{Phil_MemStr}%
  \BibitemOpen
  \bibfield  {author} {\bibinfo {author} {\bibfnamefont {P.}~\bibnamefont
  {Taranto}}, \bibinfo {author} {\bibfnamefont {F.~A.}\ \bibnamefont
  {Pollock}}, \ and\ \bibinfo {author} {\bibfnamefont {K.}~\bibnamefont
  {Modi}},\ }\href {http://arxiv.org/abs/1907.12583} {\bibfield  {journal}
  {\bibinfo  {journal} {arXiv:1907.12583}\ } (\bibinfo {year}
  {2019}{\natexlab{c}})}\BibitemShut {NoStop}%
\bibitem [{\citenamefont {Giarmatzi}\ and\ \citenamefont
  {Costa}(2020)}]{arXiv:1811.03722}%
  \BibitemOpen
  \bibfield  {author} {\bibinfo {author} {\bibfnamefont {C.}~\bibnamefont
  {Giarmatzi}}\ and\ \bibinfo {author} {\bibfnamefont {F.}~\bibnamefont
  {Costa}},\ }\href {http://arxiv.org/abs/1811.03722} {\bibfield  {journal}
  {\bibinfo  {journal} {arXiv:1811.03722}\ } (\bibinfo {year}
  {2020})}\BibitemShut {NoStop}%
\bibitem [{Note16()}]{Note16}%
  \BibitemOpen
  \bibinfo {note} {This tensor, in general, is also a quantum comb, where the
  bond represents information fed forward through an ancillary
  system.}\BibitemShut {Stop}%
\bibitem [{\citenamefont {Strasberg}\ and\ \citenamefont
  {D\'{\i}az}(2019)}]{PhysRevA.100.022120}%
  \BibitemOpen
  \bibfield  {author} {\bibinfo {author} {\bibfnamefont {P.}~\bibnamefont
  {Strasberg}}\ and\ \bibinfo {author} {\bibfnamefont {M.~G.}\ \bibnamefont
  {D\'{\i}az}},\ }\href {\doibase 10.1103/PhysRevA.100.022120} {\bibfield
  {journal} {\bibinfo  {journal} {Phys. Rev. A}\ }\textbf {\bibinfo {volume}
  {100}},\ \bibinfo {pages} {022120} (\bibinfo {year} {2019})}\BibitemShut
  {NoStop}%
\bibitem [{\citenamefont {Milz}\ \emph
  {et~al.}(2020{\natexlab{b}})\citenamefont {Milz}, \citenamefont {Egloff},
  \citenamefont {Taranto}, \citenamefont {Theurer}, \citenamefont {Plenio},
  \citenamefont {Smirne},\ and\ \citenamefont {Huelga}}]{milz_when_2020}%
  \BibitemOpen
  \bibfield  {author} {\bibinfo {author} {\bibfnamefont {S.}~\bibnamefont
  {Milz}}, \bibinfo {author} {\bibfnamefont {D.}~\bibnamefont {Egloff}},
  \bibinfo {author} {\bibfnamefont {P.}~\bibnamefont {Taranto}}, \bibinfo
  {author} {\bibfnamefont {T.}~\bibnamefont {Theurer}}, \bibinfo {author}
  {\bibfnamefont {M.~B.}\ \bibnamefont {Plenio}}, \bibinfo {author}
  {\bibfnamefont {A.}~\bibnamefont {Smirne}}, \ and\ \bibinfo {author}
  {\bibfnamefont {S.~F.}\ \bibnamefont {Huelga}},\ }\href {\doibase
  10.1103/PhysRevX.10.041049} {\bibfield  {journal} {\bibinfo  {journal} {Phys.
  Rev. X}\ }\textbf {\bibinfo {volume} {10}},\ \bibinfo {pages} {041049}
  (\bibinfo {year} {2020}{\natexlab{b}})}\BibitemShut {NoStop}%
\bibitem [{\citenamefont {Chiribella}\ \emph
  {et~al.}(2008{\natexlab{b}})\citenamefont {Chiribella}, \citenamefont
  {D'Ariano},\ and\ \citenamefont {Perinotti}}]{supermaps}%
  \BibitemOpen
  \bibfield  {author} {\bibinfo {author} {\bibfnamefont {G.}~\bibnamefont
  {Chiribella}}, \bibinfo {author} {\bibfnamefont {G.~M.}\ \bibnamefont
  {D'Ariano}}, \ and\ \bibinfo {author} {\bibfnamefont {P.}~\bibnamefont
  {Perinotti}},\ }\href {http://stacks.iop.org/0295-5075/83/i=3/a=30004}
  {\bibfield  {journal} {\bibinfo  {journal} {EPL}\ }\textbf {\bibinfo {volume}
  {83}},\ \bibinfo {pages} {30004} (\bibinfo {year}
  {2008}{\natexlab{b}})}\BibitemShut {NoStop}%
\bibitem [{\citenamefont {Chiribella}\ \emph
  {et~al.}(2008{\natexlab{c}})\citenamefont {Chiribella}, \citenamefont
  {D{'}Ariano},\ and\ \citenamefont {Perinotti}}]{chiribella_quantum_2008}%
  \BibitemOpen
  \bibfield  {author} {\bibinfo {author} {\bibfnamefont {G.}~\bibnamefont
  {Chiribella}}, \bibinfo {author} {\bibfnamefont {G.~M.}\ \bibnamefont
  {D{'}Ariano}}, \ and\ \bibinfo {author} {\bibfnamefont {P.}~\bibnamefont
  {Perinotti}},\ }\href {\doibase 10.1103/PhysRevLett.101.060401} {\bibfield
  {journal} {\bibinfo  {journal} {Phys. Rev. Lett.}\ }\textbf {\bibinfo
  {volume} {101}},\ \bibinfo {pages} {060401} (\bibinfo {year}
  {2008}{\natexlab{c}})}\BibitemShut {NoStop}%
\bibitem [{\citenamefont {Kretschmann}\ and\ \citenamefont
  {Werner}(2005)}]{kretschmann_quantum_2005}%
  \BibitemOpen
  \bibfield  {author} {\bibinfo {author} {\bibfnamefont {D.}~\bibnamefont
  {Kretschmann}}\ and\ \bibinfo {author} {\bibfnamefont {R.~F.}\ \bibnamefont
  {Werner}},\ }\href {\doibase 10.1103/PhysRevA.72.062323} {\bibfield
  {journal} {\bibinfo  {journal} {Phys. Rev. A}\ }\textbf {\bibinfo {volume}
  {72}},\ \bibinfo {pages} {062323} (\bibinfo {year} {2005})}\BibitemShut
  {NoStop}%
\bibitem [{\citenamefont {Caruso}\ \emph {et~al.}(2014)\citenamefont {Caruso},
  \citenamefont {Giovannetti}, \citenamefont {Lupo},\ and\ \citenamefont
  {Mancini}}]{caruso_quantum_2014}%
  \BibitemOpen
  \bibfield  {author} {\bibinfo {author} {\bibfnamefont {F.}~\bibnamefont
  {Caruso}}, \bibinfo {author} {\bibfnamefont {V.}~\bibnamefont {Giovannetti}},
  \bibinfo {author} {\bibfnamefont {C.}~\bibnamefont {Lupo}}, \ and\ \bibinfo
  {author} {\bibfnamefont {S.}~\bibnamefont {Mancini}},\ }\href {\doibase
  10.1103/RevModPhys.86.1203} {\bibfield  {journal} {\bibinfo  {journal} {Rev.
  Mod. Phys.}\ }\textbf {\bibinfo {volume} {86}},\ \bibinfo {pages} {1203}
  (\bibinfo {year} {2014})}\BibitemShut {NoStop}%
\bibitem [{\citenamefont {Portmann}\ \emph {et~al.}(2017)\citenamefont
  {Portmann}, \citenamefont {Matt}, \citenamefont {Maurer}, \citenamefont
  {Renner},\ and\ \citenamefont {Tackmann}}]{portmann_causal_2015}%
  \BibitemOpen
  \bibfield  {author} {\bibinfo {author} {\bibfnamefont {C.}~\bibnamefont
  {Portmann}}, \bibinfo {author} {\bibfnamefont {C.}~\bibnamefont {Matt}},
  \bibinfo {author} {\bibfnamefont {U.}~\bibnamefont {Maurer}}, \bibinfo
  {author} {\bibfnamefont {R.}~\bibnamefont {Renner}}, \ and\ \bibinfo {author}
  {\bibfnamefont {B.}~\bibnamefont {Tackmann}},\ }\href {\doibase
  10.1109/TIT.2017.2676805} {\bibfield  {journal} {\bibinfo  {journal} {IEEE
  Trans. Inf. Theory}\ }\textbf {\bibinfo {volume} {63}},\ \bibinfo {pages}
  {3277} (\bibinfo {year} {2017})}\BibitemShut {NoStop}%
\bibitem [{\citenamefont {Hardy}(2016)}]{hardy_operational_2016}%
  \BibitemOpen
  \bibfield  {author} {\bibinfo {author} {\bibfnamefont {L.}~\bibnamefont
  {Hardy}},\ }\href {http://arxiv.org/abs/1608.06940} {\bibfield  {journal}
  {\bibinfo  {journal} {arXiv:1608.06940}\ } (\bibinfo {year}
  {2016})}\BibitemShut {NoStop}%
\bibitem [{\citenamefont {Hardy}(2012)}]{hardy_operator_2012}%
  \BibitemOpen
  \bibfield  {author} {\bibinfo {author} {\bibfnamefont {L.}~\bibnamefont
  {Hardy}},\ }\href {\doibase 10.1098/rsta.2011.0326} {\bibfield  {journal}
  {\bibinfo  {journal} {Phil. Trans. R. Soc. A}\ }\textbf {\bibinfo {volume}
  {370}},\ \bibinfo {pages} {3385} (\bibinfo {year} {2012})}\BibitemShut
  {NoStop}%
\bibitem [{\citenamefont {Cotler}\ \emph {et~al.}(2018)\citenamefont {Cotler},
  \citenamefont {Jian}, \citenamefont {Qi},\ and\ \citenamefont
  {Wilczek}}]{cotler_superdensity_2017}%
  \BibitemOpen
  \bibfield  {author} {\bibinfo {author} {\bibfnamefont {J.}~\bibnamefont
  {Cotler}}, \bibinfo {author} {\bibfnamefont {C.-M.}\ \bibnamefont {Jian}},
  \bibinfo {author} {\bibfnamefont {X.-L.}\ \bibnamefont {Qi}}, \ and\ \bibinfo
  {author} {\bibfnamefont {F.}~\bibnamefont {Wilczek}},\ }\href {\doibase
  10.1007/JHEP09(2018)093} {\bibfield  {journal} {\bibinfo  {journal} {J. High
  Energy Phys.}\ }\textbf {\bibinfo {volume} {2018}},\ \bibinfo {pages} {93}
  (\bibinfo {year} {2018})}\BibitemShut {NoStop}%
\bibitem [{\citenamefont {Oreshkov}\ \emph {et~al.}(2012)\citenamefont
  {Oreshkov}, \citenamefont {Costa},\ and\ \citenamefont
  {Brukner}}]{OreshkovETAL2012}%
  \BibitemOpen
  \bibfield  {author} {\bibinfo {author} {\bibfnamefont {O.}~\bibnamefont
  {Oreshkov}}, \bibinfo {author} {\bibfnamefont {F.}~\bibnamefont {Costa}}, \
  and\ \bibinfo {author} {\bibfnamefont {{\v C}.}~\bibnamefont {Brukner}},\
  }\href {\doibase 10.1038/ncomms2076} {\bibfield  {journal} {\bibinfo
  {journal} {Nat. Commun.}\ }\textbf {\bibinfo {volume} {3}},\ \bibinfo {pages}
  {1092} (\bibinfo {year} {2012})}\BibitemShut {NoStop}%
\bibitem [{\citenamefont {Oreshkov}\ and\ \citenamefont
  {Giarmatzi}(2016)}]{oreshkov_causal_2016}%
  \BibitemOpen
  \bibfield  {author} {\bibinfo {author} {\bibfnamefont {O.}~\bibnamefont
  {Oreshkov}}\ and\ \bibinfo {author} {\bibfnamefont {C.}~\bibnamefont
  {Giarmatzi}},\ }\href {\doibase 10.1088/1367-2630/18/9/093020} {\bibfield
  {journal} {\bibinfo  {journal} {New J. Phys.}\ }\textbf {\bibinfo {volume}
  {18}},\ \bibinfo {pages} {093020} (\bibinfo {year} {2016})}\BibitemShut
  {NoStop}%
\bibitem [{\citenamefont {Allen}\ \emph {et~al.}(2017)\citenamefont {Allen},
  \citenamefont {Barrett}, \citenamefont {Horsman}, \citenamefont {Lee},\ and\
  \citenamefont {Spekkens}}]{PhysRevX.7.031021}%
  \BibitemOpen
  \bibfield  {author} {\bibinfo {author} {\bibfnamefont {J.-M.~A.}\
  \bibnamefont {Allen}}, \bibinfo {author} {\bibfnamefont {J.}~\bibnamefont
  {Barrett}}, \bibinfo {author} {\bibfnamefont {D.~C.}\ \bibnamefont
  {Horsman}}, \bibinfo {author} {\bibfnamefont {C.~M.}\ \bibnamefont {Lee}}, \
  and\ \bibinfo {author} {\bibfnamefont {R.~W.}\ \bibnamefont {Spekkens}},\
  }\href {\doibase 10.1103/PhysRevX.7.031021} {\bibfield  {journal} {\bibinfo
  {journal} {Phys. Rev. X}\ }\textbf {\bibinfo {volume} {7}},\ \bibinfo {pages}
  {031021} (\bibinfo {year} {2017})}\BibitemShut {NoStop}%
\bibitem [{\citenamefont {Gutoski}\ and\ \citenamefont
  {Watrous}(2007)}]{gutoski2007toward}%
  \BibitemOpen
  \bibfield  {author} {\bibinfo {author} {\bibfnamefont {G.}~\bibnamefont
  {Gutoski}}\ and\ \bibinfo {author} {\bibfnamefont {J.}~\bibnamefont
  {Watrous}},\ }in\ \href@noop {} {\emph {\bibinfo {booktitle} {Proceedings of
  the thirty-ninth annual ACM symposium on Theory of computing}}}\ (\bibinfo
  {organization} {ACM},\ \bibinfo {year} {2007})\ pp.\ \bibinfo {pages}
  {565--574}\BibitemShut {NoStop}%
\bibitem [{\citenamefont {Gutoski}\ \emph {et~al.}(2018)\citenamefont
  {Gutoski}, \citenamefont {Rosmanis},\ and\ \citenamefont
  {Sikora}}]{Gutoski2018fidelityofquantum}%
  \BibitemOpen
  \bibfield  {author} {\bibinfo {author} {\bibfnamefont {G.}~\bibnamefont
  {Gutoski}}, \bibinfo {author} {\bibfnamefont {A.}~\bibnamefont {Rosmanis}}, \
  and\ \bibinfo {author} {\bibfnamefont {J.}~\bibnamefont {Sikora}},\ }\href
  {\doibase 10.22331/q-2018-09-03-89} {\bibfield  {journal} {\bibinfo
  {journal} {{Quantum}}\ }\textbf {\bibinfo {volume} {2}},\ \bibinfo {pages}
  {89} (\bibinfo {year} {2018})}\BibitemShut {NoStop}%
\bibitem [{\citenamefont {Accardi}\ \emph {et~al.}(1982)\citenamefont
  {Accardi}, \citenamefont {Frigerio},\ and\ \citenamefont {Lewis}}]{accardi}%
  \BibitemOpen
  \bibfield  {author} {\bibinfo {author} {\bibfnamefont {L.}~\bibnamefont
  {Accardi}}, \bibinfo {author} {\bibfnamefont {A.}~\bibnamefont {Frigerio}}, \
  and\ \bibinfo {author} {\bibfnamefont {J.~T.}\ \bibnamefont {Lewis}},\ }\href
  {\doibase 10.2977/prims/1195184017} {\bibfield  {journal} {\bibinfo
  {journal} {Pub. Res. Inst. Math. Sci.}\ }\textbf {\bibinfo {volume} {18}},\
  \bibinfo {pages} {97} (\bibinfo {year} {1982})}\BibitemShut {NoStop}%
\bibitem [{\citenamefont {Luchnikov}\ \emph {et~al.}(2018)\citenamefont
  {Luchnikov}, \citenamefont {Vintskevich},\ and\ \citenamefont
  {Filippov}}]{Luchnikov2018}%
  \BibitemOpen
  \bibfield  {author} {\bibinfo {author} {\bibfnamefont {I.~A.}\ \bibnamefont
  {Luchnikov}}, \bibinfo {author} {\bibfnamefont {S.~V.}\ \bibnamefont
  {Vintskevich}}, \ and\ \bibinfo {author} {\bibfnamefont {S.~N.}\ \bibnamefont
  {Filippov}},\ }\href {https://arxiv.org/abs/1801.07418} {\bibfield  {journal}
  {\bibinfo  {journal} {arXiv:1801.07418}\ } (\bibinfo {year}
  {2018})}\BibitemShut {NoStop}%
\bibitem [{\citenamefont {Shrapnel}\ \emph
  {et~al.}(2018{\natexlab{b}})\citenamefont {Shrapnel}, \citenamefont {Costa},\
  and\ \citenamefont {Milburn}}]{costa2019}%
  \BibitemOpen
  \bibfield  {author} {\bibinfo {author} {\bibfnamefont {S.}~\bibnamefont
  {Shrapnel}}, \bibinfo {author} {\bibfnamefont {F.}~\bibnamefont {Costa}}, \
  and\ \bibinfo {author} {\bibfnamefont {G.}~\bibnamefont {Milburn}},\ }\href
  {\doibase 10.1142/S0219749918400105} {\bibfield  {journal} {\bibinfo
  {journal} {Int. J. Quantum Inf.}\ }\textbf {\bibinfo {volume} {16}},\
  \bibinfo {pages} {1840010} (\bibinfo {year}
  {2018}{\natexlab{b}})}\BibitemShut {NoStop}%
\bibitem [{\citenamefont {Luchnikov}\ \emph {et~al.}(2019)\citenamefont
  {Luchnikov}, \citenamefont {Vintskevich}, \citenamefont {Ouerdane},\ and\
  \citenamefont {Filippov}}]{Luchnikov2019L}%
  \BibitemOpen
  \bibfield  {author} {\bibinfo {author} {\bibfnamefont {I.~A.}\ \bibnamefont
  {Luchnikov}}, \bibinfo {author} {\bibfnamefont {S.~V.}\ \bibnamefont
  {Vintskevich}}, \bibinfo {author} {\bibfnamefont {H.}~\bibnamefont
  {Ouerdane}}, \ and\ \bibinfo {author} {\bibfnamefont {S.~N.}\ \bibnamefont
  {Filippov}},\ }\href {\doibase 10.1103/PhysRevLett.122.160401} {\bibfield
  {journal} {\bibinfo  {journal} {Phys. Rev. Lett.}\ }\textbf {\bibinfo
  {volume} {122}},\ \bibinfo {pages} {160401} (\bibinfo {year}
  {2019})}\BibitemShut {NoStop}%
\bibitem [{\citenamefont {Luchnikov}\ \emph {et~al.}(2020)\citenamefont
  {Luchnikov}, \citenamefont {Vintskevich}, \citenamefont {Grigoriev},\ and\
  \citenamefont {Filippov}}]{Luchnikov2019}%
  \BibitemOpen
  \bibfield  {author} {\bibinfo {author} {\bibfnamefont {I.~A.}\ \bibnamefont
  {Luchnikov}}, \bibinfo {author} {\bibfnamefont {S.~V.}\ \bibnamefont
  {Vintskevich}}, \bibinfo {author} {\bibfnamefont {D.~A.}\ \bibnamefont
  {Grigoriev}}, \ and\ \bibinfo {author} {\bibfnamefont {S.~N.}\ \bibnamefont
  {Filippov}},\ }\href {\doibase 10.1103/PhysRevLett.124.140502} {\bibfield
  {journal} {\bibinfo  {journal} {Phys. Rev. Lett.}\ }\textbf {\bibinfo
  {volume} {124}},\ \bibinfo {pages} {140502} (\bibinfo {year}
  {2020})}\BibitemShut {NoStop}%
\bibitem [{\citenamefont {Guo}\ \emph {et~al.}(2020)\citenamefont {Guo},
  \citenamefont {Modi},\ and\ \citenamefont {Poletti}}]{guochu2020}%
  \BibitemOpen
  \bibfield  {author} {\bibinfo {author} {\bibfnamefont {C.}~\bibnamefont
  {Guo}}, \bibinfo {author} {\bibfnamefont {K.}~\bibnamefont {Modi}}, \ and\
  \bibinfo {author} {\bibfnamefont {D.}~\bibnamefont {Poletti}},\ }\href
  {\doibase 10.1103/PhysRevA.102.062414} {\bibfield  {journal} {\bibinfo
  {journal} {Phys. Rev. A}\ }\textbf {\bibinfo {volume} {102}},\ \bibinfo
  {pages} {062414} (\bibinfo {year} {2020})}\BibitemShut {NoStop}%
\bibitem [{\citenamefont {Combes}\ \emph {et~al.}(2017)\citenamefont {Combes},
  \citenamefont {Kerckhoff},\ and\ \citenamefont {Sarovar}}]{slh}%
  \BibitemOpen
  \bibfield  {author} {\bibinfo {author} {\bibfnamefont {J.}~\bibnamefont
  {Combes}}, \bibinfo {author} {\bibfnamefont {J.}~\bibnamefont {Kerckhoff}}, \
  and\ \bibinfo {author} {\bibfnamefont {M.}~\bibnamefont {Sarovar}},\ }\href
  {\doibase 10.1080/23746149.2017.1343097} {\bibfield  {journal} {\bibinfo
  {journal} {Adv. Phys. X}\ }\textbf {\bibinfo {volume} {2}},\ \bibinfo {pages}
  {784} (\bibinfo {year} {2017})}\BibitemShut {NoStop}%
\bibitem [{\citenamefont {Chiribella}\ and\ \citenamefont
  {Ebler}(2016)}]{chiribella_optimal_2016}%
  \BibitemOpen
  \bibfield  {author} {\bibinfo {author} {\bibfnamefont {G.}~\bibnamefont
  {Chiribella}}\ and\ \bibinfo {author} {\bibfnamefont {D.}~\bibnamefont
  {Ebler}},\ }\href {\doibase 10.1088/1367-2630/18/9/093053} {\bibfield
  {journal} {\bibinfo  {journal} {New J. Phys.}\ }\textbf {\bibinfo {volume}
  {18}},\ \bibinfo {pages} {093053} (\bibinfo {year} {2016})}\BibitemShut
  {NoStop}%
\bibitem [{\citenamefont {Eggeling}\ \emph {et~al.}(2002)\citenamefont
  {Eggeling}, \citenamefont {Schlingemann},\ and\ \citenamefont
  {Werner}}]{eggeling_semicausal_2002}%
  \BibitemOpen
  \bibfield  {author} {\bibinfo {author} {\bibfnamefont {T.}~\bibnamefont
  {Eggeling}}, \bibinfo {author} {\bibfnamefont {D.}~\bibnamefont
  {Schlingemann}}, \ and\ \bibinfo {author} {\bibfnamefont {R.~F.}\
  \bibnamefont {Werner}},\ }\href {\doibase 10.1209/epl/i2002-00579-4}
  {\bibfield  {journal} {\bibinfo  {journal} {EPL}\ }\textbf {\bibinfo {volume}
  {57}},\ \bibinfo {pages} {782} (\bibinfo {year} {2002})}\BibitemShut
  {NoStop}%
\bibitem [{\citenamefont {Barrett}\ \emph {et~al.}(2020)\citenamefont
  {Barrett}, \citenamefont {Lorenz},\ and\ \citenamefont
  {Oreshkov}}]{barrett_quantum_2020}%
  \BibitemOpen
  \bibfield  {author} {\bibinfo {author} {\bibfnamefont {J.}~\bibnamefont
  {Barrett}}, \bibinfo {author} {\bibfnamefont {R.}~\bibnamefont {Lorenz}}, \
  and\ \bibinfo {author} {\bibfnamefont {O.}~\bibnamefont {Oreshkov}},\ }\href
  {http://arxiv.org/abs/1906.10726} {\bibfield  {journal} {\bibinfo  {journal}
  {arXiv:1906.10726}\ } (\bibinfo {year} {2020})}\BibitemShut {NoStop}%
\bibitem [{\citenamefont {Sakuldee}\ \emph {et~al.}(2018)\citenamefont
  {Sakuldee}, \citenamefont {Milz}, \citenamefont {Pollock},\ and\
  \citenamefont {Modi}}]{Sakuldee2018}%
  \BibitemOpen
  \bibfield  {author} {\bibinfo {author} {\bibfnamefont {F.}~\bibnamefont
  {Sakuldee}}, \bibinfo {author} {\bibfnamefont {S.}~\bibnamefont {Milz}},
  \bibinfo {author} {\bibfnamefont {F.~A.}\ \bibnamefont {Pollock}}, \ and\
  \bibinfo {author} {\bibfnamefont {K.}~\bibnamefont {Modi}},\ }\href
  {http://stacks.iop.org/1751-8121/51/i=41/a=414014} {\bibfield  {journal}
  {\bibinfo  {journal} {J. Phys. A}\ }\textbf {\bibinfo {volume} {51}},\
  \bibinfo {pages} {414014} (\bibinfo {year} {2018})}\BibitemShut {NoStop}%
\bibitem [{Note17()}]{Note17}%
  \BibitemOpen
  \bibinfo {note} {It can even be tested for if the ordering of events is not
  given a priori~\cite {giarmatzi_quantum_2018}}\BibitemShut {NoStop}%
\bibitem [{\citenamefont {Pearl}(2000)}]{Pearl}%
  \BibitemOpen
  \bibfield  {author} {\bibinfo {author} {\bibfnamefont {J.}~\bibnamefont
  {Pearl}},\ }\href@noop {} {\emph {\bibinfo {title} {Causality}}}\ (\bibinfo
  {publisher} {Oxford University Press},\ \bibinfo {year} {2000})\BibitemShut
  {NoStop}%
\bibitem [{\citenamefont {Carmichael}(1993)}]{carmichael_open_1993}%
  \BibitemOpen
  \bibfield  {author} {\bibinfo {author} {\bibfnamefont {H.}~\bibnamefont
  {Carmichael}},\ }\href {\doibase 10.1007/978-3-540-47620-7} {\emph {\bibinfo
  {title} {An {Open} {Systems} {Approach} to {Quantum} {Optics}}}},\ Lecture
  {Notes} in {Physics} {Monographs}\ (\bibinfo  {publisher} {Springer-Verlag},\
  \bibinfo {address} {Berlin},\ \bibinfo {year} {1993})\BibitemShut {NoStop}%
\bibitem [{\citenamefont {Guarnieri}\ \emph {et~al.}(2014)\citenamefont
  {Guarnieri}, \citenamefont {Smirne},\ and\ \citenamefont
  {Vacchini}}]{Guarnieri2014}%
  \BibitemOpen
  \bibfield  {author} {\bibinfo {author} {\bibfnamefont {G.}~\bibnamefont
  {Guarnieri}}, \bibinfo {author} {\bibfnamefont {A.}~\bibnamefont {Smirne}}, \
  and\ \bibinfo {author} {\bibfnamefont {B.}~\bibnamefont {Vacchini}},\ }\href
  {\doibase 10.1103/PhysRevA.90.022110} {\bibfield  {journal} {\bibinfo
  {journal} {Phys. Rev. A}\ }\textbf {\bibinfo {volume} {90}},\ \bibinfo
  {pages} {022110} (\bibinfo {year} {2014})}\BibitemShut {NoStop}%
\bibitem [{\citenamefont {Morris}\ \emph {et~al.}(2019)\citenamefont {Morris},
  \citenamefont {Pollock},\ and\ \citenamefont {Modi}}]{Morris2019}%
  \BibitemOpen
  \bibfield  {author} {\bibinfo {author} {\bibfnamefont {J.}~\bibnamefont
  {Morris}}, \bibinfo {author} {\bibfnamefont {F.~A.}\ \bibnamefont {Pollock}},
  \ and\ \bibinfo {author} {\bibfnamefont {K.}~\bibnamefont {Modi}},\ }\href
  {https://arxiv.org/abs/1902.07980} {\bibfield  {journal} {\bibinfo  {journal}
  {arXiv:1902.07980}\ } (\bibinfo {year} {2019})}\BibitemShut {NoStop}%
\bibitem [{\citenamefont {Hou}\ \emph {et~al.}(2011)\citenamefont {Hou},
  \citenamefont {Yi}, \citenamefont {Yu},\ and\ \citenamefont {Oh}}]{hou2011}%
  \BibitemOpen
  \bibfield  {author} {\bibinfo {author} {\bibfnamefont {S.~C.}\ \bibnamefont
  {Hou}}, \bibinfo {author} {\bibfnamefont {X.~X.}\ \bibnamefont {Yi}},
  \bibinfo {author} {\bibfnamefont {S.~X.}\ \bibnamefont {Yu}}, \ and\ \bibinfo
  {author} {\bibfnamefont {C.~H.}\ \bibnamefont {Oh}},\ }\href {\doibase
  10.1103/PhysRevA.83.062115} {\bibfield  {journal} {\bibinfo  {journal} {Phys.
  Rev. A}\ }\textbf {\bibinfo {volume} {83}},\ \bibinfo {pages} {062115}
  (\bibinfo {year} {2011})}\BibitemShut {NoStop}%
\bibitem [{\citenamefont {Lu}\ \emph {et~al.}(2010)\citenamefont {Lu},
  \citenamefont {Wang},\ and\ \citenamefont {Sun}}]{PhysRevA.82.042103}%
  \BibitemOpen
  \bibfield  {author} {\bibinfo {author} {\bibfnamefont {X.-M.}\ \bibnamefont
  {Lu}}, \bibinfo {author} {\bibfnamefont {X.}~\bibnamefont {Wang}}, \ and\
  \bibinfo {author} {\bibfnamefont {C.~P.}\ \bibnamefont {Sun}},\ }\href
  {\doibase 10.1103/PhysRevA.82.042103} {\bibfield  {journal} {\bibinfo
  {journal} {Phys. Rev. A}\ }\textbf {\bibinfo {volume} {82}},\ \bibinfo
  {pages} {042103} (\bibinfo {year} {2010})}\BibitemShut {NoStop}%
\bibitem [{\citenamefont {Mazzola}\ \emph {et~al.}(2012)\citenamefont
  {Mazzola}, \citenamefont {Rodr{\'\i}guez-Rosario}, \citenamefont {Modi},\
  and\ \citenamefont {Paternostro}}]{mazzola2012dynamical}%
  \BibitemOpen
  \bibfield  {author} {\bibinfo {author} {\bibfnamefont {L.}~\bibnamefont
  {Mazzola}}, \bibinfo {author} {\bibfnamefont {C.~A.}\ \bibnamefont
  {Rodr{\'\i}guez-Rosario}}, \bibinfo {author} {\bibfnamefont {K.}~\bibnamefont
  {Modi}}, \ and\ \bibinfo {author} {\bibfnamefont {M.}~\bibnamefont
  {Paternostro}},\ }\href {http://dx.doi.org/10.1103/PhysRevA.86.010102}
  {\bibfield  {journal} {\bibinfo  {journal} {Phys. Rev. A}\ }\textbf {\bibinfo
  {volume} {86}},\ \bibinfo {pages} {010102} (\bibinfo {year}
  {2012})}\BibitemShut {NoStop}%
\bibitem [{\citenamefont {Rajagopal}\ \emph {et~al.}(2010)\citenamefont
  {Rajagopal}, \citenamefont {Usha~Devi},\ and\ \citenamefont
  {Rendell}}]{PhysRevA.82.042107}%
  \BibitemOpen
  \bibfield  {author} {\bibinfo {author} {\bibfnamefont {A.~K.}\ \bibnamefont
  {Rajagopal}}, \bibinfo {author} {\bibfnamefont {A.~R.}\ \bibnamefont
  {Usha~Devi}}, \ and\ \bibinfo {author} {\bibfnamefont {R.~W.}\ \bibnamefont
  {Rendell}},\ }\href {\doibase 10.1103/PhysRevA.82.042107} {\bibfield
  {journal} {\bibinfo  {journal} {Phys. Rev. A}\ }\textbf {\bibinfo {volume}
  {82}},\ \bibinfo {pages} {042107} (\bibinfo {year} {2010})}\BibitemShut
  {NoStop}%
\bibitem [{\citenamefont {Rodr\'{\i}guez-Rosario}\ \emph
  {et~al.}(2012)\citenamefont {Rodr\'{\i}guez-Rosario}, \citenamefont {Modi},
  \citenamefont {Mazzola},\ and\ \citenamefont
  {Aspuru-Guzik}}]{rodriguez2012unification}%
  \BibitemOpen
  \bibfield  {author} {\bibinfo {author} {\bibfnamefont {C.~A.}\ \bibnamefont
  {Rodr\'{\i}guez-Rosario}}, \bibinfo {author} {\bibfnamefont {K.}~\bibnamefont
  {Modi}}, \bibinfo {author} {\bibfnamefont {L.}~\bibnamefont {Mazzola}}, \
  and\ \bibinfo {author} {\bibfnamefont {A.}~\bibnamefont {Aspuru-Guzik}},\
  }\href {http://iopscience.iop.org/article/10.1209/0295-5075/99/20010}
  {\bibfield  {journal} {\bibinfo  {journal} {EPL}\ }\textbf {\bibinfo {volume}
  {99}},\ \bibinfo {pages} {20010} (\bibinfo {year} {2012})}\BibitemShut
  {NoStop}%
\bibitem [{\citenamefont {Luo}\ \emph {et~al.}(2012)\citenamefont {Luo},
  \citenamefont {Fu},\ and\ \citenamefont {Song}}]{PhysRevA.86.044101}%
  \BibitemOpen
  \bibfield  {author} {\bibinfo {author} {\bibfnamefont {S.}~\bibnamefont
  {Luo}}, \bibinfo {author} {\bibfnamefont {S.}~\bibnamefont {Fu}}, \ and\
  \bibinfo {author} {\bibfnamefont {H.}~\bibnamefont {Song}},\ }\href {\doibase
  10.1103/PhysRevA.86.044101} {\bibfield  {journal} {\bibinfo  {journal} {Phys.
  Rev. A}\ }\textbf {\bibinfo {volume} {86}},\ \bibinfo {pages} {044101}
  (\bibinfo {year} {2012})}\BibitemShut {NoStop}%
\bibitem [{\citenamefont {He}\ \emph {et~al.}(2017)\citenamefont {He},
  \citenamefont {Zeng}, \citenamefont {Li}, \citenamefont {Wang},\ and\
  \citenamefont {Yao}}]{zhihe2017}%
  \BibitemOpen
  \bibfield  {author} {\bibinfo {author} {\bibfnamefont {Z.}~\bibnamefont
  {He}}, \bibinfo {author} {\bibfnamefont {H.-S.}\ \bibnamefont {Zeng}},
  \bibinfo {author} {\bibfnamefont {Y.}~\bibnamefont {Li}}, \bibinfo {author}
  {\bibfnamefont {Q.}~\bibnamefont {Wang}}, \ and\ \bibinfo {author}
  {\bibfnamefont {C.}~\bibnamefont {Yao}},\ }\href {\doibase
  10.1103/PhysRevA.96.022106} {\bibfield  {journal} {\bibinfo  {journal} {Phys.
  Rev. A}\ }\textbf {\bibinfo {volume} {96}},\ \bibinfo {pages} {022106}
  (\bibinfo {year} {2017})}\BibitemShut {NoStop}%
\bibitem [{\citenamefont {Fanchini}\ \emph {et~al.}(2014)\citenamefont
  {Fanchini}, \citenamefont {Karpat}, \citenamefont {\ifmmode~\mbox{\c{C}}\else
  \c{C}\fi{}akmak}, \citenamefont {Castelano}, \citenamefont {Aguilar},
  \citenamefont {Far\'{\i}as}, \citenamefont {Walborn}, \citenamefont
  {Ribeiro},\ and\ \citenamefont {de~Oliveira}}]{fanchini2014}%
  \BibitemOpen
  \bibfield  {author} {\bibinfo {author} {\bibfnamefont {F.~F.}\ \bibnamefont
  {Fanchini}}, \bibinfo {author} {\bibfnamefont {G.}~\bibnamefont {Karpat}},
  \bibinfo {author} {\bibfnamefont {B.}~\bibnamefont
  {\ifmmode~\mbox{\c{C}}\else \c{C}\fi{}akmak}}, \bibinfo {author}
  {\bibfnamefont {L.~K.}\ \bibnamefont {Castelano}}, \bibinfo {author}
  {\bibfnamefont {G.~H.}\ \bibnamefont {Aguilar}}, \bibinfo {author}
  {\bibfnamefont {O.~J.}\ \bibnamefont {Far\'{\i}as}}, \bibinfo {author}
  {\bibfnamefont {S.~P.}\ \bibnamefont {Walborn}}, \bibinfo {author}
  {\bibfnamefont {P.~H.~S.}\ \bibnamefont {Ribeiro}}, \ and\ \bibinfo {author}
  {\bibfnamefont {M.~C.}\ \bibnamefont {de~Oliveira}},\ }\href {\doibase
  10.1103/PhysRevLett.112.210402} {\bibfield  {journal} {\bibinfo  {journal}
  {Phys. Rev. Lett.}\ }\textbf {\bibinfo {volume} {112}},\ \bibinfo {pages}
  {210402} (\bibinfo {year} {2014})}\BibitemShut {NoStop}%
\bibitem [{\citenamefont {Bylicka}\ \emph {et~al.}(2014)\citenamefont
  {Bylicka}, \citenamefont {Chru{\'s}ci{\'n}ski},\ and\ \citenamefont
  {Maniscalco}}]{bylicka2014}%
  \BibitemOpen
  \bibfield  {author} {\bibinfo {author} {\bibfnamefont {B.}~\bibnamefont
  {Bylicka}}, \bibinfo {author} {\bibfnamefont {D.}~\bibnamefont
  {Chru{\'s}ci{\'n}ski}}, \ and\ \bibinfo {author} {\bibfnamefont
  {S.}~\bibnamefont {Maniscalco}},\ }\href {\doibase 10.1038/srep05720}
  {\bibfield  {journal} {\bibinfo  {journal} {Sci. Rep.}\ }\textbf {\bibinfo
  {volume} {4}},\ \bibinfo {pages} {5720} (\bibinfo {year} {2014})}\BibitemShut
  {NoStop}%
\bibitem [{\citenamefont {Pineda}\ \emph {et~al.}(2016)\citenamefont {Pineda},
  \citenamefont {Gorin}, \citenamefont {Davalos}, \citenamefont {Wisniacki},\
  and\ \citenamefont {Garc\'{\i}a-Mata}}]{pineda2016}%
  \BibitemOpen
  \bibfield  {author} {\bibinfo {author} {\bibfnamefont {C.}~\bibnamefont
  {Pineda}}, \bibinfo {author} {\bibfnamefont {T.}~\bibnamefont {Gorin}},
  \bibinfo {author} {\bibfnamefont {D.}~\bibnamefont {Davalos}}, \bibinfo
  {author} {\bibfnamefont {D.~A.}\ \bibnamefont {Wisniacki}}, \ and\ \bibinfo
  {author} {\bibfnamefont {I.}~\bibnamefont {Garc\'{\i}a-Mata}},\ }\href
  {\doibase 10.1103/PhysRevA.93.022117} {\bibfield  {journal} {\bibinfo
  {journal} {Phys. Rev. A}\ }\textbf {\bibinfo {volume} {93}},\ \bibinfo
  {pages} {022117} (\bibinfo {year} {2016})}\BibitemShut {NoStop}%
\bibitem [{\citenamefont {{Usha Devi}}\ \emph {et~al.}(2011)\citenamefont
  {{Usha Devi}}, \citenamefont {Rajagopal},\ and\ \citenamefont
  {Sudha}}]{PhysRevA.83.022109}%
  \BibitemOpen
  \bibfield  {author} {\bibinfo {author} {\bibfnamefont {A.~R.}\ \bibnamefont
  {{Usha Devi}}}, \bibinfo {author} {\bibfnamefont {A.~K.}\ \bibnamefont
  {Rajagopal}}, \ and\ \bibinfo {author} {\bibnamefont {Sudha}},\ }\href
  {\doibase 10.1103/PhysRevA.83.022109} {\bibfield  {journal} {\bibinfo
  {journal} {Phys. Rev. A}\ }\textbf {\bibinfo {volume} {83}},\ \bibinfo
  {pages} {022109} (\bibinfo {year} {2011})}\BibitemShut {NoStop}%
\bibitem [{\citenamefont {{Usha Devi}}\ \emph {et~al.}(2012)\citenamefont
  {{Usha Devi}}, \citenamefont {Rajagopal}, \citenamefont {Shenoy},\ and\
  \citenamefont {Rendell}}]{rajagopal}%
  \BibitemOpen
  \bibfield  {author} {\bibinfo {author} {\bibfnamefont {A.~R.}\ \bibnamefont
  {{Usha Devi}}}, \bibinfo {author} {\bibfnamefont {A.~K.}\ \bibnamefont
  {Rajagopal}}, \bibinfo {author} {\bibfnamefont {S.}~\bibnamefont {Shenoy}}, \
  and\ \bibinfo {author} {\bibfnamefont {R.~W.}\ \bibnamefont {Rendell}},\
  }\href {http://file.scirp.org/Html/23056.html} {\bibfield  {journal}
  {\bibinfo  {journal} {J. Quant. Info. Sci.}\ }\textbf {\bibinfo {volume}
  {2}},\ \bibinfo {pages} {47} (\bibinfo {year} {2012})}\BibitemShut {NoStop}%
\bibitem [{\citenamefont {Chru\'{s}ci\'{n}ski}\ and\ \citenamefont
  {Maniscalco}(2014)}]{sabri}%
  \BibitemOpen
  \bibfield  {author} {\bibinfo {author} {\bibfnamefont {D.}~\bibnamefont
  {Chru\'{s}ci\'{n}ski}}\ and\ \bibinfo {author} {\bibfnamefont
  {S.}~\bibnamefont {Maniscalco}},\ }\href {\doibase
  10.1103/PhysRevLett.112.120404} {\bibfield  {journal} {\bibinfo  {journal}
  {Phys. Rev. Lett.}\ }\textbf {\bibinfo {volume} {112}},\ \bibinfo {pages}
  {120404} (\bibinfo {year} {2014})}\BibitemShut {NoStop}%
\bibitem [{\citenamefont {Lindblad}(1980)}]{Lindblad1980}%
  \BibitemOpen
  \bibfield  {author} {\bibinfo {author} {\bibfnamefont {G.}~\bibnamefont
  {Lindblad}},\ }\href@noop {} {\bibfield  {journal} {\bibinfo  {journal}
  {(unpublished), Stockholm}\ } (\bibinfo {year} {1980})}\BibitemShut {NoStop}%
\bibitem [{\citenamefont {Arenz}\ \emph {et~al.}(2015)\citenamefont {Arenz},
  \citenamefont {Hillier}, \citenamefont {Fraas},\ and\ \citenamefont
  {Burgarth}}]{PhysRevA.92.022102}%
  \BibitemOpen
  \bibfield  {author} {\bibinfo {author} {\bibfnamefont {C.}~\bibnamefont
  {Arenz}}, \bibinfo {author} {\bibfnamefont {R.}~\bibnamefont {Hillier}},
  \bibinfo {author} {\bibfnamefont {M.}~\bibnamefont {Fraas}}, \ and\ \bibinfo
  {author} {\bibfnamefont {D.}~\bibnamefont {Burgarth}},\ }\href {\doibase
  10.1103/PhysRevA.92.022102} {\bibfield  {journal} {\bibinfo  {journal} {Phys.
  Rev. A}\ }\textbf {\bibinfo {volume} {92}},\ \bibinfo {pages} {022102}
  (\bibinfo {year} {2015})}\BibitemShut {NoStop}%
\bibitem [{\citenamefont {Burgarth}\ \emph {et~al.}(2021)\citenamefont
  {Burgarth}, \citenamefont {Facchi}, \citenamefont {Ligab\`o},\ and\
  \citenamefont {Lonigro}}]{arXiv:2009.10605}%
  \BibitemOpen
  \bibfield  {author} {\bibinfo {author} {\bibfnamefont {D.}~\bibnamefont
  {Burgarth}}, \bibinfo {author} {\bibfnamefont {P.}~\bibnamefont {Facchi}},
  \bibinfo {author} {\bibfnamefont {M.}~\bibnamefont {Ligab\`o}}, \ and\
  \bibinfo {author} {\bibfnamefont {D.}~\bibnamefont {Lonigro}},\ }\href
  {\doibase 10.1103/PhysRevA.103.012203} {\bibfield  {journal} {\bibinfo
  {journal} {Phys. Rev. A}\ }\textbf {\bibinfo {volume} {103}},\ \bibinfo
  {pages} {012203} (\bibinfo {year} {2021})}\BibitemShut {NoStop}%
\bibitem [{\citenamefont {Hsieh}\ \emph {et~al.}(2019)\citenamefont {Hsieh},
  \citenamefont {Su},\ and\ \citenamefont {Goan}}]{PhysRevA.100.012120}%
  \BibitemOpen
  \bibfield  {author} {\bibinfo {author} {\bibfnamefont {Y.-Y.}\ \bibnamefont
  {Hsieh}}, \bibinfo {author} {\bibfnamefont {Z.-Y.}\ \bibnamefont {Su}}, \
  and\ \bibinfo {author} {\bibfnamefont {H.-S.}\ \bibnamefont {Goan}},\ }\href
  {\doibase 10.1103/PhysRevA.100.012120} {\bibfield  {journal} {\bibinfo
  {journal} {Phys. Rev. A}\ }\textbf {\bibinfo {volume} {100}},\ \bibinfo
  {pages} {012120} (\bibinfo {year} {2019})}\BibitemShut {NoStop}%
\bibitem [{\citenamefont {Milz}\ \emph
  {et~al.}(2020{\natexlab{c}})\citenamefont {Milz}, \citenamefont {Spee},
  \citenamefont {Xu}, \citenamefont {Pollock}, \citenamefont {Modi},\ and\
  \citenamefont {G{\"u}hne}}]{milz_genuine_2020}%
  \BibitemOpen
  \bibfield  {author} {\bibinfo {author} {\bibfnamefont {S.}~\bibnamefont
  {Milz}}, \bibinfo {author} {\bibfnamefont {C.}~\bibnamefont {Spee}}, \bibinfo
  {author} {\bibfnamefont {Z.-P.}\ \bibnamefont {Xu}}, \bibinfo {author}
  {\bibfnamefont {F.~A.}\ \bibnamefont {Pollock}}, \bibinfo {author}
  {\bibfnamefont {K.}~\bibnamefont {Modi}}, \ and\ \bibinfo {author}
  {\bibfnamefont {O.}~\bibnamefont {G{\"u}hne}},\ }\href
  {http://arxiv.org/abs/2011.09340} {\bibfield  {journal} {\bibinfo  {journal}
  {arXiv:2011.09340}\ } (\bibinfo {year} {2020}{\natexlab{c}})}\BibitemShut
  {NoStop}%
\bibitem [{\citenamefont {Verstraete}\ \emph {et~al.}(2004)\citenamefont
  {Verstraete}, \citenamefont {Garc\'{\i}a-Ripoll},\ and\ \citenamefont
  {Cirac}}]{VerstraeteCirac2004}%
  \BibitemOpen
  \bibfield  {author} {\bibinfo {author} {\bibfnamefont {F.}~\bibnamefont
  {Verstraete}}, \bibinfo {author} {\bibfnamefont {J.~J.}\ \bibnamefont
  {Garc\'{\i}a-Ripoll}}, \ and\ \bibinfo {author} {\bibfnamefont {J.~I.}\
  \bibnamefont {Cirac}},\ }\href {\doibase 10.1103/PhysRevLett.93.207204}
  {\bibfield  {journal} {\bibinfo  {journal} {Phys. Rev. Lett.}\ }\textbf
  {\bibinfo {volume} {93}},\ \bibinfo {pages} {207204} (\bibinfo {year}
  {2004})}\BibitemShut {NoStop}%
\bibitem [{\citenamefont {Zwolak}\ and\ \citenamefont
  {Vidal}(2004)}]{zwolak_mixed-state_2004}%
  \BibitemOpen
  \bibfield  {author} {\bibinfo {author} {\bibfnamefont {M.}~\bibnamefont
  {Zwolak}}\ and\ \bibinfo {author} {\bibfnamefont {G.}~\bibnamefont {Vidal}},\
  }\href {\doibase 10.1103/PhysRevLett.93.207205} {\bibfield  {journal}
  {\bibinfo  {journal} {Phys. Rev. Lett.}\ }\textbf {\bibinfo {volume} {93}},\
  \bibinfo {pages} {207205} (\bibinfo {year} {2004})}\BibitemShut {NoStop}%
\bibitem [{\citenamefont {Yang}\ \emph {et~al.}(2018)\citenamefont {Yang},
  \citenamefont {Binder}, \citenamefont {Narasimhachar},\ and\ \citenamefont
  {Gu}}]{yang_matrix_2018}%
  \BibitemOpen
  \bibfield  {author} {\bibinfo {author} {\bibfnamefont {C.}~\bibnamefont
  {Yang}}, \bibinfo {author} {\bibfnamefont {F.~C.}\ \bibnamefont {Binder}},
  \bibinfo {author} {\bibfnamefont {V.}~\bibnamefont {Narasimhachar}}, \ and\
  \bibinfo {author} {\bibfnamefont {M.}~\bibnamefont {Gu}},\ }\href {\doibase
  10.1103/PhysRevLett.121.260602} {\bibfield  {journal} {\bibinfo  {journal}
  {Phys. Rev. Lett.}\ }\textbf {\bibinfo {volume} {121}},\ \bibinfo {pages}
  {260602} (\bibinfo {year} {2018})}\BibitemShut {NoStop}%
\bibitem [{\citenamefont {Bridgeman}\ and\ \citenamefont
  {Chubb}(2017)}]{MPSreview}%
  \BibitemOpen
  \bibfield  {author} {\bibinfo {author} {\bibfnamefont {J.~C.}\ \bibnamefont
  {Bridgeman}}\ and\ \bibinfo {author} {\bibfnamefont {C.~T.}\ \bibnamefont
  {Chubb}},\ }\href {http://stacks.iop.org/1751-8121/50/i=22/a=223001}
  {\bibfield  {journal} {\bibinfo  {journal} {J. Phys. A}\ }\textbf {\bibinfo
  {volume} {50}},\ \bibinfo {pages} {223001} (\bibinfo {year}
  {2017})}\BibitemShut {NoStop}%
\bibitem [{\citenamefont {Or{\'u}s}(2014)}]{orus_practical_2014}%
  \BibitemOpen
  \bibfield  {author} {\bibinfo {author} {\bibfnamefont {R.}~\bibnamefont
  {Or{\'u}s}},\ }\href {\doibase 10.1016/j.aop.2014.06.013} {\bibfield
  {journal} {\bibinfo  {journal} {Ann. Phys.}\ }\textbf {\bibinfo {volume}
  {349}},\ \bibinfo {pages} {117} (\bibinfo {year} {2014})}\BibitemShut
  {NoStop}%
\bibitem [{\citenamefont {Prior}\ \emph {et~al.}(2010)\citenamefont {Prior},
  \citenamefont {Chin}, \citenamefont {Huelga},\ and\ \citenamefont
  {Plenio}}]{prior_efficient_2010}%
  \BibitemOpen
  \bibfield  {author} {\bibinfo {author} {\bibfnamefont {J.}~\bibnamefont
  {Prior}}, \bibinfo {author} {\bibfnamefont {A.~W.}\ \bibnamefont {Chin}},
  \bibinfo {author} {\bibfnamefont {S.~F.}\ \bibnamefont {Huelga}}, \ and\
  \bibinfo {author} {\bibfnamefont {M.~B.}\ \bibnamefont {Plenio}},\ }\href
  {\doibase 10.1103/PhysRevLett.105.050404} {\bibfield  {journal} {\bibinfo
  {journal} {Phys. Rev. Lett.}\ }\textbf {\bibinfo {volume} {105}},\ \bibinfo
  {pages} {050404} (\bibinfo {year} {2010})}\BibitemShut {NoStop}%
\bibitem [{\citenamefont {Schr{\"o}der}\ and\ \citenamefont
  {Chin}(2016)}]{schroder_simulating_2016}%
  \BibitemOpen
  \bibfield  {author} {\bibinfo {author} {\bibfnamefont {F.~A. Y.~N.}\
  \bibnamefont {Schr{\"o}der}}\ and\ \bibinfo {author} {\bibfnamefont {A.~W.}\
  \bibnamefont {Chin}},\ }\href {\doibase 10.1103/PhysRevB.93.075105}
  {\bibfield  {journal} {\bibinfo  {journal} {Phys. Rev. B}\ }\textbf {\bibinfo
  {volume} {93}},\ \bibinfo {pages} {075105} (\bibinfo {year}
  {2016})}\BibitemShut {NoStop}%
\bibitem [{\citenamefont {Wall}\ \emph {et~al.}(2016)\citenamefont {Wall},
  \citenamefont {Safavi-Naini},\ and\ \citenamefont
  {Rey}}]{wall_simulating_2016}%
  \BibitemOpen
  \bibfield  {author} {\bibinfo {author} {\bibfnamefont {M.~L.}\ \bibnamefont
  {Wall}}, \bibinfo {author} {\bibfnamefont {A.}~\bibnamefont {Safavi-Naini}},
  \ and\ \bibinfo {author} {\bibfnamefont {A.~M.}\ \bibnamefont {Rey}},\ }\href
  {\doibase 10.1103/PhysRevA.94.053637} {\bibfield  {journal} {\bibinfo
  {journal} {Phys. Rev. A}\ }\textbf {\bibinfo {volume} {94}},\ \bibinfo
  {pages} {053637} (\bibinfo {year} {2016})}\BibitemShut {NoStop}%
\bibitem [{\citenamefont {Strathearn}\ \emph {et~al.}(2018)\citenamefont
  {Strathearn}, \citenamefont {Kirton}, \citenamefont {Kilda}, \citenamefont
  {Keeling},\ and\ \citenamefont {Lovett}}]{strathearn_efficient_2018}%
  \BibitemOpen
  \bibfield  {author} {\bibinfo {author} {\bibfnamefont {A.}~\bibnamefont
  {Strathearn}}, \bibinfo {author} {\bibfnamefont {P.}~\bibnamefont {Kirton}},
  \bibinfo {author} {\bibfnamefont {D.}~\bibnamefont {Kilda}}, \bibinfo
  {author} {\bibfnamefont {J.}~\bibnamefont {Keeling}}, \ and\ \bibinfo
  {author} {\bibfnamefont {B.~W.}\ \bibnamefont {Lovett}},\ }\href {\doibase
  10.1038/s41467-018-05617-3} {\bibfield  {journal} {\bibinfo  {journal} {Nat.
  Commun.}\ }\textbf {\bibinfo {volume} {9}},\ \bibinfo {pages} {3322}
  (\bibinfo {year} {2018})}\BibitemShut {NoStop}%
\bibitem [{\citenamefont {J\o{}rgensen}\ and\ \citenamefont
  {Pollock}(2019)}]{Jorgensen2019}%
  \BibitemOpen
  \bibfield  {author} {\bibinfo {author} {\bibfnamefont {M.~R.}\ \bibnamefont
  {J\o{}rgensen}}\ and\ \bibinfo {author} {\bibfnamefont {F.~A.}\ \bibnamefont
  {Pollock}},\ }\href {\doibase 10.1103/PhysRevLett.123.240602} {\bibfield
  {journal} {\bibinfo  {journal} {Phys. Rev. Lett.}\ }\textbf {\bibinfo
  {volume} {123}},\ \bibinfo {pages} {240602} (\bibinfo {year}
  {2019})}\BibitemShut {NoStop}%
\bibitem [{\citenamefont {Figueroa-Romero}\ \emph {et~al.}(2019)\citenamefont
  {Figueroa-Romero}, \citenamefont {Modi},\ and\ \citenamefont
  {Pollock}}]{Romero2018}%
  \BibitemOpen
  \bibfield  {author} {\bibinfo {author} {\bibfnamefont {P.}~\bibnamefont
  {Figueroa-Romero}}, \bibinfo {author} {\bibfnamefont {K.}~\bibnamefont
  {Modi}}, \ and\ \bibinfo {author} {\bibfnamefont {F.~A.}\ \bibnamefont
  {Pollock}},\ }\href {\doibase 10.22331/q-2019-04-30-136} {\bibfield
  {journal} {\bibinfo  {journal} {{Quantum}}\ }\textbf {\bibinfo {volume}
  {3}},\ \bibinfo {pages} {136} (\bibinfo {year} {2019})}\BibitemShut {NoStop}%
\bibitem [{\citenamefont {Figueroa-Romero}\ \emph {et~al.}(2021)\citenamefont
  {Figueroa-Romero}, \citenamefont {Pollock},\ and\ \citenamefont
  {Modi}}]{Romero2020}%
  \BibitemOpen
  \bibfield  {author} {\bibinfo {author} {\bibfnamefont {P.}~\bibnamefont
  {Figueroa-Romero}}, \bibinfo {author} {\bibfnamefont {F.~A.}\ \bibnamefont
  {Pollock}}, \ and\ \bibinfo {author} {\bibfnamefont {K.}~\bibnamefont
  {Modi}},\ }\href {https://arxiv.org/abs/2004.07620} {\bibfield  {journal}
  {\bibinfo  {journal} {Commun. Phys.}\ } (\bibinfo {year} {2021})}\BibitemShut
  {NoStop}%
\bibitem [{\citenamefont {Modi}\ \emph {et~al.}(2010)\citenamefont {Modi},
  \citenamefont {Paterek}, \citenamefont {Son}, \citenamefont {Vedral},\ and\
  \citenamefont {Williamson}}]{arXiv:0911.5417}%
  \BibitemOpen
  \bibfield  {author} {\bibinfo {author} {\bibfnamefont {K.}~\bibnamefont
  {Modi}}, \bibinfo {author} {\bibfnamefont {T.}~\bibnamefont {Paterek}},
  \bibinfo {author} {\bibfnamefont {W.}~\bibnamefont {Son}}, \bibinfo {author}
  {\bibfnamefont {V.}~\bibnamefont {Vedral}}, \ and\ \bibinfo {author}
  {\bibfnamefont {M.}~\bibnamefont {Williamson}},\ }\href@noop {} {\bibfield
  {journal} {\bibinfo  {journal} {Phys. Rev. Lett.}\ }\textbf {\bibinfo
  {volume} {104}},\ \bibinfo {pages} {080501} (\bibinfo {year}
  {2010})}\BibitemShut {NoStop}%
\bibitem [{\citenamefont {Taranto}(2020)}]{TarantoThesis}%
  \BibitemOpen
  \bibfield  {author} {\bibinfo {author} {\bibfnamefont {P.}~\bibnamefont
  {Taranto}},\ }\href {\doibase 10.1142/S0219749919410028} {\bibfield
  {journal} {\bibinfo  {journal} {Int. J. Quantum Inf.}\ }\textbf {\bibinfo
  {volume} {18}},\ \bibinfo {pages} {1941002} (\bibinfo {year}
  {2020})}\BibitemShut {NoStop}%
\bibitem [{Note18()}]{Note18}%
  \BibitemOpen
  \bibinfo {note} {If an element $\protect \mathbf {A}_{\protect \mathbf
  {x}_j}$ is linearly dependent on $\protect \mathbf {A}_{\protect \mathbf
  {x}_k}$ and $\protect \mathbf {A}_{\protect \mathbf {x}_\ell }$, then either
  the conditional future and past processes are the same for both outcomes $k$
  and $\ell $ or outcome $j$ does yield conditional independence.}\BibitemShut
  {Stop}%
\bibitem [{\citenamefont {Hayden}\ \emph {et~al.}(2004)\citenamefont {Hayden},
  \citenamefont {Jozsa}, \citenamefont {Petz},\ and\ \citenamefont
  {Winter}}]{Hayden2004}%
  \BibitemOpen
  \bibfield  {author} {\bibinfo {author} {\bibfnamefont {P.}~\bibnamefont
  {Hayden}}, \bibinfo {author} {\bibfnamefont {R.}~\bibnamefont {Jozsa}},
  \bibinfo {author} {\bibfnamefont {D.}~\bibnamefont {Petz}}, \ and\ \bibinfo
  {author} {\bibfnamefont {A.}~\bibnamefont {Winter}},\ }\href {\doibase
  10.1007/s00220-004-1049-z} {\bibfield  {journal} {\bibinfo  {journal}
  {Commun. Math. Phys.}\ }\textbf {\bibinfo {volume} {246}},\ \bibinfo {pages}
  {359} (\bibinfo {year} {2004})}\BibitemShut {NoStop}%
\bibitem [{\citenamefont {Petz}(1986)}]{Petz1986}%
  \BibitemOpen
  \bibfield  {author} {\bibinfo {author} {\bibfnamefont {D.}~\bibnamefont
  {Petz}},\ }\href {https://projecteuclid.org:443/euclid.cmp/1104115260}
  {\bibfield  {journal} {\bibinfo  {journal} {Commun. Math. Phys.}\ }\textbf
  {\bibinfo {volume} {105}},\ \bibinfo {pages} {123} (\bibinfo {year}
  {1986})}\BibitemShut {NoStop}%
\bibitem [{\citenamefont {Petz}(2003)}]{Petz2003}%
  \BibitemOpen
  \bibfield  {author} {\bibinfo {author} {\bibfnamefont {D.}~\bibnamefont
  {Petz}},\ }\href {\doibase 10.1142/S0129055X03001576} {\bibfield  {journal}
  {\bibinfo  {journal} {Rev. Math. Phys.}\ }\textbf {\bibinfo {volume} {15}},\
  \bibinfo {pages} {79} (\bibinfo {year} {2003})}\BibitemShut {NoStop}%
\bibitem [{\citenamefont {Ruskai}(2002)}]{Ruskai2002}%
  \BibitemOpen
  \bibfield  {author} {\bibinfo {author} {\bibfnamefont {M.~B.}\ \bibnamefont
  {Ruskai}},\ }\href {\doibase 10.1063/1.1497701} {\bibfield  {journal}
  {\bibinfo  {journal} {J. Math. Phys.}\ }\textbf {\bibinfo {volume} {43}},\
  \bibinfo {pages} {4358} (\bibinfo {year} {2002})}\BibitemShut {NoStop}%
\bibitem [{\citenamefont {Ibinson}\ \emph {et~al.}(2008)\citenamefont
  {Ibinson}, \citenamefont {Linden},\ and\ \citenamefont
  {Winter}}]{Ibinson2008}%
  \BibitemOpen
  \bibfield  {author} {\bibinfo {author} {\bibfnamefont {B.}~\bibnamefont
  {Ibinson}}, \bibinfo {author} {\bibfnamefont {N.}~\bibnamefont {Linden}}, \
  and\ \bibinfo {author} {\bibfnamefont {A.}~\bibnamefont {Winter}},\ }\href
  {\doibase 10.1007/s00220-007-0362-8} {\bibfield  {journal} {\bibinfo
  {journal} {Commun. Math. Phys.}\ }\textbf {\bibinfo {volume} {277}},\
  \bibinfo {pages} {289} (\bibinfo {year} {2008})}\BibitemShut {NoStop}%
\bibitem [{\citenamefont {Wilde}(2015)}]{Wilde2015}%
  \BibitemOpen
  \bibfield  {author} {\bibinfo {author} {\bibfnamefont {M.~M.}\ \bibnamefont
  {Wilde}},\ }\href {\doibase 10.1098/rspa.2015.0338} {\bibfield  {journal}
  {\bibinfo  {journal} {Proc. Royal Soc. A}\ }\textbf {\bibinfo {volume}
  {471}},\ \bibinfo {pages} {2182} (\bibinfo {year} {2015})}\BibitemShut
  {NoStop}%
\bibitem [{\citenamefont {Sutter}\ \emph {et~al.}(2017)\citenamefont {Sutter},
  \citenamefont {Berta},\ and\ \citenamefont {Tomamichel}}]{Sutter2017}%
  \BibitemOpen
  \bibfield  {author} {\bibinfo {author} {\bibfnamefont {D.}~\bibnamefont
  {Sutter}}, \bibinfo {author} {\bibfnamefont {M.}~\bibnamefont {Berta}}, \
  and\ \bibinfo {author} {\bibfnamefont {M.}~\bibnamefont {Tomamichel}},\
  }\href {\doibase 10.1007/s00220-016-2778-5} {\bibfield  {journal} {\bibinfo
  {journal} {Commun. Math. Phys.}\ }\textbf {\bibinfo {volume} {352}},\
  \bibinfo {pages} {37} (\bibinfo {year} {2017})}\BibitemShut {NoStop}%
\bibitem [{Note19()}]{Note19}%
  \BibitemOpen
  \bibinfo {note} {The duals are, for example, all positive, if the blocking
  instrument consists of elements with Choi that are orthogonal projectors. In
  this case, $\protect \widetilde {\protect \mathcal {W}}_{M\rightarrow FM}$ is
  a proper recovery map and the QCMI vanishes.}\BibitemShut {Stop}%
\bibitem [{Note20()}]{Note20}%
  \BibitemOpen
  \bibinfo {note} {S stands for a scattering matrix, L for jump operators, and
  H for Hamiltonians}\BibitemShut {NoStop}%
\bibitem [{\citenamefont {Parthasarathy}(1992)}]{Parthasarathy}%
  \BibitemOpen
  \bibfield  {author} {\bibinfo {author} {\bibfnamefont {K.}~\bibnamefont
  {Parthasarathy}},\ }\href@noop {} {\emph {\bibinfo {title} {An introduction
  to quantum stochastic calculus}}},\ Monographs in Mathematics. 85.\ (\bibinfo
  {address} {Basel: Birkh{\"a}user Verlag},\ \bibinfo {year}
  {1992})\BibitemShut {NoStop}%
\bibitem [{\citenamefont {Daley}(2014)}]{qtraj}%
  \BibitemOpen
  \bibfield  {author} {\bibinfo {author} {\bibfnamefont {A.~J.}\ \bibnamefont
  {Daley}},\ }\href {\doibase 10.1080/00018732.2014.933502} {\bibfield
  {journal} {\bibinfo  {journal} {Adv. Phys.}\ }\textbf {\bibinfo {volume}
  {63}},\ \bibinfo {pages} {77} (\bibinfo {year} {2014})}\BibitemShut {NoStop}%
\bibitem [{\citenamefont {Nurdin}(2019)}]{Nurdin2019}%
  \BibitemOpen
  \bibfield  {author} {\bibinfo {author} {\bibfnamefont {H.~I.}\ \bibnamefont
  {Nurdin}},\ }\enquote {\bibinfo {title} {Quantum stochastic processes and the
  modelling of quantum noise},}\ in\ \href {\doibase
  10.1007/978-1-4471-5102-9_100160-1} {\emph {\bibinfo {booktitle}
  {Encyclopedia of Systems and Control}}},\ \bibinfo {editor} {edited by\
  \bibinfo {editor} {\bibfnamefont {J.}~\bibnamefont {Baillieul}}\ and\
  \bibinfo {editor} {\bibfnamefont {T.}~\bibnamefont {Samad}}}\ (\bibinfo
  {publisher} {Springer London},\ \bibinfo {address} {London},\ \bibinfo {year}
  {2019})\ pp.\ \bibinfo {pages} {1--8}\BibitemShut {NoStop}%
\bibitem [{\citenamefont {Zurek}(2009)}]{darwin}%
  \BibitemOpen
  \bibfield  {author} {\bibinfo {author} {\bibfnamefont {W.~H.}\ \bibnamefont
  {Zurek}},\ }\href {\doibase 10.1038/nphys1202} {\bibfield  {journal}
  {\bibinfo  {journal} {Nature Physics}\ }\textbf {\bibinfo {volume} {5}},\
  \bibinfo {pages} {181} (\bibinfo {year} {2009})}\BibitemShut {NoStop}%
\bibitem [{\citenamefont {Heyl}\ \emph {et~al.}(2013)\citenamefont {Heyl},
  \citenamefont {Polkovnikov},\ and\ \citenamefont
  {Kehrein}}]{PhysRevLett.110.135704}%
  \BibitemOpen
  \bibfield  {author} {\bibinfo {author} {\bibfnamefont {M.}~\bibnamefont
  {Heyl}}, \bibinfo {author} {\bibfnamefont {A.}~\bibnamefont {Polkovnikov}}, \
  and\ \bibinfo {author} {\bibfnamefont {S.}~\bibnamefont {Kehrein}},\ }\href
  {\doibase 10.1103/PhysRevLett.110.135704} {\bibfield  {journal} {\bibinfo
  {journal} {Phys. Rev. Lett.}\ }\textbf {\bibinfo {volume} {110}},\ \bibinfo
  {pages} {135704} (\bibinfo {year} {2013})}\BibitemShut {NoStop}%
\bibitem [{\citenamefont {Turner}\ \emph {et~al.}(2018)\citenamefont {Turner},
  \citenamefont {Michailidis}, \citenamefont {Abanin}, \citenamefont {Serbyn},\
  and\ \citenamefont {Papi{\'c}}}]{scar}%
  \BibitemOpen
  \bibfield  {author} {\bibinfo {author} {\bibfnamefont {C.~J.}\ \bibnamefont
  {Turner}}, \bibinfo {author} {\bibfnamefont {A.~A.}\ \bibnamefont
  {Michailidis}}, \bibinfo {author} {\bibfnamefont {D.~A.}\ \bibnamefont
  {Abanin}}, \bibinfo {author} {\bibfnamefont {M.}~\bibnamefont {Serbyn}}, \
  and\ \bibinfo {author} {\bibfnamefont {Z.}~\bibnamefont {Papi{\'c}}},\ }\href
  {\doibase 10.1038/s41567-018-0137-5} {\bibfield  {journal} {\bibinfo
  {journal} {Nature Physics}\ }\textbf {\bibinfo {volume} {14}},\ \bibinfo
  {pages} {745} (\bibinfo {year} {2018})}\BibitemShut {NoStop}%
\bibitem [{\citenamefont {Skinner}\ \emph {et~al.}(2019)\citenamefont
  {Skinner}, \citenamefont {Ruhman},\ and\ \citenamefont
  {Nahum}}]{PhysRevX.9.031009}%
  \BibitemOpen
  \bibfield  {author} {\bibinfo {author} {\bibfnamefont {B.}~\bibnamefont
  {Skinner}}, \bibinfo {author} {\bibfnamefont {J.}~\bibnamefont {Ruhman}}, \
  and\ \bibinfo {author} {\bibfnamefont {A.}~\bibnamefont {Nahum}},\ }\href
  {\doibase 10.1103/PhysRevX.9.031009} {\bibfield  {journal} {\bibinfo
  {journal} {Phys. Rev. X}\ }\textbf {\bibinfo {volume} {9}},\ \bibinfo {pages}
  {031009} (\bibinfo {year} {2019})}\BibitemShut {NoStop}%
\bibitem [{\citenamefont {Viola}\ \emph
  {et~al.}(1999{\natexlab{a}})\citenamefont {Viola}, \citenamefont {Knill},\
  and\ \citenamefont {Lloyd}}]{PhysRevLett.82.2417}%
  \BibitemOpen
  \bibfield  {author} {\bibinfo {author} {\bibfnamefont {L.}~\bibnamefont
  {Viola}}, \bibinfo {author} {\bibfnamefont {E.}~\bibnamefont {Knill}}, \ and\
  \bibinfo {author} {\bibfnamefont {S.}~\bibnamefont {Lloyd}},\ }\href
  {\doibase 10.1103/PhysRevLett.82.2417} {\bibfield  {journal} {\bibinfo
  {journal} {Phys. Rev. Lett.}\ }\textbf {\bibinfo {volume} {82}},\ \bibinfo
  {pages} {2417} (\bibinfo {year} {1999}{\natexlab{a}})}\BibitemShut {NoStop}%
\bibitem [{\citenamefont {Viola}\ \emph
  {et~al.}(1999{\natexlab{b}})\citenamefont {Viola}, \citenamefont {Lloyd},\
  and\ \citenamefont {Knill}}]{PhysRevLett.83.4888}%
  \BibitemOpen
  \bibfield  {author} {\bibinfo {author} {\bibfnamefont {L.}~\bibnamefont
  {Viola}}, \bibinfo {author} {\bibfnamefont {S.}~\bibnamefont {Lloyd}}, \ and\
  \bibinfo {author} {\bibfnamefont {E.}~\bibnamefont {Knill}},\ }\href
  {\doibase 10.1103/PhysRevLett.83.4888} {\bibfield  {journal} {\bibinfo
  {journal} {Phys. Rev. Lett.}\ }\textbf {\bibinfo {volume} {83}},\ \bibinfo
  {pages} {4888} (\bibinfo {year} {1999}{\natexlab{b}})}\BibitemShut {NoStop}%
\bibitem [{\citenamefont {Arenz}\ \emph {et~al.}(2018)\citenamefont {Arenz},
  \citenamefont {Burgarth}, \citenamefont {Facchi},\ and\ \citenamefont
  {Hillier}}]{Arenz2017}%
  \BibitemOpen
  \bibfield  {author} {\bibinfo {author} {\bibfnamefont {C.}~\bibnamefont
  {Arenz}}, \bibinfo {author} {\bibfnamefont {D.}~\bibnamefont {Burgarth}},
  \bibinfo {author} {\bibfnamefont {P.}~\bibnamefont {Facchi}}, \ and\ \bibinfo
  {author} {\bibfnamefont {R.}~\bibnamefont {Hillier}},\ }\href {\doibase
  10.1063/1.5016495} {\bibfield  {journal} {\bibinfo  {journal} {J. Math.
  Phys.}\ }\textbf {\bibinfo {volume} {59}},\ \bibinfo {pages} {032203}
  (\bibinfo {year} {2018})}\BibitemShut {NoStop}%
\bibitem [{\citenamefont {Addis}\ \emph {et~al.}(2015)\citenamefont {Addis},
  \citenamefont {Ciccarello}, \citenamefont {Cascio}, \citenamefont {Palma},\
  and\ \citenamefont {Maniscalco}}]{addis-dd}%
  \BibitemOpen
  \bibfield  {author} {\bibinfo {author} {\bibfnamefont {C.}~\bibnamefont
  {Addis}}, \bibinfo {author} {\bibfnamefont {F.}~\bibnamefont {Ciccarello}},
  \bibinfo {author} {\bibfnamefont {M.}~\bibnamefont {Cascio}}, \bibinfo
  {author} {\bibfnamefont {G.~M.}\ \bibnamefont {Palma}}, \ and\ \bibinfo
  {author} {\bibfnamefont {S.}~\bibnamefont {Maniscalco}},\ }\href {\doibase
  10.1088/1367-2630/17/12/123004} {\bibfield  {journal} {\bibinfo  {journal}
  {New J. Phys.}\ }\textbf {\bibinfo {volume} {17}},\ \bibinfo {pages} {123004}
  (\bibinfo {year} {2015})}\BibitemShut {NoStop}%
\bibitem [{\citenamefont {Burgarth}\ \emph
  {et~al.}(2020{\natexlab{a}})\citenamefont {Burgarth}, \citenamefont {Facchi},
  \citenamefont {Fraas},\ and\ \citenamefont {Hillier}}]{arXiv:1904.03627}%
  \BibitemOpen
  \bibfield  {author} {\bibinfo {author} {\bibfnamefont {D.}~\bibnamefont
  {Burgarth}}, \bibinfo {author} {\bibfnamefont {P.}~\bibnamefont {Facchi}},
  \bibinfo {author} {\bibfnamefont {M.}~\bibnamefont {Fraas}}, \ and\ \bibinfo
  {author} {\bibfnamefont {R.}~\bibnamefont {Hillier}},\ }\href
  {https://www.arXiv.org/abs/1904.03627} {\bibfield  {journal} {\bibinfo
  {journal} {arXiv:1904.03627}\ } (\bibinfo {year}
  {2020}{\natexlab{a}})}\BibitemShut {NoStop}%
\bibitem [{\citenamefont {Zanardi}(1999)}]{zanardi}%
  \BibitemOpen
  \bibfield  {author} {\bibinfo {author} {\bibfnamefont {P.}~\bibnamefont
  {Zanardi}},\ }\href {\doibase 10.1016/S0375-9601(99)00365-5} {\bibfield
  {journal} {\bibinfo  {journal} {Phys. Lett. A}\ }\textbf {\bibinfo {volume}
  {258}},\ \bibinfo {pages} {77} (\bibinfo {year} {1999})}\BibitemShut
  {NoStop}%
\bibitem [{\citenamefont {Lidar}\ \emph {et~al.}(1998)\citenamefont {Lidar},
  \citenamefont {Chuang},\ and\ \citenamefont {Whaley}}]{PhysRevLett.81.2594}%
  \BibitemOpen
  \bibfield  {author} {\bibinfo {author} {\bibfnamefont {D.~A.}\ \bibnamefont
  {Lidar}}, \bibinfo {author} {\bibfnamefont {I.~L.}\ \bibnamefont {Chuang}}, \
  and\ \bibinfo {author} {\bibfnamefont {K.~B.}\ \bibnamefont {Whaley}},\
  }\href {\doibase 10.1103/PhysRevLett.81.2594} {\bibfield  {journal} {\bibinfo
   {journal} {Phys. Rev. Lett.}\ }\textbf {\bibinfo {volume} {81}},\ \bibinfo
  {pages} {2594} (\bibinfo {year} {1998})}\BibitemShut {NoStop}%
\bibitem [{\citenamefont {Lidar}\ and\ \citenamefont {Brun}(2013)}]{lidarQEC}%
  \BibitemOpen
  \bibfield  {author} {\bibinfo {author} {\bibfnamefont {D.~A.}\ \bibnamefont
  {Lidar}}\ and\ \bibinfo {author} {\bibfnamefont {T.~A.}\ \bibnamefont
  {Brun}},\ }\href@noop {} {\emph {\bibinfo {title} {Quantum Error
  Correction}}}\ (\bibinfo  {publisher} {Cambridge University Press},\ \bibinfo
  {year} {2013})\BibitemShut {NoStop}%
\bibitem [{\citenamefont {Paz-Silva}\ \emph {et~al.}(2012)\citenamefont
  {Paz-Silva}, \citenamefont {Rezakhani}, \citenamefont {Dominy},\ and\
  \citenamefont {Lidar}}]{PhysRevLett.108.080501}%
  \BibitemOpen
  \bibfield  {author} {\bibinfo {author} {\bibfnamefont {G.~A.}\ \bibnamefont
  {Paz-Silva}}, \bibinfo {author} {\bibfnamefont {A.~T.}\ \bibnamefont
  {Rezakhani}}, \bibinfo {author} {\bibfnamefont {J.~M.}\ \bibnamefont
  {Dominy}}, \ and\ \bibinfo {author} {\bibfnamefont {D.~A.}\ \bibnamefont
  {Lidar}},\ }\href {\doibase 10.1103/PhysRevLett.108.080501} {\bibfield
  {journal} {\bibinfo  {journal} {Phys. Rev. Lett.}\ }\textbf {\bibinfo
  {volume} {108}},\ \bibinfo {pages} {080501} (\bibinfo {year}
  {2012})}\BibitemShut {NoStop}%
\bibitem [{\citenamefont {Haase}\ \emph {et~al.}(2018)\citenamefont {Haase},
  \citenamefont {Vetter}, \citenamefont {Unden}, \citenamefont {Smirne},
  \citenamefont {Rosskopf}, \citenamefont {Naydenov}, \citenamefont {Stacey},
  \citenamefont {Jelezko}, \citenamefont {Plenio},\ and\ \citenamefont
  {Huelga}}]{PhysRevLett.121.060401}%
  \BibitemOpen
  \bibfield  {author} {\bibinfo {author} {\bibfnamefont {J.~F.}\ \bibnamefont
  {Haase}}, \bibinfo {author} {\bibfnamefont {P.~J.}\ \bibnamefont {Vetter}},
  \bibinfo {author} {\bibfnamefont {T.}~\bibnamefont {Unden}}, \bibinfo
  {author} {\bibfnamefont {A.}~\bibnamefont {Smirne}}, \bibinfo {author}
  {\bibfnamefont {J.}~\bibnamefont {Rosskopf}}, \bibinfo {author}
  {\bibfnamefont {B.}~\bibnamefont {Naydenov}}, \bibinfo {author}
  {\bibfnamefont {A.}~\bibnamefont {Stacey}}, \bibinfo {author} {\bibfnamefont
  {F.}~\bibnamefont {Jelezko}}, \bibinfo {author} {\bibfnamefont {M.~B.}\
  \bibnamefont {Plenio}}, \ and\ \bibinfo {author} {\bibfnamefont {S.~F.}\
  \bibnamefont {Huelga}},\ }\href {\doibase 10.1103/PhysRevLett.121.060401}
  {\bibfield  {journal} {\bibinfo  {journal} {Phys. Rev. Lett.}\ }\textbf
  {\bibinfo {volume} {121}},\ \bibinfo {pages} {060401} (\bibinfo {year}
  {2018})}\BibitemShut {NoStop}%
\bibitem [{\citenamefont {Burgarth}\ \emph
  {et~al.}(2020{\natexlab{b}})\citenamefont {Burgarth}, \citenamefont {Facchi},
  \citenamefont {Nakazato}, \citenamefont {Pascazio},\ and\ \citenamefont
  {Yuasa}}]{Burgarth2020quantumzenodynamics}%
  \BibitemOpen
  \bibfield  {author} {\bibinfo {author} {\bibfnamefont {D.}~\bibnamefont
  {Burgarth}}, \bibinfo {author} {\bibfnamefont {P.}~\bibnamefont {Facchi}},
  \bibinfo {author} {\bibfnamefont {H.}~\bibnamefont {Nakazato}}, \bibinfo
  {author} {\bibfnamefont {S.}~\bibnamefont {Pascazio}}, \ and\ \bibinfo
  {author} {\bibfnamefont {K.}~\bibnamefont {Yuasa}},\ }\href {\doibase
  10.22331/q-2020-07-06-289} {\bibfield  {journal} {\bibinfo  {journal}
  {{Quantum}}\ }\textbf {\bibinfo {volume} {4}},\ \bibinfo {pages} {289}
  (\bibinfo {year} {2020}{\natexlab{b}})}\BibitemShut {NoStop}%
\bibitem [{\citenamefont {Caruso}\ \emph {et~al.}(2012)\citenamefont {Caruso},
  \citenamefont {Montangero}, \citenamefont {Calarco}, \citenamefont {Huelga},\
  and\ \citenamefont {Plenio}}]{caruso_coherent_2012}%
  \BibitemOpen
  \bibfield  {author} {\bibinfo {author} {\bibfnamefont {F.}~\bibnamefont
  {Caruso}}, \bibinfo {author} {\bibfnamefont {S.}~\bibnamefont {Montangero}},
  \bibinfo {author} {\bibfnamefont {T.}~\bibnamefont {Calarco}}, \bibinfo
  {author} {\bibfnamefont {S.~F.}\ \bibnamefont {Huelga}}, \ and\ \bibinfo
  {author} {\bibfnamefont {M.~B.}\ \bibnamefont {Plenio}},\ }\href {\doibase
  10.1103/PhysRevA.85.042331} {\bibfield  {journal} {\bibinfo  {journal} {Phys.
  Rev. A}\ }\textbf {\bibinfo {volume} {85}},\ \bibinfo {pages} {042331}
  (\bibinfo {year} {2012})}\BibitemShut {NoStop}%
\bibitem [{\citenamefont {Plenio}\ \emph {et~al.}(1999)\citenamefont {Plenio},
  \citenamefont {Huelga}, \citenamefont {Beige},\ and\ \citenamefont
  {Knight}}]{plenio_cavity-loss-induced_1999}%
  \BibitemOpen
  \bibfield  {author} {\bibinfo {author} {\bibfnamefont {M.~B.}\ \bibnamefont
  {Plenio}}, \bibinfo {author} {\bibfnamefont {S.~F.}\ \bibnamefont {Huelga}},
  \bibinfo {author} {\bibfnamefont {A.}~\bibnamefont {Beige}}, \ and\ \bibinfo
  {author} {\bibfnamefont {P.~L.}\ \bibnamefont {Knight}},\ }\href {\doibase
  10.1103/PhysRevA.59.2468} {\bibfield  {journal} {\bibinfo  {journal} {Phys.
  Rev. A}\ }\textbf {\bibinfo {volume} {59}},\ \bibinfo {pages} {2468}
  (\bibinfo {year} {1999})}\BibitemShut {NoStop}%
\bibitem [{\citenamefont {Verstraete}\ \emph {et~al.}(2009)\citenamefont
  {Verstraete}, \citenamefont {Wolf},\ and\ \citenamefont
  {Cirac}}]{verstraete_quantum_2009}%
  \BibitemOpen
  \bibfield  {author} {\bibinfo {author} {\bibfnamefont {F.}~\bibnamefont
  {Verstraete}}, \bibinfo {author} {\bibfnamefont {M.~M.}\ \bibnamefont
  {Wolf}}, \ and\ \bibinfo {author} {\bibfnamefont {J.~I.}\ \bibnamefont
  {Cirac}},\ }\href {\doibase 10.1038/nphys1342} {\bibfield  {journal}
  {\bibinfo  {journal} {Nat. Phys.}\ }\textbf {\bibinfo {volume} {5}},\
  \bibinfo {pages} {633} (\bibinfo {year} {2009})}\BibitemShut {NoStop}%
\bibitem [{\citenamefont {Caruso}\ \emph {et~al.}(2009)\citenamefont {Caruso},
  \citenamefont {Chin}, \citenamefont {Datta}, \citenamefont {Huelga},\ and\
  \citenamefont {Plenio}}]{caruso_highly_2009}%
  \BibitemOpen
  \bibfield  {author} {\bibinfo {author} {\bibfnamefont {F.}~\bibnamefont
  {Caruso}}, \bibinfo {author} {\bibfnamefont {A.~W.}\ \bibnamefont {Chin}},
  \bibinfo {author} {\bibfnamefont {A.}~\bibnamefont {Datta}}, \bibinfo
  {author} {\bibfnamefont {S.~F.}\ \bibnamefont {Huelga}}, \ and\ \bibinfo
  {author} {\bibfnamefont {M.~B.}\ \bibnamefont {Plenio}},\ }\href {\doibase
  10.1063/1.3223548} {\bibfield  {journal} {\bibinfo  {journal} {J. Chem.
  Phys.}\ }\textbf {\bibinfo {volume} {131}},\ \bibinfo {pages} {105106}
  (\bibinfo {year} {2009})}\BibitemShut {NoStop}%
\bibitem [{\citenamefont {Caruso}\ \emph
  {et~al.}(2010{\natexlab{b}})\citenamefont {Caruso}, \citenamefont {Huelga},\
  and\ \citenamefont {Plenio}}]{caruso_noise-enhanced_2010}%
  \BibitemOpen
  \bibfield  {author} {\bibinfo {author} {\bibfnamefont {F.}~\bibnamefont
  {Caruso}}, \bibinfo {author} {\bibfnamefont {S.~F.}\ \bibnamefont {Huelga}},
  \ and\ \bibinfo {author} {\bibfnamefont {M.~B.}\ \bibnamefont {Plenio}},\
  }\href {\doibase 10.1103/PhysRevLett.105.190501} {\bibfield  {journal}
  {\bibinfo  {journal} {Phys. Rev. Lett.}\ }\textbf {\bibinfo {volume} {105}},\
  \bibinfo {pages} {190501} (\bibinfo {year} {2010}{\natexlab{b}})}\BibitemShut
  {NoStop}%
\bibitem [{\citenamefont {Kastoryano}\ \emph {et~al.}(2013)\citenamefont
  {Kastoryano}, \citenamefont {Wolf},\ and\ \citenamefont
  {Eisert}}]{kastoryano_precisely_2013}%
  \BibitemOpen
  \bibfield  {author} {\bibinfo {author} {\bibfnamefont {M.~J.}\ \bibnamefont
  {Kastoryano}}, \bibinfo {author} {\bibfnamefont {M.~M.}\ \bibnamefont
  {Wolf}}, \ and\ \bibinfo {author} {\bibfnamefont {J.}~\bibnamefont
  {Eisert}},\ }\href {\doibase 10.1103/PhysRevLett.110.110501} {\bibfield
  {journal} {\bibinfo  {journal} {Phys. Rev. Lett.}\ }\textbf {\bibinfo
  {volume} {110}},\ \bibinfo {pages} {110501} (\bibinfo {year}
  {2013})}\BibitemShut {NoStop}%
\bibitem [{\citenamefont {Chin}\ \emph {et~al.}(2012)\citenamefont {Chin},
  \citenamefont {Huelga},\ and\ \citenamefont
  {Plenio}}]{PhysRevLett.109.233601}%
  \BibitemOpen
  \bibfield  {author} {\bibinfo {author} {\bibfnamefont {A.~W.}\ \bibnamefont
  {Chin}}, \bibinfo {author} {\bibfnamefont {S.~F.}\ \bibnamefont {Huelga}}, \
  and\ \bibinfo {author} {\bibfnamefont {M.~B.}\ \bibnamefont {Plenio}},\
  }\href {\doibase 10.1103/PhysRevLett.109.233601} {\bibfield  {journal}
  {\bibinfo  {journal} {Phys. Rev. Lett.}\ }\textbf {\bibinfo {volume} {109}},\
  \bibinfo {pages} {233601} (\bibinfo {year} {2012})}\BibitemShut {NoStop}%
\bibitem [{\citenamefont {Am-Shallem}\ and\ \citenamefont
  {Kosloff}(2014)}]{weakfieldphase}%
  \BibitemOpen
  \bibfield  {author} {\bibinfo {author} {\bibfnamefont {M.}~\bibnamefont
  {Am-Shallem}}\ and\ \bibinfo {author} {\bibfnamefont {R.}~\bibnamefont
  {Kosloff}},\ }\href {\doibase 10.1063/1.4890822} {\bibfield  {journal}
  {\bibinfo  {journal} {J. Chem. Phys.}\ }\textbf {\bibinfo {volume} {141}},\
  \bibinfo {pages} {044121} (\bibinfo {year} {2014})}\BibitemShut {NoStop}%
\bibitem [{\citenamefont {Liu}\ \emph {et~al.}(2020)\citenamefont {Liu},
  \citenamefont {Sun}, \citenamefont {Liu}, \citenamefont {Li}, \citenamefont
  {Guo}, \citenamefont {Hamedani~Raja}, \citenamefont {Lyyra},\ and\
  \citenamefont {Piilo}}]{liu_experimental_2020}%
  \BibitemOpen
  \bibfield  {author} {\bibinfo {author} {\bibfnamefont {Z.-D.}\ \bibnamefont
  {Liu}}, \bibinfo {author} {\bibfnamefont {Y.-N.}\ \bibnamefont {Sun}},
  \bibinfo {author} {\bibfnamefont {B.-H.}\ \bibnamefont {Liu}}, \bibinfo
  {author} {\bibfnamefont {C.-F.}\ \bibnamefont {Li}}, \bibinfo {author}
  {\bibfnamefont {G.-C.}\ \bibnamefont {Guo}}, \bibinfo {author} {\bibfnamefont
  {S.}~\bibnamefont {Hamedani~Raja}}, \bibinfo {author} {\bibfnamefont
  {H.}~\bibnamefont {Lyyra}}, \ and\ \bibinfo {author} {\bibfnamefont
  {J.}~\bibnamefont {Piilo}},\ }\href {\doibase 10.1103/PhysRevA.102.062208}
  {\bibfield  {journal} {\bibinfo  {journal} {Phys. Rev. A}\ }\textbf {\bibinfo
  {volume} {102}},\ \bibinfo {pages} {062208} (\bibinfo {year}
  {2020})}\BibitemShut {NoStop}%
\bibitem [{\citenamefont {Huelga}\ \emph {et~al.}(2012)\citenamefont {Huelga},
  \citenamefont {Rivas},\ and\ \citenamefont
  {Plenio}}]{PhysRevLett.108.160402}%
  \BibitemOpen
  \bibfield  {author} {\bibinfo {author} {\bibfnamefont {S.~F.}\ \bibnamefont
  {Huelga}}, \bibinfo {author} {\bibfnamefont {A.}~\bibnamefont {Rivas}}, \
  and\ \bibinfo {author} {\bibfnamefont {M.~B.}\ \bibnamefont {Plenio}},\
  }\href {\doibase 10.1103/PhysRevLett.108.160402} {\bibfield  {journal}
  {\bibinfo  {journal} {Phys. Rev. Lett.}\ }\textbf {\bibinfo {volume} {108}},\
  \bibinfo {pages} {160402} (\bibinfo {year} {2012})}\BibitemShut {NoStop}%
\bibitem [{\citenamefont {Burgarth}\ \emph {et~al.}(2014)\citenamefont
  {Burgarth}, \citenamefont {Facchi}, \citenamefont {Giovannetti},
  \citenamefont {Nakazato}, \citenamefont {Pascazio},\ and\ \citenamefont
  {Yuasa}}]{compen}%
  \BibitemOpen
  \bibfield  {author} {\bibinfo {author} {\bibfnamefont {D.~K.}\ \bibnamefont
  {Burgarth}}, \bibinfo {author} {\bibfnamefont {P.}~\bibnamefont {Facchi}},
  \bibinfo {author} {\bibfnamefont {V.}~\bibnamefont {Giovannetti}}, \bibinfo
  {author} {\bibfnamefont {H.}~\bibnamefont {Nakazato}}, \bibinfo {author}
  {\bibfnamefont {S.}~\bibnamefont {Pascazio}}, \ and\ \bibinfo {author}
  {\bibfnamefont {K.}~\bibnamefont {Yuasa}},\ }\href {\doibase
  10.1038/ncomms6173} {\bibfield  {journal} {\bibinfo  {journal} {Nat.
  Commun.}\ }\textbf {\bibinfo {volume} {5}},\ \bibinfo {pages} {5173}
  (\bibinfo {year} {2014})}\BibitemShut {NoStop}%
\bibitem [{\citenamefont {Norris}\ \emph {et~al.}(2016)\citenamefont {Norris},
  \citenamefont {Paz-Silva},\ and\ \citenamefont
  {Viola}}]{PhysRevLett.116.150503}%
  \BibitemOpen
  \bibfield  {author} {\bibinfo {author} {\bibfnamefont {L.~M.}\ \bibnamefont
  {Norris}}, \bibinfo {author} {\bibfnamefont {G.~A.}\ \bibnamefont
  {Paz-Silva}}, \ and\ \bibinfo {author} {\bibfnamefont {L.}~\bibnamefont
  {Viola}},\ }\href {\doibase 10.1103/PhysRevLett.116.150503} {\bibfield
  {journal} {\bibinfo  {journal} {Phys. Rev. Lett.}\ }\textbf {\bibinfo
  {volume} {116}},\ \bibinfo {pages} {150503} (\bibinfo {year}
  {2016})}\BibitemShut {NoStop}%
\bibitem [{\citenamefont {Paz-Silva}\ \emph {et~al.}(2017)\citenamefont
  {Paz-Silva}, \citenamefont {Norris},\ and\ \citenamefont
  {Viola}}]{PhysRevA.95.022121}%
  \BibitemOpen
  \bibfield  {author} {\bibinfo {author} {\bibfnamefont {G.~A.}\ \bibnamefont
  {Paz-Silva}}, \bibinfo {author} {\bibfnamefont {L.~M.}\ \bibnamefont
  {Norris}}, \ and\ \bibinfo {author} {\bibfnamefont {L.}~\bibnamefont
  {Viola}},\ }\href {\doibase 10.1103/PhysRevA.95.022121} {\bibfield  {journal}
  {\bibinfo  {journal} {Phys. Rev. A}\ }\textbf {\bibinfo {volume} {95}},\
  \bibinfo {pages} {022121} (\bibinfo {year} {2017})}\BibitemShut {NoStop}%
\bibitem [{\citenamefont {Strasberg}\ and\ \citenamefont
  {Esposito}(2018)}]{PhysRevLett.121.040601}%
  \BibitemOpen
  \bibfield  {author} {\bibinfo {author} {\bibfnamefont {P.}~\bibnamefont
  {Strasberg}}\ and\ \bibinfo {author} {\bibfnamefont {M.}~\bibnamefont
  {Esposito}},\ }\href {\doibase 10.1103/PhysRevLett.121.040601} {\bibfield
  {journal} {\bibinfo  {journal} {Phys. Rev. Lett.}\ }\textbf {\bibinfo
  {volume} {121}},\ \bibinfo {pages} {040601} (\bibinfo {year}
  {2018})}\BibitemShut {NoStop}%
\bibitem [{\citenamefont {Chru\ifmmode \acute{s}\else
  \'{s}\fi{}ci\ifmmode~\acute{n}\else \'{n}\fi{}ski}\ \emph
  {et~al.}(2017)\citenamefont {Chru\ifmmode \acute{s}\else
  \'{s}\fi{}ci\ifmmode~\acute{n}\else \'{n}\fi{}ski}, \citenamefont
  {Macchiavello},\ and\ \citenamefont {Maniscalco}}]{PhysRevLett.118.080404}%
  \BibitemOpen
  \bibfield  {author} {\bibinfo {author} {\bibfnamefont {D.}~\bibnamefont
  {Chru\ifmmode \acute{s}\else \'{s}\fi{}ci\ifmmode~\acute{n}\else
  \'{n}\fi{}ski}}, \bibinfo {author} {\bibfnamefont {C.}~\bibnamefont
  {Macchiavello}}, \ and\ \bibinfo {author} {\bibfnamefont {S.}~\bibnamefont
  {Maniscalco}},\ }\href {\doibase 10.1103/PhysRevLett.118.080404} {\bibfield
  {journal} {\bibinfo  {journal} {Phys. Rev. Lett.}\ }\textbf {\bibinfo
  {volume} {118}},\ \bibinfo {pages} {080404} (\bibinfo {year}
  {2017})}\BibitemShut {NoStop}%
\bibitem [{\citenamefont {Benedetti}\ \emph {et~al.}(2014)\citenamefont
  {Benedetti}, \citenamefont {Paris},\ and\ \citenamefont
  {Maniscalco}}]{benedetti_non-markovianity_2014}%
  \BibitemOpen
  \bibfield  {author} {\bibinfo {author} {\bibfnamefont {C.}~\bibnamefont
  {Benedetti}}, \bibinfo {author} {\bibfnamefont {M.~G.~A.}\ \bibnamefont
  {Paris}}, \ and\ \bibinfo {author} {\bibfnamefont {S.}~\bibnamefont
  {Maniscalco}},\ }\href {\doibase 10.1103/PhysRevA.89.012114} {\bibfield
  {journal} {\bibinfo  {journal} {Phys. Rev. A}\ }\textbf {\bibinfo {volume}
  {89}},\ \bibinfo {pages} {012114} (\bibinfo {year} {2014})}\BibitemShut
  {NoStop}%
\bibitem [{\citenamefont {Schindler}\ \emph {et~al.}(2013)\citenamefont
  {Schindler}, \citenamefont {M\"uller}, \citenamefont {Nigg}, \citenamefont
  {Barreiro}, \citenamefont {Martinez}, \citenamefont {Hennrich}, \citenamefont
  {Monz}, \citenamefont {Diehl}, \citenamefont {Zoller},\ and\ \citenamefont
  {Blatt}}]{Schindler}%
  \BibitemOpen
  \bibfield  {author} {\bibinfo {author} {\bibfnamefont {P.}~\bibnamefont
  {Schindler}}, \bibinfo {author} {\bibfnamefont {M.}~\bibnamefont {M\"uller}},
  \bibinfo {author} {\bibfnamefont {D.}~\bibnamefont {Nigg}}, \bibinfo {author}
  {\bibfnamefont {J.~T.}\ \bibnamefont {Barreiro}}, \bibinfo {author}
  {\bibfnamefont {E.~A.}\ \bibnamefont {Martinez}}, \bibinfo {author}
  {\bibfnamefont {M.}~\bibnamefont {Hennrich}}, \bibinfo {author}
  {\bibfnamefont {T.}~\bibnamefont {Monz}}, \bibinfo {author} {\bibfnamefont
  {S.}~\bibnamefont {Diehl}}, \bibinfo {author} {\bibfnamefont
  {P.}~\bibnamefont {Zoller}}, \ and\ \bibinfo {author} {\bibfnamefont
  {R.}~\bibnamefont {Blatt}},\ }\href {\doibase 10.1038/nphys2630} {\bibfield
  {journal} {\bibinfo  {journal} {Nat. Phys.}\ }\textbf {\bibinfo {volume}
  {9}},\ \bibinfo {pages} {361} (\bibinfo {year} {2013})}\BibitemShut {NoStop}%
\bibitem [{\citenamefont {Bylicka}\ \emph {et~al.}(2013)\citenamefont
  {Bylicka}, \citenamefont {Chru\'sci\'nski},\ and\ \citenamefont
  {Maniscalco}}]{arXiv:1301.2585}%
  \BibitemOpen
  \bibfield  {author} {\bibinfo {author} {\bibfnamefont {B.}~\bibnamefont
  {Bylicka}}, \bibinfo {author} {\bibfnamefont {D.}~\bibnamefont
  {Chru\'sci\'nski}}, \ and\ \bibinfo {author} {\bibfnamefont {S.}~\bibnamefont
  {Maniscalco}},\ }\href {https://arxiv.org/abs/1301.2585} {\bibfield
  {journal} {\bibinfo  {journal} {arXiv:1301.2585}\ } (\bibinfo {year}
  {2013})}\BibitemShut {NoStop}%
\bibitem [{\citenamefont {Rosset}\ \emph {et~al.}(2018)\citenamefont {Rosset},
  \citenamefont {Buscemi},\ and\ \citenamefont {Liang}}]{PhysRevX.8.021033}%
  \BibitemOpen
  \bibfield  {author} {\bibinfo {author} {\bibfnamefont {D.}~\bibnamefont
  {Rosset}}, \bibinfo {author} {\bibfnamefont {F.}~\bibnamefont {Buscemi}}, \
  and\ \bibinfo {author} {\bibfnamefont {Y.-C.}\ \bibnamefont {Liang}},\ }\href
  {\doibase 10.1103/PhysRevX.8.021033} {\bibfield  {journal} {\bibinfo
  {journal} {Phys. Rev. X}\ }\textbf {\bibinfo {volume} {8}},\ \bibinfo {pages}
  {021033} (\bibinfo {year} {2018})}\BibitemShut {NoStop}%
\bibitem [{\citenamefont {Bylicka}\ \emph {et~al.}(2016)\citenamefont
  {Bylicka}, \citenamefont {Tukiainen}, \citenamefont {Chru{\'s}ci{\'n}ski},
  \citenamefont {Piilo},\ and\ \citenamefont
  {Maniscalco}}]{bylicka_thermodynamic_2016}%
  \BibitemOpen
  \bibfield  {author} {\bibinfo {author} {\bibfnamefont {B.}~\bibnamefont
  {Bylicka}}, \bibinfo {author} {\bibfnamefont {M.}~\bibnamefont {Tukiainen}},
  \bibinfo {author} {\bibfnamefont {D.}~\bibnamefont {Chru{\'s}ci{\'n}ski}},
  \bibinfo {author} {\bibfnamefont {J.}~\bibnamefont {Piilo}}, \ and\ \bibinfo
  {author} {\bibfnamefont {S.}~\bibnamefont {Maniscalco}},\ }\href {\doibase
  10.1038/srep27989} {\bibfield  {journal} {\bibinfo  {journal} {Sci. Rep.}\
  }\textbf {\bibinfo {volume} {6}},\ \bibinfo {pages} {27989} (\bibinfo {year}
  {2016})}\BibitemShut {NoStop}%
\bibitem [{\citenamefont {Berk}\ \emph {et~al.}(2021)\citenamefont {Berk},
  \citenamefont {Garner}, \citenamefont {Yadin}, \citenamefont {Modi},\ and\
  \citenamefont {Pollock}}]{berk}%
  \BibitemOpen
  \bibfield  {author} {\bibinfo {author} {\bibfnamefont {G.~D.}\ \bibnamefont
  {Berk}}, \bibinfo {author} {\bibfnamefont {A.~J.~P.}\ \bibnamefont {Garner}},
  \bibinfo {author} {\bibfnamefont {B.}~\bibnamefont {Yadin}}, \bibinfo
  {author} {\bibfnamefont {K.}~\bibnamefont {Modi}}, \ and\ \bibinfo {author}
  {\bibfnamefont {F.~A.}\ \bibnamefont {Pollock}},\ }\href
  {https://arxiv.org/abs/1907.07003} {\bibfield  {journal} {\bibinfo  {journal}
  {Quantum}\ } (\bibinfo {year} {2021})}\BibitemShut {NoStop}%
\bibitem [{\citenamefont {Singha~Roy}\ and\ \citenamefont
  {Bae}(2019)}]{singha_roy_information-theoretic_2019}%
  \BibitemOpen
  \bibfield  {author} {\bibinfo {author} {\bibfnamefont {S.}~\bibnamefont
  {Singha~Roy}}\ and\ \bibinfo {author} {\bibfnamefont {J.}~\bibnamefont
  {Bae}},\ }\href {\doibase 10.1103/PhysRevA.100.032303} {\bibfield  {journal}
  {\bibinfo  {journal} {Phys. Rev. A}\ }\textbf {\bibinfo {volume} {100}},\
  \bibinfo {pages} {032303} (\bibinfo {year} {2019})}\BibitemShut {NoStop}%
\bibitem [{\citenamefont {Abiuso}\ and\ \citenamefont
  {Giovannetti}(2019)}]{abiuso_non-markov_2019}%
  \BibitemOpen
  \bibfield  {author} {\bibinfo {author} {\bibfnamefont {P.}~\bibnamefont
  {Abiuso}}\ and\ \bibinfo {author} {\bibfnamefont {V.}~\bibnamefont
  {Giovannetti}},\ }\href {\doibase 10.1103/PhysRevA.99.052106} {\bibfield
  {journal} {\bibinfo  {journal} {Phys. Rev. A}\ }\textbf {\bibinfo {volume}
  {99}},\ \bibinfo {pages} {052106} (\bibinfo {year} {2019})}\BibitemShut
  {NoStop}%
\bibitem [{\citenamefont {Strasberg}(2019{\natexlab{a}})}]{Strasberg2018}%
  \BibitemOpen
  \bibfield  {author} {\bibinfo {author} {\bibfnamefont {P.}~\bibnamefont
  {Strasberg}},\ }\href {\doibase 10.1103/PhysRevE.100.022127} {\bibfield
  {journal} {\bibinfo  {journal} {Phys. Rev. E}\ }\textbf {\bibinfo {volume}
  {100}},\ \bibinfo {pages} {022127} (\bibinfo {year}
  {2019}{\natexlab{a}})}\BibitemShut {NoStop}%
\bibitem [{\citenamefont {Strasberg}\ and\ \citenamefont
  {Winter}(2019)}]{Strasberg2019-1}%
  \BibitemOpen
  \bibfield  {author} {\bibinfo {author} {\bibfnamefont {P.}~\bibnamefont
  {Strasberg}}\ and\ \bibinfo {author} {\bibfnamefont {A.}~\bibnamefont
  {Winter}},\ }\href {\doibase 10.1103/PhysRevE.100.022135} {\bibfield
  {journal} {\bibinfo  {journal} {Phys. Rev. E}\ }\textbf {\bibinfo {volume}
  {100}},\ \bibinfo {pages} {022135} (\bibinfo {year} {2019})}\BibitemShut
  {NoStop}%
\bibitem [{\citenamefont
  {Strasberg}(2019{\natexlab{b}})}]{strasberg_repeated_2019}%
  \BibitemOpen
  \bibfield  {author} {\bibinfo {author} {\bibfnamefont {P.}~\bibnamefont
  {Strasberg}},\ }\href {\doibase 10.1103/PhysRevLett.123.180604} {\bibfield
  {journal} {\bibinfo  {journal} {Phys. Rev. Lett.}\ }\textbf {\bibinfo
  {volume} {123}},\ \bibinfo {pages} {180604} (\bibinfo {year}
  {2019}{\natexlab{b}})}\BibitemShut {NoStop}%
\bibitem [{\citenamefont {Strasberg}(2020)}]{strasberg_thermodynamics_2020}%
  \BibitemOpen
  \bibfield  {author} {\bibinfo {author} {\bibfnamefont {P.}~\bibnamefont
  {Strasberg}},\ }\href {\doibase 10.22331/q-2020-03-02-240} {\bibfield
  {journal} {\bibinfo  {journal} {Quantum}\ }\textbf {\bibinfo {volume} {4}},\
  \bibinfo {pages} {240} (\bibinfo {year} {2020})}\BibitemShut {NoStop}%
\bibitem [{\citenamefont {Chru\ifmmode \acute{s}\else
  \'{s}\fi{}ci\ifmmode~\acute{n}\else \'{n}\fi{}ski}\ \emph
  {et~al.}(2010)\citenamefont {Chru\ifmmode \acute{s}\else
  \'{s}\fi{}ci\ifmmode~\acute{n}\else \'{n}\fi{}ski}, \citenamefont
  {Kossakowski},\ and\ \citenamefont {Pascazio}}]{PhysRevA.81.032101}%
  \BibitemOpen
  \bibfield  {author} {\bibinfo {author} {\bibfnamefont {D.}~\bibnamefont
  {Chru\ifmmode \acute{s}\else \'{s}\fi{}ci\ifmmode~\acute{n}\else
  \'{n}\fi{}ski}}, \bibinfo {author} {\bibfnamefont {A.}~\bibnamefont
  {Kossakowski}}, \ and\ \bibinfo {author} {\bibfnamefont {S.}~\bibnamefont
  {Pascazio}},\ }\href {\doibase 10.1103/PhysRevA.81.032101} {\bibfield
  {journal} {\bibinfo  {journal} {Phys. Rev. A}\ }\textbf {\bibinfo {volume}
  {81}},\ \bibinfo {pages} {032101} (\bibinfo {year} {2010})}\BibitemShut
  {NoStop}%
\bibitem [{\citenamefont {Chru\ifmmode \acute{s}\else
  \'{s}\fi{}ci\ifmmode~\acute{n}\else \'{n}\fi{}ski}\ and\ \citenamefont
  {Wudarski}(2015)}]{PhysRevA.91.012104}%
  \BibitemOpen
  \bibfield  {author} {\bibinfo {author} {\bibfnamefont {D.}~\bibnamefont
  {Chru\ifmmode \acute{s}\else \'{s}\fi{}ci\ifmmode~\acute{n}\else
  \'{n}\fi{}ski}}\ and\ \bibinfo {author} {\bibfnamefont {F.~A.}\ \bibnamefont
  {Wudarski}},\ }\href {\doibase 10.1103/PhysRevA.91.012104} {\bibfield
  {journal} {\bibinfo  {journal} {Phys. Rev. A}\ }\textbf {\bibinfo {volume}
  {91}},\ \bibinfo {pages} {012104} (\bibinfo {year} {2015})}\BibitemShut
  {NoStop}%
\bibitem [{\citenamefont {Banchi}\ \emph {et~al.}(2017)\citenamefont {Banchi},
  \citenamefont {Burgarth},\ and\ \citenamefont
  {Kastoryano}}]{PhysRevX.7.041015}%
  \BibitemOpen
  \bibfield  {author} {\bibinfo {author} {\bibfnamefont {L.}~\bibnamefont
  {Banchi}}, \bibinfo {author} {\bibfnamefont {D.}~\bibnamefont {Burgarth}}, \
  and\ \bibinfo {author} {\bibfnamefont {M.~J.}\ \bibnamefont {Kastoryano}},\
  }\href {\doibase 10.1103/PhysRevX.7.041015} {\bibfield  {journal} {\bibinfo
  {journal} {Phys. Rev. X}\ }\textbf {\bibinfo {volume} {7}},\ \bibinfo {pages}
  {041015} (\bibinfo {year} {2017})}\BibitemShut {NoStop}%
\bibitem [{\citenamefont {Figueroa-Romero}\ \emph {et~al.}(2020)\citenamefont
  {Figueroa-Romero}, \citenamefont {Modi},\ and\ \citenamefont
  {Pollock}}]{Romero2019}%
  \BibitemOpen
  \bibfield  {author} {\bibinfo {author} {\bibfnamefont {P.}~\bibnamefont
  {Figueroa-Romero}}, \bibinfo {author} {\bibfnamefont {K.}~\bibnamefont
  {Modi}}, \ and\ \bibinfo {author} {\bibfnamefont {F.~A.}\ \bibnamefont
  {Pollock}},\ }\href {\doibase 10.1103/PhysRevE.102.032144} {\bibfield
  {journal} {\bibinfo  {journal} {Phys. Rev. E}\ }\textbf {\bibinfo {volume}
  {102}},\ \bibinfo {pages} {032144} (\bibinfo {year} {2020})}\BibitemShut
  {NoStop}%
\bibitem [{\citenamefont {Popescu}\ \emph {et~al.}(2006)\citenamefont
  {Popescu}, \citenamefont {Short},\ and\ \citenamefont
  {Winter}}]{Popescu2006}%
  \BibitemOpen
  \bibfield  {author} {\bibinfo {author} {\bibfnamefont {S.}~\bibnamefont
  {Popescu}}, \bibinfo {author} {\bibfnamefont {A.~J.}\ \bibnamefont {Short}},
  \ and\ \bibinfo {author} {\bibfnamefont {A.}~\bibnamefont {Winter}},\ }\href
  {\doibase 10.1038/nphys444} {\bibfield  {journal} {\bibinfo  {journal} {Nat.
  Phys.}\ }\textbf {\bibinfo {volume} {2}},\ \bibinfo {pages} {754} (\bibinfo
  {year} {2006})}\BibitemShut {NoStop}%
\bibitem [{\citenamefont {Linden}\ \emph {et~al.}(2009)\citenamefont {Linden},
  \citenamefont {Popescu}, \citenamefont {Short},\ and\ \citenamefont
  {Winter}}]{PhysRevE.79.061103}%
  \BibitemOpen
  \bibfield  {author} {\bibinfo {author} {\bibfnamefont {N.}~\bibnamefont
  {Linden}}, \bibinfo {author} {\bibfnamefont {S.}~\bibnamefont {Popescu}},
  \bibinfo {author} {\bibfnamefont {A.~J.}\ \bibnamefont {Short}}, \ and\
  \bibinfo {author} {\bibfnamefont {A.}~\bibnamefont {Winter}},\ }\href
  {\doibase 10.1103/PhysRevE.79.061103} {\bibfield  {journal} {\bibinfo
  {journal} {Phys. Rev. E}\ }\textbf {\bibinfo {volume} {79}},\ \bibinfo
  {pages} {061103} (\bibinfo {year} {2009})}\BibitemShut {NoStop}%
\bibitem [{\citenamefont {Masanes}\ \emph {et~al.}(2013)\citenamefont
  {Masanes}, \citenamefont {Roncaglia},\ and\ \citenamefont
  {Ac\'{\i}n}}]{Masanes2013}%
  \BibitemOpen
  \bibfield  {author} {\bibinfo {author} {\bibfnamefont {L.}~\bibnamefont
  {Masanes}}, \bibinfo {author} {\bibfnamefont {A.~J.}\ \bibnamefont
  {Roncaglia}}, \ and\ \bibinfo {author} {\bibfnamefont {A.}~\bibnamefont
  {Ac\'{\i}n}},\ }\href {\doibase 10.1103/PhysRevE.87.032137} {\bibfield
  {journal} {\bibinfo  {journal} {Phys. Rev. E}\ }\textbf {\bibinfo {volume}
  {87}},\ \bibinfo {pages} {032137} (\bibinfo {year} {2013})}\BibitemShut
  {NoStop}%
\bibitem [{\citenamefont {D'Alessio}\ \emph {et~al.}(2016)\citenamefont
  {D'Alessio}, \citenamefont {Kafri}, \citenamefont {Polkovnikov},\ and\
  \citenamefont {Rigol}}]{DAlessio2016}%
  \BibitemOpen
  \bibfield  {author} {\bibinfo {author} {\bibfnamefont {L.}~\bibnamefont
  {D'Alessio}}, \bibinfo {author} {\bibfnamefont {Y.}~\bibnamefont {Kafri}},
  \bibinfo {author} {\bibfnamefont {A.}~\bibnamefont {Polkovnikov}}, \ and\
  \bibinfo {author} {\bibfnamefont {M.}~\bibnamefont {Rigol}},\ }\href
  {\doibase 10.1080/00018732.2016.1198134} {\bibfield  {journal} {\bibinfo
  {journal} {Adv Phys}\ }\textbf {\bibinfo {volume} {65}},\ \bibinfo {pages}
  {239} (\bibinfo {year} {2016})}\BibitemShut {NoStop}%
\bibitem [{\citenamefont {Plenio}\ and\ \citenamefont
  {Virmani}(2007)}]{PhysRevLett.99.120504}%
  \BibitemOpen
  \bibfield  {author} {\bibinfo {author} {\bibfnamefont {M.~B.}\ \bibnamefont
  {Plenio}}\ and\ \bibinfo {author} {\bibfnamefont {S.}~\bibnamefont
  {Virmani}},\ }\href {\doibase 10.1103/PhysRevLett.99.120504} {\bibfield
  {journal} {\bibinfo  {journal} {Phys. Rev. Lett.}\ }\textbf {\bibinfo
  {volume} {99}},\ \bibinfo {pages} {120504} (\bibinfo {year}
  {2007})}\BibitemShut {NoStop}%
\bibitem [{\citenamefont {Plenio}\ and\ \citenamefont
  {Virmani}(2008)}]{virmani}%
  \BibitemOpen
  \bibfield  {author} {\bibinfo {author} {\bibfnamefont {M.~B.}\ \bibnamefont
  {Plenio}}\ and\ \bibinfo {author} {\bibfnamefont {S.}~\bibnamefont
  {Virmani}},\ }\href {\doibase 10.1088/1367-2630/10/4/043032} {\bibfield
  {journal} {\bibinfo  {journal} {New. J. Phys.}\ }\textbf {\bibinfo {volume}
  {10}},\ \bibinfo {pages} {043032} (\bibinfo {year} {2008})}\BibitemShut
  {NoStop}%
\bibitem [{\citenamefont {Haikka}\ \emph {et~al.}(2013)\citenamefont {Haikka},
  \citenamefont {McEndoo},\ and\ \citenamefont
  {Maniscalco}}]{PhysRevA.87.012127}%
  \BibitemOpen
  \bibfield  {author} {\bibinfo {author} {\bibfnamefont {P.}~\bibnamefont
  {Haikka}}, \bibinfo {author} {\bibfnamefont {S.}~\bibnamefont {McEndoo}}, \
  and\ \bibinfo {author} {\bibfnamefont {S.}~\bibnamefont {Maniscalco}},\
  }\href {\doibase 10.1103/PhysRevA.87.012127} {\bibfield  {journal} {\bibinfo
  {journal} {Phys. Rev. A}\ }\textbf {\bibinfo {volume} {87}},\ \bibinfo
  {pages} {012127} (\bibinfo {year} {2013})}\BibitemShut {NoStop}%
\bibitem [{\citenamefont {Haikka}\ \emph {et~al.}(2012)\citenamefont {Haikka},
  \citenamefont {Goold}, \citenamefont {McEndoo}, \citenamefont {Plastina},\
  and\ \citenamefont {Maniscalco}}]{PhysRevA.85.060101}%
  \BibitemOpen
  \bibfield  {author} {\bibinfo {author} {\bibfnamefont {P.}~\bibnamefont
  {Haikka}}, \bibinfo {author} {\bibfnamefont {J.}~\bibnamefont {Goold}},
  \bibinfo {author} {\bibfnamefont {S.}~\bibnamefont {McEndoo}}, \bibinfo
  {author} {\bibfnamefont {F.}~\bibnamefont {Plastina}}, \ and\ \bibinfo
  {author} {\bibfnamefont {S.}~\bibnamefont {Maniscalco}},\ }\href {\doibase
  10.1103/PhysRevA.85.060101} {\bibfield  {journal} {\bibinfo  {journal} {Phys.
  Rev. A}\ }\textbf {\bibinfo {volume} {85}},\ \bibinfo {pages} {060101}
  (\bibinfo {year} {2012})}\BibitemShut {NoStop}%
\bibitem [{\citenamefont {Cosco}\ and\ \citenamefont
  {Maniscalco}(2018)}]{PhysRevA.98.053608}%
  \BibitemOpen
  \bibfield  {author} {\bibinfo {author} {\bibfnamefont {F.}~\bibnamefont
  {Cosco}}\ and\ \bibinfo {author} {\bibfnamefont {S.}~\bibnamefont
  {Maniscalco}},\ }\href {\doibase 10.1103/PhysRevA.98.053608} {\bibfield
  {journal} {\bibinfo  {journal} {Phys. Rev. A}\ }\textbf {\bibinfo {volume}
  {98}},\ \bibinfo {pages} {053608} (\bibinfo {year} {2018})}\BibitemShut
  {NoStop}%
\bibitem [{\citenamefont {Giarmatzi}\ and\ \citenamefont
  {Costa}(2018)}]{giarmatzi_quantum_2018}%
  \BibitemOpen
  \bibfield  {author} {\bibinfo {author} {\bibfnamefont {C.}~\bibnamefont
  {Giarmatzi}}\ and\ \bibinfo {author} {\bibfnamefont {F.}~\bibnamefont
  {Costa}},\ }\href {\doibase 10.1038/s41534-018-0062-6} {\bibfield  {journal}
  {\bibinfo  {journal} {npj Quantum Inf.}\ }\textbf {\bibinfo {volume} {4}},\
  \bibinfo {pages} {1} (\bibinfo {year} {2018})}\BibitemShut {NoStop}%
\bibitem [{\citenamefont {Chiribella}\ and\ \citenamefont
  {Ebler}(2019)}]{chiribella_quantum_2019}%
  \BibitemOpen
  \bibfield  {author} {\bibinfo {author} {\bibfnamefont {G.}~\bibnamefont
  {Chiribella}}\ and\ \bibinfo {author} {\bibfnamefont {D.}~\bibnamefont
  {Ebler}},\ }\href {\doibase 10.1038/s41467-019-09383-8} {\bibfield  {journal}
  {\bibinfo  {journal} {Nat. Commun.}\ }\textbf {\bibinfo {volume} {10}},\
  \bibinfo {pages} {1472} (\bibinfo {year} {2019})}\BibitemShut {NoStop}%
\bibitem [{\citenamefont {Chiribella}\ \emph {et~al.}(2013)\citenamefont
  {Chiribella}, \citenamefont {D'Ariano}, \citenamefont {Perinotti},\ and\
  \citenamefont {Valiron}}]{chiribella_quantum_2013}%
  \BibitemOpen
  \bibfield  {author} {\bibinfo {author} {\bibfnamefont {G.}~\bibnamefont
  {Chiribella}}, \bibinfo {author} {\bibfnamefont {G.~M.}\ \bibnamefont
  {D'Ariano}}, \bibinfo {author} {\bibfnamefont {P.}~\bibnamefont {Perinotti}},
  \ and\ \bibinfo {author} {\bibfnamefont {B.}~\bibnamefont {Valiron}},\ }\href
  {\doibase 10.1103/PhysRevA.88.022318} {\bibfield  {journal} {\bibinfo
  {journal} {Phys. Rev. A}\ }\textbf {\bibinfo {volume} {88}},\ \bibinfo
  {pages} {022318} (\bibinfo {year} {2013})}\BibitemShut {NoStop}%
\bibitem [{\citenamefont {Procopio}\ \emph {et~al.}(2015)\citenamefont
  {Procopio}, \citenamefont {Moqanaki}, \citenamefont {Ara{\'u}jo},
  \citenamefont {Costa}, \citenamefont {Alonso~Calafell}, \citenamefont {Dowd},
  \citenamefont {Hamel}, \citenamefont {Rozema}, \citenamefont {Brukner},\ and\
  \citenamefont {Walther}}]{procopio_experimental_2015}%
  \BibitemOpen
  \bibfield  {author} {\bibinfo {author} {\bibfnamefont {L.~M.}\ \bibnamefont
  {Procopio}}, \bibinfo {author} {\bibfnamefont {A.}~\bibnamefont {Moqanaki}},
  \bibinfo {author} {\bibfnamefont {M.}~\bibnamefont {Ara{\'u}jo}}, \bibinfo
  {author} {\bibfnamefont {F.}~\bibnamefont {Costa}}, \bibinfo {author}
  {\bibfnamefont {I.}~\bibnamefont {Alonso~Calafell}}, \bibinfo {author}
  {\bibfnamefont {E.~G.}\ \bibnamefont {Dowd}}, \bibinfo {author}
  {\bibfnamefont {D.~R.}\ \bibnamefont {Hamel}}, \bibinfo {author}
  {\bibfnamefont {L.~A.}\ \bibnamefont {Rozema}}, \bibinfo {author}
  {\bibfnamefont {v.}~\bibnamefont {Brukner}}, \ and\ \bibinfo {author}
  {\bibfnamefont {P.}~\bibnamefont {Walther}},\ }\href {\doibase
  10.1038/ncomms8913} {\bibfield  {journal} {\bibinfo  {journal} {Nat.
  Commun.}\ }\textbf {\bibinfo {volume} {6}},\ \bibinfo {pages} {7913}
  (\bibinfo {year} {2015})}\BibitemShut {NoStop}%
\bibitem [{\citenamefont {Rubino}\ \emph {et~al.}(2017)\citenamefont {Rubino},
  \citenamefont {Rozema}, \citenamefont {Feix}, \citenamefont {Ara{\'u}jo},
  \citenamefont {Zeuner}, \citenamefont {Procopio}, \citenamefont {Brukner},\
  and\ \citenamefont {Walther}}]{rubino_experimental_2017}%
  \BibitemOpen
  \bibfield  {author} {\bibinfo {author} {\bibfnamefont {G.}~\bibnamefont
  {Rubino}}, \bibinfo {author} {\bibfnamefont {L.~A.}\ \bibnamefont {Rozema}},
  \bibinfo {author} {\bibfnamefont {A.}~\bibnamefont {Feix}}, \bibinfo {author}
  {\bibfnamefont {M.}~\bibnamefont {Ara{\'u}jo}}, \bibinfo {author}
  {\bibfnamefont {J.~M.}\ \bibnamefont {Zeuner}}, \bibinfo {author}
  {\bibfnamefont {L.~M.}\ \bibnamefont {Procopio}}, \bibinfo {author}
  {\bibfnamefont {{\v C}.}~\bibnamefont {Brukner}}, \ and\ \bibinfo {author}
  {\bibfnamefont {P.}~\bibnamefont {Walther}},\ }\href {\doibase
  10.1126/sciadv.1602589} {\bibfield  {journal} {\bibinfo  {journal} {Sci.
  Adv.}\ }\textbf {\bibinfo {volume} {3}},\ \bibinfo {pages} {e1602589}
  (\bibinfo {year} {2017})}\BibitemShut {NoStop}%
\bibitem [{\citenamefont {Ebler}\ \emph {et~al.}(2018)\citenamefont {Ebler},
  \citenamefont {Salek},\ and\ \citenamefont
  {Chiribella}}]{ebler_enhanced_2018}%
  \BibitemOpen
  \bibfield  {author} {\bibinfo {author} {\bibfnamefont {D.}~\bibnamefont
  {Ebler}}, \bibinfo {author} {\bibfnamefont {S.}~\bibnamefont {Salek}}, \ and\
  \bibinfo {author} {\bibfnamefont {G.}~\bibnamefont {Chiribella}},\ }\href
  {\doibase 10.1103/PhysRevLett.120.120502} {\bibfield  {journal} {\bibinfo
  {journal} {Phys. Rev. Lett.}\ }\textbf {\bibinfo {volume} {120}},\ \bibinfo
  {pages} {120502} (\bibinfo {year} {2018})}\BibitemShut {NoStop}%
\bibitem [{\citenamefont {Goswami}\ \emph {et~al.}(2018)\citenamefont
  {Goswami}, \citenamefont {Giarmatzi}, \citenamefont {Kewming}, \citenamefont
  {Costa}, \citenamefont {Branciard}, \citenamefont {Romero},\ and\
  \citenamefont {White}}]{goswami_indefinite_2018}%
  \BibitemOpen
  \bibfield  {author} {\bibinfo {author} {\bibfnamefont {K.}~\bibnamefont
  {Goswami}}, \bibinfo {author} {\bibfnamefont {C.}~\bibnamefont {Giarmatzi}},
  \bibinfo {author} {\bibfnamefont {M.}~\bibnamefont {Kewming}}, \bibinfo
  {author} {\bibfnamefont {F.}~\bibnamefont {Costa}}, \bibinfo {author}
  {\bibfnamefont {C.}~\bibnamefont {Branciard}}, \bibinfo {author}
  {\bibfnamefont {J.}~\bibnamefont {Romero}}, \ and\ \bibinfo {author}
  {\bibfnamefont {A.~G.}\ \bibnamefont {White}},\ }\href {\doibase
  10.1103/PhysRevLett.121.090503} {\bibfield  {journal} {\bibinfo  {journal}
  {Phys. Rev. Lett.}\ }\textbf {\bibinfo {volume} {121}},\ \bibinfo {pages}
  {090503} (\bibinfo {year} {2018})}\BibitemShut {NoStop}%
\bibitem [{\citenamefont {Goswami}\ \emph {et~al.}(2020)\citenamefont
  {Goswami}, \citenamefont {Cao}, \citenamefont {Paz-Silva}, \citenamefont
  {Romero},\ and\ \citenamefont {White}}]{goswami_communicating_2020}%
  \BibitemOpen
  \bibfield  {author} {\bibinfo {author} {\bibfnamefont {K.}~\bibnamefont
  {Goswami}}, \bibinfo {author} {\bibfnamefont {Y.}~\bibnamefont {Cao}},
  \bibinfo {author} {\bibfnamefont {G.~A.}\ \bibnamefont {Paz-Silva}}, \bibinfo
  {author} {\bibfnamefont {J.}~\bibnamefont {Romero}}, \ and\ \bibinfo {author}
  {\bibfnamefont {A.~G.}\ \bibnamefont {White}},\ }\href {\doibase
  10.1103/PhysRevResearch.2.033292} {\bibfield  {journal} {\bibinfo  {journal}
  {Phys. Rev. Research}\ }\textbf {\bibinfo {volume} {2}},\ \bibinfo {pages}
  {033292} (\bibinfo {year} {2020})}\BibitemShut {NoStop}%
\bibitem [{\citenamefont {Ara{\'u}jo}\ \emph {et~al.}(2017)\citenamefont
  {Ara{\'u}jo}, \citenamefont {Feix}, \citenamefont {Navascu{\'e}s},\ and\
  \citenamefont {Brukner}}]{araujo_purification_2017}%
  \BibitemOpen
  \bibfield  {author} {\bibinfo {author} {\bibfnamefont {M.}~\bibnamefont
  {Ara{\'u}jo}}, \bibinfo {author} {\bibfnamefont {A.}~\bibnamefont {Feix}},
  \bibinfo {author} {\bibfnamefont {M.}~\bibnamefont {Navascu{\'e}s}}, \ and\
  \bibinfo {author} {\bibfnamefont {{\v C}.}~\bibnamefont {Brukner}},\ }\href
  {\doibase 10.22331/q-2017-04-26-10} {\bibfield  {journal} {\bibinfo
  {journal} {Quantum}\ }\textbf {\bibinfo {volume} {1}},\ \bibinfo {pages} {10}
  (\bibinfo {year} {2017})}\BibitemShut {NoStop}%
\end{thebibliography}%
\end{document}